\documentclass[11pt]{article}

\usepackage[margin=1in]{geometry}
\usepackage[mathscr]{eucal}
\usepackage[title]{appendix}
\usepackage{amsmath,amsthm,amssymb}
\usepackage{bbm}
\usepackage{enumitem}
\setlist{topsep=0pt, leftmargin=*}
\usepackage{authblk}
\usepackage{mathtools}
\usepackage[ruled,vlined]{algorithm2e}

\usepackage[sans]{dsfont}
\usepackage{xcolor}

\SetKwComment{Comment}{/* }{ */}
\usepackage{booktabs} 
\usepackage{hyperref}
\hypersetup{pdfborder = {0 0 0},colorlinks=true,allcolors=blue}
\usepackage{url}
\usepackage{natbib}
\usepackage{breqn}
\usepackage{footnote}
\usepackage{graphicx}
\usepackage{tipa}
\usepackage{subfig}
\usepackage{subcaption}
\usepackage{upgreek}
\usepackage{bm}
\usepackage{calrsfs}
\usepackage{subfiles}

\usepackage{mwe}
\usepackage{dutchcal}
\usepackage[T1]{fontenc}
\usepackage[utf8]{inputenc}
\usepackage{textcomp}

\DeclareMathSymbol{\invques}{\mathord}{operators}{`>}
\DeclareUnicodeCharacter{00BF}{\tmquestiondown}
\DeclareRobustCommand{\tmquestiondown}{%
  \ifmmode\invques\else\textquestiondown\fi
}

\newtheorem{theorem}{Theorem}

\newtheorem{lemma}{Lemma}[section]

\newtheorem{assumption}{Assumption}

\numberwithin{theorem}{section}
\numberwithin{lemma}{section}
\newtheorem{innercustomgeneric}{\customgenericname}
\providecommand{\customgenericname}{}

\SetKwRepeat{Do}{do}{while}

\SetKw{Continue}{continue}
\SetKw{Break}{break}

\newcommand{\scr}[1]{\mathscr{#1}}

\newcommand{\wh}[1]{\widehat{#1}}

\DeclareMathAlphabet{\pazocal}{OMS}{zplm}{m}{n}
\newcommand{\ca}[1]{\pazocal{#1}}
\newcommand{\mca}[1]{\mathcal{#1}}

\newcommand{\bX}{\mathbf{X}}

\newcommand{\sign}{\operatorname{sgn}}
\newcommand{\bsign}{\boldsymbol{sgn}}
\newcommand{\pta}{\partial}

\newcommand{\bs}{B_S}

\newcommand{\E}{\mathbb{E}}
\renewcommand{\P}{\mathbb{P}}

\newcommand{\bS}{\mathbf{S}}

\newcommand{\bt}{\mathbf{t}}

\newcommand{\btau}{\boldsymbol{\tau}}

\newcommand{\Indicator}{\mathbbm{1}}
\newcommand{\reals}{\mathbb{R}}

\newcommand{\scrC}{\mathscr{C}}

\DeclareMathOperator*{\argmax}{arg\,max}

\newcommand{\bp}{\mathbf{p}}
\newcommand{\bT}{\mathbf{T}}

\renewcommand{\citet}{\cite}

\title{Robust Inference for the Direct Average Treatment Effect with Treatment Assignment Interference}

\author{
    Matias D. Cattaneo\thanks{Department of Operations Research and Financial Engineering, Princeton University.} \qquad \qquad
    Yihan He\thanks{Department of Operations Research and Financial Engineering, Princeton University.} \qquad \qquad
    Ruiqi (Rae) Yu\thanks{Department of Operations Research and Financial Engineering, Princeton University.}
}

\begin{document}

\maketitle

\begin{abstract}%
    This paper develops methods for uncertainty quantification in causal inference settings with random network interference. We study the large-sample distributional properties of the classical difference-in-means H\'ajek treatment effect estimator, and propose a robust inference procedure for the (conditional) direct average treatment effect. Our framework allows for cross-unit interference in both the outcome equation and the treatment assignment mechanism. Drawing from statistical physics, we introduce a novel Ising model to capture complex dependencies in treatment assignment, and derive three results. First, we establish a Berry–Esseen-type distributional approximation that holds pointwise in the degree of interference induced by the Ising model. This approximation recovers existing results in the absence of treatment interference, and highlights the fragility of inference procedures that do not account for the presence of interference in treatment assignment. Second, we establish a uniform distributional approximation for the H\'ajek estimator and use it to develop robust inference procedures that remain valid uniformly over all interference regimes allowed by the model. Third, we propose a novel resampling method to implement the robust inference procedure and validate its performance through Monte Carlo simulations. A key technical innovation is the introduction of a conditional i.i.d. Gaussianization that may have broader applications. We also discuss extensions and generalizations of our results.

\end{abstract}

\textit{Keywords}: causal inference under interferece; Ising Model; Distribution Theory; Robust Inference

\thispagestyle{empty} 
\clearpage

\section{Introduction}

We study the large-sample distributional properties of the classical difference-in-means H\'ajek estimator for the direct average treatment effect, and propose a robust inference procedure for the (conditional) direct average treatment effect in the presence of cross-unit interference in both the outcome equation and treatment assignment mechanism. This problem arises in a variety of casusal inference settings, including social networks, medical trials, and socio-spatial studies, and has received renewed attention. Recent contributions include \cite{auerbach2025local}, \cite{Bhattacharya-Sen_2025_AOS}, \cite{hu2022average}, \cite{lee2025efficient}, \cite{leung2022causal}, \cite{li2022random}, \cite{ogburn2024causal}, \cite{vazquez2023identification}, and references therein. See \cite{Hernan-Robins_2020_Book} for a modern textbook introduction to causal inference.

A central challenge in causal inference with interference is that units can affect one another in arbitrary ways, complicating identification and inference. To discipline these interactions, it is common to model units as vertices in a network, where edges encode potential interference relationships. While early research assumed the network was fixed and observed, recent advances allow for random and latent (unobserved) network structures (see Assumption \ref{assump:network} below). To manage this added complexity, \cite{li2022random} restricted the degree of interference, especially in the potential outcomes (e.g., by assuming \textit{anonymity} or \textit{exchangeability}, see Assumption~\ref{assump:pout}), while typically assuming a random (independent) treatment assignment. Notable recent exceptions include \cite{ogburn2024causal} and \cite{Bhattacharya-Sen_2025_AOS}, who allow for observed-covariate-based interference in treatment assignment. We contribute to this emerging literature by uncovering a rich family of limiting distributions, and establishing uniformly valid inference procedures across different interference regimes in the treatment assignment mechanism.

To model interference in treatment assignment, we introduce a class of Ising-based assignment mechanisms inspired by statistical physics \citep{ellis2006entropy}. Specifically, we assume that the vector of binary treatment assignments $\mathbf{T} = (T_1,\ldots,T_n)^\top \in \{0,1\}^n$ follows the distribution
\begin{align}\label{ist}
    \mathbb{P}_{\beta}(\mathbf{T} = \mathbf{t}) \propto \exp\left( \frac{\beta}{n} \sum_{i \neq j} (2t_i - 1)(2t_j - 1) \right),
\end{align}
where $\bt = (t_1,\ldots,t_n)^\top \in \{0,1\}^n$, and the parameter $\beta \geq 0$ governs the degree of cross-unit dependence in treatment assignment (see Assumption~\ref{assump:ta}). This model incorporates potential interference effects by allowing treatment assignments to interact globally, capturing the equilibrium distribution of treatment assignments under network influence, analogous to a system in thermal equilibrium with inverse temperature parameter $\beta$. When $\beta = 0$, the model reduces to the classical independent equiprobable treatment assignment. As $\beta$ increases, treatment assignments become more dependent, modeling correlated behaviors such as peer effects or strategic interactions. This equiprobable Ising model allows us to study how interference in assignment affects inference procedures proposed in the literature, and to evaluate their validity and robustness as a function of the unknown dependence parameter $\beta$.

To streamline the analysis, and due to some technical issues, we focus on the moderate interference regime $\beta\in[0,1]$; see Section \ref{sec:Extensions} for extensions and generalizations. Our first contribution is to derive the large-sample distribution of the H\'ajek estimator in this setting. Theorem~\ref{thm: pointwise BE} provides a Berry-Esseen bound for the estimator, that is, a distributional approximation in Kolmogorov distance with explicit convergence rates. The closest antecedent is \cite{li2022random}, who considered the same causal inference model under independent treatment assignment ($\beta = 0$), and established a Gaussian approximation for the Hájek estimator. Our results show that:
\begin{enumerate}
    \item For $\beta \in [0,1)$, the limiting distribution remains Gaussian, but the asymptotic variance includes an additional term capturing dependence in treatment assignment. The asymptotic variance is increasing and unbounded in $\beta$, encompassing the known special case $\beta = 0$.    
    \item At $\beta = 1$, the limiting distribution becomes non-Gaussian. Therefore, there is a discontinuity in the asymptotic distributional behavior of the Hájek estimator.
\end{enumerate}
These results highlight a fragility in the inference methods proposed by \cite{li2022random}, which are not robust to the presence of interference in the treatment assignment. Since their Gaussian approximation becomes invalid when $\beta>0$, the associated inference procedures, while valid at $\beta = 0$, can perform poorly for moderate or large interference in the treatment assignment.

The lack of uniform validity (in $\beta$) poses a major challenge for developing robust inference procedures in the presence of interference in the treatment assignment. \cite{mukherjee2018global} showed that no consistent estimator exists for $\beta\in[0,1)$, making plug-in inference procedures infeasible, even pointwise in $\beta\in[0,1)$. To overcome these challenges, Theorem \ref{thm: uniform BE} establishes a uniform in $\beta\in[0,1]$ distributional approximation for the H\'ajek estimator, along with a parallel result for the Maximum Pseudo-Likelihood estimator (MPLE) for $\beta$. Our uniform distributional approximation depends on a localization parameter that smoothly interpolates between the Gaussian and non-Gaussian regimes.

Building on these results, we construct uniformly valid confidence sets for the (conditional) direct average treatment effect $\tau_n$. More precisely, using a Bonferroni-based correction that aggregates inference across $\beta$-regimes, we construct both infeasible (Theorem~\ref{adaptiveband}) and feasible (Theorem~\ref{thm:implement}) prediction intervals, generically denoted by ${\ca C}_n(\alpha)$, such that
\begin{align*}
    \liminf_{n\to\infty} \inf_{\beta\in[0,1]} \P_\beta[\tau_n \in {\ca C}_n(\alpha)] \geq 1-\alpha,
\end{align*}
for $\alpha\in[0,1]$. The intervals ${\ca C}_n(\alpha)$ are based on the H\'ajek estimator and a resampling procedure designed to capturing capture uncertainty due to the latent network structure. To our knowledge, this is the first feasible inference procedure for this setting, and the first to provide robust uniform validity over $\beta \in [0,1]$, including in the known special case $\beta = 0$. We also present a simulation study demonstrating the performance of our proposed methods.

Our methodological contributions are twofold. First, we introduce the Ising equiprobable treatment assignment model, which generalizes independent assignment to allow nonparametric, structured interference. Second, we develop a new robust and feasible inference procedure for the (conditional) direct average treatment effect, uniformly valid across a wide class of treatment dependence structures. This method combines novel uniform distributional approximations, a hierarchical confidence region construction, and a new resampling-based variance estimator. See \cite{hudgens2008toward}, \cite{tchetgen2012causal}, \cite{manski2013identification}, and references therein, for overviews of the literature.

On the technical side, our work also contributes to the applied probability and statistical mechanics literature. Interference in assignment induces strong dependence across units, complicating traditional asymptotic arguments. For example, as shown in Theorem~\ref{thm: pointwise BE}, both concentration rates and limit distributions depend discontinuously on $\beta\in[0,1]$. To address this technical challenge, we introduce a \textit{De-Finetti Machine}, a new theoretical approach that leverages exchangeability in the Ising model to conditionally Gaussianize the assignment vector. This method transforms the problem into a conditional i.i.d. setting, enabling Berry-Esseen approximations even in the presence of strong dependence. Our main technical result is given in Lemma \ref{lem:be}, which generalizes results from \cite{chatterjee2011nonnormal} and \cite{eichelsbacher2010stein}, and provides an alternative to Stein's method. This technique may be useful in other settings involving exchangeable or dependent random variables. Furthermore, our uniform distributional approximations also contribute to the growing literature on the probabilistic properties of the Ising model, and may be of independent interest in statistics, econometrics, and beyond.

\section{Setup}\label{sec:setup}

We consider a random potential outcome framework under network interference. For each unit $i\in[n]=\{1,2,\cdots,n\}$, let $Y_i(t;\mathbf{t}_{-i})$ denote its random potential outcome when assigned to treatment level $t\in\{0,1\}$ while the other units are assigned to treatment levels $\mathbf{t}_{-i}\in\{0,1\}^{n-1}$. The vector of observed random treatment assignments for the $n$ units is $\mathbf{T}=(T_i:i\in[n])$, and $\mathbf{T}_{-i}$ denotes the associated random treatment assignment vector excluding $T_i$. Thus, the observed data is $(Y_i,T_i:i\in[n])$ with $Y_i = (1-T_i) Y_i(0;\mathbf{T}_{-i}) + T_i Y_i(1;\mathbf{T}_{-i})$ for each $i\in[n]$.

Interference among the $n$ units is modelled via a latent network characterized by an undirected random graph $G(\mathbf{V},\mathbf{E})$ with vertex set $\mathbf{V}=[n]$ and (random) adjacency matrix $\mathbf{E}=(E_{ij}:(i,j)\in[n]\times[n])\in\{0,1\}^{n\times n}$. The following assumption restricts this random graph structure.

\begin{assumption}[Network Structure]\label{assump:network} The random network $\mathbf{E}$ satisfies:
    For all $1\leq i \leq j \leq n$ and $\rho_n\in(0,1]$, $E_{ii} = 0$, $E_{ij}=E_{ji}$, and
    $E_{ij} = \Indicator(\xi_{ij} \leq \min\{1,\rho_n G(U_i,U_j)\})$,
    where $G:[0,1]^2\mapsto\mathbb{R}_{+}$ is symmetric, continuous and positive on $[0,1]^2$, $\mathbf{U} = (U_i:i\in[n])$ are i.i.d. $\mathsf{Uniform}[0,1]$ random variables, $\boldsymbol{\Xi} = (\xi_{ij}: (i,j) \in [n] \times [n], i < j)$ are i.i.d $\mathsf{Uniform}[0,1]$ random variables.
    Finally, $\mathbf{U}$ and $\boldsymbol{\Xi}$ are independent.
\end{assumption}

This assumption corresponds to the \emph{sparse graphon model} of \cite{bickel2009nonparametric}. The parameter $\rho_n$ controls the sparsity of the network, and will play an important role in our theoretical results. The variable $U_i$ is a \emph{latent} heterogenous property of the $i$th unit, and $G(U_i,U_j)$ measures similarity between traits of $U_i$ and $U_j$. This allows for a stochastic model for the edge formation.

Building on the underlying random graph structure, the following assumption imposes discipline on the interference entering the outcome equation.

\begin{assumption}[Exchangeable Smooth Potential Outcomes Model]\label{assump:pout}
    For all $i\in[n]$, $Y_i(T_i; \mathbf{T}_{-i}) = f_i(T_i; M_i/N_i)$
    where $M_i=\sum_{j \neq i} E_{ij} T_j$, $N_i=\sum_{j \neq i} T_j$, and $\mathbf{f} = (f_i:i\in[n])$ are i.i.d random functions.
    In addition, for all $i\in[n]$ and some integer $p\geq 4$, $\max_{1 \leq i \leq n}\max_{t\in\{0,1\}}|\partial_2^{(p)}f_i(t, \cdot)| \leq C$ for some $C$ not depending on $n$ and $\beta$.
    Finally, $\mathbf{f}$ is independent to $\boldsymbol{\Xi}$.
\end{assumption}

This second assumption imposes two main restrictions on the potential outcomes. First, a dimension reduction is assumed via the underlying network structure (Assumption \ref{assump:network}), making the potential outcomes for each unit $i\in[n]$ a function of only their own treatment assignment and the fraction of other treated units among their (connected) peers. Second, the potential outcomes are assumed to be smooth as a function of the fraction of treated peers, thereby ruling out certain types of outcome variables (e.g., binary or similarly limited dependent variable models). Assumption \ref{assump:pout} explicitly parametrizes the smoothness level $p$ because, together with the the sparsity parameter $\rho_n$ in Assumption \ref{assump:network}, it will play an important role in our theoretical results.

To complete the causal inference model, the following assumption restricts the treatment assignment distribution. We propose an Ising model from statistical physics \citep{ellis2006entropy}.

\begin{assumption}[Ising Equiprobable Treatment Assignment]\label{assump:ta}
    The treatment assignment mechanism follows a Curie-Weiss distribution:
    \begin{equation}\label{eq: curie-weiss}
        \mathbb{P}_\beta(\mathbf{T}=\mathbf{t})
        = \frac{1}{C_{\beta}} \exp\bigg( \frac{\beta}{n}\sum_{1\leq i\neq j\leq n}(2t_i-1)(2t_j-1)\bigg),
    \end{equation}
    where $\mathbf{t}\in\{0,1\}^n$, $\beta\in[0,1]$, and $C_\beta$ is determined by the condition $\sum_{\mathbf{t}}\P_\beta(\mathbf{T} = \mathbf{t}) = 1$.
\end{assumption}

This model encodes a nonparametric class of equiprobable, possibly dependent treatment assignment mechanisms. Assumption \ref{assump:ta} implies $\P_\beta(T_i = 1) = 1/2$ for $i\in[n]$ and all $\beta\geq0$, but allows for correlation in treatment assignment as controlled by $\beta$. When $\beta=0$, treatment assignment becomes independent across units, and thus the assignment mechanism reduces to the canonical (equiprobable) randomized allocation. For $\beta\in[0,1]$, the Ising mechanism induces positive pairwise correlations, capturing social interdependence phenomena like peer influence \citep{lipowski2017phase} characteristic of observational settings.

We study the distributional properties of, and proposed uncertainty quantification methods based on, the classical H\'ajek estimator
\begin{align}\label{eq:Hajek Estimator}
    \wh \tau_n = \frac{\sum_{i=1}^n T_i Y_i}{\sum_{i = 1}^n T_i} - \frac{\sum_{i=1}^n(1-T_i)Y_i}{\sum_{i=1}^n(1-T_i)},
\end{align}
which is commonly used in causal inference, both with and without interference. In particular, \cite{li2022random} studied the asymptotic properties of $\wh \tau_n$ under Assumptions \ref{assump:network}--\ref{assump:ta} with $\beta=0$, and showed that
\begin{align}\label{eq:Li-Wager CLT}
    \sqrt{n}(\wh \tau_n - \tau_n) \rightsquigarrow \mathsf{N}(0, \kappa_2),
    \qquad
    \kappa_s= \E[(R_i - \E[R_i] + Q_i)^s],
\end{align}
where $\rightsquigarrow$ denotes weak convergence as $n\to\infty$, $\mathsf{N}(0, \sigma^2)$ denotes a mean-zero Gaussian distribution with variance $\sigma^2$, the centering is the (conditional) direct average treatment effect
\begin{align}\label{eq:(conditional) direct average treatment effect}
    \tau_{n} = 
        \frac{1}{n}\sum_{i=1}^n \E\big[Y_i(1;\mathbf{T}_{-i}) - Y_i(0;\mathbf{T}_{-i}) \big| f_i(\cdot), \mathbf{E} \big],
\end{align}
and the random variables
\begin{align*}
    R_i = f_i \Big(1,\frac{1}{2}\Big) + f_i\Big(0,\frac{1}{2}\Big)
    \qquad\text{and}\qquad 
    Q_i = \E \Big[\frac{G(U_i,U_j)}{\E[G(U_i,U_j)|U_j]}\Big(f_j^{\prime}\Big(1,\frac{1}{2}\Big) - f_j^{\prime}\Big(0,\frac{1}{2}\Big)\Big)\Big|U_i\Big]
\end{align*}
depend on the underlying stochastic potential outcomes and network structure. The (conditional) direct average treatment effect in \eqref{eq:(conditional) direct average treatment effect} is a \textit{predictand}, not an \textit{estimand}, because it is a random variable that needs not to settle to a non-random probability limit under the assumptions imposed. Consequently, we propose prediction intervals for $\tau_{n}$ for uncertainty quantification.

\section{Distribution Theory}\label{sec:dist}

We present pointwise and uniform in $\beta\in[0,1]$ distributional approximations for $\wh \tau_n$. We also study pointwise and uniform distributional approximations for the Maximum Pseudo-Likelihood Estimator (MPLE) for $\beta$. We employ a novel Berry-Esseen bound for the Ising model, which may be of independent interest. Proofs and technical details are given in the supplemental appendix.

\subsection{Pointwise Distributional Approximation}

Our first result is a Berry-Esseen bound for the H\'ajek estimator, pointwise in $\beta\in[0,1]$. Let $\P_\beta$ denote a probability distribution under parametrization $\beta$.

\begin{theorem}[Pointwise Distribution Theory]\label{thm: pointwise BE}
    Suppose Assumptions~\ref{assump:network}, \ref{assump:pout}, and \ref{assump:ta} hold. Then,
    \begin{align*}
        \sup_{t \in \reals} \big|\P_\beta[\wh \tau_n - \tau_n \leq t] - L_n(t;\beta,\kappa_1,\kappa_2)\big|
        = O\Big(\frac{\log n}{\sqrt{n \rho_n}} + \mathtt{r}_{n,\beta}\Big),
    \end{align*}
    where $L_n(\cdot;\beta,\kappa_1,\kappa_2)$ and $\mathtt{r}_{n,\beta}$ are as follows.
    \begin{enumerate}
        \item High temperature: If $\beta \in [0,1)$, then
                \begin{align}\label{eq:law-beta-high}
                    L_n(t;\beta,\kappa_1,\kappa_2) = \P\Big[n^{-1/2}\Big(\kappa_2 + \kappa_1^2 \frac{\beta}{1 - \beta}\Big)^{1/2}\mathsf{Z} \leq t\Big]
                \end{align}
              with $\mathsf{Z} \thicksim \mathsf{N}(0,1)$, and $\mathtt{r}_{n,\beta} = \frac{\sqrt{n \log n}}{(n \rho_n)^{(p+1)/2}}$.
        \item Critical temperature: If $\beta = 1$, then
                \begin{align}\label{eq:law-beta-critical}
                    L_n(t;\beta,\kappa_1,\kappa_2) = \P\big[n^{-1/4} \kappa_1 \mathsf{W}_0 \leq t\big]
                \end{align}
              with
              \begin{align*}
                  \P[\mathsf{W}_c \leq w]
                  = \frac{\int_{-\infty}^w \exp (-\frac{x^4}{12}-\frac{c x^2}{2})dx}{\int_{-\infty}^{\infty}\exp(-\frac{x^4}{12}-\frac{c x^2}{2})dx}, \qquad w \in \reals,\quad c \in \reals_+,
              \end{align*}
              and $\mathtt{r}_{n,\beta} = \frac{\log^3 n}{n^{1/4}} + \frac{n^{1/4}\sqrt{\log n}}{(n \rho_n)^{(p+1)/2}}$.
    \end{enumerate}
\end{theorem}

In the high temperature regime, $\sqrt{n}(\widehat{\tau}_n - \tau_n)$ is asymptotically normal with variance $\kappa_2 + \kappa_1^2 \beta/(1 - \beta)$. Thus, our result recovers \eqref{eq:Li-Wager CLT} when $\beta = 0$, but the asymptotic variance is strictly larger (increasing in $\beta\in[0,1)$) unless $\kappa_1=0$ (no randomness from the underlying network) when $\beta \in (0,1)$. In the critical temperature regime ($\beta = 1$), the limiting distribution is non-Gaussian. The distinct asymptotic behaviors of $\widehat{\tau}_n$ across the interference regimes mirror the phase transition phenomena observed in the Ising model's magnetization $\mathcal{m} = \frac{1}{n}\sum_{i = 1}^n (2 T_i - 1)$. Theorem~\ref{thm: pointwise BE} highlights critical challenges in uncertainty quantification due to the fact that $\kappa_1$ and $\kappa_2$ are unknown quantities, and the interference regime parameter $\beta\in[0,1]$ is also unknown.

\cite{bhattacharya2018inference} established an impossibility result showing that no consistent estimator for $\beta$ exists in the high-temperature regime. Due to the existence of the normalizing constant $C_\beta$ in \eqref{eq: curie-weiss}, maximum likelihood estimation is also computationally prohibitive. However, the conditional distribution of $T_i$ given the rest of treatments adopts a closed form solution and can be optimized efficiently. Define \( W_i = 2T_i - 1 \), \( \mathbf{W}_{-i} = \{W_j: j \in [n], j \neq i\} \), and $\mca m_i = n^{-1}\sum_{j \neq i} W_j$. The MPLE for $\beta$ is
\begin{align*}
   \widehat{\beta}_n
   &= \argmax_{\beta \in [0,1]} \sum_{i \in [n]} \log \mathbb{P}_{\beta}[ W_i | \mathbf{W}_{-i} ]
    = \argmax_{\beta \in [0,1]} \sum_{i \in [n]} -\log \bigg( \frac{1}{2}W_i \tanh(\beta \mca m_i) + \frac{1}{2}\bigg).
\end{align*}
We show in the supplementary appendix (Lemma SA-8) that the limiting distribution of $\widehat{\beta}_n$ also depends on the regime $\beta\in[0,1]$. For $\beta \in [0,1)$, $1 - \wh \beta_n \rightsquigarrow (1 - \beta)\max\{(\chi_1^2)^{-1},0\}$, thereby ruling out consistent estimation, as already anticipated by \cite{bhattacharya2018inference}. For $\beta = 1$, $\sqrt{n}(\wh{\beta}_n - 1) \rightsquigarrow \min\{\mathsf{W}_0^2/3 - 1/\mathsf{W}_0^2,1\}$, where $\mathsf{W}_0$ is given in Theorem~\ref{thm: pointwise BE}.

\subsection{Uniform Distributional Approximation}

From Theorem~\ref{thm: pointwise BE}, for all $\beta\in[0,1)$, the limiting variance of $\sqrt{n}(\widehat{\tau}_n - \tau_n)$ is $\kappa_2 + \kappa_1^2 \beta / (1 - \beta)$. Thus, when $\kappa_1 \neq 0$, the asymptotic variance diverges as $\beta$ approaches the critical value $\beta=1$. In contrast, Theorem~\ref{thm: pointwise BE} shows that when $\beta=1$ the limiting variance of $n^{1/4}(\widehat{\tau}_n - \tau_n)$ is finite. This discrepancy indicates a lack of uniform (in $\beta$) validity for the two distributional approximations. The asymptotic distribution of $\wh{\beta}_n - 1$ exhibits a discontinuity at $\beta = 1$, also highlighting the need for a distributional approximation that is uniform in $\beta$ for valid inference across all regimes.

We establish a uniform distributional approximation based on the drifting sequence $\beta_n = 1 + c n^{-1/2}$. This sequence follows the \textit{knife-edge} rate, ensuring that the resulting approximating laws for $\widehat{\tau}_n$ and $\widehat{\beta}_n$ smoothly interpolate between the different pointwise distributional approximations indexed by $\beta\in[0,1]$.

\begin{theorem}[Uniform Distribution Theory]\label{thm: uniform BE}
    Suppose Assumptions~\ref{assump:network}, \ref{assump:pout} and \ref{assump:ta} hold, and define $c_{\beta,n} = \sqrt{n}(1 - \beta)$. Then, the following distributional approximations hold.
    \begin{enumerate}
        \item H\'ajek Estimator:
        \begin{align*}
            \lim_{n\to\infty}\sup_{0 \leq \beta \leq 1}\sup_{t \in \reals}
            \big| \P_{\beta}[\wh \tau_n - \tau_n \leq t] - H_n(t;\kappa_1,\kappa_2,c_{\beta,n}) \big| = 0,
        \end{align*}
        where            
        \begin{align*}
            H_n(t;\kappa_1,\kappa_2,c_{\beta,n})
            = \P\big[ n^{-1/2}\kappa_2^{1/2}\mathsf{Z} + \beta^{1/2} n^{-1/4}\kappa_1 \mathsf{W}_{c_{\beta,n}}\leq t \big]
        \end{align*}
        with
        $\mathsf{Z} \thicksim \mathsf{N}(0,1)$ independent of $\mathsf{W}_c$.

        \item MPLE for $\beta$:
        \begin{align*}
            \lim_{n\to\infty}\sup_{0 \leq \beta \leq 1}\sup_{t \in \reals}
            \Big| \P_{\beta}\big[1 - \wh \beta_n \leq t\big]
                - \P\big[\min\{\max \{\mathsf{T}_{c_{\beta,n},n}^{-2} - \mathsf{T}_{c_{\beta,n},n}^2/(3n),0\},1\} \leq t\big]
            \Big| = 0
        \end{align*}
        where $\mathsf{T}_{c,n} = \mathsf{Z} + n^{1/4} \mathsf{W}_{c}$.      
    \end{enumerate}
\end{theorem}

Theorem~\ref{thm: uniform BE} establishes that $H_n(t;\kappa_1,\kappa_2,c_{\beta,n})$ uniformly approximates the distribution of \( \widehat{\tau}_n - \tau_n \) in both the high-temperature and critical-temperature regimes. Under the \emph{knife-edge} scaling, the leading term $n^{-1/2} \kappa_2^{1/2} \mathsf{Z}$ becomes negligible, and the typical \emph{knife-edge} representation retains only the second term $\beta^{1/2} n^{-1/4} \kappa_1 \mathsf{W}_c$. However, when $\beta\in[0,1)$ is fixed and $c_{\beta,n} = \sqrt{n}(1 - \beta) \to \infty$, $\mathsf{W}_{c_{\beta,n}}$ approximates \( n^{-1/4} \mathsf{N}(0, (1 - \beta)^{-1}) \), making both terms comparable in order. Consequently, we retain both terms in the distributional approximation. In the supplementary appendix (Lemma SA-5), we show that when $\beta$ is fixed and $c_{\beta,n} = \sqrt{n}(1 - \beta)$, we have \( \sup_{t \in \mathbb{R}} | H_n(t;\kappa_1,\kappa_2,c_{\beta,n}) - L_n(t;\kappa_1,\kappa_2,\beta) | \to 0 \). The same ideas apply to $1 - \wh{\beta}_n$.

\subsection{Main Technical Contribution}\label{sec:tech}

Our main results, Theorem \ref{thm: pointwise BE} and Theorem \ref{thm: uniform BE}, rely on a novel Berry-Esseen distributional approximation for Curie-Weiss magnetization with independent multipliers given in the following lemma. The lemma is self-contained because it may be of independent interest in other settings, and its proof and related technical details are given in the supplemental appendix.

\begin{lemma}[Ising Berry-Esseen Bound]\label{lem:be}
    Let $\P_\beta[\mathbf{W} = \mathbf{w}] \propto \exp(\frac{\beta}{n}\sum_{i \neq j} w_i w_j)$, where $\mathbf{W}=(W_1,\cdots,W_n)^\top\in \{-1,1\}^n$, $\mathbf{w} = (w_1,\cdots,w_n)^\top \in \{-1,1\}^n$, $\beta \geq 0$, and $(X_1,\cdots,X_n)$ are i.i.d. with $\E[|X_i|^3] < \infty$, and independent of $\mathbf{W}$. Then, the following results hold.
    \begin{enumerate}
        \item $\sup_{t \in \reals} \big|\P_\beta(n^{-1}\sum_{i=1}^n X_i W_i \leq t) - L_n(t;\E[X_i],\E[X_i^2],\beta)\big| = O(n^{-1/2} (1 + \Indicator(\beta=1) \log^3 n))$, where $L_n$ is given in Theorem~\ref{thm: pointwise BE}.

        \item $\sup_{\beta \in [0,1]}\sup_{t \in \reals} \big|\P(n^{-1}\sum_{i=1}^n X_i W_i \leq t) - H_n(t;\E[X_i],\E[X_i^2],c_{\beta,n})\big| = O(n^{-1/2}\log^3 n)$, where $c_{\beta,n} = \sqrt{n}(\beta - 1)$, and $H_n$ is given in Theorem~\ref{thm: uniform BE}.
    \end{enumerate}
\end{lemma}

This lemma generalizes the Berry-Esseen bounds for Curie-Weiss magnetization $n^{-1}\sum_{i = 1}^n W_i$, $W_i = 2 T_i - 1$, obtained by \cite{chatterjee2011nonnormal} and \cite{eichelsbacher2010stein}, to allow for multipliers $\bX_i$'s independent to $T_i$'s. Our rate differs from theirs only in a logarithmic term, while allowing for fairly general random weights $X_i$ with bounded third moment. 

Our proof strategy differs significantly from \cite{chatterjee2011nonnormal} and \cite{eichelsbacher2010stein}. They rely on the existence of an exchangeable pair: for $A = n^{-1} \sum_{i = 1}^n W_i$, the existence of another random variable $A^{\prime}$ such that $(A, A^{\prime})$ and $(A^{\prime}, A)$ have the same distribution. The limiting distribution of $A$ can be studied using Stein's method characterized by $g$, if $\E[A - A^{\prime}|A] = g(A) + r(A)$ where $g(A)$ is the leading term and $r(A)$ is negligible. See Equation (1.1) in \citet{chatterjee2011nonnormal}, and Equation (1.3) in \citet{eichelsbacher2010stein}. However, we have not succeed in finding such an exchangeable pair construction. The Glauber dynamics construction used in both previous works does not work here, due to the fact that $\E[X_i^{\prime} (2T_i^{\prime} -1)|\bX_{-i}, \bT_{-i}] \approx$ $\E[X_i] \tanh(2 n^{-1} \sum_{i = 1}^n T_i - 1)$, not preserving the multiplicative structure $n^{-1} \sum_{i  =1}^n X_i (2 T_i - 1)$, where $X_i^{\prime}$ and $T_i^{\prime}$ are random variables following the conditional distribution of $X_i$ given $\bX_{-i}$, and $T_i$ given $\bT_{-i}$, respectively. In contrast, our method is based on the \textit{de Finetti's Lemma}, which ensures that the Ising spins $W_i = 2 T_i - 1$ are i.i.d., conditional on a latent variable $\mathsf{U}_n$. Hence, we can apply an i.i.d Berry-Esseen result condition on $\mathsf{U}_n$, and characterize the conditional mean and variance based on distribution of $\mathsf{U}_n$. The density of $\mathsf{U}_n$ at $u \in \reals$ is proportional to $\exp (- 1/2 u^2 + n \log \cosh( \sqrt{\beta/n} u))$, and then show that $\mathsf{U}_n$ is close to $\mathsf{N}(0,(1 - \beta)^{-1})$ when $\beta \in [0,1)$, and $\mathsf{U}_n$ is close to $n^{1/4} \mathsf{W}_0$ when $\beta = 1$. See the supplementary appendix (Section SA-2) for details.

\section{Uniform Inference}

Building on Theorem \ref{thm: uniform BE}, we present prediction intervals for $\tau_n$ that are valid uniformly over $\beta\in[0,1]$. First, we discuss an infeasible construction assuming $\kappa_1$ and $\kappa_2$ are known, thereby focusing on uniformity over $\beta$. Then, we present feasible prediction intervals for the (conditional) direct average treatment effect. Finally, we present a simulation study.

\subsection{Known Random Graph Structure}

This section addresses inference when the interference regime parameter $\beta$ is unknown, but $\kappa_1$ and $\kappa_2$ are known. We propose a conservative prediction interval based on the following Bonferroni-correction procedure. In particular, in the first step, a uniform confidence interval for $\beta$ is constructed under the \emph{knife-edge} approximation, and in the second step, we choose the largest quantile for $\wh \tau_n - \tau_n$ among all $\beta$'s in the confidence interval. The quantile chosen is also based on the \emph{knife-edge} approximation. 

The following algorithm summarizes the infeasible, uniform-in-$\beta$ inference procedure.

\medskip\begin{algorithm}[H]
    \caption{Infeasible Prediction Interval for $\tau_n$}
    \label{alg: infeasible}
    \SetAlgoLined
    \KwIn{Treatments and outcomes $(T_i,Y_i)_{i\in[n]}$, MPLE-estimator $\wh \beta_n$, an upper bound $K_n$ such that $\kappa_2 \leq K_n^2$, confidence level parameters $\alpha_1, \alpha_2 \in (0,1)$.}
    \KwOut{An $(1 - \alpha_1 - \alpha_2)$ prediction interval $\widetilde{\ca C}(\alpha_1, \alpha_2)$ for $\tau_n$.}
    \BlankLine
    Get the maximum pseudo-likelihood estimator $\wh \beta_n$ of $\beta$\; 
    \smallskip
    Define the $(1 - \alpha_1)$-confidence region given by $\ca I(\alpha_1)=\{\beta \in [0,1]:1 - \wh\beta_n\in [\mathtt{q}, \infty)\}$, where $\mathtt{q} = \inf\{q: \P[\min\{\max \{ \mathsf{T}_{c_{\beta,n},n}^{-2} - \mathsf{T}_{c_{\beta,n},n}^2/(3n),0\},1\} \leq q ] \geq \alpha_1\}$\;
    \smallskip
    Take $\mathtt{U} = \sup_{\beta \in \ca I(\alpha_1)} H_n(1 - \frac{\alpha_2}{2};K_n, K_n,c_{\beta,n})$, $\mathtt{L} = \inf_{\beta \in \ca I(\alpha_1)} H_{n}(\frac{\alpha_2}{2};K_n, K_n, c_{\beta,n})$.
    \BlankLine
    \Return{$\widetilde{\ca C}(\alpha_1,\alpha_2) = [\wh \tau_n + \mathtt{L}, \wh \tau_n + \mathtt{U}]$}.
\end{algorithm}\medskip

Validity of our proposed infeasible, uniform-in-$\beta$ prediction interval for $\tau_n$ is established in the following theorem.

\begin{theorem}[Infeasible Uniform Inference]\label{adaptiveband}
    Suppose Assumptions~\ref{assump:network}, \ref{assump:pout} and \ref{assump:ta} hold, and let $K_n$ be a sequence such that $\kappa_2 \leq K_n^2$. Then, the prediction interval $\widetilde{\ca C}(\alpha_1,\alpha_2)$ given in Algorithm~\ref{alg: infeasible} satisfies
    $\liminf_{n \to \infty} \inf_{\beta \in [0,1]} \P_{\beta}[\tau_n \in \widetilde{\ca C}(\alpha_1,\alpha_2)] \geq 1 - \alpha_1 - \alpha_2$.
\end{theorem}

Algorithm~\ref{alg: infeasible} can be implemented without the knowledge of the degree of interference in the treatment assignment, but requires knowledge of of the underlying random graph structure.

\subsection{Feasible Prediction Interval}\label{sec:implementation}

The unknown parameters $\kappa_1$ and $\kappa_2$ capture moments of the underlying random graph structure. Building on \cite{lin2020theoretical}, we propose a resampling method for consistent estimation of those parameters under an additional nonparametric assumption on the outcome equation.

\begin{assumption}\label{poutconserv}
    Suppose $f_i(\cdot,\cdot) = f(\cdot,\cdot) + \varepsilon_i$, where $f(t,\cdot)$ is 4-times continuously differentiable on $[0,1]$ for $t \in \{0,1\}$, and $(\varepsilon_i: 1 \leq i \leq n)$ are i.i.d  and independent of $\mathbf{E}$ and $\mathbf{T}$, with $\E[\varepsilon_i] = 0$ and $\E[|\varepsilon_i|^{2 + \nu}] < \infty$ for some $\nu > 0$.
\end{assumption}

This assumption allows for nonparametric learning the non-random regression function $f$. In the supplementary appendix (Section SA-5), we provide one example of such learner, but here we remain agnostic and thus present high-level conditions. This step aims to find a consistent estimate for both the function $f$ and its derivative $\frac{\pta f(\cdot,x)}{\pta x}$.

We propose the following novel algorithm for estimating $\kappa_2$ based on resampling methods.

\medskip\begin{algorithm}[H]
    \caption{Estimation of $\kappa_2$}
    \label{alg:conservative}
    \KwIn{Treatments and outcomes $(T_i,Y_i)_{i\in[n]}$, realized graph $\mathbf{E}$, non-parametric learner $\wh f$ of $f$.}
    \KwOut{An upper bound $\wh K_n$ for $\kappa_2$.}
    
    \BlankLine
    
    Generate a new sample $(T_i^*: 1 \leq i \leq n)$ with $\beta = 0$\;
    
    Take $M_j^{\ast} = \sum_{l \neq j} E_{jl} T_l^{\ast}$, $N_j^{\ast} = \sum_{l \neq j} E_{jl}$, $M_{j,(i)}^{\ast} = \sum_{l \neq i,j} E_{jl} T_l^{\ast}$, $N_{j,(i)}^{\ast} = \sum_{l \neq i,j} E_{jl}$\;

    Take $\wh \varepsilon_i = Y_i - T_i \wh f(1,\frac{\sum_{j \neq i}E_{ij}T_j}{\sum_{j \neq i}T_j}) - (1 - T_i)  \wh f(0,\frac{\sum_{j \neq i}E_{ij}T_j}{\sum_{j \neq i}T_j}) $\;
    
    Take  $\tau_{(i)}^a  = n^{-1} \sum_{j \neq i} 2 T_j^\ast (\widehat{f}(1,\frac{M_j^{\ast}}{N_j^{\ast}}) + \wh \varepsilon_j) - 2 (1 - T_j^\ast) (\widehat{f}(0,\frac{M_j^{\ast}}{N_j^{\ast}}) + \wh \varepsilon_j)$, and $\tau_{(i)}^b = n^{-1} \sum_{j \in [n]} 2 T_j^\ast (\widehat{f}(1,\frac{M_{j,(i)}^{\ast}}{N_{j,(i)}^{\ast}}) + \wh \varepsilon_j) - 2 (1 - T_i^\ast) (\widehat{f}(0,\frac{M_{j,(i)}^{\ast}}{N_{j,(i)}^{\ast}}) + \wh \varepsilon_j)$\;
    
    Take $\overline{\tau}^a = n^{-1} \sum_{i \in [n]} \tau_{(i)}^a$, $\overline{\tau}^b = n^{-1} \sum_{i \in [n]} \tau_{(i)}^b$, and 
    $\wh K_n = n \sum_{i \in [n]} (\tau_{(i)}^a - \overline{\tau}^a + \tau_{(i)}^b - \overline{\tau}^b)^2.$
    
    \Return $\wh K_n$.
\end{algorithm}\medskip

Our inference procedure consists of three steps. In step $1$, we estimate $f$ non-parametrically by $\wh f$.  In step $2$, we construct \emph{two} types of plug-in and leave-one-out estimators, denoted by $\{\tau^a_{(i)}\}_{i \in [n]}$ and $\{\tau^b_{(i)}\}_{i \in [n]}$ respectively. $\tau^a_{(i)}$ accounts for the randomness from flipping $i$-th unit's own treatment. $\tau^b_{(i)}$ accounts for randomness from flipping $j$-th unit's treatment, where $j$ is a neighbor of $i$. In Step $3$, we form our final variance estimator using the resampling based treatment effect estimators similar to the i.i.d. case. The supplementary appendix (Lemma SA-18, SA-19) presents formal results on the convergence guarantees for these estimators.

We outline the construction of our proposed feasible, robust prediction interval for $\tau_n$ in the following algorithm.

\medskip\begin{algorithm}[H]
    \caption{Feasible Prediction Interval for $\tau_n$}
    \label{alg:data-driven}
    \KwIn{Treatments and outcomes $(T_i,Y_i)_{i\in[n]}$, realized graph $\mathbf{E}$, non-parametric learner $\wh f$ of $f$.}
    \KwOut{A data-driven $(1 - \alpha_1 - \alpha_2)$ prediction interval $\wh{\ca C}(\alpha_1, \alpha_2)$ for $\tau_n$.}
    
    \BlankLine
    
    Get $\wh K_n$ from Algorithm~\ref{alg:conservative} using the treatments and outcomes $(T_i,Y_i)_{i\in[n]}$, the realized random graph $\mathbf{E}$, a non-parametric learner $\wh f$ for $f$\;
    
    Get $\wh{\ca C}(\alpha_1,\alpha_2)$ from Algorithm~\ref{alg: infeasible} given $(T_i,Y_i)_{i \in [n]}$ and $\wh K_n$.
    
    \Return $\wh{\ca C}(\alpha_1,\alpha_2)$.
\end{algorithm}\medskip

The next theorem establishes the validity of our feasible, robust prediction interval for $\tau_n$.

\begin{theorem}[Feasible Robust Prediction Interval]\label{thm:implement}
    Suppose Assumptions~\ref{assump:network}, \ref{assump:pout}, \ref{assump:ta}, and \ref{poutconserv} hold. Suppose the non-parametric learner $\wh f$ satisfies $\wh f(\ell,\cdot) \in C_2([0,1])$, and $|\wh f(\ell,\pi_*) - f(\ell,\pi_*)| = o_{\P}(1)$, $|\partial_2 \wh f(\ell,\pi_*) - \partial_2 f(\ell,\pi_*)| = o_{\P}(1)$, for $\ell \in \{0,1\}$. If $n \rho_n^3 \to \infty$, then the prediction interval $\wh{\ca C}(\alpha_1,\alpha_2)$ given in Algorithm~\ref{alg:data-driven} satisfies
    \begin{align*}
        \liminf_{n \to \infty} \sup_{\beta \in [0,1]} \P_{\beta}\big[\tau_n \in \wh{\ca C}(\alpha_1,\alpha_2)\big] \geq 1 - \alpha_1 - \alpha_2.
    \end{align*}
\end{theorem}

\subsection{Simulations}\label{sec:simulations}

We exhibit the finite sample performance of our robust inference procedure. Take $(U_i: 1 \leq i \leq n)$ i.i.d $\operatorname{Uniform}([0,1])$-distributed, graph function $G(\cdot, \cdot) \equiv 0.5$ and density $\rho_n = 0.5$. The Ising-treatments satisfy Assumption~\ref{assump:ta} with various $n$ and $\beta$. $Y_i$ has data generating process $Y_i = \Indicator(T_i = 1) f(1,M_i/N_i) + \Indicator(T_i = 0) f(0,M_i/N_i) + \varepsilon_i$, with $f(x_1,x_2) = x_1^2 + x_1(x_2 + 1)^2, (x_1, x_2) \in \reals^2$ and $(\varepsilon_i: 1 \leq i \leq n)$ are i.i.d $N(0,0.05)$ noise terms independent to $((U_i,T_i): 1 \leq i \leq n)$. The Monte-Carlo simulations are repeated with $5000$ iterations and look at the $1 - \alpha$ prediction interval with $\alpha = 0.1$. 

\begin{figure}[t]
    \centering
    \begin{minipage}[b]{0.35\textwidth}
        \centering
        \includegraphics[width=\textwidth]{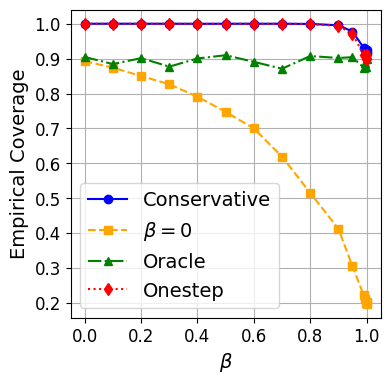}
        \textbf{(a)} {\footnotesize Empirical Coverage Across $\beta$}    
    \end{minipage}
    \hspace{2em}
    \vspace{0.5em}
    \begin{minipage}[b]{0.35\textwidth}  
        \centering 
        \includegraphics[width=\textwidth]{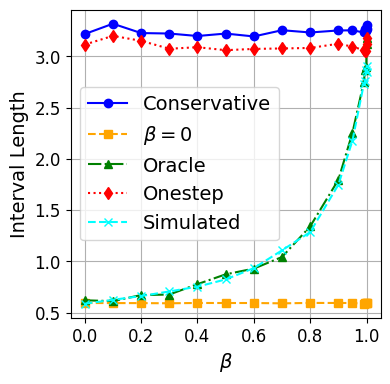}
        \textbf{(b)} {\footnotesize Interval Length Across $\beta$}    
    \end{minipage}
    
    \vspace{-0.5em} % Reduce space between rows

    \begin{minipage}[b]{0.35\textwidth}   
        \centering 
        \includegraphics[width=\textwidth]{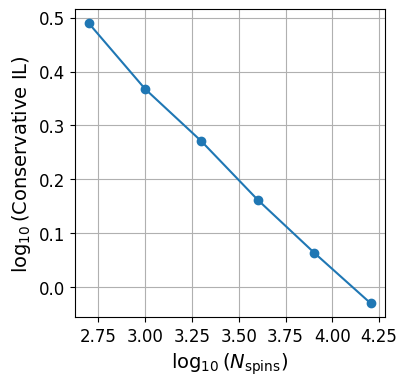}
        \textbf{(c)} {\footnotesize \texttt{Conserv} Interval Length vs $n$}
    \end{minipage}
    \hspace{2em}
    \vspace{0.5em}
    \begin{minipage}[b]{0.35\textwidth}   
        \centering 
        \includegraphics[width=\textwidth]{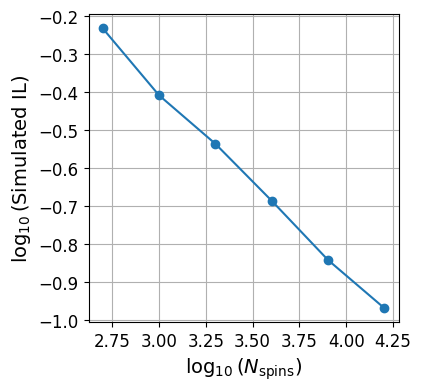}
        \textbf{(d)} {\footnotesize \texttt{Simulated} Interval Length vs $n$}  
    \end{minipage}
    
    \vspace{-1em} % Reduce space before caption

    \caption{(a) and (b) are empirical coverages and interval lengths of four methods across $\beta \in [0,1]$: \texttt{Conservative} and \texttt{Onestep} remain conservative except when $\beta$ is close to $1$; \texttt{Oracle} has empirical coverage close to $1 - \alpha$ and interval length close to the true interval length from Monte-Carlo simulation; the approach of plugging in $\beta = 0$ becomes invalid as $\beta$ deviates from zero. (c) shows \texttt{Conserv} interval length $\propto n^{-0.34}$. (d) shows \texttt{Simulated} interval length $\propto n^{-0.52}$.}
    \label{fig:jackmatrix}
\end{figure}

Figure~\ref{fig:jackmatrix} (a) and (b) demonstrate the empirical coverage and interval length against $\beta$, while fixing $n = 500$. To compare multiple methods, \texttt{conserv} stands for Algorithm~\ref{alg:data-driven}, "$\beta = 0$" stands for using the formula from Theorem~\ref{thm: pointwise BE}, \texttt{Oracle} stands for using the law $n^{-1/2} \wh{\kappa}_2^{1/2} \mathsf{Z} + n^{-1/4} \wh{\kappa}_1 \mathsf{W}_{c_{\beta,n}}$ from Theorem~\ref{thm: uniform BE} with $c_{\beta,n} = \sqrt{n}(1 - \beta)$ assumed to be known, and \texttt{Onestep} stands for Algorithm~\ref{alg: infeasible} but taking the first step confidence interval ${\ca I}(\alpha_1)$ to be the full range $[0,1]$ instead. For interval length, \texttt{Simulated} stands for the true interval length from Monte-Carlo simulations. \texttt{Conservative} and \texttt{Onestep} remain conservative except when $\beta$ is close to $1$, due to the second step in Algorithm~\ref{alg: infeasible} taking maximum quantile from $\beta \in {\ca I}(\alpha_1)$; \texttt{Oracle} has empirical coverage close to $1 - \alpha$ and interval length close to the true interval length from Monte-Carlo simulation; the approach of plugging in $\beta = 0$ becomes invalid as $\beta$ deviates from zero. Figure~\ref{fig:jackmatrix} (c) and (d) demonstrate log-log plots of interval length against sample size, fixing $\beta = 0$. While the Monte-Carlo interval length \texttt{Simulated} interval length $\propto n^{-0.52}$, consistent with the $\sqrt{n}$-convergence with $\beta = 0$, \texttt{Conserv} has interval length $\propto n^{-0.34}$, an effect of taking the maximum quantile among $\beta \in {\ca I}(\alpha_1)$. Full replication codes and data are publicly available at the GitHub repository (\url{https://github.com/ruiiiiqi/network_interference}).

\section{Extensions and Generalizations}\label{sec:Extensions}

We present pointwise distributional approximations under more general interference regimes in treatment assignment. These results demonstrate additional technical complications: developing valid (pointwise and uniform in $\beta$) inference procedures in the settings considered in this section is substantially more challenging, and thus beyond the scope of this paper.

\subsection{Low Temperature Treatment Assignment}

The low temperature regime corresponds to $\beta > 1$ in Assumption~\ref{assump:ta}. In this case, the H\'ajek estimator concentrates in probability around a different (conditional) direct treatment effect, which now also depends on which side of the half line the random variable $\sign(\mca m) = \sign(2 n^{-1}\sum_{i = 1}^n T_i - 1)$ lies on, where $\sign(x) = +$ if $x \geq 0$, and $\sign(x) = - $ if $x < 0$. This is due to the convergence of $M_i/N_i$ to a two-point distribution depending on $\sign(\mca m)$. We thus define
\begin{align*}
    \tau_{n,\ell} = \frac{1}{n}\sum_{i = 1}^n \E[Y_i(1;\mathbf{T}_{-i}) - Y_i(0;\mathbf{T}_{-i})|f_i(\cdot),\mathbf{E},\sign(\mca m ) = \ell],
    \qquad \ell\in\{-,+\},
\end{align*}
which is a new causal predictand in the context of our causal inference model with interference. 

The following lemma gives an analogue of Theorem \ref{thm: pointwise BE} in the low temperature regime.

\begin{lemma}\label{lem:dist low temp}
    Under the Assumptions~\ref{assump:network}, \ref{assump:pout} and \ref{assump:ta} hold with $\beta > 1$,  then
    \begin{align*}
        \sup_{t \in \reals} \max_{\ell \in \{-,+\}}\big|\P(\wh \tau_n - \tau_{n,\ell} \leq t|\sign(\mca m) = \ell) - L_n(t;\beta,\kappa_{1,\ell},\kappa_{2,\ell})\big|
        = O\Big(\sqrt{\frac{n \log n}{(n \rho_n)^{p+1}}} + \frac{\log n}{\sqrt{n \rho_n}}\Big),
    \end{align*}
    where
    \begin{align*}
        L_{n}(t;\beta,\kappa_{1,\ell},\kappa_{2,\ell})
        = \P\Big[ n^{-1/2}\Big(\kappa_{2,\ell} (1 - \pi_*^2) + \kappa_{1,\ell}^2 \frac{\beta (1 - \pi_*^2)}{1 - \beta (1  - \pi_*^2)}\Big)^{1/2}\mathsf{Z} \leq t \Big],
    \end{align*}
    with $\kappa_{s,\ell}= \E[(R_{i,\ell} - \E[R_{i,\ell}] +  Q_{i,\ell})^s]$, 
    \begin{gather*}
        R_{i,\ell} = (1 + \pi_{\ell})^{-1} f_i \Big(1,\frac{\pi_{\ell} + 1}{2}\Big) + (1 - \pi_{\ell})^{-1} f_i\Big(0,\frac{\pi_{\ell} + 1}{2}\Big), \\
        Q_{i,\ell} = \E \Big[\frac{G(U_i,U_j)}{\E[G(U_i,U_j)|U_j]} \frac{1}{2} \Big(f_j^{\prime}\Big(1,\frac{\pi_{\ell} + 1}{2}\Big) - f_j^{\prime}\Big(0,\frac{\pi_{\ell} + 1}{2}\Big)\Big)\Big|U_i\Big],
    \end{gather*}
    $\pi_*$ the positive root of $x = \tanh(\beta x)$, $\pi_+ = 1/2 + \pi_*/2$, $\pi_- = 1/2 - \pi_*/2$, and $\mathsf{Z} \thicksim \mathsf{N}(0,1)$ independent of $\mca m$.
\end{lemma}

Due to the change of centering, establishing uniformly valid inference is substantially more challenging in this context.

\subsection{Asymmetric Treatment Assignment} 

The following assumption allows for unequal treatment assignment probabilities, and thereby providing another (partial) generalization of Assumption \ref{assump:ta}.

\begin{assumption}[Ising Asymmetric Treatment Assignment]\label{assump:ta-unequal}
    Let $\beta\in\mathbb{R}_+$ and let $h\in\mathbb{R}\setminus\{0\}$ denote,
    respectively, the interaction strength and the external field of a
    Curie–Weiss model.  
    The treatment vector
    $\mathbf{T}=(T_1,\dots,T_n)^\top\in\{0,1\}^n$ is assumed to follow the
    joint distribution
    \begin{align*}
      \mathbb{P}_{\beta,h} (\mathbf{T}=\mathbf{t})
       = \frac{1}{C_n(\beta,h)}
            \exp\Big(
               \frac{\beta}{n}
               \sum_{1\le i<j\le n}(2t_i-1)(2t_j-1)
               +
               h\sum_{i=1}^{n}(2t_i-1)
            \Big),
    \end{align*}
    where $\mathbf{t}=(t_1,\ldots,t_n)\in\{0,1\}^n$, and $C_n(\beta,h)$ is the normalizing constant.
\end{assumption}

The external field $h$ induces unequal marginal treatment probabilities, so that $\mathbb{P}_{\beta,h}(T_i=1)=(1+\pi)/2$ with $\pi\in(-1,1)$ satisfying $\pi=\tanh(\beta\pi+h)$. Setting $h=0$ recovers the equiprobable mechanism of Assumption~\ref{assump:ta}.

The following lemma gives an analogue of Theorem \ref{thm: pointwise BE} for the generalized treatment assignment mechanism.

\begin{lemma}\label{lem:dist unequal}
    Suppose Assumptions~\ref{assump:network}, \ref{assump:pout} and \ref{assump:ta-unequal} hold. Then,
    \begin{align*}
        \sup_{t \in \reals} \big|\P[\wh \tau_n - \tau_n \leq t] - L_n(t;\beta,h,\kappa_1,\kappa_2)\big|
        = O\Big(\frac{\log n}{\sqrt{n \rho_n}} + \sqrt{\frac{n \log n}{(n \rho_n)^{p+1}}} \Big),
    \end{align*}
    where
    \begin{align*}
        L_n(t;\beta,h,\kappa_1,\kappa_2) = \P\Big[n^{-1/2}\Big(\kappa_2 (1 - \pi^2) + \kappa_1^2 \frac{\beta (1 - \pi^2)^2}{1 - \beta (1 - \pi^2)}\Big)^{1/2}\mathsf{Z} \leq t\Big]
    \end{align*}
    with $\mathsf{Z} \thicksim \mathsf{N}(0,1)$, $\pi$ the unique solution to $x = \tanh(\beta x + h)$,
    \begin{gather*}
        R_i = (1 + \pi)^{-1} f_i \Big(1,\frac{\pi + 1}{2}\Big) + (1 - \pi)^{-1} f_i\Big(0,\frac{\pi + 1}{2}\Big), \\
        Q_i = \E \Big[\frac{G(U_i,U_j)}{\E[G(U_i,U_j)|U_j]} \frac{1}{2} \Big(f_j^{\prime}\Big(1,\frac{\pi + 1}{2}\Big) - f_j^{\prime}\Big(0,\frac{\pi + 1}{2}\Big)\Big)\Big|U_i\Big],
    \end{gather*}
    and $\kappa_s= \E[(R_i - \E[R_i] + Q_i)^s]$.
\end{lemma}

This results demonstrates that the H\'ajek estimator is $\sqrt{n}$-consistent and asymptotically Gaussian, with a limiting distribution generalizing our previous result for the high temperature case $h = 0$ and $ \beta \in [0,1)$. Unfortunately, establishing a general distributional approximation that is valid (poitwise and) uniform over $(h,\beta)\in\mathbb{R} \times \mathbb{R}_+$ is quite challenging.

\subsection{Block Treatment Assignment}

We have assumed that treatment assignments interfere with each other through a fully connected network, but this assumption can be relaxed by imposing a block structure in the underlying latent network.

\begin{assumption}[Ising Block Treatment Assignment]\label{assump:ta block}
    For a fixed $K \in \mathbb{N}$, suppose $[n]$ has a partition $[n] = \sqcup_{k = 1}^K \scrC_k$. For each partition $k=1,\ldots,K$, assume the exists a positive vector $\bp = (p_1, \cdots, p_K)$ such that the sample sizes $n_k = \sum_{i = 1}^n \Indicator(i \in \mathscr{C}_k)$ satisfy $n_k/n = p_k + O(n^{-1/2})$ as $n \to \infty$. With parameters $(\beta_k, h_k) \in \reals_+ \times \reals$ for $k=1,\ldots,K$, the Ising block treatment assignment mechanism is
    \begin{align*}
        \P_{\boldsymbol{\beta},\boldsymbol{h}}(\bT = \bt)
        = \prod_{k = 1}^K \P_{\boldsymbol{\beta},\boldsymbol{h}}\big[(T_i, i \in \scrC_k) = (t_i, i \in \scrC_k)\big]
    \end{align*}
    with $\boldsymbol{\beta} = (\beta_1,\ldots,\beta_K)^\top$, $\boldsymbol{h} = (h_1,\ldots,h_K)^\top$,
    \begin{align*}
        \P_{\boldsymbol{\beta},\boldsymbol{h}}\big[(T_i, i \in \scrC_k) = (t_i, i \in \scrC_k)\big]
        \propto \exp \Big(\frac{\beta_k}{n} \sum_{i,j \in \scrC_k, i < j}(2 t_i - 1) (2 t_j - 1) + h_k \sum_{i \in \scrC_k}  (2 t_i - 1) \Big).
    \end{align*}
\end{assumption}

Denote the collections of high temperature or nonzero external field blocks by $\scr{H} = \{k \in [K]: h_k = 0, \beta_k \in [0,1) \text{ or } h_k \neq 0\}$, the collections of critical temperature blocks by $\scr{C} = \{k \in [K]: h_k = 0, \beta_k = 1\}$, and the collections of low temperature blocks by $\scr{L} = \{k \in [K]: h_k = 0, \beta_k > 1\}$. For block $k \in \scr{H} \cup \scr{C}$, $\pi_k$ denotes the unique solution to $x = \tanh(\beta_k x + h_k)$. For block $k \in \scr{L}$, $\pi_{k,+}$ and $\pi_{k,-}$ denote the unique positive and negative solutions to $x = \tanh(\beta_k x + h_k)$, respectively. Due to the potential existence of low temperature blocks, we use $\bsign$ to collect the average spins in all low temperature blocks, and fill in the positions for high and critical temperature blocks with zeros:
\begin{align*}
    \bsign = (\sign(\mca m_1) \Indicator(1 \in \mathscr{L}), \cdots, \sign(\mca m_K) \Indicator(K \in \mathscr{L})).
\end{align*}
Let $\mathscr{S}$ denote the collection of all possible configurations of $\bsign$:
\begin{align*}
    \mathscr{S} = \{(s_k)_{1 \leq k \leq K}: s_k = - \text{ or } + \text{ if } k \in \mathscr{L}, s_k = 0 \text{ otherwise}\}.
\end{align*}
Finally, we denote the conditional fixed point of block $k$ based on $\bsign = \bs$ by
\begin{align*}
    \pi_{k,(\bs)} = 
    \begin{cases}
        \pi_k, & \text{ if } k \in \mathscr{H} \cup \mathscr{C}, \\
        \pi_{k,s_k}, & \text{ if } k \in \mathscr{L},
    \end{cases}
    \qquad 1 \leq k \leq K, \bs \in \mathscr{S}.
\end{align*}

In this setting, we consider the blockwise (conditional) average treatment effects $\btau_n = (\tau_{n,1},\cdots,\tau_{n,K})^\top$, where
\begin{align*}
    \tau_{n,k} = \frac{1}{n_k} \sum_{i \in \mathscr{C}_k}\mathbb{E}\big[Y_i(1;\mathbf{T}_{-i}) - Y_i(0;\mathbf{T}_{-i}) \big| f_i(\cdot), \mathbf{E}, \bsign  \big], \qquad k = 1, \cdots, K,
\end{align*}
and the associated $K$-dimensional H\'ajek estimator is $\wh\btau_n = (\wh\tau_{n,1},\cdots,\wh\tau_{n,K})^\top$, where
\begin{align*}
     \widehat{\tau}_{n,k}=  \frac{\sum_{i \in \mathscr{C}_k} T_i Y_i}{\sum_{i \in \mathscr{C}_k} T_i} - \frac{\sum_{i \in \mathscr{C}_k} (1 - T_i) Y_i}{\sum_{i \in \mathscr{C}_k} (1 - T_i)}, \qquad 
     k = 1, \cdots, K.
\end{align*}

The following theorem gives an analogue of Theorem \ref{thm: pointwise BE} in this generalized setting. Let $\mathscr{R}$ be the collection of all hyperrectangles in $\mathbb{R}^K$. 

\begin{lemma}\label{lem:dist block}
Suppose Assumptions~\ref{assump:network}, \ref{assump:pout} and \ref{assump:ta block} hold. Then,
\begin{align*}
    &\max_{\mathbf{s} \in \mathscr{S}} \sup_{A \in \mathscr{R}}
    \big| \P(\wh{\btau}_n  - \btau_n \in A|  \bsign = \mathbf{s})
          - L_n(A; \boldsymbol{\beta},\boldsymbol{h})
    \big|\\
    &\qquad\qquad\qquad = O\Big(\frac{n^{1/2}}{(n \rho_n)^{(p+1)/2}} + \Big(\frac{\log^7 n}{n}\Big)^{1/6} + \mathtt{r}_{n, \boldsymbol{\beta}, \boldsymbol{h}} \Big),
\end{align*}
where
\begin{align*}
    L_n(A; \boldsymbol{\beta},\boldsymbol{h})
    &= \P\Big(n^{-1/2} \Big(\sum_{k = 1}^K \E[\bS_{k,i,(\bs)} \bS_{k,i,(\bs)}^\top] (1 - \pi_{k,(\bs)}^2) p_k^2\Big)^{1/2} \mathsf{Z}_K \\
    & \qquad \qquad + n^{-1/2} \sum_{k \in \mathscr{H} \cup \mathscr{L}} p_k \sigma_{k,(\bs)} \E[\bS_{k,i,(\bs)}] \mathsf{Z}_{(k)}\\
    & \qquad \qquad + n^{-1/4} \sum_{k \in \mathscr{C}} p_k \E[\bS_{k,i,(\bs)}] \mathsf{R}_{(k)} \in A\Big)
\end{align*}
with $\mathsf{Z}_K \thicksim \mathsf{N}(\mathbf{0},\mathbf{I})$, where $\mathbf{0}$ is a $K$-dimensional vector of zeros and $\mathbf{I}$ is the $K\times K$ identity matrix, $\mathsf{Z}_{(k)} \thicksim \mathsf{N}(0,1)$ for $k \in \mathscr{H} \cup \mathscr{L}$ independent, $\mathsf{R}_{(k)}$ for $k \in \scrC$ independent random variables with distribution function
\begin{align*}
    F_0(t) = \frac{\int_{-\infty}^t \exp(-z^4/12)d z}{\int_{-\infty}^{\infty} \exp(-z^4/12)d z},\qquad  t \in \reals,
\end{align*}
and $\mathsf{Z}_K$, $(\mathsf{Z}_{(k)}: k \in \mathscr{H} \cup \mathscr{L})$ and $(\mathsf{R}_{(k)}: k \in \mathscr{C})$ mutually independent. In addition, $\bS_{l,i,(\bs)} = (S_{1,l,i,(\bs)}, \cdots, S_{K,l,i,(\bs)})^\top$ with $S_{k,l,i,(\bs)} = Q_{i,(\bs)} + \Indicator(k = l) p_k^{-1}(R_{i,l,(\bs)} - \E[R_{i,l,(\bs)}])$, $1 \leq k, l \leq K$ and $1 \leq i \leq n$,
\begin{align*}
    R_{i,l,(\bs)} & = \frac{g_i(1, \overline{\pi}_{(\bs)})}{1 + \pi_{l,(\bs)}} + \frac{g_i(-1, \overline{\pi}_{(\bs)})}{1 - \pi_{l,(\bs)}},
    \qquad 
    \overline{\pi}_{(\bs)} = \sum_{k = 1}^K p_k \pi_{k,(\bs)}, \\
    Q_{i,(\bs)} & = \E \Big[\frac{G(U_i, U_j)}{\E[G(U_i, U_j)|U_j]} (g_j^{\prime}(1, \overline{\pi}_{(\bs)}) - g_j^{\prime}(-1, \overline{\pi}_{(\bs)}))\Big| U_i\Big],
\end{align*}
and
\begin{align*}
    \mathtt{r}_{n, \boldsymbol{\beta}, \boldsymbol{h}}
    = \begin{cases}
        \sqrt{\log n} (n \rho_n)^{-1/2} &  \text{if } \sum_{k = 1}^K \Indicator(k \in \mathscr{C}) = 0\\
        \sqrt{\log n} (n \rho_n^2)^{-1/4} & \text{if } \sum_{k = 1}^K \Indicator(k \in \mathscr{C}) > 0
    \end{cases}
    .
\end{align*}
\end{lemma}

If there is no low temperature block, then $\bsign = (0, \cdots, 0)$ almost surely, and $\mathsf{S}$ is the singleton set containing $(0, \cdots, 0)$. Then, the result reduces to the unconditional distributional approximation. More generally, when comparing to the single-block results, Lemma \ref{lem:dist block} shows that an \emph{overflow} behavior happens: as long as there is at least one low temperature block, the Hajek estimators for all blocks converge to the conditional estimand given low temperature block average Ising spin signs. Moreover, the distribution approximation error depends on whether there is at least one low temperature block. This is because unit $i$'s potential outcome $Y_i(T_i; \bT_{-i}) = f_i(T_i; M_i/N_i)$ depends on units from all blocks through $M_i/N_i$. Finally, the additional, slower rate $\log(n)^{7/6} n^{-1/6}$ comes from a multivariate conditional i.i.d Berry-Esseen result (see Section~\ref{sec:tech}) with a degenerate covariance matrix, via \citet[Theorem 2.1]{chernozhukov2017central}.

% Acknowledgments---Will not appear in anonymized version
\noindent \textbf{Acknowledgements.} Cattaneo gratefully acknowledges financial support from the National Science Foundation through DMS-2210561 and SES-2241575. 

\section*{Supplementary material}\label{SM}

The supplementary appendix presents more general theoretical results encompassing those discussed in the paper, and presents the proofs of those general results.

\bibliographystyle{plain}
\bibliography{bib.bib}

\begin{thebibliography}{10}

\bibitem{auerbach2025local}
Eric Auerbach, Hongchang Guo, and Max Tabord-Meehan.
\newblock The local approach to causal inference under network interference.
\newblock {\em arXiv preprint arXiv:2105.03810}, 2025.

\bibitem{bhattacharya2018inference}
Bhaswar~B Bhattacharya and Sumit Mukherjee.
\newblock Inference in ising models.
\newblock {\em Bernoulli}, 24(1):493--525, 2018.

\bibitem{Bhattacharya-Sen_2025_AOS}
Sohom Bhattacharya and Subhabrata Sen.
\newblock Causal effect estimation under network interference with mean-field
  methods.
\newblock {\em Annals of Statistics}, 2025.

\bibitem{bickel2009nonparametric}
Peter~J Bickel and Aiyou Chen.
\newblock A nonparametric view of network models and newman--girvan and other
  modularities.
\newblock {\em Proceedings of the National Academy of Sciences},
  106(50):21068--21073, 2009.

\bibitem{chatterjee2011nonnormal}
Sourav Chatterjee and Qi-Man Shao.
\newblock Nonnormal approximation by stein’s method of exchangeable pairs
  with application to the curie--weiss model.
\newblock {\em Annals of Applied Probability}, 21(2):464--483, 2011.

\bibitem{chernozhukov2017central}
Victor Chernozhukov, Denis Chetverikov, and Kengo Kato.
\newblock Central limit theorems and bootstrap in high dimensions.
\newblock {\em Annals of Probability}, 45(4):2309 -- 2352, 2017.

\bibitem{eichelsbacher2010stein}
Peter Eichelsbacher and Matthias Loewe.
\newblock Stein's method for dependent random variables occuring in statistical
  mechanics.
\newblock {\em Electronic Journal of Probability}, 15:962 -- 988, 2010.

\bibitem{ellis2006entropy}
Richard~S Ellis.
\newblock {\em Entropy, Large Deviations, and Statistical Mechanics}.
\newblock Taylor \& Francis, 2006.

\bibitem{Hernan-Robins_2020_Book}
Miguel~A. Hernán and James~M. Robins.
\newblock {\em Causal Inference: What If}.
\newblock Boca Raton: Chapman \& Hall/CRC, 2020.

\bibitem{hu2022average}
Yuchen Hu, Shuangning Li, and Stefan Wager.
\newblock Average direct and indirect causal effects under interference.
\newblock {\em Biometrika}, 109(4):1165--1172, 2022.

\bibitem{hudgens2008toward}
Michael~G Hudgens and M~Elizabeth Halloran.
\newblock Toward causal inference with interference.
\newblock {\em Journal of the American Statistical Association},
  103(482):832--842, 2008.

\bibitem{lee2025efficient}
Chanhwa Lee, Donglin Zeng, and Michael~G Hudgens.
\newblock Efficient nonparametric estimation of stochastic policy effects with
  clustered interference.
\newblock {\em Journal of the American Statistical Association},
  120(549):382--39, 2025.

\bibitem{leung2022causal}
Michael~P Leung.
\newblock Causal inference under approximate neighborhood interference.
\newblock {\em Econometrica}, 90(1):267--293, 2022.

\bibitem{li2022random}
Shuangning Li and Stefan Wager.
\newblock Random graph asymptotics for treatment effect estimation under
  network interference.
\newblock {\em Annals of Statistics}, 50(4):2334--2358, 2022.

\bibitem{lin2020theoretical}
Qiaohui Lin, Robert Lunde, and Purnamrita Sarkar.
\newblock On the theoretical properties of the network jackknife.
\newblock In {\em International Conference on Machine Learning}, pages
  6105--6115. PMLR, 2020.

\bibitem{lipowski2017phase}
Adam Lipowski, Dorota Lipowska, and Ant{\'o}nio~Luis Ferreira.
\newblock Phase transition and power-law coarsening in an ising-doped voter
  model.
\newblock {\em Physical Review E}, 96(3):032145, 2017.

\bibitem{manski2013identification}
Charles~F Manski.
\newblock Identification of treatment response with social interactions.
\newblock {\em The Econometrics Journal}, 16(1):S1--S23, 2013.

\bibitem{mukherjee2018global}
Rajarshi Mukherjee, Sumit Mukherjee, and Ming Yuan.
\newblock Global testing against sparse alternatives under ising models.
\newblock {\em Annals of Statistics}, 46(5):2062--2093, 2018.

\bibitem{ogburn2024causal}
Elizabeth~L Ogburn, Oleg Sofrygin, Ivan Diaz, and Mark~J Van~der Laan.
\newblock Causal inference for social network data.
\newblock {\em Journal of the American Statistical Association},
  119(545):597--611, 2024.

\bibitem{tchetgen2012causal}
Eric J~Tchetgen Tchetgen and Tyler~J VanderWeele.
\newblock On causal inference in the presence of interference.
\newblock {\em Statistical methods in medical research}, 21(1):55--75, 2012.

\bibitem{vazquez2023identification}
Gonzalo Vazquez-Bare.
\newblock Identification and estimation of spillover effects in randomized
  experiments.
\newblock {\em Journal of Econometrics}, 237(1):105237, 2023.

\end{thebibliography}


\begin{thebibliography}{10}

\bibitem{bhattacharya2018inference}
Bhaswar~B Bhattacharya and Sumit Mukherjee.
\newblock Inference in ising models.
\newblock {\em Bernoulli}, 24(1):493--525, 2018.

\bibitem{bleistein1975asymptotic}
Norman Bleistein and Richard~A Handelsman.
\newblock {\em Asymptotic Expansions of Integrals}.
\newblock Ardent Media, 1975.

\bibitem{chatterjee2010spin}
Sourav Chatterjee.
\newblock Spin glasses and stein’s method.
\newblock {\em Probability Theory and Related Fields}, 148(3):567--600, 2010.

\bibitem{chernozhukov2017central}
Victor Chernozhukov, Denis Chetverikov, and Kengo Kato.
\newblock Central limit theorems and bootstrap in high dimensions.
\newblock {\em Annals of Probability}, 45(4):2309 -- 2352, 2017.

\bibitem{dagan2021learning}
Yuval Dagan, Constantinos Daskalakis, Nishanth Dikkala, and Anthimos~Vardis
  Kandiros.
\newblock Learning ising models from one or multiple samples.
\newblock In {\em Proceedings of the 53rd Annual ACM SIGACT Symposium on Theory
  of Computing}, pages 161--168, 2021.

\bibitem{diaconis1988recent}
Persi Diaconis.
\newblock Recent progress on de finetti’s notions of exchangeability.
\newblock {\em Bayesian statistics}, 3(111-125):13--14, 1988.

\bibitem{diaconis1980finetti}
Persi Diaconis and David Freedman.
\newblock de finetti's theorem for markov chains.
\newblock {\em The Annals of Probability}, pages 115--130, 1980.

\bibitem{eichelsbacher2010stein}
Peter Eichelsbacher and Matthias Loewe.
\newblock Stein's method for dependent random variables occuring in statistical
  mechanics.
\newblock {\em Electronic Journal of Probability}, 15:962 -- 988, 2010.

\bibitem{ellis1978statistics}
Richard~S Ellis and Charles~M Newman.
\newblock The statistics of curie-weiss models.
\newblock {\em Journal of Statistical Physics}, 19(2):149--161, 1978.

\bibitem{friedli2017statistical}
Sacha Friedli and Yvan Velenik.
\newblock {\em Statistical Mechanics of Lattice Systems: A Concrete
  Mathematical Introduction}.
\newblock Cambridge University Press, 2017.

\bibitem{li2022random}
Shuangning Li and Stefan Wager.
\newblock Random graph asymptotics for treatment effect estimation under
  network interference.
\newblock {\em Annals of Statistics}, 50(4):2334--2358, 2022.

\bibitem{vershynin2018high}
Roman Vershynin.
\newblock {\em High-Dimensional Probability: An Introduction with Applications
  in Data Science}, volume~47.
\newblock Cambridge university press, 2018.

\bibitem{wainwright2019high}
Martin~J Wainwright.
\newblock {\em High-Dimensional Statistics: A Non-asymptotic Viewpoint},
  volume~48.
\newblock Cambridge university press, 2019.

\end{thebibliography}

\end{document}

\maketitle

\begin{abstract}
\noindent This supplemental appedix contains general theoretical results encompassing those discussed in the main paper, includes proofs of those general results, and discusses additional methodological and technical results.\bigskip
\end{abstract}

\clearpage

\tableofcontents

\clearpage

\section{Notations}

For $n \in \mathbb{N}$, $[n] = \{1, \cdots, n\}$. For reals sequences $a_n = o(b_n)$ if $\limsup_{n\to\infty} \frac{|a_n|}{|b_n|} = 0$,  $|a_n| \lesssim |b_n|$ if there exists some constant $C$ and $N > 0$ such that $n > N$ implies $|a_n| \leq C |b_n|$. For sequences of random variables $a_n = o_{\P}(b_n)$ if $\operatorname{plim}_{n \rightarrow \infty}\frac{|a_n|}{|b_n|} = 0$, $a_n = O_{\P}(b_n)$ if $\limsup_{M \rightarrow \infty} \limsup_{n \rightarrow \infty} \P[|\frac{a_n}{b_n}| \geq M] = 0$. For positive real sequences $a_n \ll b_n$ if $a_n = o(b_n)$. For a sequence of real-valued random variables $X_n$, we say $X_n = O_{\psi_p}(r_n)$ if there exists $N \in \mathbb{N}$ and $M > 0$ such that $\norm{X_n}_{\psi_p} \leq M r_n$ for all $n \geq N$, where $\norm{\cdot}_{\psi_p}$ is the Orlicz norm w.r.p $\psi_p(x) = \exp(x^p) - 1$. We say $X_n = O_{\psi_p, tc}(r_n)$, $tc$ stands for tail control, if there exists $N \in \mathbb{N}$ and $M > 0$ such that for all $n \geq N$ and $t > 0$, $\P(|X_n| \geq t) \leq 2 n\exp(-(t/(Mr_n))^{p}) + M n^{-1/2}$. 

For a vector $\bv \in \reals^k$, the Euclidean norm is $\norm{\bv} = (\sum_{i = 1}^k \bv_i^2)^{1/2}$,and  the infinity norm is $\norm{\bv}_{\infty} = \max_{1 \leq i \leq k} |v_i|$. For a matrix $A = (a_{ij})_{i \in [m], j \in [n]} \in \reals^{m \times n}$, the operator norm is $\norm{A} = \norm{A}_2 = \sup_{\norm{\bx} = 1} \norm{A\bx}$, the maximum absolute column sum norm is $\norm{A}_1 = \sup_{1 \leq j \leq n} \sum_{i = 1}^m |a_{ij}|$, and the Frobenius norm is $\norm{A}_F = \sqrt{\sum_{i = 1}^m \sum_{j = 1}^n a_{ij}^2}$. For sets $A$ and $B$, denote by $A \Delta B$ the set difference $(A \setminus B) \cup (B \setminus A)$. 

$\sign$ denotes the function such that $\sign(x) = +$ if $x \geq 0$, and $\sign(x) = -$ otherwise. $\Phi(x)$ denotes the standard Gaussian cumulative distribution function. For $\boldsymbol{\mu} \in \reals^{k  \times k}$ and $\boldsymbol{\Sigma} \in \reals^{k \times k}$, $\mathsf{N}(\boldsymbol{\mu}, \boldsymbol{\Sigma})$ denotes the multivariate normal distribution with mean $\boldsymbol{\mu}$ and covariance matrix $\boldsymbol{\Sigma}$.

\section{Curie-Weiss Magnetization with Independent Multipliers}\label{sa-sec:berry-esseen}

For notational simplicity, we consider $$\bW = (W_i)_{1 \leq i \leq n}, \qquad W_i = 2 T_i - 1, 1 \leq i \leq n.$$ And we consider a more general setting compare to Assumption 3 in the main paper.

\begin{assumption}[Curie-Weiss]\label{assump-one-block}
For $\beta \geq 0$ and $h \in \reals$, suppose $\bW = (W_i)_{1 \leq i \leq n}$ are such that for some $C_{\beta,h} \in \reals$, 
\begin{align}\label{sa-eq:curie-weiss}
    \P_{\beta,h}(\bW = \bw) = C_{\beta,h}^{-1} \exp \bigg(\frac{\beta}{n}\sum_{1 \leq i < j \leq n} w_i w_j + h \sum_{i = 1}^n w_i\bigg), \qquad \bw = (w_1, \cdots, w_n) \in \{-1,1\}^n,
\end{align}
where $C_{\beta,h}$ is a normalizing constant.
\end{assumption}

The Curie-Weiss model has a phase transition phenomena in different regimes. Let $\mca m  = n^{-1}\sum_{i = 1}^n W_i$.
\begin{enumerate}
    \item High temperature or non-zero external field $\mathcal{A}_H = \{(\beta,h) \in \reals_+ \times \reals: h = 0, 0 \leq \beta < 1 \text{ or } h \neq 0\}$: $\mca m$ concentrates around $\pi$, where $\pi$ is the unique solution to $x = \tanh(\beta x + h)$. In particular, $\mca m = \pi + \Theta_{\P}(n^{-1/2})$. Moreover, $\mathcal{A}_H = \mathcal{A}_{H,1} \sqcup \mathcal{A}_{H,2}$, where $\mathcal{A}_{H,1} = \{(\beta, h) \in \reals^+ \times \reals: h = 0, 0 \leq \beta < 1\}$ and $\mathcal{A}_{H,2} = \{(\beta, h) \in \reals^+ \times \reals: h \neq 0\}$.
    \item Critical temperature $\mathcal{A}_{C} = \{(1,0)\}$: $\mca m$ concentrates around $\pi$, where $\pi$ is the unique solution to $x = \tanh(\beta x + h)$. In particular, $\mca m = \pi + \Theta_{\P}(n^{-1/4})$.
    \item Low temperature regime $\mathcal{A}_{R} = \{(\beta,h) \in \reals_+ \times \reals: h = 0, \beta > 1\}$: $\mca m$ concentrates on the set $\{\pi_-,\pi_+\}$, with $\pi_-$ and $\pi_+$ the unique negative and positive solutions to $x = \tanh(\beta x)$, respectively. In particular, condition on $\sign(\mca m) = \ell$, $\mca m = \pi_{\ell} + \Theta_{\P}(n^{-1/2})$.
\end{enumerate}

In the main paper, we focus on $(\beta, h)$ in $\mathscr{H}_1$. But for this section, we provide the results for all of $\mathscr{H}$, $\mathscr{C}$ and $\mathscr{L}$.

Suppose $\bX = (X_1,\cdots,X_n)$ has i.i.d components such that $\E \left[|X_1|^3\right] < \infty$ independent to $\bW$. The goal is to study the limiting distribution and the rate of convergence for 
\begin{align*}
    \mca g_n = n^{-1}\sum_{i = 1}^n X_i (W_i - \pi).
\end{align*}
The magnetization $n^{-1}\sum_{i = 1}^n (W_i - \pi)$ has been studied using Stein's method \citep{eichelsbacher2010stein}, \citep{chatterjee2010spin}. Due to the multipliers, the Stein's method can not be directly applied for $\mca g_n$. We use a novel strategy based on the following de Finetti's lemma to show Berry Essseen results.

\begin{lemma}[de Finetti's Theorem]\label{sa-lem: definetti} There exists a latent variable $\mathsf{U}_n$ with density
\begin{align*}
    f_{\mathsf{U}_n}(u) = I_{\mathsf{U}_n}^{-1}\exp \bigg(- \frac{1}{2} u^2 + n \log \cosh \bigg( \sqrt{\frac{\beta}{n}} u + h\bigg) \bigg), 
\end{align*}
where $I_{\mathsf{U}_n} = \int_{-\infty}^{\infty} \exp (- \frac{1}{2} u^2 + n \log \cosh ( \sqrt{\frac{\beta}{n}} u + h)) d u$, such that $W_1,\cdots,W_n$ are i.i.d condition on $\mathsf{U}_n$. 
\end{lemma}

The de Finetti's theorem for exchangable sequences of random variable is a classical result \citep{diaconis1988recent, diaconis1980finetti, ellis1978statistics}. For completeness, we include a short proof for the Curie-Weiss model in Section~\ref{sa-sec: berry esseen proof}.

\begin{lemma}\label{sa-lem:sub-gaussian}
Take $\mathsf{U}_n$ to be the latent variable from Lemma~\ref{sa-lem: definetti} and $\mathsf{W}_n = n^{-\frac{1}{4}}\mathsf{U}_n$. Then
\begin{enumerate}
    \item High-temparature case: Suppose $h \neq 0$ or $h = 0, \beta < 1$. Then $\norm{\mathsf{U}_n - \E[\mathsf{U}_n]}_{\psi_2} \lesssim 1$.
    \item Critical-temparature case: Suppose $h = 0$ and $\beta = 1$. Then $\norm{\mathsf{U}_n}_{\psi_2} \lesssim n^{1/4}$.
    \item Low-temparature case: Suppose $h = 0$ and $\beta > 1$. Then condition on $\mathsf{U}_n \in \mathcal{C}_l$, $\norm{\mathsf{U}_n - \E[\mathsf{U}_n|\sign(\mathsf{U}_n) = \ell]}_{\psi_2} \lesssim 1$. 
    \item Drifting sequence case: Suppose $h = 0$, $\beta = 1 - c n^{-\frac{1}{2}}, c \in \reals^+$. Then $\norm{\mathsf{U}_n}_{\psi_2} \leq \mathtt{C} n^{1/4}$ for large enough $n$ with $\mathtt{C}$ not depending on $\beta$.
\end{enumerate}
\end{lemma}

Fix $\beta > 0$. We characterize the limiting distribution of $n^{-1}\sum_{i = 1}^n W_i X_i$  and the rate of convergence as $n \rightarrow \infty$ in the following lemma. In particular, we will see that the limiting distribution changes from a Gaussian distribution under high temperature, to a non-Gaussian distribution under critical temperature, to a Gaussian mixture under low temperature.

\begin{lemma}[Fixed Temperature Berry-Esseen]\label{sa-lem:fixed-temp-be}
Recall $\mca g_n = n^{-1}\sum_{i = 1}^n X_i (W_i - \pi)$.
\begin{enumerate}
    \item When $\beta < 1$ and $h = 0$ or $h \neq 0$, 
    \begin{align*}
        \sup_{t\in\bb R}|\bb P_{\beta,h}(n^{\frac{1}{2}} \Big(\E[X_i^2](1 - \pi^2) + \E[X_i]^2\frac{\beta(1 - \pi^2)}{1 - \beta (1 - \pi^2)}\Big)^{-\frac{1}{2}}  \mca g_{n} \leq t)  -\Phi(t)|  =O(n^{-\frac{1}{2}}).
    \end{align*}
    \item When $\beta = 1$ and $h = 0$, denote $F_0(t) = \frac{\int_{-\infty}^t \exp(-z^4/12)d z}{\int_{-\infty}^{\infty} \exp(-z^4/12)d z}, t \in \reals$, then
    \begin{align*}
        \sup_{t\in\bb R}|\bb P_{\beta,h}(n^{\frac{1}{4}} \E[X_i]^{-1} \mca g_n \leq t)-F_0(t)| =O((\log n)^3n^{-\frac{1}{2}}).
    \end{align*}
    \item When $\beta > 1$ and $h = 0$, denote $\mca g_{n,\ell} = \frac{1}{n}\sum_{i = 1}^n X_i (W_i - \pi_\ell)$, then
    \begin{align*}
        \sup_{t\in\bb R}|\bb P_{\beta,h}(n^{\frac{1}{2}} \Big(\E[X_i^2](1 - \pi_\ell^2) + \E[X_i]^2\frac{\beta(1 - \pi_\ell^2)}{1 - \beta (1 - \pi_\ell^2)}\Big)^{-\frac{1}{2}}  \mca g_{n,\ell} \leq t| & \sign(\mca m) = \ell)  -\Phi(t)| \\
        & =O(n^{-\frac{1}{2}}), \quad t \in \{ -,+\}.
    \end{align*}
\end{enumerate}
\end{lemma}

\begin{remark}\label{sa-remarK: conditional concentration under low temp}
Lemma~\ref{sa-lem:sub-gaussian}(3) and Lemma~\ref{sa-lem:fixed-temp-be}(3) together implies when $h = 0, \beta > 1$, condition on $\sign(\mca m) = \ell$, $\norm{n^{-1/2} \mathsf{U}_n - \pi_{\ell}}_{\psi_2} \lesssim n^{-1/2}$.
\end{remark}

\begin{lemma}[Size-Dependent Temperature Berry-Esseen when $h = 0$]\label{sa-lem: localization to singularity}
Suppose $\mathsf{Z}$ is a standard Gaussian random variable.
(1) Suppose $\beta_n = 1 + c n^{-\frac{1}{2}}$ and $h = 0$, where $c < 0$. Then
\begin{align*}
    \sup_{t \in \reals}\bigg|\P_{\beta_n, h}(n^{\frac{1}{4}} \mca g_n \leq t) - \P ( n^{-\frac{1}{4}}\E[X_i^2]^{\frac{1}{2}}\mathsf{Z} +\beta_n^{\frac{1}{2}}\E[X_i]\mathsf{W}_c \leq t)\bigg| = O((\log n)^{3}n^{-\frac{1}{2}}),
\end{align*}
where $O(\cdot)$ is up to a universal constant, and recall from Theorem 3.1 in the main paper that $\mathsf{W}_c$ is a random variable independent to $\mathsf{Z}$ with cummulative distribution function
\begin{align*}
  \P[\mathsf{W}_c \leq w]
  = \frac{\int_{-\infty}^w \exp (-\frac{x^4}{12}-\frac{c x^2}{2})dx}{\int_{-\infty}^{\infty}\exp(-\frac{x^4}{12}-\frac{c x^2}{2})dx}, \qquad w \in \reals,\quad c \in \reals_+.
\end{align*}

(2) Suppose $\beta_n = 1 + c n^{-1/2}$ and $h = 0$, where $c > 0$. Then
\begin{align*}
    \sup_{c \in \reals^+}\sup_{t \in \reals}\bigg|\P_{1 + cn^{-1/2}, h} (n^{\frac{1}{4}} \mca g_n \leq t| \mca m \in \ca I_{c,n,\ell}) - \P ( n^{-\frac{1}{4}}\E[X_i^2]^{\frac{1}{2}}\mathsf{Z} +\beta_n^{\frac{1}{2}}\E[X_i]\mathsf{W}_{c,n} \leq t & |\mathsf{W}_{c,n} \in \ca I_{c,n,\ell})\bigg| \\
    & = O((\log n)^{3}n^{-\frac{1}{2}}),
\end{align*}
where with $v_{n,+}$ and $v_{n,-}$ the positive and negative solutions to $x = \tanh(\beta_n x)$, and 
\begin{align*}
    a_{c,n} & =  v_{n,+}^2 - c n^{-1/2}, \\
    b_{c,n} & = 2 (1 + c n^{-1/2} - v_{n,+}^2) v_{n,+}^2, \\
    c_{c,n} & = 2 (1 + c n^{-1/2} - v_{n,+}^2) (1 + c n^{-1/2} - 3 v_{n,+}^2),
\end{align*}
where $\mathsf{W}_{c,n}$ is a random variable taking values in $\reals$ with density at $w \in \reals$ proportional to $\exp(-h_{c,n}(w))$ independent to $\mathsf{Z}$, 
\begin{align*}
    & h_{c,n}(w) = \frac{\sqrt{n}a_{c,n}}{2} (w - n^{1/4} v_{n,\sign(w)})^2 + \frac{n^{1/4}b_{c,n}}{6}  (w - n^{1/4} v_{n, \sign(w)})^3 + \frac{c_{c,n}}{24}(w - n^{1/4} v_{n,\sign(w)})^4,
\end{align*}
and $\ca I_{c,n,-} = (-\infty,K_{c,n,-})$ and $\ca I_{c,n,+} = (K_{c,n,+},\infty)$ such that $\E[\mathsf{W}_{c,n}|\mathsf{W}_{c,n} \in \ca I_{c,n,\ell}] = n^{1/4} v_{c,n,\ell}$ for $\ell \in \{-,+\}$, and $O(\cdot)$ is up to a universal constant.
\end{lemma}

\begin{remark}
In (2), we consider drifting from the low temperature regime to the critical temperature regime. In Lemma~\ref{sa-lem:fixed-temp-be} we show $\mca g_n$ concentrates on the conditional means given $\sign(\mca m)$ in the low temperature regime, whereas it concentrates on the unconditional mean in the critical temperature regime. The drifting region $\ca I_{c,n,\ell}$ captures this effect. $\ca I_{c,n,\ell} = (-\infty, 0)$ or $(0,\infty)$ when $c = 0$, and $\ca I_{c,n,\ell} = \reals$ when $c = \infty$.
\end{remark}

\begin{lemma}[$\sqrt{n}$-sequence is knife-edge]\label{sa-lem: knife-edge}
Suppose $h = 0$. (1) Suppose $|\beta_n - 1| = o(n^{-\frac{1}{2}})$, then 
\begin{align*}
    \sup_{t \in \reals}\bigg|\P_{\beta_n, h}(n^{\frac{1}{4}} \mca g_n \leq t) - \P (\E[X_i]\mathsf{W}_0 \leq t)\bigg| = o(1).
\end{align*}

(2) Suppose $1 - \beta_n \gg n^{-\frac{1}{2}}$, then
\begin{align*}
    \sup_{t \in \reals}\bigg|\P_{\beta_n, h}(\V[\mca g_n]^{-\frac{1}{2}} \mca g_n \leq t) - \Phi(t)\bigg| = o(1).
\end{align*}

(3) Suppose $\beta_n - 1 \gg n^{-\frac{1}{2}}$, then for $\ell \in \{-,+\}$,
\begin{align*}
    \sup_{t \in \reals} \bigg|\P_{\beta_n, h} \Big(\V[\mca g_n|\mca m \in \ca I_{\ell}])^{-\frac{1}{2}}(\mca g_n - \E[\mca g_n|\mca m \in \ca I_{\ell}]) \leq t \Big) - \Phi(t) \bigg| = o(1),
\end{align*}
where $\ca I_+ = [0,\infty)$ and $\ca I_- = (-\infty,0)$. 
\end{lemma}

\begin{lemma}[Fixed Temperature Berry-Esseen  with Multivariate Multiplier]\label{lem: mult-be}
Suppose $\bW$ satisfies Assumption~\ref{assump-one-block}, and $\bX_1, \cdots, \bX_n$ are i.i.d random vectors taking values in $\reals^d$, independent to $\bW$. Suppose there exists some constant $b > 0$ such that $\E[X_{ij}^2] \geq b$ for all $j = 1, \cdots, d$, and for some sequence of constants $B_n \geq 1$, $|X_{ij}| \leq B_n$ for all $i = 1, \cdots, n$ and $j = 1, \cdots, d$. Let $\mathcal{R}$ be the collection of all hyperrectangles in $\reals^d$. 

\begin{enumerate}
    \item When $\beta < 1$ and $h =  0$ or $h \neq 0$, 
    \begin{align*}
        \sup_{A \in \mathcal{R}} \Big|\bb P_{\beta,h} \Big(\frac{1}{n} \sum_{i = 1}^n \bX_i (W_i - \pi) \in A \Big)  - \P(n^{-1/2}\bSigma^{1/2}\mathsf{Z}_d + n^{-1/2}\boldsymbol{\eta} \mathsf{Z} \in A) \Big| = O\Big(\Big(\frac{B_n^2 \log (n)^7}{n}\Big)^{1/6}\Big),
    \end{align*}
    where $\bSigma = (1 - \pi^2) \E[\bX_i \bX_i^{\top}]$, $\boldsymbol{\eta} = (\frac{\beta(1 - \pi^2)^2}{1 - \beta(1 - \pi^2)})^{1/2} \E[\bX_i]$, and $\mathsf{Z}_d \sim \mathsf{N}(\mathbf{0}, \mathbf{I}_{d \times d})$ independent to $\mathsf{Z} \sim \mathsf{N}(0,1)$.
    \item When $\beta = 1$ and $h =  0$, 
    \begin{align*}
        \sup_{A \in \mathcal{R}} \Big|\bb P_{\beta,h} \Big(\frac{1}{n} \sum_{i = 1}^n \bX_i W_i \in A \Big)-\P(n^{-1/2}\bSigma^{1/2}\mathsf{Z}_d + n^{-1/4}\E[\bX_i] \mathsf{R} \in A) \Big| = O\Big(\Big(\frac{B_n^2 \log (n)^7}{n}\Big)^{1/6}\Big).
    \end{align*}
    where $\mathsf{R}$ be a random variable with cummulative distribution function $F_0(t) = \frac{\int_{-\infty}^t \exp(-z^4/12)d z}{\int_{-\infty}^{\infty} \exp(-z^4/12)d z}$, $t \in \reals$, independent to $\mathsf{Z}_d$.
    \item When $\beta > 1$ and $h =  0$, for $\ell = -,+$, 
    \begin{align*}
        \sup_{A \in \mathcal{R}} \Big|\bb P_{\beta,h} \Big(\frac{1}{n} \sum_{i = 1}^n \bX_i (W_i - \pi_{\ell}) \in A \Big| \sign(\mca m) = \ell \Big)  - \P(n^{-1/2}\bSigma^{1/2}\mathsf{Z}_d & + n^{-1/2} \boldsymbol{\eta} \mathsf{Z} \in A) \Big| \\
        & = O\Big(\Big(\frac{B_n^2 \log (n)^7}{n}\Big)^{1/6}\Big),
    \end{align*}
    where $\bSigma = (1 - \pi_{+}^2) \E[\bX_i \bX_i^{\top}]$,  $\boldsymbol{\eta} = 
    (\frac{\beta(1 - \pi_{+}^2)^2}{1 - \beta(1 - \pi_{+}^2)})^{1/2} \E[\bX_i]$.
\end{enumerate}
\end{lemma}

\subsection{Proof Sketch of Lemma~\ref{sa-lem: localization to singularity}}\label{sa-sec: proof sketch}

The magnetization $n^{-1}\sum_{i = 1}^n W_i$ has been studied using Stein's method \citep{eichelsbacher2010stein,chatterjee2010spin}. Due to the multipliers, the Stein's method can not be directly applied to $n^{-1}\sum_{i = 1}^n X_i W_i$. We use a proof strategy based on the \textit{de Finetti's Lemma} in Lemma~\ref{sa-lem: definetti}: There exists a latent variable $\mathsf{U}_n$ such that $W_1,\cdots,W_n$ are i.i.d condition on $\mathsf{U}_n$. Moreover, the density of $\mathsf{U}_n$ satisfies $f_{\mathsf{U}_n}(u) \propto \exp (- 1/2 u^2 + n \log \cosh( \sqrt{\beta/n} u)), u \in \reals$.

We provide a proof sketch of Lemma~\ref{sa-lem: localization to singularity} (1) only. Throughout, take $c_{n, \beta} = \sqrt{n}(\beta - 1)$.

\begin{center}
\textbf{Step 1: Conditional Berry-Esseen.} 
\end{center}

$W_i$'s are i.i.d condition on $\mathsf{U}_n$ with 
\begin{align*}
    e(\mathsf{U}_n) & = \E \left[X_i W_i| \mathsf{U}_n\right] =  \E \left[X_i\right] \tanh( \sqrt{\beta/n} \mathsf{U}_n), \\
    v(\mathsf{U}_n) & = \V \left[X_i W_i| \mathsf{U}_n \right] =  \E \left [X_i^2\right]
    - \E \left[X_i\right]^2 \tanh^2 ( \sqrt{\beta/n} \mathsf{U}_n).
\end{align*}
Apply Berry-Esseen Theorem conditional on $\mathsf{U}_n$, and take $\mathsf{Z} \sim \mathsf{N}(0,1)$ independent to $\mathsf{U}_n$,
\begin{align*}
    \sup_{t \in \reals}\Big|\P\Big(\frac{1}{n}\sum_{i = 1}^n X_i W_i \leq t\Big|\mathsf{U_n}\Big) - \P( \sqrt{v(\mathsf{U}_n)}\mathsf{Z} + \sqrt{n}e(\mathsf{U}_n)\leq t|\mathsf{U_n})\Big|\leq \mathtt{C} \E \left[|X_i|^3\right] v(\mathsf{U}_n) n^{-1/2}.
\end{align*}
Lemma 2 in the supplementary material shows $\norm{\mathsf{U}_n}_{\psi_2} \leq \mathtt{C}n^{1/4}$, hence by concentration arguments, $$\sup_{t \in \reals}|\P(n^{-1}\sum_{i = 1}^n X_i W_i \leq t) - \P( \sqrt{v(\mathsf{U}_n)}\mathsf{Z} + \sqrt{n}e(\mathsf{U}_n)\leq t)|\leq K n^{-1/2}.$$

\begin{center}
\textbf{Step 2: Non-Gaussian Approximation for $n^{-\frac{1}{4}}\mathsf{U}_n$.}
\end{center} 

Consider $\mathsf{W}_n = n^{-1/4}\mathsf{U}_n$. By a change of variable from $\mathsf{U}_n$ and Taylor expand what is inside the exponent, we show $\mathsf{W}_n$ has density satisfying
\begin{align*}  
    f_{\mathsf{W}_n}(w) \propto \exp (-\frac{c_{\beta,n}}{2} w^2 - \frac{\beta_n^2}{12} w^4 + g(w)\beta_n^3 n^{-\frac{1}{2}}w^6),
\end{align*}
where $g$ is a bounded smooth function. We show based on sub-Gaussianity of $\mathsf{W}_{n}$, with an upper bound of sub-Gaussian norm not depending on $\beta$, that the sixth order term is negligible and $$\sup_{t \in \reals}|\P(\mathsf{W}_n \leq t) - \P(\mathsf{W} \leq t)| = O(\log^3 n n^{-1/2}),$$ where $\mathsf{W}$ has density proportional to $\exp (-c_{\beta,n}/2 w^2 - \beta_n^2  w^4 / 12)$.

\begin{center}
\textbf{Step 3: Concentration Arguments.}
\end{center}
Since $\mathsf{Z}$ is independent to $(\mathsf{U}_n, \mathsf{W}_n)$, we use data processing inequality and the previous two steps to show $n^{-1}\sum_{i = 1}^n X_i W_i$ is close to $n^{-1/4}v(n^{1/4}\mathsf{W}_{c_{\beta,n}})^{1/2}\mathsf{Z} + n^{1/4}e(n^{1/4}\mathsf{W}_{c_{\beta,n}}))$. Lemma 2 in the supplementary appendix imply $\norm{\mathsf{W}_{c_{\beta,n}}}_{\psi_2} \leq \mathtt{K}$. By Taylor expanding $e(\cdot)$ and $v(\cdot)$ at $0$, we show $n^{1/4}e(\mathsf{U}_n)$ is close to $\E[X_i] \mathsf{W}_{c_{\beta,n}}$ and $n^{-1/4}\sqrt{v(\mathsf{U}_n)}\mathsf{Z}$ is close to $n^{-1/4}v(n^{1/4}\mathsf{W}_{c_{\beta,n}})^{1/2}\mathsf{Z}$.

% \textbf{Step 4: Non-Gaussian Approximation for $n^{\frac{1}{4}}e(n^{\frac{1}{4}}\mathsf{W}_c)$.} Taylor expanding $\tanh$ at zero, we have $n^{1/4}e(n^{1/4}\mathsf{W}_c)) =\E[X_i] n^{\frac{1}{4}}\tanh(n^{-\frac{1}{4}}\mathsf{W}_c) = \E[X_i] [\mathsf{W}_c -  O (\frac{\mathsf{W}_c^2}{3\sqrt{n}})]$. We can then show by concentration of $\mathsf{W}_c$ that $\dKS \left(n^{1/4}e(n^{1/4}\mathsf{W})), \E[X_i]\mathsf{W} \right) = O(n^{-1/2})$.

% \textbf{Step 5: Stabilization of Variance.} Denote $\mathsf{U_c} = n^{1/4}\mathsf{W}_c$. Then we have $\dKS(v(\mathsf{U}_c)^{1/2} \mathsf{Z} + \sqrt{n}e(\mathsf{U}_c)), \E[v(\mathsf{U}_c)]^{1/2}Z + \sqrt{n}e(\mathsf{U}_c))$ can be upper bounded by $$\sup_{t \in \reals} \E \left[ \left|\phi \left(\frac{t - \sqrt{n}e(\mathsf{U}_c)}{v^{\ast}(\mathsf{U}_c)^{1/2}} \right)(t - \sqrt{n}e(\mathsf{U}_c))\left(v(\mathsf{U}_c)^{-1/2} - \E[v(\mathsf{U}_c)]^{-1/2} \right)\right|\right],$$ where $v^{\ast}(\mathsf{U}_c)$ is some quantity between $\E[v(\mathsf{U}_c)]$ and $v(\mathsf{U}_c)$. Again Taylor expanding $v$ and use concentration of $\mathsf{U}_c$, we can get $\dKS(n^{-\frac{1}{4}}v(n^{\frac{1}{4}}\mathsf{W})^{\frac{1}{2}}\mathsf{Z} + n^{\frac{1}{4}}e(n^{\frac{1}{4}}\mathsf{W}),n^{-\frac{1}{4}}\E[X_i^2]^{\frac{1}{2}} \mathsf{Z} + \E[X_i] \mathsf{W})) \leq \mathtt{K} \frac{\log n}{\sqrt{n}}.$

% The conclusion follows from combining the five steps.

\section{Pseudo-Likelihood Estimator for Curie-Weiss Regimes}\label{sa-sec:mple}

\begin{lemma}[No Consistent Variance Estimator]\label{sa-lem:var-inconsist}
Suppose Assumptions 1,2,3 in the main paper hold. Then there is no consistent estimator of $n \V[\widehat{\tau}_n - \tau_n]$.
\end{lemma}

The pseudo-likelihood estimator for Curie-Weiss regime with $h = 0$ is given by
\begin{align*}
   \nonumber \wh\beta & = \argmax_{\beta}\sum_{i\in [n]}\log\bb P_{\beta}\lef(W_i|W_{-i}\rig)\\
    &=\argmax_{\beta} \sum_{i \in [n]}-\log\bl\frac{W_i\tanh(\beta n^{-1}\sum_{j\neq i}W_j)+1}{2}\br.
\end{align*}

\begin{lemma}[Fixed Temperature Distribution Approximation]\label{sa-lem: fixed temp MPLE}
(1) If $\beta \in [0,1)$ and $h = 0$, then
\begin{align*}
    \wh \beta \overset{d}{\to} \max\bigg\{1 - \frac{1 - \beta}{\chi^2(1)},0\bigg\}.
\end{align*}
(2) If $\beta = 1$ and $h = 0$, then
\begin{align*}
    n^{\frac{1}{2}}(1 - \wh\beta)&\overset{d}{\to}  \max \bigg\{\frac{1}{\mathsf{W}_0^2} - \frac{\mathsf{W}_0^2}{3},0\bigg\}.
\end{align*}
(3) If $\beta > 1$ and $h = 0$, we define an unrestricted pseud-likelihood estimator,
\begin{align*}
   \nonumber \betaur = \argmax_{\beta \in \reals} \log \mathbb{P}_{\beta} \left( W_i \mid \mathbf{W}_{-i} \right) = \sum_{i \in [n]} -\log \bigg( \frac{1}{2}W_i \tanh(\beta \mca m_i) + \frac{1}{2}\bigg).
\end{align*}
Then
\begin{align*}
    \sup_{t \in \reals}|\P(n^{1/2}(\betaur - \beta) \leq t |\mca m \in \ca I_\ell) -  \P((\frac{1 - \beta(1 - \pi_{\ell}^2)}{1 - \pi_{\ell}^2})^{1/2}\mathsf{Z} \leq t)| = o(1).
\end{align*}
\end{lemma}

\begin{lemma}[Drifting Temperature Distribution Approximation]\label{sa-lem: drifting MPLE}
For any $\beta \in [0,1]$ and $h = 0$, define $c_{\beta,n} = \sqrt{n}(1 - \beta)$, and suppose $z_{\beta,n}$ is a random variable such that
\begin{align*}
    \P(z_{\beta,n} \leq t)  = \P(\mathsf{Z} + n^{\frac{1}{4}}\mathsf{W}_{c_{\beta,n}} \leq t), \qquad t \in \reals.
\end{align*}
then
\begin{align*}
    \sup_{\beta \in [0,1]}\sup_{t \in \reals}|\P (1 - \wh \beta \leq t) - \P (\min\{\max\{z_{\beta,n}^{-2} - \frac{1}{3n} z_{\beta,n}^2,0\},1\} \leq t)| = o(1).
\end{align*}
\end{lemma}
\section{Stochastic Linearization}\label{sa-sec:stochlin}

Recall $\bW = (W_1, \cdots, W_n)$ satisfies Assumption~\ref{assump-one-block}. And for notational simplicity, let $g_i$ be the function such that
\begin{align*}
    g_i(x,y) = f_i\Big(\frac{1}{2}x+\frac{1}{2}, \frac{1}{2}y + \frac{1}{2}\Big), \qquad x \in \{-1,1\}, y \in [-1,1].
\end{align*}
We denote $M_i = \sum_{j \neq i} E_{ij} W_i$, $N_i = \sum_{j \neq i} E_{ij}$. Then
\begin{align*}
    g_i(T_i, \mathbf{T}_{-i}) = f_i\Big(T_i,\frac{\sum_{j \neq i}E_{ij} T_i}{\sum_{j \neq i}E_{ij}}\Big) = g_i\Big(W_i, \frac{M_i}{N_i}\Big).
\end{align*}
Recall our definition of regimes: High temperature regime $\mathcal{A}_H = \{(\beta,h) \in [0,\infty) \times \reals: h \neq 0 \text{ or } h = 0, \beta < 1\}$, critical temperature regime $\mathcal{A}_C = \{(1, 0)\}$, and low temperature regime $\mathcal{A}_L = \{(\beta,h) \in [0,\infty) \times \reals: h = 0, \beta > 1\}$. Define the following rates that will be used in the convergence analysis:
\begin{align*}
    \balpha = \begin{cases}
        1/2, & \text{ if } (\beta, h) \in \mathcal{A}_H \cup \mathcal{A}_L, \\
        3/4, & \text{ if } (\beta, h) \in \mathcal{A}_C,
    \end{cases}
    \qquad 
    \bgamma = \begin{cases}
        1/2, & \text{ if } (\beta, h) \in \mathcal{A}_H \cup \mathcal{A}_L, \\
        1/4, & \text{ if } (\beta, h) \in \mathcal{A}_C.
    \end{cases}
\end{align*}
and
\begin{align*}
    \pgamma = \begin{cases}
        1/2, & \text{ if } (\beta, h) \in \mathcal{A}_H \cup \mathcal{A}_L, \\
        1/4, & \text{ if } (\beta, h) \in \mathcal{A}_C,
    \end{cases}
    \qquad
    \psi_{\beta,h}(x) = \begin{cases}
        \exp(x^2) - 1, & \text{ if } (\beta, h) \in \mathcal{A}_H \cup \mathcal{A}_L, \\
         \exp(x^4) - 1, & \text{ if } (\beta, h) \in \mathcal{A}_C.
    \end{cases}
\end{align*}
Throughout Section~\ref{sa-sec:stochlin}, we work with $(\beta, h) \in \mathcal{A}_H \cup \mathcal{A}_C$, and let $\pi$ be the unique solution to $x = \tanh(\beta x + h)$. Then \cite[Section 2.5.2]{friedli2017statistical} implies $\E[W_i] = \pi + O(n^{-1})$. Let $\mca m = n^{-1} \sum_{i= 1}^n W_i$ and $\mca m_i = n^{-1} \sum_{j \neq i} W_j$.

\subsection{The Unbiased Estimator}\label{sa-sec:unbiased}
Denote $p_i = \bb P_{\beta, h}(W_i = 1; \bW_{-i}) = \left(\exp \left(-2\beta \mca m_i - 2 h\right) + 1\right)^{-1}$. We propose an unbiased estimator given by
\begin{align*}
    \wh\tau_{n,\text{UB}} = \frac{1}{n} \sum_{i = 1}^n \bigg[\frac{T_i Y_i}{p_i} - \frac{(1 - T_i) Y_i}{1 - p_i}\bigg]
    .
\end{align*}

\begin{lemma}[Unbiased Estimator]\label{sa-lem:unbiased}
$\wh\tau_{n,\text{UB}}$ is an unbiased estimator for $\tau_n$ in the sense that,
\begin{align*}
    \E[\tauub|\mathbf{E},(f_i)_{i \in [n]}] = \tau_n.
\end{align*}
\end{lemma}
We will show the followings have weak limits:
\begin{align*}
    n^{-\balpha} \sum_{i = 1}^n \left[\frac{T_i Y_i}{p_i} - \frac{(1 - T_i) Y_i}{1 - p_i} - \tau_n\right].
\end{align*}
W.l.o.g, we analyse the error for treated data, the error for control data follows in the same way. First, decompose by
\begin{gather*}
     n^{-\balpha} \sum_{i = 1}^n \left[\frac{T_i Y_i}{p_i} - \frac{(1 - T_i)Y_i}{1 - p_i}\right] = \Delta_1 + \Delta_2, \\
    \Delta_1  = n^{-\balpha} \sum_{i = 1}^n \bigg[ \frac{T_i}{p_i} Y_i (1, \pi) - \frac{1 - T_i}{1 - p_i} g_i( -1,\pi)\bigg], \\ 
    \Delta_2 = n^{-\balpha} \sum_{i = 1}^n \left[\frac{T_i}{p_i}\Big( g_i \Big(1, \frac{M_i}{N_i}\Big) - g_i \Big(1, \pi \Big) \Big) - \frac{1 - T_i}{1 - p_i}\Big(g_i \Big(-1, \frac{M_i}{N_i}\Big) - g_i \Big(-1, \pi \Big) \Big)\right].
\end{gather*}

\begin{lemma}\label{sa-lem: approx delta_1} Suppose Assumption \ref{assump-one-block}, and Assumptions 2, and 3 from the main paper hold with $(\beta, h) \in \mathcal{A}_H \cup \mathcal{A}_C$. Then 
\begin{align*}
    \Delta_1 - \E[\Delta_1|\mathbf{E},(g_i)_{i \in [n]}] 
    =  n^{-\balpha}\sum_{i = 1}^n \Big(\frac{g_i(1,\pi)}{1 + \pi} + \frac{g_i(-1,\pi)}{1 - \pi} & - \beta \mathtt{d}\Big) \left(W_i - \pi\right)  \\
    + O_{\psi_{2},tc}(\sqrt{\log n}n^{-\bgamma}),
\end{align*}
where $\mathtt{d} = (1 - \pi)\E[g_i(1,\pi)] +  (1 + \pi)\E[g_i(-1,\pi)]$.

% (2) When $h = 0$ and $\beta \in [0,1]$, 
% \begin{align*}
%     \Delta_1 - \E[\Delta_1|\mathbf{E},(f_i)_{i \in [n]}] 
%     =  n^{-\balpha}\sum_{i = 1}^n \Big(\frac{g_i(1,\pi)}{1 + \pi} + \frac{g_i(-1,\pi)}{1 - \pi} & - \beta \mathtt{d}\Big) \left(W_i - \pi\right)  \\
%     + O_{\psi_{2},tc}(\sqrt{\log n}n^{-1/4}),
% \end{align*}
% where the constants in $O_{\psi_{2},tc}(\cdot)$ can be taken to not depend on $\beta$ or $n$.
\end{lemma}

Now consider $\Delta_2$.  Since $\frac{T_i}{p_i} = \frac{T_i - p_i}{p_i} + 1$, we have the decomposition,
\begin{equation}\label{delta_2 decomposition}
\begin{aligned}
    \Delta_2 = &  n^{-\balpha} \sum_{i = 1}^n \frac{T_i}{p_i} \left[g_i \left(1, \frac{M_i}{N_i} \right) - g_i \left(1, \pi \right)\right] = \Delta_{2,1} + \Delta_{2,2} + \Delta_{2,3} 
\end{aligned}
\end{equation}
where
\begin{align*}
    \Delta_{2,1} & = n^{-\balpha} \sum_{i =1}^n g_i^{\prime} (1,  \pi) \bigg(\frac{M_i}{N_i} -  \pi \bigg), \\
    \Delta_{2,2} & = n^{-\balpha} \sum_{i = 1}^n \frac{T_i - p_i}{p_i} g_i^{\prime}\left( 1, \pi\right) \left(\frac{M_i}{N_i} -  \pi \right), \\
    \Delta_{2,3} & = n^{-\balpha}\sum_{i = 1}^n  \frac{T_i Y_i^{\prime \prime} \left(1, \eta_i^{\ast}\right)}{2 p_i} \left(\frac{M_i}{N_i} -  \pi\right)^2
\end{align*}
where $\eta_i^{\ast}$ is some random quantity between $ \frac{M_i}{N_i}$ and $\pi$. Define $b_i = \sum_{j \neq i} \frac{E_{ij}}{N_j} Y_j^{\prime}\left(1, \pi \right)$. Then by reordering the terms,
\begin{align*}
    \Delta_{2,1} = n^{-\balpha}\sum_{i = 1}^n b_i \left(W_i - \pi \right).
\end{align*}

\begin{lemma}\label{lem: delta_2,2}
Assumption \ref{assump-one-block}, and Assumptions 2, and 3 from the main paper hold with $(\beta, h) \in \mathcal{A}_H \cup \mathcal{A}_C$. Then condition on $\mathbf{U}$ such that $A(\mathbf{U}) \in \mathcal{A} = \{A \in \bb R^{n \times n}: \min_{i \in [n]} \sum_{j \neq i}A_{ij} \geq 32 \log n\}$,
\begin{align*}
    \Delta_{2,2} = O_{\psi_{2},tc}\bigg(\log n \max_{i \in [n]}\E[N_i|\mathbf{U}]^{-1/2}\bigg) + O_{\psi_{\beta,\gamma},tc}(\sqrt{\log n} n^{-\bgamma}).
\end{align*}
% (2) When $h = 0$ and $\beta \in [0,1]$, condition on $A(\mathbf{U}) \in \mathcal{A}$,
% \begin{align*}
%     \Delta_{2,2} = O_{\psi_{2},tc}\bigg(\log n \max_{i \in [n]}\E[N_i|\mathbf{U}]^{-1/2}\bigg) + O_{\psi_1,tc}(\sqrt{\log n} n^{-1/4}),
% \end{align*}
% where the constants in $O_{\psi_{2},tc}$ and $O_{\psi_1,tc}$ can be taken to not depend on $\beta$ and $n$.
\end{lemma}

% \begin{lemma}\label{lem: delta_{2,3}}
%  Suppose Assumption~\ref{Graph-1}, \ref{Pout-1} and \ref{assump-one-block} hold. If $g_i(1, \cdot)$ and $g_i(-1,\cdot)$ are $p+1$-times continuously differentiable, $p \geq 3$, then
% \begin{align*}
%     \Delta_{2,3} - \E[\Delta_{2,3}|\mathbf{E}] = O_{\P} \left( \frac{B}{\sqrt{ \min_i N_i}} + \frac{B n^{\frac{1}{2}- \balpha}\sqrt{\sum_i \sum_j \sum_{k \neq i,j}E_{ik}E_{jk}}}{\sqrt{n \min_i N_i^3}} + \frac{B}{n^{\bgamma}} + \frac{B n^{\bgamma}}{(\min_i N_i)^{\frac{p+1}{2}}}\right).
% \end{align*}
% \end{lemma}

For the term $\Delta_{2,3}$, we further decompose it into two parts:
\begin{align*}
    \Delta_{2,3} = \Delta_{2,3,1} + \Delta_{2,3,2},
\end{align*}
where
\begin{align*}
    \Delta_{2,3,1} & = n^{-\balpha} \sum_{i = 1}^n  \left[g_i \left(1, \frac{M_i}{N_i} \right) - g_i \left(1, \pi \right) - g_i^{\prime} \left(1, \pi \right) \left(\frac{M_i}{N_i} - \pi \right)\right], \\
    \Delta_{2,3,2} & = n^{-\balpha} \sum_{i = 1}^n \frac{1}{2} \frac{W_i - \E[W_i | \bW_{-i}]}{p_i}\left[g_i \left(1, \frac{M_i}{N_i} \right) - g_i \left(1, \pi \right) - g_i^{\prime} \left(1, \pi \right) \left(\frac{M_i}{N_i} - \pi \right)\right].
\end{align*}

% \begin{lemma}\label{lem: delta_{2,3}}
%  Suppose Assumption~\ref{Graph-1}, \ref{Pout-1} and \ref{assump-one-block} hold. If $g_i(1, \cdot)$ and $g_i(-1,\cdot)$ are $p+1$-times continuously differentiable, $p \geq 3$, then
% \begin{align*}
%     \Delta_{2,3} - \E[\Delta_{2,3}|\mathbf{E}] = O_{\P} \left( \frac{B}{\sqrt{ \min_i N_i}} + \frac{B n^{\frac{1}{2}- \balpha}\sqrt{\sum_i \sum_j \sum_{k \neq i,j}E_{ik}E_{jk}}}{\sqrt{n \min_i N_i^3}} + \frac{B}{n^{\bgamma}} + \frac{B n^{\bgamma}}{(\min_i N_i)^{\frac{p+1}{2}}}\right).
% \end{align*}
% \end{lemma}

\begin{lemma}\label{lem:delta_231}
 Assumption \ref{assump-one-block}, and Assumptions 2, and 3 from the main paper hold with $(\beta, h) \in \mathcal{A}_H \cup \mathcal{A}_C$. Then condition on $\mathbf{U}$ such that $A(\mathbf{U}) \in \mathcal{A} = \{A \in \bb R^{n \times n}: \min_{i \in [n]} \sum_{j \neq i}A_{ij} \geq 32 \log n\}$,
\begin{align*}
    & \Delta_{2,3,1} - \E[\Delta_{2,3,1}|\mathbf{E},(f_i)_{i \in [n]}] \\
    = & O_{\psi_{\mathtt{p}_{\beta,h}/2}}(n^{-\bgamma}) + O_{\psi_{\beta,h},tc}(\max_i \E[N_i|\mathbf{U}]^{-1/2}) + O_{\psi_1, tc}(n^{-1/2}) \\
    & \qquad + O_{\psi_2,tc}(n^{\frac{1}{2} - \balpha}\max \E[N_i|\mathbf{U}]^{-1/2}).
\end{align*}
\end{lemma}

\begin{lemma}\label{lem: delta_2,3,2}
 Assumption \ref{assump-one-block}, and Assumptions 2, and 3 from the main paper hold with $(\beta, h) \in \mathcal{A}_H \cup \mathcal{A}_C$. If $g_i(1, \cdot)$ and $g_i(-1,\cdot)$ are $4$-times continuously differentiable, then condition on $\mathbf{U}$ such that $A(\mathbf{U}) \in \mathcal{A}$,
\begin{align*}
    & \Delta_{2,3,2} - \E[\Delta_{2,3,2}|\mathbf{E},(f_i)_{i \in [n]}] \\
    =  &  O_{\psi_{\pgamma/2},tc}((\log n)^{-1/\pgamma}n^{-2\bgamma}) + O_{\psi_{1},tc}((\log n)^{-1/\pgamma} (\min_i\E[N_i|\mathbf{U}])^{-1}) \\
    & + O_{\psi_1, tc} \left(n^{1/2 - \balpha} \left( \frac{\max_i \E[N_i|\mathbf{U}]^3}{\min_i \E[N_i|\mathbf{U}]^4}\right)^{1/2} \right) + O_{\psi_{2/(p+1)},tc}\left(n^{\bgamma} (\min_i \E[N_i|\mathbf{U}]^{-(p+1)/2}) \right).
\end{align*}
\end{lemma}

\subsection{Hajek Estimator}

\begin{lemma}\label{sa-lem: hajek}
Assumption \ref{assump-one-block}, and Assumptions 2, and 3 from the main paper hold with $(\beta, h) \in \mathcal{A}_H \cup \mathcal{A}_C$. Then
\begin{align*}
    \wh{\tau}_n - \wh{\tau}_{n,\text{UB}} 
     = - \bigg(\frac{\E[g_i(1,\pi)]}{\pi +  1} + \frac{\E[g_i(-1,\pi)]}{1 - \pi} \bigg)(1 - \beta(1  - \pi^2))(\mca m - \pi) + O_{\psi_1}(n^{-2\bgamma}).
\end{align*}
\end{lemma}

\subsection{Stochastic Linearization}

\begin{lemma}\label{sa-lem:lin-fixed-temp}
Suppose Assumptions \ref{assump-one-block}, and Assumptions 2, and 3 from the main paper hold with $(\beta, h) \in \mathcal{A}_H \cup \mathcal{A}_C$. Define
\begin{align*}
    R_i = \frac{g_i(1,\pi)}{1 + \pi} + \frac{g_i(-1,\pi)}{1 - \pi}, \qquad Q_i = \E [\frac{G(U_i,U_j)}{\E[G(U_i,U_j)|U_j]}(g_j^{\prime}(1,\pi) - g_j^{\prime}(-1,\pi))|U_i].
\end{align*}

Then,
\begin{align*}
    \sup_{t \in \reals} \big|\P_{\beta,h}(\wh \tau_n - \tau_n \leq t) - \P_{\beta,h}(\frac{1}{n}\sum_{i =1}^n (R_i - \E[R_i] + Q_i)(W_i - \pi) \leq t)\big|
    = O\Big(\frac{\log n}{\sqrt{n \rho_n}} + \mathtt{r}_{n,\beta}\Big),
\end{align*}
where $\mathtt{r}_{n,\beta} = \sqrt[4]{n}\sqrt{\log n}(n \rho_n)^{-\frac{p+1}{2}}$ if $\beta = 1, h = 0$; and $\sqrt{n \log n}(n \rho_n)^{-\frac{p+1}{2}}$ if $\beta < 1$ or $h \neq 0$.
\end{lemma}

\begin{lemma}\label{sa-lem:lin-stoch-unif}
Suppose Assumption \ref{assump-one-block}, and Assumptions 2, and 3 from the main paper with $h = 0$, $\beta \in [0,1]$. Define
\begin{align*}
    R_i = g_i(1,0) + g_i(-1,0), \qquad Q_i = \E [\frac{G(U_i,U_j)}{\E[G(U_i,U_j)|U_j]}(g_j^{\prime}(1,0) - g_j^{\prime}(-1,0))|U_i].
\end{align*}
Then,
\begin{align*}
    \sup_{\beta \in [0,1]}\sup_{t \in \reals} \big|\P_{\beta,h}(\wh \tau_n - \tau_n \leq t) - \P_{\beta,h}(\frac{1}{n}\sum_{i =1}^n (R_i - \E[R_i] + Q_i) W_i \leq t)\big| = o(1).
\end{align*}
\end{lemma}
\section{Jacknife-Assisted Variance Estimation}\label{sec: jacknife}

\begin{lemma}\label{sa-lem:jacknife}
Suppose Assumptions 1,2,3,4 from the main paper hold with $h = 0$, and $n \rho_n^3 \rightarrow \infty$ as $n \rightarrow \infty$. Suppose the non-parametric learner $\wh f$ satisfies $\wh f(\ell,\cdot) \in C_2([0,1])$, and $|\wh f(\ell,\frac{1}{2}) - f(\ell,\frac{1}{2})| = o_\P(1)$, $|\partial_2 \wh f(\ell,\frac{1}{2}) - \partial_2 f(\ell,\frac{1}{2})| = o_\P(1)$, for $\ell \in \{0,1\}$, where the rate in $o_\P(\cdot)$ does not depend on $\beta$. Suppose $\wh K_n$ is the jacknife estimator from Algorithm 2. Then
\begin{align*}
    \wh K_n = \E[(R_i - \E[R_i] + Q_i)^2] + o_\P(1),
\end{align*}
where the rate in $o_\P(1)$ also does not depend on $\beta$.
\end{lemma}

Here we give a local-polynomial based learner $\wh f$ that satisfies requirements of Lemma~\ref{sa-lem:jacknife} (hence Theorem 4 in the main paper.)

\begin{lemma}\label{sa-lem:lp}
Use a local polynomial estimator to fit the potential outcome functions: Take
\begin{align*}
    \widehat{f}(1,x) & := \widehat{\gamma}_0 + \widehat{\gamma}_1 x,  \\
    (\widehat{\gamma}_0, \widehat{\gamma}_1) &:= \argmin_{\gamma_0,\gamma_1} \sum_{i = 1}^n \Big(Y_i - \gamma_0 - \gamma_1 \frac{M_i}{N_i} \Big)^2 K_h \Big(\frac{M_i}{N_i}\Big)\Indicator(T_i = 1),
\end{align*}
where $K_h(\cdot) = h^{-1}K(\cdot/h)$ where $K$ is a kernel function, $h$ is the optimal bandwidth. Then $\widehat{f}(1,0) = f(1,0) + o_{\P}(1), \partial_2\widehat{f}(1,0) = \partial_2 f(1,0) + o_{\P}(1)$, the same for control group. Moreover, the rate of convergence can be made not depending on $\beta$.
\end{lemma}

\section{Additional Distributional Results}
\label{sa-sec: add}
This section presents the additional distributional results in the appendix. We continue to use the notations defined at the beginning of Section~\ref{sa-sec:stochlin}.

\subsection{Low Temperature Treatment Assignment}

Recall we consider a conditional estimand given by 
\begin{align*}
    \tau_{n,\ell} = \frac{1}{n}\sum_{i = 1}^n \E[Y_i(1;\mathbf{T}_{-i}) - Y_i(0;\mathbf{T}_{-i})|f_i(\cdot),\mathbf{E},\sign(\mca m ) = \ell],
    \qquad \ell\in\{-,+\},
\end{align*}
where $\sign(\mca m) = \sign(2 n^{-1}\sum_{i = 1}^n T_i - 1)$. Let $\pi_*$ be the positive root of $x = \tanh(\beta x)$, and take $\pi_+ = 1/2 + \pi_*/2$, $\pi_- = 1/2 - \pi_*/2$.

\begin{lemma}\label{sa-lem:lin-low}
Suppose Assumptions \ref{assump-one-block}, and Assumptions 2, and 3 from the main paper hold with $\beta > 1$ and $h = 0$. Define
\begin{align*}
    R_{i,\ell} = \frac{g_i(1,\pi_\ell)}{1 + \pi_{\ell}} + \frac{g_i(-1,\pi_\ell)}{1 - \pi_{\ell}}, \quad Q_{i,\ell} = \E \Big[\frac{G(U_i,U_j)}{\E[G(U_i,U_j)|U_j]}(g_j^{\prime}(1,\pi_{\ell}) - g_j^{\prime}(-1,\pi_{\ell}))\Big|U_i\Big], \quad \ell \in \{-,+\}.
\end{align*}
Then,
\begin{align*}
    \sup_{t \in \reals} \max_{\ell \in \{-,+\}} \big|\P_{\beta,h}(\wh \tau_n - \tau_{n,\ell} \leq t | \sign(\mca m) = \ell) - \P(\frac{1}{n}\sum_{i =1}^n (R_{i,\ell}  & - \E[R_{i,\ell}] + Q_{i,\ell})( W_i - \pi_{\ell}) \leq t)\big| \\
    & = O\Big(\frac{\log n}{\sqrt{n \rho_n}} + \sqrt{n \log n}(n \rho_n)^{-\frac{p+1}{2}}\Big).
\end{align*}
\end{lemma}

\begin{lemma}\label{sa-lem:clt-low}
Suppose Assumptions \ref{assump-one-block}, and Assumptions 2, and 3 from the main paper hold with $\beta > 1$ and $h = 0$. Then
\begin{align*}
    \sup_{t \in \reals} \max_{\ell \in \{-,+\}}\big|\P_{\beta,h}(\wh \tau_n - \tau_{n,\ell} \leq t|\sign(\mca m) = \ell) - L_n(t;\beta,\kappa_{1,\ell},\kappa_{2,\ell})\big|
    = O\Big(\sqrt{\frac{n \log n}{(n \rho_n)^{p+1}}} + \frac{\log n}{\sqrt{n \rho_n}}\Big),
\end{align*}
where with $\mathsf{Z} \sim \mathsf{N}(0,1)$,
\begin{align*}
    L_{n}(t;\beta,\kappa_{1,\ell},\kappa_{2,\ell}) = \P\Big\{n^{-1/2}\Big(\kappa_{2,\ell} (1 - \pi_*^2) + \kappa_{1,\ell}^2 \frac{\beta (1 - \pi_*^2)}{1 - \beta (1  - \pi_*^2)}\Big)^{1/2}\mathsf{Z} \leq t\Big\},
\end{align*}
where $\kappa_{s,\ell} = \E[(R_{i,\ell} - \E[R_{i,\ell}] + Q_{i,\ell})^s]$ for $s = 1,2$ and $\ell = -, +$.
\end{lemma}

\subsection{Asymmetric Treatment Assignment}

Recall the following treatment assignment model from Section A.1: For $\beta \in [0,\infty)$ and $h \neq 0$, the treatment vector $\bT = (T_1, \cdots, T_n)$ satisfies a distribution on $\{0,1\}^n$ such that
\begin{align*}
    \mathbb{P}_{\beta, h}(\mathbf{T} = \mathbf{t}) \propto\exp \bigg( \frac{\beta}{n} \sum_{i < j} (2 t_i - 1) (2 t_j - 1) + h \sum_{i = 1}^n (2 t_i - 1) \bigg), \qquad \mathbf{t} \in \{0,1\}^n.
\end{align*}
Let $\pi$ be the unique solution to $x = \tanh(\beta x + h)$.

\begin{lemma}\label{sa-lem:lin-h-nonzero}
Suppose Assumptions \ref{assump-one-block}, and Assumptions 2, and 3 from the main paper hold with $\beta > 0$ and $h \neq 0$. Define
\begin{align*}
    R_i = \frac{g_i(1,\pi)}{1 + \pi} + \frac{g_i(-1,\pi)}{1 - \pi}, \qquad Q_i = \E \Big[\frac{G(U_i,U_j)}{\E[G(U_i,U_j)|U_j]}(g_j^{\prime}(1,\pi) - g_j^{\prime}(-1,\pi))\Big|U_i\Big].
\end{align*}
Then,
\begin{align*}
    \sup_{t \in \reals} \big|\P_{\beta,h}(\wh \tau_n - \tau_n \leq t) - \P(\frac{1}{n}\sum_{i =1}^n (R_i - \E[R_i] + Q_i)(W_i - \pi) \leq t)\big|
    = O\Big(\frac{\log n}{\sqrt{n \rho_n}} + \sqrt{n \log n}(n \rho_n)^{-\frac{p+1}{2}}\Big).
\end{align*}
\end{lemma}

\begin{lemma}\label{sa-thm:dist-nonzero}
Suppose Assumptions \ref{assump-one-block}, and Assumptions 2, and 3 from the main paper hold with $\beta > 0$ and $h \neq 0$. Then
\begin{align*}
    \sup_{t \in \reals} \big|\P[\wh \tau_n - \tau_n \leq t] - L_n(t;\beta,h,\kappa_1,\kappa_2)\big|
    = O\Big(\frac{\log n}{\sqrt{n \rho_n}} + \sqrt{n \log n}(n \rho_n)^{-(p+1)/2}\Big),
\end{align*}
where $L_n(\cdot;\beta,h,\kappa_1,\kappa_2)$ is as follows:
\begin{align*}
    L_n(t;\beta,h,\kappa_1,\kappa_2) = \P_{\beta,h}\Big[n^{-1/2}\Big(\kappa_2 (1 - \pi^2) + \kappa_1^2 \frac{\beta (1 - \pi^2)^2}{1 - \beta (1 - \pi^2)}\Big)^{1/2}\mathsf{Z} \leq t\Big]
\end{align*}
with $\mathsf{Z} \thicksim \mathsf{N}(0,1)$, and $\kappa_s= \E[(R_i - \E[R_i] + Q_i)^s]$ for $s = 1, 2$.
\end{lemma}

\subsection{Ising Block Treatment Assignment}

Recall our notations: For block $k$ with $h_k \neq 0$ or $h_k = 0, 0 \leq \beta_k \leq 1$, $\pi_k$ denotes the unique solution to $x = \tanh(\beta_k x + h_k)$. For block $k$ with $h_k = 0, \beta_k > 1$, $\pi_{k,+}$ and $\pi_{k,-}$ denote the unique positive and negative solutions to $x = \tanh(\beta_k x + h_k)$, respectively.

Due to the potential existence of low temperature blocks, we use $\bsign$ to collect the average spins in all low temperature blocks, and fill in the positions for high and critical temperature blocks with zeros, that is,
\begin{align*}
    \bsign = (\sign(\mca m_1) \Indicator(1 \in \mathscr{L}), \cdots, \sign(\mca m_K) \Indicator(K \in \mathscr{L})).
\end{align*}
And we use $\mathscr{S}$ to denote the collection of all possible configurations of $\bsign$, that is,
\begin{align*}
    \mathscr{S} = \{(s_k)_{1 \leq k \leq K}: s_k = - \text{ or } + \text{ if } k \in \mathscr{L}, s_k = 0 \text{ otherwise}\}.
\end{align*}
Also we denote the conditional fixed point based on $\bsign = \bs$ by
\begin{align*}
    \pi_{k,(\bs)} = 
    \begin{cases}
        \pi_k, & \text{ if } k \in \mathscr{H} \cup \mathscr{C}, \\
        \pi_{k,s_k}, & \text{ if } k \in \mathscr{L},
    \end{cases}
    \qquad 1 \leq k \leq K, \bs \in \mathscr{S}.
\end{align*}
We denote by $\mathscr{R}$ the collection of all hyperrectangles in $\mathbb{R}^K$.

\begin{lemma}\label{sa-lem: mult-stoch-lin}
Suppose Assumptions 2, 3, and 6 from the main paper hold. Condition on $\bsign = \mathbf{s}$,
\begin{align*}
    & \bigg\lVert \wh{\btau}_{n} - \btau_n - \frac{1}{n} \sum_{l = 1}^K \sum_{i \in \scrC_l} \bS_{l,i,(\bs)}(W_i - \pi_{l,(\mathbf{s})}) \bigg\rVert_2 = O_{\psi_1,tc}(\mathtt{r}_n).
\end{align*}
where $\bS_{l,i,(\bs)} = (S_{1,l,i,(\bs)}, \cdots, S_{K,l,i,(\bs)})^{\Trans}$, where
\begin{align*}
    S_{k,l,i,(\bs)} = Q_{i,(\bs)} + \Indicator(k = l) p_k^{-1}(R_{i,l,(\bs)} - \E[R_{i,l,(\bs)}]), \qquad 1 \leq k, l \leq K, 1 \leq i \leq n,
\end{align*}
with $\overline{\pi}_{(\bs)} = \sum_{k = 1}^K p_k \pi_{k,(\bs)}$, 
\begin{align*}
    R_{i,l,(\bs)} & = \frac{g_i(1, \overline{\pi}_{(\bs)})}{1 + \pi_{l,(\bs)}} + \frac{g_i(-1, \overline{\pi}_{(\bs)})}{1 - \pi_{l,(\bs)}}, \\
    Q_{i,(\bs)} & = \E \Big[\frac{G(U_i, U_j)}{\E[G(U_i, U_j)|U_j]} (g_j^{\prime}(1, \overline{\pi}_{(\bs)}) - g_j^{\prime}(-1, \overline{\pi}_{(\bs)}))\Big| U_i\Big].
\end{align*}
and $\mathtt{r}_n = \sqrt{\log n} \max_{1 \leq k \leq K} n^{-\bgammak}(n \rho_n)^{-1/2} + (n \rho_n)^{-(p+1)/2}$.
\end{lemma}

\begin{lemma}\label{sa-lem:clt-block}
Suppose Assumptions 2, 3, and 6 from the main paper hold. Condition on $\bsign = \mathbf{s}$, we have
\begin{align*}
    \max_{\mathbf{s} \in \mathscr{S}} \sup_{A \in \mathscr{R}} & | \P_{\boldsymbol{\beta}, \boldsymbol{h}}(\wh{\btau}_n  - \btau_n \in A|  \bsign = \mathbf{s}) - \\
    & \P(n^{-1/2} \bSigma_{(\bs)}^{1/2} \mathsf{Z}_K + n^{-1/2} \sum_{k \in \mathscr{H} \cup \mathscr{L}} p_k \sigma_{k,(\bs)} \E[\bS_{k,i,(\bs)}] \mathsf{Z}_{(k)} + n^{-1/4} \sum_{k \in \mathscr{C}} p_k \E[\bS_{k,i,(\bs)}] \mathsf{R}_{(k)} \in A)| \\
    & \qquad \quad = O(n^{1/2} \mathtt{r}_n + (\log n)^{7/6} n^{-1/6}),
\end{align*}
where $\mathsf{Z}_K \sim \mathsf{N}(\mathbf{0},\mathbf{I}_{K \times K})$, $\mathsf{Z}_{(k)} \sim \mathsf{N}(0,1)$ for $k \in \mathscr{H} \cup \mathscr{L}$, and $\mathsf{R}_{(k)}$ has cummulative distribution function $F_0(t) = \frac{\int_{-\infty}^t \exp(-z^4/12)d z}{\int_{-\infty}^{\infty} \exp(-z^4/12)d z}, t \in \reals,$ for $k \in \mathscr{C}$, with $\mathsf{Z}_K$, $\mathsf{Z}_{(k)}, k \in \mathscr{H} \cup \mathscr{L}$ and $\mathsf{R}_{(k)}, k \in \mathscr{C}$ mutually independent, and 
\begin{align*}
    \bSigma_{(\bs)} = (\sum_{k = 1}^K \E[\bS_{k,i,(\bs)} \bS_{k,i,(\bs)}^\top] (1 - \pi_{k,(\bs)}^2) p_k^2)^{1/2}.
\end{align*}
\end{lemma}

\begin{remark}\label{sa-remark: multi-block}
If there is no low temperature block, then $\bsign = (0, \cdots, 0)$ almost surely, and $\mathsf{S}$ is the singleton set containing $(0, \cdots, 0)$. Hence the result reduces to the unconditional distributional approximation.
\end{remark}

\section{Proofs: Main Paper}

\subsection{Proof of Theorem 3.1}

The conclusion follows from the stochastic linearization result in Lemma~\ref{sa-lem:lin-fixed-temp}, and the Berry-Esseen result for Curie-Weiss magnetization with independent multipliers in Lemma~\ref{sa-lem:fixed-temp-be} (1) and (2).

\subsection{Proof of Theorem 3.2}

The conclusion for Hajek estimator follows from the stochastic linearization result in Lemma~\ref{sa-lem:lin-fixed-temp}, and the (uniform in $\beta$) Berry-Esseen result for Curie-Weiss magnetization with independent multipliers in Lemma~\ref{sa-lem: localization to singularity} (1).

The conclusion for MPLE follows from Lemma~\ref{sa-lem: drifting MPLE}.

\subsection{Proof of Lemma 3.1}

The conclusion follows from Lemma~\ref{sa-lem:fixed-temp-be} and Lemma~\ref{sa-lem: localization to singularity}.

\subsection{Proof of Theorem 4.1} 

The uniform approximation for $\sqrt{n}(\wh \beta_n - 1)$ established in Lemma~\ref{sa-lem: drifting MPLE} implies $$\inf_{\beta}\P_\beta(\beta \in \ca I(\alpha_1)) \geq \inf_{\beta}\P_\beta(\sqrt{n}(1 - \beta) \geq \mca q) \geq  1 - \alpha_1 + o_\P(1).$$
where $\mca q$ is the $\alpha_1$ quantile of $\min\{\max \{\mathsf{T}_{c_{\beta,n},n}^{-2} - \mathsf{T}_{c_{\beta,n},n}^2/(3n),0\},1\}$.

Then by a Bonferroni correction argument, the second step coverage can be lower bounded by
\begin{align*}
    \inf_{\beta \in [0,1]}\P_\beta(\tau_n \in \wh{\ca C}(\alpha_1,\alpha_2))
    & \geq \inf_{\beta \in [0,1]}\P_\beta(\tau_n \in \wh{\ca C}(\alpha_1,\alpha_2), \beta \in \ca I(\alpha_1)) - \P_\beta(\beta \notin \ca I(\alpha_1)).
\end{align*}
Observe that the event $\tau_n \in \wh{\ca C}(\alpha_1,\alpha_2)$ conincides with the event $\wh \tau_n - \tau_n \in [\mathtt{L}, \mathtt{U}]$, where $\mathtt{U} = \sup_{\beta \in \ca I(\alpha_1)} H_n(1 - \frac{\alpha_2}{2};K_n, K_n,c_{\beta,n})$, $\mathtt{L} = \inf_{\beta \in \ca I(\alpha_1)} H_{n}(\frac{\alpha_2}{2};K_n, K_n, c_{\beta,n})$.
Hence
\begin{align*}
    & \inf_{\beta \in [0,1]}\P_\beta(\tau_n \in \wh{\ca C}(\alpha_1,\alpha_2), \beta \in \ca I(\alpha_1)) \\
    \geq & \inf_{\beta \in [0,1]}\P_\beta(\wh \tau_n - \tau_n \in [ H_{n}(\frac{\alpha_2}{2};K_n, K_n, c_{\beta,n}), H_n(1 - \frac{\alpha_2}{2};K_n, K_n,c_{\beta,n})], \beta \in \ca I(\alpha_1)) \\
    \geq & \inf_{\beta \in [0,1]}\P_\beta(\wh \tau_n - \tau_n \in [ H_{n}(\frac{\alpha_2}{2};K_n, K_n, c_{\beta,n}), H_n(1 - \frac{\alpha_2}{2};K_n, K_n,c_{\beta,n})]) - \P_\beta(\beta \in \ca I(\alpha_1)).
\end{align*}
Theorem 2 shows that the quantiles of the distributions of $\wh \tau_n - \tau_n$ can be uniformly approximated by quantiles from $H_n(\cdot;\kappa_1, \kappa_2,c_{\beta,n})$, if $\kappa_1$ and $\kappa_2$ are correctly specified, and the confidence interval is conservative, if we use upper bound $\mathtt{K}_n$ for $\kappa_1$ and $\kappa_2$. The conclusion then follows. 

\subsection{Proof of Theorem 4.2}

The conclusion follows from Theorem 4.1 and Lemma~\ref{sa-lem:jacknife}.

\subsection{Proof of Lemma 5.1}

The conclusion follows from Lemma~\ref{sa-lem:clt-low}.

\subsection{Proof of Lemma 5.2}

The conclusion follows from Lemma~\ref{sa-thm:dist-nonzero}.

\subsection{Proof of Lemma 5.3}

The conclusion follows from Lemma~\ref{sa-lem:clt-block}.

\section{Proofs: Section~\ref{sa-sec:berry-esseen}}\label{sa-sec: berry esseen proof}

\subsection{Proof of Lemma~\ref{sa-lem: definetti}}

Using Gaussian integral identity $\exp(v^2/2) = \frac{1}{\sqrt{2 \pi}} \int_{-\infty}^{\infty} \exp \left( - u^2/2 + uv\right) du$, 
\begin{align*}
    \P \left(\bW = \bw \right) & = \int_{-\infty}^{\infty} \frac{\exp \left( \left(\sqrt{\frac{\beta}{n}} u + h \right)\left(\sum_{i = 1}^n w_i\right) \right)}{2^n \exp \left(n \log \cosh \left( \sqrt{\frac{\beta}{n}} u + h\right) \right)} f_{\mathsf{U}_n}(u) d u.
\end{align*}

\subsection{Proof of Lemma~\ref{sa-lem:sub-gaussian}}

    Our proof is divided according to the different temperature regimes.
    \begin{center}
        \textbf{ The High Temperature Regime. } 
    \end{center}
    We introduce the handy notation given by $F(v):=-\frac{1}{2}v^2+\log\cosh(\sqrt{\beta}v+h)$.
    For the high temperature regime, we note that the term in the exponential can be expanded across its global minimum $v^*$ (which satisfies the first order stationary point condition given by $v^*=\sqrt{\beta}\tanh(\sqrt{\beta}v^*+h)$) by
    \begin{align*}
        F(v)&=F(v^*)+F^{\prime}(v^*)(v-v^*)+\frac{1}{2}F^{(2)}(v^*)(v-v^*)^2+O((v-v^*)^3)\\
        &=F(v^*)-\frac{1}{2}(1-\beta\sech^2(\sqrt{\beta}v^*+h))(v-v^*)^2+O((v-v^*)^3).
    \end{align*}
    Therefore, to obtain the limit of the expectation, we note that by the Laplace method given similar to the proof of Lemma~\ref{sa-lem:fixed-temp-be} and the definition of $\mathsf{V}_n:=n^{-1/2}\mathsf{U}_n$:
    \begin{align*}
        \bb E[\mathsf{V}_n]=\frac{\int_{\bb R}v\exp\lef(-nF(v)\rig)dv}{\int_{\bb R}\exp(-nF(v))dv}=v^*(1+O(n^{-1})).
    \end{align*}
    % Then we also note that
    % \begin{align*}
    %     \int_{\bb R}\exp\bl-\frac{1}{2}u^2&+n\log\cosh\lef(\sqrt{\beta/n}u+h\rig)\br du\\
    %     &=\int_{(n^{})}\exp\bl-\frac{1}{2}(1-\beta\sech^2(h))u^2+\sqrt{\beta n}\tanh(h)u+O(n^{-\frac{1}{2}})u^3\br du
    % \end{align*}
    Then, we note that for $\ell\in\bb N$, when $h=0$ and $\beta<1$ we use the Laplace method again to obtain that for all $\ell\in\bb N$,
    \begin{align*}
       \bb E\lef[(\mathsf{V}_n-\bb E[\mathsf{V}_n])^{2\ell}\rig]&=\frac{\int_{\bb R} (v-v^*)^{2\ell}\exp(-n(F(v)-F(v^*)))dv}{\int_{\bb R}\exp(-n(F(v)-F(v^*)))dv}(1+O(n^{-1}))\\
       &=\frac{1}{\sqrt{\pi}}\bl\frac{2}{n(1-\beta\sech^2(\sqrt{\beta}v^*+h))}\br^{\ell}\Gamma\bl\frac{2\ell+1}{2}\br(1+O(n^{-1})).
    \end{align*}
    Then we can obtain that for all $t\in\bb R$, we have
    \begin{align*}
        \bb E[\exp(t(\mathsf{V}_n-\bb E[\mathsf{V}_n]))]&=\sum_{\ell=0}^{\infty}\frac{t^{\ell}}{\ell!}\bb E[(\mathsf{V}_n-\bb E[\mathsf{V}_n])^{\ell}]=\sum_{\ell=0}^{\infty}\frac{t^{2\ell}}{(2\ell)!}\bb E[(\mathsf{V}_n-\bb E[\mathsf{V}_n])^{2\ell}]\\
        &\leq\exp\bl\frac{(1+o(1))t^2}{2n(1-\beta\sech^2(\sqrt{\beta}v^*+h))}\br,
    \end{align*}
    which alternatively implies that
    \begin{align}\label{eq: subG high-temp}
        \Vert \mathsf{U}_n-\bb E[\mathsf{U}_n]\Vert_{\psi_2}=n^{1/2}\Vert \mathsf{V}_n-\bb E[\mathsf{V}_n]\Vert_{\psi_2}\leq (1+o(1))(1-\beta\sech^2(\sqrt{\beta}v^*+h))^{\frac{1}{2}}.
        \end{align}
        \begin{center}
            \textbf{The Critical Temperature Regime.}
        \end{center}
        Then we study the critical temperature regime with $\beta=1$. Note that one has $\bb E[\mathsf{U}_n]=0$ and for all $\ell\in\bb N$ we have
        \begin{align*}\textbf{}
            F(v)&=F(0)+F^\prime(0)v+\frac{1}{2}F^{(2)}(0)v^2+\frac{1}{6}F^{(3)}(0)v^3+\frac{1}{24}F^{(4)}(0)v^4+O(v^5)\\
            &=F(0)+\frac{1}{12}v^4+O(v^5).
        \end{align*}
        Then we can obtain that $\ell\in\bb N$,
        \begin{align*}
            \bb E\lef[V^{2\ell}_n\rig]&
            =\frac{\int_{\bb R}v^{2\ell}\exp(-nF(v))dv}{\int_{\bb R}\exp(-nF(v))dv}=(1+o(1))\cdot 2^{\ell-\frac{1}{2}}\cdot 3^{\frac{\ell}{2}+\frac{1}{4}}\frac{\Gamma\lef(\frac{\ell}{2}+\frac{1}{4}\rig)}{\Gamma(1/4)}\\
            &\leq(1+o(1))\frac{1}{\sqrt \pi}\bl \frac{2^{3/2}\cdot 3^{3/4}\Gamma(3/4)}{n^{1/2}\Gamma(1/4)}\br^{\ell}\Gamma\lef(\frac{2\ell+1}{2}\rig).
        \end{align*}
        And we immediately obtain that
        \begin{align*}
            \bb E\lef[\exp(t\mathsf{V}_n)\rig]&=\sum_{\ell=0}^{\infty}\frac{t^{\ell}\bb E[\mathsf{V}_n^{2\ell}]}{\Gamma(1+\ell)}\leq\sum_{\ell=0}^{\infty}\frac{1+o(1)}{\Gamma(1+2\ell)}\frac{1}{\sqrt \pi}\bl \frac{2^{1/2}\cdot 3^{3/4}\sqrt 2\Gamma(3/4)}{n^{1/2}\Gamma(1/4)}\br^{\ell}\Gamma\lef(\frac{2\ell+1}{2}\rig)t^{\ell}\\
            &\leq\exp\bl\frac{1+o(1)}{2}t^2\bl\frac{2^{3/2}\cdot 3^{3/4}\Gamma(3/4)}{n^{1/2}\Gamma(1/4)}\br\br ,
        \end{align*}
        which finally leads to 
        \begin{align}\label{eq: subG critc-temp}
            \Vert \mathsf{V}_n\Vert_{\psi_2}\leq (1+o(1))\sqrt{\frac{2^{1/2}\cdot 3^{3/4}\Gamma(3/4)}{n^{1/2}\Gamma(1/4)}}.
        \end{align}
        \begin{center}
            \textbf{ The Low Temperature Regime. }
        \end{center}
        We shall note that at the low temperature regime the function $F(v)$ has two symmetric global minima $v_1>0>v_2$, satisfying
        \begin{align*}
            F^\prime(v_1)=F^\prime(v_2)=0\quad\Rightarrow\quad v_{\ell}=\sqrt{\beta}\tanh(\sqrt{\beta}v_{\ell}+h)\quad\text{ for }\ell\in\{1,2\}. 
        \end{align*}
        Then we can check that by the Laplace method, for all $t>0$ (following the path given by the high temperature regime) we have
        \begin{align*}
            \bb E[\exp(t(\mathsf{V}_n-\bb E[\mathsf{V}_n|\mathsf{V}_n>0]))|\mathsf{V}_n>0]&=\frac{\int_{[0,\infty)}\exp\lef(t(v-v_1)-nF(v)\rig)dv}{\int_{[0,\infty)}\exp(-nF(v))dv}\\
            &=\exp\bl\frac{(1+o(1))t^2}{2n(1-\sqrt{\beta}\sech^2(\sqrt{\beta}v_1))}\br.
        \end{align*}
        Then we similarly obtain that
$            \bb E[\exp(t(\mathsf{V}_n-\bb E[\mathsf{V}_n|\mathsf{V}_n<0]))|\mathsf{V}_n<0]=\exp\lef(\frac{(1+o(1))t^2}{2n(1-\sqrt{\beta}\sech^2(\sqrt{\beta}v_1))}\rig)$. Hence we obtain that
\begin{align}\label{eq: subG low-temp}
    \nonumber \Vert \mathsf{V}_n-\bb E[\mathsf{V}_n|\mathsf{V}_n<0]|\mathsf{V}_n<0\Vert_{\psi_2} & =\Vert \mathsf{V}_n-\bb E[\mathsf{V}_n|\mathsf{V}_n>0]|\mathsf{V}_n>0\Vert_{\psi_2}\\
    & \leq (1+o(1))(1-\beta\sech^2(\sqrt{\beta}v_1))^{\frac{1}{2}}.
\end{align}
\begin{center}
    \textbf{ The Drifting Sequence Case. }
\end{center}
Then we consider the drifting case. 

First consider $\beta= 1-cn^{-\frac{1}{2}}$ with $c\in\bb R^+$ and $\beta \geq 0$. We will show that for any fixed $n$, $\norm{W_n}_{\psi_2}$ is increasing in $\beta$ when $\beta \in [0,1]$. This will imply that in the drifting case, $\norm{W_n}_{\psi_2}$ will be no larger than its value at the critical regime.

For a comparison argument, denote $F_{\beta}(v) = - \frac{1}{2}v^2 + \log \cosh(\sqrt{\beta} v)$. Let $0 < \beta_1 < \beta_2 \leq 1$. Then
\begin{align*}
    \frac{\exp(n F_{\beta_2}(v))}{\exp(n F_{\beta_1}(v))} = \exp (n \log \cosh(\sqrt{\beta_2} v ) - n \log \cosh(\sqrt{\beta_1} v)),
\end{align*}
where 
\begin{align*}
    \frac{d}{d v} \frac{\cosh(\sqrt{\beta_2} v)}{\cosh(\sqrt{\beta_1} v)} = \frac{(\sqrt{\beta_2} - \sqrt{\beta_1})\sinh((\sqrt{\beta_2} - \sqrt{\beta_1})v)}{\cosh^2(\sqrt{\beta_1} v)} > 0.
\end{align*}
Hence for any $n \in \mathbb{N}$ and $t > 0$, \begin{align*}
    \P_\beta(|W_n| \geq t) = 2 \frac{\int_t^\infty\exp(n F_\beta(v))d v}{\int_0^\infty \exp(n F_\beta(v))d v}
\end{align*}
increases as $\beta \in [0,1]$ increases. This shows that $\norm{W_n}_{\psi_2}$ increases as $\beta \in [0,1]$ increases. Together with Equation~\eqref{eq: subG critc-temp}, we have under $\beta_n = 1 - \frac{c}{\sqrt{n}}$, $0 \leq c \leq \sqrt{n}$,
\begin{align*}
    \norm{\mathsf{V}_n}_{\psi_2} \leq (1+o(1))\sqrt{\frac{2^{1/2}\cdot 3^{3/4}\Gamma(3/4)}{n^{1/2}\Gamma(1/4)}},
\end{align*}
where $o(\cdot)$ is by an absolute constant.

Then we consider $\beta=1+cn^{-\frac{1}{2}}$. We shall note that under this situation it is not hard to check that
\begin{align*}
    \bb E[\exp(t\mathsf{V}_n)]&=\frac{1}{2}\lef(\bb E[\exp(t\mathsf{V}_n)|\mathsf{V}_n>0]+\bb E[\exp(t\mathsf{V}_n)|\mathsf{V}_n<0]\rig)\\
    &=\frac{1}{2}\lef(\bb E[\exp(t(\mathsf{V}_n-v_+))|\mathsf{V}_n>0]\exp(tv_{+})+\bb E[\exp(t(\mathsf{V}_n-v_{-}))|\mathsf{V}_n<0]\exp(tv_{-})\rig).
\end{align*}
Then, under this case we have by Taylor expanding $F$ at $0$ and the fact that $\sup_{v \in \reals}|F^{(5)}(v)| < \infty$,
\begin{align*}
    f_{\mathsf{V}_n}(v)\propto\sum_{l \in \{-,+\}} \Indicator(v \in \mathcal{C}_l)\exp\bl-cn^{\frac{1}{2}}(v-v_{l})^2-\frac{\sqrt{3c}}{3}n^{\frac{3}{4}}(v-v_{l})^3-\frac{1}{12}n(v-v_{l})^4-O(n(v-v_{l})^5)\br.
\end{align*}
Before we start to upper bound the moments, we first use the fact that  $v_+=O(n^{-1/4})$ to obtain that
\begin{align*}
    \int_{(-v_+,0)}v^{2\ell}\exp\lef(-\sqrt{3c}v^3\rig)dv\leq n^{-\frac{1}{4}}v_{+}^{2\ell}\exp(-\sqrt{3c}n^{-1/4})=O\lef(n^{-1/4-\ell/2}\rig).
\end{align*}
Then we obtain that
\begin{align*}
    \bb E&[(\mathsf{V}_n-v_+)^{2\ell}|\mathsf{V}_n>0]
    =n^{-\frac{\ell}{2}}\frac{\int_{(-v_+,+\infty)}v^{2\ell}\exp\lef(-cv^2-\frac{\sqrt{3c}}{3}v^3-\frac{1}{12}v^4\rig)dv}{\int_{(-v_+,+\infty)}\exp\lef(-cv^2-\frac{\sqrt{3c}}{3}v^3-\frac{1}{12}v^4\rig)dv}(1+o(1))\\
    &\leq n^{-\frac{\ell}{2}}(1+o(1))\frac{\int_{\bb R}v^{2\ell}\exp(-3cv^2)dv+\int_{(-v_{+},+\infty)}v^{2\ell}\exp(-\sqrt{3c}v^3)dv+\int_{\bb R}v^{2\ell}\exp(-\frac{1}{4}v^4)dv}{\int_{(-v_{+},+\infty)}\exp\lef(-cv^2-\frac{\sqrt{3c}}{3}v^3-\frac{1}{12}v^4\rig)dv}\\
    &= n^{-\frac{\ell}{2}}(1+o(1))\frac{\int_{\bb R}v^{2\ell}\exp(-3cv^2)dv+\int_{\bb R^+}v^{2\ell}\exp(-\sqrt{3c}v^3)dv+\int_{\bb R}v^{2\ell}\exp(-\frac{1}{4}v^4)dv}{\int_{(-v_{+},+\infty)}\exp\lef(-cv^2-\frac{\sqrt{3c}}{3}v^3-\frac{1}{12}v^4\rig)dv}+O(n^{-1/4-\ell/2})\\
    &=n^{-\frac{\ell}{2}}(1+o(1))\bl\mca C_3\bl\frac{1}{3c}\br^{\ell}\Gamma\bl\ell+\frac{1}{2}\br+\mca C_4(3c)^{-\frac{\ell}{3}}\Gamma\bl\frac{2\ell}{3}+\frac{1}{3}\br +\mca C_52^{\ell}\Gamma\bl\frac{\ell}{2}+\frac{1}{4}\br\br,
\end{align*}
with 
$\mca C_3:=\frac{(3c)^{-1/2}}{3\int_{(-v_+,+\infty)}\exp\lef(-cv^2-\frac{\sqrt{3c}}{3}v^3-\frac{1}{12}v^4\rig)dv}$, $\mca C_4=\frac{1}{9\int_{(-v_+,+\infty)}\exp\lef(-cv^2-\frac{\sqrt{3c}}{3}v^3-\frac{1}{12}v^4\rig)dv}$,\\ and $\mca C_5=\frac{2^{-3/2}}{\int_{(-v_+,+\infty)}\exp\lef(-cv^2-\frac{\sqrt{3c}}{3}v^3-\frac{1}{12}v^4\rig)dv}$.
Therefore, we can simply use the definition of the m.g.f. to obtain that
\begin{align*}
    \bb E[\exp(t^2&(\mathsf{V}_n-v_+)^2)|\mathsf{V}_n>0]=\sum_{\ell=0}^{\infty}\frac{t^{{2\ell}}\bb E[(\mathsf{V}_n-v_+)^{2\ell}|\mathsf{V}_n>0]}{\Gamma(2\ell+1)}\\
    &\leq\sum_{\ell=0}^{\infty}\frac{(1+o(1))n^{-\ell/2}t^{2\ell}}{\Gamma(2\ell+1)}\bl\mca C_3\bl\frac{1}{3c}\br^{\ell}\Gamma\bl\ell+\frac{1}{2}\br+\mca C_4(3c)^{-\frac{\ell}{3}}\Gamma\bl\frac{2\ell}{3}+\frac{1}{3}\br +\mca C_52^{\ell}\Gamma\bl\frac{\ell}{2}+\frac{1}{4}\br\br\\
    &\leq\sum_{\ell=0}^\infty\frac{(1+o(1))n^{-\ell/2}t^{2\ell}}{\Gamma(2\ell+1)}\bl \mca C_3({3c})^{-1}\Gamma\bl\frac{3}{2}\br+\mca C_4(3c)^{-1/3}\Gamma(1)+2\mca C_5 \Gamma\bl\frac{3}{4}\br\br^{\ell}\Gamma\bl\frac{2\ell+1}{2}\br\\
    &\leq (1-2t^2n^{1/2}/\sigma^2)^{-\frac{1}{2}},\qquad\sigma:=\bl \mca C_3({3c})^{-1}\Gamma\bl\frac{3}{2}\br+\mca C_4(3c)^{-1/3}\Gamma(1)+2\mca C_5 \Gamma\bl\frac{3}{4}\br\br^{\frac{1}{2}}.
\end{align*}
Then we use the fact that $\bb E[\mathsf{V}_n|\mathsf{V}_n>0]=v_+$ to obtain that (here we use proposition 2.5.2 in \citep{vershynin2018high})
\begin{align*}
    \bb E[\exp(t(\mathsf{V}_n-v_+))|\mathsf{V}_n>0]\leq\exp\lef( 18e^2n^{-1/2}\sigma^2t^2\rig).
\end{align*}
Similarly one obtains that
$    \bb E[\exp(t(\mathsf{V}_n-v_-))|\mathsf{V}_n<0]\leq\exp(18e^2n^{-1/2}\sigma^2t^2)$.
And hence
\begin{align*}
    \bb E[\exp(t\mathsf{V}_n)]\leq\frac{1}{2}\lef(\exp(tv_+)+\exp(-tv_+)\rig)\exp(18e^2n^{-1/2}\sigma^2t^2)\leq\exp\lef(\frac{1}{2}t^2v_{+}^2\rig).
\end{align*}

\subsection{Proof for Lemma~\ref{sa-lem:fixed-temp-be} High Temperature}

We will leverage the representation of $\bW$ as a mixture of independent Bernouli random variables after conditioning on some latent variable $\mathsf{U}_n$. We take $\mathsf{U}_n$ to be a random variable with density
\begin{align}\label{sa-eq:latent}
    f_{\mathsf{U}_n}(u) = \frac{\exp \left(- \frac{1}{2} u^2 + n \log \cosh \left( \sqrt{\frac{\beta}{n}} u + h\right) \right)}{\int_{-\infty}^{\infty} \exp \left(- \frac{1}{2} v^2 + n \log \cosh \left( \sqrt{\frac{\beta}{n}} v + h\right) \right) d v} 
\end{align}
Using Gaussian integral identity $\exp(v^2/2) = \frac{1}{\sqrt{2 \pi}} \int_{-\infty}^{\infty} \exp \left( - u^2/2 + uv\right) du$, 
\begin{align}\label{eq: mixture of indep}
    \P \left(\bW = \bw \right) & = \int_{-\infty}^{\infty} \frac{\exp \left( \left(\sqrt{\frac{\beta}{n}} u + h \right)\left(\sum_{i = 1}^n w_i\right) \right)}{2^n \exp \left(n \log \cosh \left( \sqrt{\frac{\beta}{n}} u + h\right) \right)} f_{\mathsf{U}_n}(u) d u.
\end{align}
Hence condition on $\mathsf{U}_n$, $\bW_i$ are i.i.d Bernouli with $\P(W_i = 1|\mathsf{U}_n) = \frac{1}{2} (\tanh (\sqrt{\frac{\beta}{n}} \mathsf{U}_n + h) + 1)$, and
\begin{align}\label{eq: conditional mean and var on Un}
    e(\mathsf{U}_n)  & = \E \left[X_i (W_i - \pi) | \mathsf{U}_n\right] \\ 
    & =  \E \left[X_i\right] \bigg(\tanh \bigg( \sqrt{\frac{\beta}{n}} \mathsf{U}_n + h\bigg) - \pi \bigg), 
\end{align}
\begin{align*}
    v(\mathsf{U}_n) &  = \V \left[X_i (W_i - \pi) | \mathsf{U}_n \right] \\
    & =  \E \left [X_i^2\right]\left\{\frac{(1 - \pi)^2}{2} (\tanh (\sqrt{\frac{\beta}{n}} \mathsf{U}_n + h ) + 1) + \frac{(1 + \pi)^2}{2} (1 - \tanh (\sqrt{\frac{\beta}{n}} \mathsf{U}_n + h ) )\right\} \\
    & \phantom{ v(\mathsf{U}_n)  =} - \E \left[X_i\right]^2 \left(\tanh \left( \sqrt{\frac{\beta}{n}} \mathsf{U}_n + h\right) - \pi \right)^2 \\
    & \geq  \V[X_i] \min \left\{ (1  - \pi)^2, (1 + \pi)^2\right\} =: C_2 \V[X_i], 
\end{align*}
Moreover,
\begin{align*}
    & \E \left[\left|X_i^3 (W_i - \pi)^3\right| |\mathsf{U}_n \right] \leq \E \left[|X_i|^3 \right] \max \left\{(1 - \pi)^3, (1 + \pi)^3 \right\}  =: C_3 \E \left[ |X_i|^3\right].
\end{align*}

\paragraph*{Step 1: Conditional Berry-Esseen} Apply Berry-Esseen Theorem conditional on $\mathsf{U}_n$, 
\begin{align*}
    \sup_{u \in \reals}\sup_{t \in \reals} \left|\P \left(G_n \leq t | \mathsf{U}_n = u \right) -  \Phi \left(\frac{t - \sqrt{n} \E \left[X_i (W_i - \pi) | \mathsf{U}_n = u\right]}{\V \left[X_i (W_i - \pi) | \mathsf{U}_n = u\right]^{1/2}} \right)\right| \leq 3 \frac{C_3 \E \left[|X_i|^3\right]}{C_2^{\frac{3}{2}}\V\left[X_i\right]^{\frac{3}{2}}} n^{-1/2}.
\end{align*}
Take $Z \sim N(0,1)$ independent to $\bW$ and $X_i$'s. $\mathsf{U}_n$ is sub-Gaussian by Equation~\ref{eq: subG high-temp}, hence
 \begin{align*}
    \dKS\left(G_n, v(\mathsf{U}_n)^{1/2}Z + \sqrt{n}e(\mathsf{U}_n)\right) & = \sup_{t \in \reals} \left|\int_{-\infty}^{\infty} ( \P \left(G_n \leq t \middle| \mathsf{U}_n = u \right) -  \Phi\left(\frac{t - \sqrt{n}e(\mathsf{U}_n)}{v(\mathsf{U}_n)^{1/2}}\right)) f_{\mathsf{U}_n}(u) d u\right| \\
    & \leq 3 \frac{C_3 \E \left[|X_i|^3\right]}{C_2^{\frac{3}{2}}\E\left[X_i^2\right]^{\frac{3}{2}}} n^{-1/2}.
 \end{align*}
 
 \paragraph*{Step 2: Stabilization of Variance} By independence between $\mathsf{U}_n$ and $Z$, we have
 \begin{align*}
     & \dKS\left(v(\mathsf{U}_n)^{1/2}Z + \sqrt{n}e(\mathsf{U}_n)), \E[v(\mathsf{U}_n)]^{1/2}Z + \sqrt{n}e(\mathsf{U}_n)\right) \\
     = & \sup_{t \in \reals}\E \left[\Phi \left(\frac{t - \sqrt{n}e(\mathsf{U}_n)}{v(\mathsf{U}_n)^{1/2}}\right) - \Phi \left(\frac{t - \sqrt{n}e(\mathsf{U}_n)}{\E[v(\mathsf{U}_n)]^{1/2}}\right) \right]\\
     \leq & \sup_{t \in \reals} \E \left[ \left|\phi \left(\frac{t - \sqrt{n}e(\mathsf{U}_n)}{v^{\ast}(\mathsf{U}_n)^{1/2}} \right)(t - \sqrt{n}e(\mathsf{U}_n))\left(v(\mathsf{U}_n)^{-1/2} - \E[v(\mathsf{U}_n)]^{-1/2} \right)\right|\right],
 \end{align*}
 where $v^{\ast}(\mathsf{U}_n)$ is some quantity between $\E[v(\mathsf{U}_n)]$ and $v(\mathsf{U}_n)$, and by Equation~\ref{eq: conditional mean and var on Un}, $v^{\ast}(\mathsf{U}_n) \geq C_2 \V[X_i]$. It follows from  boundedness of $v(\mathsf{U}_n)$ and Lipshitzness of $\tanh$ in the expression of $v(\mathsf{U}_n)$  that
 \begin{align*}
    & \dKS\left(v(\mathsf{U}_n)^{1/2}Z + \sqrt{n}e(\mathsf{U}_n)), \E[v(\mathsf{U}_n)]^{1/2}Z + \sqrt{n}e(\mathsf{U}_n)\right) \\
    \leq & \sup_{t \in \reals} \sup_{u \in \reals}\left|\phi \left(\frac{t - \sqrt{n}e(u)}{\sqrt{\E[X_i^2]2(\pi^2 + 1)}} \right)(t - \sqrt{n}e(u))\right|\frac{1}{2 \sqrt{C_2 \V[X_i]}}\E \left[\left|v(\mathsf{U}_n) - \E[v(\mathsf{U}_n)] \right|\right] = O(n^{-1/2}).
 \end{align*}

 \paragraph*{Step 3: Reduction Through TV-distance Inequality}
 \begin{align*}
     & \dKS\left(\E[v(\mathsf{U}_n)]^{1/2}Z + \sqrt{n}e(\mathsf{U}_n), \E[v(\mathsf{U}_n)]^{1/2}Z + \sqrt{n}e(b_n + \mathsf{U})\right) \\
     \leq & \dTV\left(\E[v(\mathsf{U}_n)]^{1/2}Z + \sqrt{n}e(\mathsf{U}_n), \E[v(\mathsf{U}_n)]^{1/2}Z + \sqrt{n}e(b_n + \mathsf{U})\right) \\
     \stackrel{(2)}{\leq} & \dTV \left(e(\mathsf{U}_n), e(\mathsf{U} + b_n) \right) \stackrel{(3)}{\leq} \dTV \left(\mathsf{U}_n, \mathsf{U} + b_n \right),
 \end{align*}
 where $b_n = \sqrt{n}v_0$. The first inequality is by relation between KS- and TV-distances. For the second inequality, denote $X = \sqrt{n}e(\mathsf{U}_n)$, $Y = \sqrt{n}e(b_n + \mathsf{U})$. Denote by $f_X, f_Y, f_Z$ the Lebesgue density of $X, Y, Z$ respectively. Then using $Z \indep X$ and $Z \indep Y$, by data processing inequality,
 \begin{align*}
    \dTV \left(Z + X, Z + Y \right) \leq \dTV(X, Y).
 \end{align*}
 Above proves inequality (2). Inequality (3) is by scale-invariance of TV distance and data processing inequality. 
 
 \paragraph*{Step 4: Gaussian Approximation for $\mathsf{U}_n$}
Consider $\mathsf{V}_n = n^{-1/2} \mathsf{U}_n$. Then
 \begin{align*}
    f_{\mathsf{V}_n}(v) \propto \exp \left(- \frac{1}{2} n v^2 + n \log \cosh \left(\sqrt{\beta}v + h \right) \right) =: \exp \left( - n \phi(v) \right),
 \end{align*}
 where $\phi(v) = - \frac{1}{2} v^2 + \log \cosh (\sqrt{\beta} v + h)$. $\phi$ is maximized at $v_0$ that solves
 \begin{align}\label{eq: v_0}
    v_0 = \sqrt{\beta}\tanh \left(\sqrt{\beta} v_0 + h \right).
 \end{align}
We will approximate the integral of $f_{\mathsf{V}_n}$ by Laplace method. We will introduce constants $c_0, c_1$ and $c_2$ that only depends on $\beta$ and $h$. By Equation (5.1.21) in \cite{bleistein1975asymptotic},
\begin{align*}
    \int_{-\infty}^{\infty} \exp \left(-n \phi(v)\right) d v & = \sqrt{\frac{2 \pi}{n \phi^{\prime \prime}(v_0)}} \exp \left(- n \phi(v_0) \right) + O \left(\frac{\exp(-n \phi(v_0))}{n^{3/2}} \right) \\
    & = \sqrt{\frac{2 \pi}{n \phi^{\prime \prime}(v_0)}} \exp \left(- n \phi(v_0) \right) \left[1 + O(n^{-1})\right],
\end{align*}
where the $O(n^{-1})$ term only depends on $n$ and $\phi$. It follows that
\begin{align*}
    f_{\mathsf{V}_n}(v) = \sqrt{\frac{n \phi^{\prime\prime}(v_0)}{2 \pi}}\exp \left(- n \phi(v) + n \phi(v_0) \right)\left[1 + O(n^{-1}) \right].
\end{align*}
Then by a change of variable and the fact that $O(n^{-1})$ term does not depend on $v$,
\begin{align}\label{eq: density of Un}
    f_{\mathsf{U}_n}(u) = \sqrt{\frac{\phi^{\prime \prime}(v_0)}{2 \pi}} \exp \left( - n \phi(n^{-1/2}u) + n \phi(n^{-1/2}u_0) \right)[1 + O(n^{-1})].
\end{align}
Taylor expanding $\phi$ at $v_0 = n^{-1/2}u_0$ and using $\phi^{\prime}(v_0) = 0$, we get
\begin{align}\label{eq: aprx Un}
    -n \phi(n^{-1/2}u) + n \phi(n^{-1/2}u_0) & = - \frac{\phi^{\prime \prime}(v_0)}{2} (u - u_0)^2 -  \tanh(\sqrt{\beta}v_{\ast} + h) \sech^2(\sqrt{\beta}v_{\ast} + h)\frac{(u - u_0)^3}{3 \sqrt{n}}\\
    & = - \frac{1}{2} \left(1 - \beta + v_0^2 \right)(u-u_0)^2 -  \tanh(\sqrt{\beta}v_{\ast} + h) \sech^2(\sqrt{\beta}v_{\ast} + h)\frac{(u - u_0)^3}{3 \sqrt{n}},
\end{align}
where $v_{\ast}$ is some quantity between $v_0$ and $n^{-1/2}u$. Now take $b_n = u_0 = \sqrt{n}v_0$ and take $\mathsf{U} \sim N(0, (1 - \beta + v_0^2)^{-1})$, we have
\begin{align*}
    \dTV(\mathsf{U}_n, b_n + \mathsf{U}) = &  \int_{-\infty}^{\infty}\left|f_{\mathsf{U}_n}(u) - f_{b_n + U}(u) \right|d u \\
    \leq & \int_{-\infty}^{\infty} \sqrt{\frac{\phi^{\prime \prime}(v_0)}{2 \pi}} \exp \left( -\frac{1}{2}(1 - \beta + v_0^2)(u - u_0)^2 \right)\\
    & \cdot \left[\exp \left(-  \tanh(\sqrt{\beta}v_{\ast}(u) + h) \sech^2(\sqrt{\beta}v_{\ast}(u) + h)\frac{(u - u_0)^3}{3 \sqrt{n}}\right) - 1 \right]d u \left[1 + O(n^{-1}) \right],
\end{align*}
where $v^{\ast}(u)$ is some random quantity between $v_0 = n^{-1/2}u_0$ and $n^{-1/2} u$. We will show that we can restrict the analysis to the region $[u_0 - c_0\sqrt{\log n}, u_0 + c_0 \sqrt{\log n}]$, which is where the bulk of mass lies. Since $\mathsf{U} \sim N(u_0, (1 - \beta + v_0^2)^{-1})$, for some constant $c$ only depending on $\beta$ and $h$, $\P \left(\left|b_n + \mathsf{U} - u_0 \right| \geq c \sqrt{\log n} \right) \leq n^{-1}$. Using a change of variable and concavity of $\phi$,
\begin{align*}
    & \P \left( \left|\mathsf{U}_n - u_0\right| \geq c \sqrt{\log n} \right) = \P \left(\left| \mathsf{V}_n - v_0\right| \geq c \sqrt{\frac{\log n}{n}} \right) \\
    = & \int_{\reals \setminus [v_0 - \sqrt{\frac{\log n}{n}}, v_0 + \sqrt{\frac{\log n}{n}}]} \sqrt{\frac{n \phi^{\prime \prime}(v_0)}{2 \pi}} \exp \left(- n (\phi(v) - \phi(v_0))\right) \left[1 + O(n^{-1}) \right] d v \\
    \leq & \int_{\reals \setminus [v_0 - \sqrt{\frac{\log n}{n}}, v_0 + \sqrt{\frac{\log n}{n}}]} \sqrt{\frac{n \phi^{\prime \prime}(v_0)}{2 \pi}} \exp \left(- n c_1 (v - v_0)^2)\right) \left[1 + O(n^{-1}) \right] d v \\
    \leq & \int_{\reals \setminus [-\sqrt{\log n}, \sqrt{\log n}]}  \sqrt{\frac{\phi^{\prime \prime}(v_0)}{2 \pi}} \exp \left(- c_1 s^2)\right) \left[1 + O(n^{-1}) \right] d s = O(n^{-1}).
\end{align*}
In the third line we used the fact that $\phi(v_0 + t) - \phi(v_0) = \int_{0}^t \phi^{\prime}(v_0 + s) d s$ and the first derivative is bounded by
\begin{align}\label{eq: concave phi}
    |\phi^{\prime}(v_0 +s)| & = |v_0 + s - \sqrt{\beta} \tanh\left(\sqrt{\beta} v_0 + \sqrt{\beta} s + h \right)| \\
    & = |s +  \sqrt{\beta}\tanh\left(\sqrt{\beta} v_0 + h\right) - \sqrt{\beta}\tanh \left( \sqrt{\beta} v_0 + \sqrt{\beta} s + h\right)| \\
    & \geq |s| \left(1 - \sech^2(w_0) \right),
\end{align}
where $w_0$ is the solution to $\tanh(\sqrt{\beta} v_0 + h) - \tanh(w_0) = (\sqrt{\beta}v_0 + h - w_0) \sech^2(w_0)$. It follows that $\phi(v) - \phi(v_0) \leq - \frac{1}{2} (1 - \sech(w_0)^2)(v - v_0)^2$. Using boundedness of $\tanh$ and $\sech$ and the Lipschitzness of $\exp$ when restricted to $[-1,1]$, we have
\begin{align*}
    & \dTV(\mathsf{U}_n, b_n + \mathsf{U}) \\
    \leq & \int_{u_0 - c \sqrt{\log n}}^{u_0 + c\sqrt{\log n}} \sqrt{\frac{\phi^{\prime \prime}(v_0)}{2 \pi}} \exp \left( -\frac{1}{2}(1 - \beta + v_0^2)(u - u_0)^2 \right)\\
    & \cdot \left[\exp \left(-  \tanh(\sqrt{\beta}v_{\ast}(u) + h) \sech^2(\sqrt{\beta}v_{\ast}(u) + h)\frac{(u - u_0)^3}{3 \sqrt{n}}\right) - 1 \right]d u \left[1 + O(n^{-1}) \right] + O(n^{-1}) \\
    \leq & \int_{u_0 - c \sqrt{\log n}}^{u_0 + c\sqrt{\log n}} \sqrt{\frac{\phi^{\prime \prime}(v_0)}{2 \pi}} \exp \left( -\frac{1}{2}(1 - \beta + v_0^2)(u - u_0)^2 \right) c_2 \frac{|u - u_0|^3}{\sqrt{n}}d u \left[1 + O(n^{-1}) \right] + O(n^{-1}) \\
    = & O(n^{-1/2}).
\end{align*}
\paragraph*{Step 5: Gaussian Approximation for $\sqrt{n}e(b_n + \mathsf{U})$} In this step, we will show that $\sqrt{n}e(b_n + \mathsf{U})$ can be well-approximated by $\sqrt{\beta} \sech^2(\sqrt{\beta} v_0 + h)\mathsf{U}$ and hence $\frac{1}{\sqrt{n}} \sum_{i = 1}^n X_i(W_i - \pi)$ can be well-approximated by a Gaussian.
\begin{align*}
    & \dKS \left(\E[v(\mathsf{U}_n)]^{1/2}Z + \sqrt{n}e(b_n + \mathsf{U}), \E[v(\mathsf{U}_n)]^{1/2}Z + \sqrt{\beta} \sech^2(\sqrt{\beta} v_0 + h)\mathsf{U}\right) \\
    \leq & \sup_{t \in \reals} \E \left[\Phi\left( \frac{t - \sqrt{n}e(b_n + \mathsf{U})}{\E[v(\mathsf{U}_n)]^{1/2}} \right) - \Phi\left( \frac{t - \sqrt{\beta}\sech^2(\sqrt{\beta}v_0 + h)U}{\E[v(\mathsf{U}_n)]^{1/2}} \right)\right] \\
    \leq & \frac{\infnorm{\phi}}{\E[v(\mathsf{U}_n)^{1/2}]} \E \left[\left|\sqrt{n}e(b_n + \mathsf{U}) - \sqrt{\beta} \sech^2(\sqrt{\beta}v_0 + h)\mathsf{U} \right|\right]
\end{align*}
Since $\dKS(\mathsf{U}_n, \mathsf{U}) = O(n^{-1/2})$ and $\pi = \E \left[\tanh \left( \sqrt{\frac{\beta}{n}}\mathsf{U}_n + h \right)\right]$, Taylor expanding $\tanh$ at $\sqrt{\beta}v_0 + h$, 
\begin{align*}
    \sqrt{n}e(b_n + \mathsf{U}) & =\E[X_i] \sqrt{n} \left[ \tanh\left(\sqrt{\frac{\beta}{n}}(\sqrt{n} v_0 + \mathsf{U}) + h\right) - \E \left[\tanh\left(\sqrt{\frac{\beta}{n}}(\sqrt{n} v_0 + \mathsf{U}) + h\right)\right]\right] \\
    & = \E[X_i]\sqrt{\beta}\sech^2(\sqrt{\beta}v_0 + h)\mathsf{U} + O \left(\frac{\beta}{\sqrt{n}}\mathsf{U}^2 \right) + O(n^{-1/2}) \\
    & = \E[X_i] \sqrt{\beta}(1 - \frac{v_0^2}{\beta})\mathsf{U} +  O \left(\frac{\beta}{\sqrt{n}}\mathsf{U}^2 \right) + O(n^{-1/2}),
\end{align*}
 It follows that $\E \left[\left|\sqrt{n}e(b_n + \mathsf{U}) - \sqrt{\beta} (1 - \frac{v_0^2}{\beta})\mathsf{U} \right|\right] = O(n^{-1/2})$ and hence
\begin{align*}
    \dKS \left(\E[v(\mathsf{U}_n)]^{1/2}Z + \sqrt{n}e(b_n + \mathsf{U}), \E[v(\mathsf{U}_n)]^{1/2}Z +  \E[X_i] \sqrt{\beta}(1 - \frac{v_0^2}{\beta})\mathsf{U}\right) = O(n^{-1/2}).
\end{align*}
Recall $\mathsf{U} \sim N(0, (1 - \beta + v_0^2)^{-1})$, hence $\E[X_i] \sqrt{\beta}(1 - \frac{v_0^2}{\beta})\mathsf{U} \sim N(0, \E[X_i]^2\frac{(\beta - v_0)^2}{\beta (1 - \beta + v_0^2)})$. Moreover,
\begin{align*}
    \E[v(\mathsf{U}_n)] & = \E[\E[X_i^2]\E[(W_i -\pi)^2|\mathsf{U}_n]] - \E[\E[X_i]^2 \E[W_i - \pi|\mathsf{U}_n]^2] \\
    & = \E[X_i^2](1 - \pi^2) - \E[X_i]^2(\E[W_i|\mathsf{U}_n]^2 - \pi^2) \\
    & = \E[X_i^2](1 - \pi^2) + O(n^{-1/2}),
\end{align*}
where the last line is because $\E[W_i|\mathsf{U}_n] = \tanh(\sqrt{\beta/n}\mathsf{U}_n)$ and $\mathsf{U}_n$ is sub-Gaussian.
Since $Z \indep \mathsf{U}$,  
\begin{align*}
    \dKS \bigg(\E[v(\mathsf{U}_n)]^{1/2}Z +  \E[X_i] \sqrt{\beta}(1 - \frac{v_0^2}{\beta})U, N(0, \E[X_i^2](1 - \pi^2) + \E[X_i]^2 \frac{(\beta - v_0^2)^2}{\beta (1 - \beta + v_0^2)})\bigg) = O(n^{-1/2}).
\end{align*}
Combining the previous five steps, we get
\begin{align*}
    \dKS \bigg(\frac{1}{\sqrt{n}}\sum_{i = 1}^n X_i (W_i -\pi), N\left(0, \E[X_i^2](1 - \pi^2) + \E[X_i]^2 \frac{(\beta - v_0^2)^2}{\beta (1 - \beta + v_0^2)}\right) \bigg) = O(n^{-1/2}).
\end{align*}

\subsection{Proof for Lemma~\ref{sa-lem:fixed-temp-be} Critical Temperature}

Throughout the proof, we denote by $\mathtt{C}$ an absolute constant, and $\mathtt{K}$ a constant that only depends on the distribution of $X_i$. The proofs for the critical temperature case will have a similar structure as the proof for the high temperature case, based the same $\mathsf{U}_n$ defined in Equation~\eqref{sa-eq:latent}. 

\smallskip

\noindent \textbf{Step 1: Conditional Berry-Esseen.}

\smallskip

The same argument as in the high-temperature case gives
\begin{align*}
& \dKS\left(\mca g_n, v(\mathsf{U}_n)^{1/2}\mathsf{Z} + \sqrt{n}e(\mathsf{U}_n)\right) \leq \mathtt{K} n^{-1/2}.
\end{align*}

\noindent \textbf{Step 2: Approximation for $\mathsf{U}_n$.}

\smallskip

Take $\mathsf{W}$ to be a random variable with density function
\begin{align*}  
    f_{\mathsf{W}}(z) = \frac{\sqrt{2}}{3^{1/4}\Gamma(\frac{1}{4})} \exp \left(-\frac{1}{12}z^4\right), \quad z \in \reals,
\end{align*}
independent to $\mathsf{Z}$. Take $\mathsf{W}_n = n^{-1/4}\mathsf{U}_n$ and $\mathsf{V}_n = n^{-1/2}\mathsf{U}_n$. Again $f_{\mathsf{V}_n}(v) \propto \exp(-n \phi(v))$, 
where $\phi(v) := - \frac{1}{2} v^2 + \log \cosh (v)$. In particular, $\phi^{(v)}(0) = 0$ for all $0 \leq v \leq 3$, and $\phi^{(4)}(0) = -2 < 0$, $\phi^{(5)}(0) = 0$, $\phi^{(6)}(0) = 16 > 0$. Example 5.2.1 in \cite{bleistein1975asymptotic} leads to 
$$f_{\mathsf{V}_n}(v) = n^{\frac{1}{4}}\frac{\sqrt{2}}{3^{\frac{1}{4}}\Gamma(\frac{1}{4})}\exp(n \phi(v) - n \phi(0))(1 + o(1)),$$
which implies $f_{\mathsf{W}_n}(w) = f_{W}(w)(1 + o(1))$. Results in \cite{bleistein1975asymptotic} do not give a rate, however. We will use a more cumbersome approach to obtain a slightly sub-optimal rate.
 
By a change of variable, $f_{\mathsf{W}_n}(w) = \frac{h_n(w)}{\int_{-\infty}^{\infty}h_n(u)du}$, where $h_n$ can be written as
\begin{align*}
    h_n(w) = \exp \left( - \frac{\sqrt{n}}{2} w^2 + n \log \cosh \left(n^{-\frac{1}{4}} w \right) \right) = \exp \left(- \frac{1}{12} w^4 + g(w) n^{-\frac{1}{2}}w^6 \right).
\end{align*}
The last equality follows from Taylor expanding the term in $\exp(\cdot)$ at $w = 0$, and $g$ is some bounded function. 
\begin{align*}
    \int_{-10\sqrt{\log n}}^{10\sqrt{\log n}} h_n(w) d w & = I_n (1 + O((\log n)^3 n^{-\frac{1}{2}})), \quad I_n := \int_{-10\sqrt{\log n}}^{10\sqrt{\log n}} \exp\bigg(- \frac{1}{12}w^4\bigg) d w
\end{align*}
Moreover, $\int_{[-10\sqrt{\log n}, 10\sqrt{\log n}]^c} h_n(w) d w  = O(n^{-1/2}) = I_n [1 + O(n^{-\frac{1}{2}})]$. Hence for denominator, we have $\int_{-\infty}^{\infty} h_n(w)dw =I_n[1 + O((\log n)^3 n^{-\frac{1}{2}})]$. It follows that
\begin{align*}
   &  
    \dTV(\mathsf{W}_n, \mathsf{W}) \\
    \lesssim &  \int_{-10\sqrt{\log n}}^{10\sqrt{\log n}} I_n^{-1} \exp \bigg(- \frac{1}{12} w^4 \bigg) n^{-\frac{1}{2}}w^6 d w + \int_{-10\sqrt{\log n}}^{10\sqrt{\log n}}I_n^{-1} O((\log n)^3 n^{-\frac{1}{2}})  d w \\
    & + P(|\mathsf{W}_n| \geq 10\sqrt{\log n}) + \P(|\mathsf{W}| \geq 10\sqrt{\log n}) \\
    = & O((\log n)^3 n^{-\frac{1}{2}}).
\end{align*}

\noindent \textbf{Step 3: Data Processing Inequality.}

\smallskip

We can use data processing inequality to get
 \begin{align*}
    & \dKS\left(v(\mathsf{U}_n)^{1/2}\mathsf{Z} + \sqrt{n}e(\mathsf{U}_n),v(n^{1/4}\mathsf{W})^{1/2}\mathsf{Z} + \sqrt{n}e(n^{1/4}\mathsf{W})\right)
    \leq \dTV\left(\mathsf{W}_n, \mathsf{W}\right) = O(n^{-1/2}).
\end{align*}

\paragraph*{Step 4: Non-Gaussian Approximation for $n^{\frac{1}{4}}e(n^{\frac{1}{4}}\mathsf{W})$}

\begin{align*}
    n^{1/4}e(n^{1/4}\mathsf{W})) & =\E[X_i] n^{\frac{1}{4}}\tanh\left(n^{-\frac{1}{4}}\mathsf{W}\right) 
    = \E[X_i] \left[\mathsf{W} -  O \left(\frac{\mathsf{W}^2}{3\sqrt{n}} \right)\right],
\end{align*}
where we have use the fact that $\tanh^{(2)}(0) = 0$.
Hence there exists $C >0$ such that for $n$ large enough, for any $t > 0$,
\begin{align*}
    \P\left(\E[X_i] \left[\mathsf{W} + C \frac{\mathsf{W}^2}{\sqrt{n}} \right] \leq t\right)
    \leq \P \left(n^{1/4}e(n^{1/4}\mathsf{W})) \leq t\right) \leq \P\left(\E[X_i] \left[\mathsf{W} - C \frac{\mathsf{W}^2}{\sqrt{n}} \right] \leq t\right). \tag{0}
\end{align*}
We have showed that there exists $c > 0$ such that 
\begin{align*}
    \P(|\mathsf{W}| \geq c \sqrt{\log n}) \leq n^{-1/2}, \tag{1}
\end{align*}
in which case $\mathsf{W}^2/\sqrt{n} \leq 1$ for large enough $n$. Hence for large enough $n$ if $t/\E[X_i] > c \sqrt{\log n} + 1$, then
\begin{align*}
    \P \left(\mathsf{W} + C \frac{\mathsf{W}^2}{\sqrt{n}} \leq \frac{t}{\E[X_i]}, |\mathsf{W}| \leq c \sqrt{\log n} \right) - \P \left( \mathsf{W} \leq \frac{t}{\E[X_i]}, |\mathsf{W}| \leq c \sqrt{\log n}\right) = 0. \tag{2}
\end{align*}
If $0 < t/\E[X_i] < c\sqrt{\log n} + 1$, then
\begin{align*}  
    & \left|\P \left(\mathsf{W} + \frac{\mathsf{W}^2}{\sqrt{n}} \leq \frac{t}{\E[X_i]}, |\mathsf{W}| \leq c \sqrt{\log n} \right) - \P \left( \mathsf{W} \leq \frac{t}{\E[X_i]}, |\mathsf{W}| \leq c \sqrt{\log n}\right)\right| \\
    \leq & \P \left( \frac{t}{\E[X_i]} \leq \mathsf{W} \leq \frac{1 - \sqrt{1 - 4 n^{-1/2} t/\E[X_i]}}{2 n^{-1/2}} , |\mathsf{W}| \leq c \sqrt{\log n}\right). 
\end{align*}
Now we study $g(x; \alpha) = (1 - \sqrt{1 - 4 x \alpha})/(2x), x >0$. Then $\sup_{\alpha \leq \frac{1}{4}} \sup_{0 \leq x \leq \frac{1}{2}}|\theta^{\prime}(x; \alpha)| \leq 2$ and $g(0;\alpha) = \alpha$. Since for large enough $n$, $0 < t/\E[X_i] < c\sqrt{\log n} + 1 \leq \frac{1}{4}$ and $0 \leq n^{-1/2} \leq \frac{1}{2}$, we have $\frac{1 - \sqrt{1 - 4 n^{-1/2} t/\E[X_i]}}{2 n^{-1/2}} \leq t/\E[X_i] + 2 n^{-1/2}$. Hence if $0 < t/\E[X_i] < c\sqrt{\log n} + 1$, 
\begin{align*}
    & \left|\P \left(\mathsf{W} + \frac{\mathsf{W}^2}{\sqrt{n}} \leq \frac{t}{\E[X_i]}, |\mathsf{W}| \leq c \sqrt{\log n} \right) - \P \left( \mathsf{W} \leq \frac{t}{\E[X_i]}, |\mathsf{W}| \leq c \sqrt{\log n}\right)\right| = O(n^{-1/2}). \tag{3}
\end{align*}
Combining (1), (2), (3), 
\begin{align*}
    \sup_{t > 0}\left|\P \left(\mathsf{W} + \frac{\mathsf{W}^2}{\sqrt{n}} \leq \frac{t}{\E[X_i]} \right) - \P \left( \mathsf{W} \leq \frac{t}{\E[X_i]}\right)\right| = O(n^{-1/2}).
\end{align*}
By similar argument, we can show 
\begin{align*}
    \sup_{t > 0}\left|\P \left(\mathsf{W}  - \frac{\mathsf{W}^2}{\sqrt{n}} \leq \frac{t}{\E[X_i]} \right) - \P \left( \mathsf{W} \leq \frac{t}{\E[X_i]}\right)\right| = O(n^{-1/2}).
\end{align*}
Noticing that $W$ and $-W$ have the same distribution, the above two inequalities also hold for $t \leq 0$. Hence it follows from (0) that
\begin{align*}
    \dKS \left(n^{1/4}e(n^{1/4}\mathsf{W})), \E[X_i] \mathsf{W} \right) = O(n^{-1/2}).
\end{align*}

\paragraph*{Step 5: Vanishing Variance Term.}

\smallskip

Denote by $f_{\mathsf{W} + n^{-1/4}\mathsf{Z}}$ the density of $\mathsf{W} + n^{-1/4}\mathsf{Z}$. Then
\begin{align*}
    f_{\mathsf{W} + n^{-1/4}\mathsf{Z}}(y) & = \int_{-\infty}^{\infty} \frac{\sqrt{2}}{3^{1/4}\Gamma(\frac{1}{4})}\exp \bigg(-\frac{1}{12}(y - x)^4\bigg)\frac{\exp (-\sqrt{n}x^2/2)}{\sqrt{2 \pi n^{-1/2}}}  d x.
\end{align*}
We will use Laplace method to show $f_{\mathsf{W} + n^{-1/4}\mathsf{\mathsf{Z}}}$ is close to $f_{\mathsf{W}}$. However, to get uniformity over $y$, we need to work harder than in the high temperature case. Define $\varphi(x) = x^2/2$ and $g_y(t) =\exp(-(t-y)^4/12)$. Consider
\begin{align*}
    I_{y,+}(\lambda) = \int_{0}^{\infty} g_{y}(t)\exp(-\lambda\varphi(t))d t, \qquad I_{y,-}(\lambda) = \int_{-\infty}^0 g_{y}(t)\exp(-\lambda \varphi(t))d t.
\end{align*}
Following Section 5.1 in \cite{bleistein1975asymptotic}, take $\tau > 0$ such that $\varphi(t) = \tau $, by a change of variable,
\begin{align*}
    I_{y,+}(\lambda) & = \exp(-\lambda \varphi(0)) \int_0^{\infty} \bigg[\frac{g_y(t)}{\varphi^{\prime}(t)}\bigg|_{t = \varphi^{-1}(\tau)}\bigg] \exp(-\lambda \tau) d \tau  = \int_{0}^{\infty} \frac{\exp(-(\sqrt{2 \tau} - y)^4/12)}{\sqrt{2 \tau}} \exp(-\lambda \tau) d \tau.
\end{align*}
To get rate of convergence uniformly in $y$, we follow the proof of Watson's Lemma but consider only up to first order term. Taylor expanding $x \mapsto \exp(-x ^4)/12$ up to first order at $y$, we have
\begin{align*}
    \frac{\exp(-(\sqrt{2 \tau} - y)^4/12)}{\sqrt{2 \tau}} & = \frac{\exp(-y^4/12)}{\sqrt{2 \tau}} + \frac{1}{3} \exp(-y^4/12)y^3 + \frac{h_y(\tau^{\ast})}{2}\sqrt{2 \tau},
\end{align*}
where $\tau^{\ast}$ is some quantity between $0$ and $\sqrt{2\tau}$ and 
\begin{align*}
    h_y(u) = - \exp (-(u - y)^4/12)(u - y)^2 + \frac{1}{9} \exp(-(u - y)^4/12)(u - y)^6.
\end{align*}
In particular, we have $\sup_{y \in \reals} \sup_{u \in \reals}|h_y(u)| < C$ for some absolute constant $C$. Then 
\begin{align*}
    \sup_{y \in \reals}\bigg|\int_{0}^{\infty} \frac{h_y(\tau^{\ast})}{2}\sqrt{2 \tau} \exp(-\lambda \tau) d \tau \bigg| \leq \frac{C}{\sqrt{2}} \Gamma\bigg(\frac{3}{2}\bigg) \lambda^{-3/2}, \quad \forall \lambda > 0.
\end{align*}
Evaluating the first two terms, we get
\begin{align*}
    \sup_{y \in \reals}\bigg|I_{y,+}(\lambda) - \sqrt{\frac{\pi}{2 \lambda}} \exp(-y^4/12) - \int_{0}^{\infty}\frac{1}{3} \exp(-y^4/12)y^3 \exp(-\lambda \tau) d \tau \bigg| \leq \frac{C}{\sqrt{2}} \Gamma \bigg(\frac{3}{2}\bigg)\lambda^{-3/2}, \forall \lambda > 0.
\end{align*}
Similarly, for $I_{y,-}$, change of variable by taking $\tau < 0$ such that $\varphi(t) = \tau$, we have
\begin{align*}
    \sup_{y \in \reals} \bigg|I_{y,-}(\lambda) - \sqrt{\frac{\pi}{2 \lambda}}\exp(-y^4/12) + \int_{0}^{\infty} \frac{1}{3}\exp(-y^4/12)y^3\exp(-\lambda \tau) d \tau \bigg| \leq \frac{C}{\sqrt{2}} \Gamma \bigg(\frac{3}{2}\bigg)\lambda^{-3/2}, \forall \lambda > 0.
\end{align*}
Combining the two parts, we get 
\begin{align*}
    \sup_{y \in \reals}\bigg|\int_{-\infty}^{\infty}g_y(t)\exp(-\lambda \varphi(t)) d t - \sqrt{\frac{2\pi}{\lambda}} \exp(-y^4/12) \bigg| \leq C \sqrt{2} \Gamma \bigg(\frac{3}{2}\bigg)\lambda^{-3/2}, \quad \forall \lambda > 0.
\end{align*}
Now take $\lambda = \sqrt{n}$ and multiply both sides by $\frac{n^{1/4}}{3^{1/4}\Gamma(\frac{1}{4})\sqrt{\pi}}$, we get
\begin{align*}
    \sup_{y \in \reals}\bigg|f_{\mathsf{W} + n^{-1/4}\mathsf{Z}}(y) - \frac{\sqrt{2}}{3^{1/4}\Gamma(\frac{1}{4})}\exp(-y^4/12)\bigg| \leq C \frac{\sqrt{2} \Gamma(\frac{3}{2})}{3^{1/4}\Gamma(\frac{1}{4})\sqrt{\pi}} n^{-1/2}.
\end{align*}
By a truncation argument, we have
\begin{align*}
    \dKS(\mathsf{W} + n^{-1/4}\mathsf{Z}, \mathsf{W}) & \leq \dTV(\mathsf{W} + n^{-1/4}\mathsf{Z}, \mathsf{W}) \\
    & = \int_{-\sqrt{\log n}}^{\sqrt{\log n}}|f_{\mathsf{W}+n^{-1/4}\mathsf{Z}}(y) - f_{\mathsf{W}}(y)|d y + \P(|\mathsf{W} + n^{-1/4}\mathsf{Z}| \geq \sqrt{\log n}) \\
    & \qquad + \P(|\mathsf{W}| \geq \sqrt{\log n}) \\
    & \leq C \sqrt{n^{-1}\log n}.
\end{align*}
Together with the fact that 
\begin{align*}
    n^{-1/4}v(n^{1/4}\mathsf{W}) & = n^{-1/4}(\E[X_i^2] - \E[X_i]^2 \tanh^2(\sqrt{\beta}n^{-1/4}\mathsf{W}))^{1/2} \\
    & = n^{-1/4}\E[X_i^2]^{1/2}(1 + O_{\psi_2}(n^{-1/4})),
\end{align*}
we know 
\begin{align*}
    \dKS(n^{-1/4}v(n^{1/4}\mathsf{W})^{1/2}\mathsf{Z} + n^{1/4} e(n^{1/4}\mathsf{W}),\mathsf{W}) = O(\sqrt{\log n} n^{-1/2}).
\end{align*}
Putting together all previous steps, we have
\begin{align*}
    \dKS(n^{1/4}\mca g_n, \E[X_i] \mathsf{W}) = O((\log n)^3 n^{-1/2}).
\end{align*}

\subsection{Proof for Lemma~\ref{sa-lem:fixed-temp-be} Low Temperature}

Throughout the proof, we denote by $\mathtt{C}$ an absolute constant, and $\mathtt{K}$ a constant that only depends on the distribution of $X_i$. The proofs are based on essentially the same argument as in the high temperature case. 

Instead of using sub-Gaussianity of $\mathsf{U}_n$, here we use $\mathsf{U}_n$ is sub-Gaussian condition on $\mathsf{U}_n \in \ca I_\ell$, $\ell \in\{-,+\}$. In particular, the previous step 2 by:

\smallskip

\noindent \textbf{Step 2: Approximation for $\mathsf{U}_n$.}

\smallskip

In case $\beta > 1$, $\phi(v) = \frac{1}{2}v^2 - \log(\cosh(\sqrt{\beta}v))$ has two global minimum $v_+$ and $v_-$, which are the two solutions of $v -\sqrt{\beta}\tanh(\sqrt{\beta}v) = 0$. We want to show $\phi^{(2)}(v_+) = \phi^{(2)}(v_-) = 1 - \beta + v_+^2 > 0$. It sufffices to show $v_+ > \sqrt{\beta - 1}$. Since $\phi^{\prime}(v) < 0$ for $v \in (0,v_+)$ and $\phi^{\prime}(v) > 0$ for $v \in (v_+,\infty)$, it suffices to show $\phi^{\prime}(\sqrt{\beta - 1}) < 0$. But
\begin{align*}
    \phi^{\prime}(\sqrt{\beta - 1}) < 0 \Leftrightarrow \sqrt{\beta - 1} - \sqrt{\beta}\tanh(\sqrt{\beta(\beta - 1)}) < 0 \Leftrightarrow \beta > 1.
\end{align*}
Hence $\phi^{(2)}(v_+) = \phi^{(2)}(v_-) > 0$. Observe that on $\ca I_- = (-\infty,0)$ and $\ca I_+ = (0,\infty)$ respectively, the absolute minimum of $\phi$ occurs at $v_-$ and $v_+$, and $\phi^{\prime}$ is non-zero on $\ca I_-$ and $\ca I_+$ except at $v_-$ and $v_+$. Hence we can apply Laplace method (Equation 5.1.21 in \cite{bleistein1975asymptotic}) sperarately on $\ca I_-$ and $\ca I_+$ to get
\begin{align*}
    & \int_{-\infty}^0 \exp (-n \phi(v)) d v = \sqrt{\frac{2 \pi}{n \phi^{(2)}(v_-)}}\exp(-n\phi(v_-))(1 + O(n^{-1})), \\ & \int_{0}^{\infty} \exp (-n \phi(v)) d v = \sqrt{\frac{2 \pi}{n \phi^{(2)}(v_+)}}\exp(-n\phi(v_+))(1 + O(n^{-1})).
\end{align*}
It follows from the definition of $f_{\mathsf{V}_n}$ and a change of variable that the density of $\mathsf{U}_n = \sqrt{n} \mathsf{V}_n$ can be approximated by
\begin{align*}
    f_{\mathsf{U}_n}(u) = \sum_{l=+,-}\Indicator(u \in \mathcal{C}_l)\sqrt{\frac{\phi^{(2)}(v_-)}{8 \pi}} \exp(- n \phi(n^{-1/2}u)+n\phi(n^{-1/2}u_l))(1 + O(n^{-1})),
\end{align*}
where $u_l = \sqrt{n} v_l, l \in \{ +,- \}$. Since $\P(\mathsf{U}_n \in \ca I_+) = \P(\mathsf{U}_n \in \ca I_-) = \frac{1}{2}$, condition on $\mathsf{U}_n \in \ca I_+$, 
\begin{align*}
    f_{\mathsf{U}_n|\mathsf{U}_n \in \ca I_+}(u) = \sqrt{\frac{\phi^{(2)}(v_+)}{2 \pi}} \exp(-n\phi(n^{-1/2}u)+n\phi(n^{-1/2}u_+))(1 + O(n^{-1})).
\end{align*}
It then follows from Equation~\ref{eq: aprx Un} that if we define $\mathsf{U}_+$ to be a random variable with density
\begin{align*}
    f_{\mathsf{U}_+}(u) =  \sqrt{\frac{1 - \beta + v_+^2}{2 \pi}} \exp (-(1 - \beta + v_+^2) (u - u_+)^2/2),
\end{align*}
then by Taylor expanding $\phi$ at $v_+ = n^{-1/2}u_+$ and a similar argument as in the proof for high temperature case,
\begin{align*}
    \dTV(\mathsf{U}_n|\mathsf{U}_n \in \ca I_+, \mathsf{U}_+) = O(n^{-1/2}).
\end{align*}
The rest follows from the same argument as in the proof for high temperature case and is sub-Gaussianity of $\mathsf{U}_n$ condition on $\mathsf{U}_n \in \ca I_\ell$, $\ell \in\{-,+\}$.

\subsection{Proof for Remark~\ref{sa-remarK: conditional concentration under low temp}}

Take $\mathsf{V}_n = n^{-1/2}\mathsf{U}_n$ and $\mathsf{Z}$ be a $\mathsf{N}(0,1)$ variable independent to $\mca m$ and $\mathsf{U}_n$. Based on the conditional mean and variance formulas in Equation~\eqref{eq: conditional mean and var on Un}, using the conditional on $\mathsf{U}_n$ Berry-Esseen bound,
\begin{align*}
    \P (\mca m < 0|\mathsf{U}_n \geq 0) & = \P(n^{-\frac{1}{4}} v(\mathsf{U}_n)^{\frac{1}{2}}\mathsf{Z} + n^{\frac{1}{4}}\tanh(\sqrt{\beta} n^{- \frac{1}{2}}\mathsf{U}_n) < 0|\mathsf{U}_n \geq 0) + O(n^{-\frac{1}{2}}).
\end{align*}
Using the fact that $v(\mathsf{U}_n)$ is bounded above and $\mathsf{Z}$ is Gaussian, and Taylor expanding $\tanh$, we get
\begin{align*}
    \P (\mca m < 0|\mathsf{U}_n \geq 0) & \leq \P (n^{\frac{1}{4}}\tanh(\sqrt{\beta} n^{-\frac{1}{2}}\mathsf{U}_n) \leq \sqrt{\log n} n^{-\frac{1}{4}} | \mathsf{U}_n \geq 0) + O(n^{-\frac{1}{2}}) \\
    & \leq \P (\mathsf{U}_n \leq \sqrt{\log n} | \mathsf{U}_n \geq 0) + O(n^{-\frac{1}{2}}).
\end{align*}
The proof of Lemma~\ref{sa-lem:fixed-temp-be} (low temperature) shows that $ \dTV(\mathsf{U}_n|\mathsf{U}_n \in \ca I_+, \mathsf{U}_+) = O(n^{-1/2})$ where $\mathsf{U}_+ \sim \mathsf{N}(\sqrt{n}\pi_+, (1 - \beta(1 - \pi_+^2))^{-1})$. Hence $\P (\mathsf{U}_n \leq \sqrt{\log n} | \mathsf{U}_n \geq 0) \lesssim \exp(-n)$. It follows that
\begin{align*}  
    \P (\mca m < 0|\mathsf{U}_n \geq 0) = O(n^{-\frac{1}{2}}).
\end{align*}
By symmetry and the fact that $\P(\sign(\mca m) = \ell) = \P(\sign(\mathsf{U}_n) = \ell) = 1/2$ for $\ell = -, +$, we know 
\begin{align*}
    \P(\{\sign(\mca m) = \ell\} \Delta\{\sign(\mathsf{U}_n) = \ell\}) = O(n^{-1/2}), \qquad \ell = -, +.
\end{align*}
The conclusion then follows from Lemma~\ref{sa-lem:sub-gaussian}(3).

\subsection{Proof for Lemma~\ref{sa-lem: localization to singularity} Drifting from High Temperature}

Throughout the proof, we denote by $\mathtt{C}$ an absolute constant, and $\mathtt{K}$ a constant that only depends on the distribution of $X_i$. 

Let $\mathsf{U}_n(c)$, $e(\mathsf{U}_n(c))$, $v(\mathsf{U}_n(c))$ be the latent variable, conditional mean, and conditional variance as previously defined when $\beta_n = 1 + c n^{-\frac{1}{2}}$, $c < 0$. For notational simplicity, we abbreviate the $c$, and call them $\mathsf{U}_n, e(\mathsf{U}_n), v(\mathsf{U}_n)$ respectively. By Lemma~\ref{sa-lem:sub-gaussian}, $\norm{\mathsf{U}_n}_{\psi_2} \leq \mathtt{C} n^{1/4}$.

\smallskip

\noindent \textbf{Step 1: Conditional Berry-Esseen.}

\smallskip

Apply Berry-Esseen Theorem conditional on $\mathsf{U}_n$ in the same way as in the high temperature case, we get
\begin{align*}
    \dKS\left(\mca g_n, v(\mathsf{U}_n)^{1/2}\mathsf{Z} + \sqrt{n}e(\mathsf{U}_n)\right) \leq \mathtt{K} n^{-1/2}.
\end{align*}

\noindent \textbf{Step 2: Non-Normal Approximation for $n^{-\frac{1}{4}}\mathsf{U}_n$.}

\smallskip

Consider $\mathsf{W}_n = n^{-1/4}\mathsf{U}_n$. Then $f_{\mathsf{W}_n}(w) = I_n(c)^{-1} h_n(w)$, with $I_n(c) = \int_{-\infty}^{\infty}h_n(w)dw$, and
\begin{align*}
    h_n(w) = \exp \left( - \frac{\sqrt{n}}{2} w^2 + n \log \cosh \left(n^{-\frac{1}{4}} \sqrt{\beta_n} w \right) \right) = \exp \left(-\frac{c}{2} w^2 - \frac{\beta_n^2}{12} w^4 + g(w)\beta_n^3 n^{-\frac{1}{2}}w^6 \right),
\end{align*}
where by smoothness of $\log(\cosh(\cdot))$, $\infnorm{\theta} \leq \mathtt{K}$. Then
\begin{align}\label{eq: drift high dem 1}
    \int_{-\mathtt{C}\sqrt{\log n}}^{\mathtt{C}\sqrt{\log n}} h_n(w) d w & = \int_{-\mathtt{C}\sqrt{\log n}}^{\mathtt{C}\sqrt{\log n}} \exp(-\frac{c}{2}w^2 - \frac{\beta_n^2}{12}w^4) d w [1 + O(\mathtt{C}^6(\log n)^3 n^{-\frac{1}{2}})] \\
    & = I(c) [1 + O(\mathtt{C}^6(\log n)^3 n^{-\frac{1}{2}})].
\end{align}
Moreover, by a change of variable and the fact that $\beta_n \leq 1$,
\begin{align*}
    I_n(c) & := \int_{-\infty}^{\infty} h_n(w) d w = n^{-\frac{1}{4}}\int_{-\infty}^{\infty} \exp \Big(- n \Big(\frac{v^2}{2} - \log \cosh(\sqrt{\beta_n v}) \Big) \Big)d v \\
    & \leq n^{-\frac{1}{4}} \int_{-\infty}^{\infty} \exp \Big(- n \Big(\frac{v^2}{2} - \log \cosh(\sqrt{v}) \Big) \Big)d v \leq \mathtt{C}.
\end{align*}
Since $\norm{\mathsf{W}_n(c)}_{\psi_2} \leq \mathtt{C}$, $I_n(c)^{-1}\int_{(-\mathtt{C}\sqrt{\log n},\mathtt{C} \sqrt{\log n})^c}h_n(w) dw \leq \mathtt{C} n^{-1/2}$. It follows that
\begin{align}\label{eq: drift high dem 2}
    \int_{(-\mathtt{C}\sqrt{\log n}, \mathtt{C} \sqrt{\log n})^c} h_n(w) dw \leq \mathtt{C} n^{-1/2}.
\end{align}
Combining Equation~\ref{eq: drift high dem 1} and \ref{eq: drift high dem 2}, we have $I_n(c) =I(c)[1 + O(\mathtt{C}^6(\log n)^3 n^{-1/2})]$. It follows that
\begin{align*}
    & \dTV(\mathsf{W}_n, \mathsf{W}) \\
    \leq & \int_{-\mathtt{C}\sqrt{\log n}}^{\mathtt{C} \sqrt{\log n}} \Big| \frac{h_n(w)}{I_n(c)} - \frac{h(w)}{I(c)}\Big| d w + \P(|\mathsf{W}_n| \geq \mathtt{C}\sqrt{\log n}) + \P(|\mathsf{W}| \geq \mathtt{C}\sqrt{\log n}) \\
    \leq &  \int_{-\mathtt{C}\sqrt{\log n}}^{\mathtt{C} \sqrt{\log n}} \Big| \frac{h_n(w) - h(w)}{I(c)}\Big| + h_n(w)\Big|\frac{1}{I(c)} - \frac{1}{I_n(c)}\Big| d w + O(n^{-\frac{1}{2}}) \\
    \leq &  \int_{-\mathtt{C}\sqrt{\log n}}^{\mathtt{C}\sqrt{\log n}} \exp \Big( - \frac{c}{2}w^2 - \frac{\beta_n^2}{12} w^4 \Big) \frac{w^6}{\sqrt{n}I(c)} d w  + \int_{-\mathtt{C}\sqrt{\log n}}^{\mathtt{C}\sqrt{\log n}}\frac{1}{I(c)} O(\mathtt{C}^6(\log n)^3 n^{-\frac{1}{2}})  d w +  O(n^{-\frac{1}{2}})\\
    \leq & \mathtt{C} (\log n)^3 n^{-1/2}.
\end{align*}

\noindent \textbf{Step 3: A Reduction through TV-distance Inequality.}

\smallskip

Since $\mathsf{Z} \indep (\mathsf{U}_n, \mathsf{W}_n)$, we can use data processing inequality to get
 \begin{align*}
   \dKS\left(n^{-\frac{1}{4}}v(\mathsf{U}_n)^{\frac{1}{2}}\mathsf{Z} + n^{\frac{1}{4}}e(\mathsf{U}_n), n^{-\frac{1}{4}}v(n^{\frac{1}{4}}\mathsf{W})^{\frac{1}{2}}\mathsf{Z} + n^{\frac{1}{4}}e(n^{\frac{1}{4}}\mathsf{W})\right) & \leq \dTV\left(\mathsf{W}_n, \mathsf{W}\right) \\
   & \leq \mathtt{C} (\log n)^3 n^{-1/2}.
\end{align*}

\noindent \textbf{Step 4: Non-Gaussian Approximation for $n^{\frac{1}{4}}e(n^{\frac{1}{4}}\mathsf{W})$.}

\smallskip

This is essentially the same as the proof for step 4 from the critical temperature case in Lemma~\ref{sa-lem:fixed-temp-be}.
\begin{align*}
    \dKS \Big(n^{1/4}e(n^{1/4}\mathsf{W}), \E[X_i] \mathsf{W}\Big) \leq \mathtt{K} \frac{\log n}{\sqrt{n}}.
\end{align*}

\noindent \textbf{Step 5: Stabilization of Variance.}

\smallskip

Using the same argument as Step 4 in the high temperature case for Lemma~\ref{sa-lem:fixed-temp-be}, and $\norm{\mathsf{W}} \leq \mathtt{K}$, 
\begin{align*}
    \dKS(n^{-\frac{1}{4}}v(n^{\frac{1}{4}}\mathsf{W})^{\frac{1}{2}}\mathsf{Z} + n^{\frac{1}{4}}e(n^{\frac{1}{4}}\mathsf{W}),n^{-\frac{1}{4}}\E[X_i^2]^{\frac{1}{2}} \mathsf{Z} + \E[X_i] \mathsf{W})) \leq \mathtt{K} \frac{\log n}{\sqrt{n}}.
\end{align*}
The conclusion then follows from putting together the previous five steps.

\subsection{Proof for Lemma~\ref{sa-lem: localization to singularity} Drifting from Low Temperature}

Consider the same $\mathsf{U}_n$ defined in Equation~\eqref{sa-eq:latent}. Recall $\phi(v) = \frac{v^2}{2} - \log \cosh(\sqrt{\beta_n}v)$, $\phi^{\prime}(v) = v - \sqrt{\beta_n} \tanh(\sqrt{\beta_n} v)$, $\phi^{(2)}(v) = 1 - \beta_n \sech^2(\sqrt{\beta_n}v)$. And we take $v_{n,+} > 0$, $v_{n,-} < 0$ to be the two solutions of $v - \sqrt{\beta_n}\tanh(\sqrt{\beta_n}v) = 0$.

\smallskip

\noindent \textbf{Step 2': Non-Normal Approximation for $n^{-\frac{1}{4}}\mathsf{U}_n$.}

\smallskip

Take $\mathsf{V}_n = n^{-1/2}\mathsf{U}_n$. Then $f_{\mathsf{V}_n}(v) \propto \exp(-n \phi(v))$. Taylor expanding $\phi^{\prime}$ at $0$, we know there exists some function $g$ that is uniformly bounded such that $\phi^{\prime}(v) = (1 - \beta_n) v + \frac{1}{3}\beta_n^2 v^3 + \beta_n^3 g(v) v^5$. Hence $$v_{n,+} = \sqrt{\frac{3(\beta_n - 1)}{\beta_n^2}} + O(\beta_n - 1) = \sqrt{3 c}n^{-1/4} +  O(n^{-1/2}).$$ Taylor expand $\tanh$ and $\sech$ at $0$, 
\begin{align*}
    \phi^{(2)}(v_{n,+}) & = 1 - \beta_n + v_{n,+}^2 \\
    & = - c n^{-1/2} + 3 c n^{-1/2}(1 + O(c n^{-1/2}))^{-2} + O((c n^{-1/2})^{5/2}) \\
    & = 2 c n^{-1/2}(1 + O(cn^{-1/2})), \\
    \phi^{(3)}(v_{n,+}) & = 2 (\beta_n - v_{n,+}^2) v_{n,+}^2\\
    & = 2 \beta_n^{3/2} \sech^2(\sqrt{\beta_n} v_{n,+}) \tanh(\sqrt{\beta_n} v_{n,+}) \\
    & = 2 (1 + O(c n^{-1/2}))(1 + O(v_{n,+}^2))(\sqrt{\beta_n}v_{n,+} + O(v_{n,+}^3)) \\
    & = 2 \sqrt{3 c} n^{-1/4} (1 + O(c n^{-1/2})), \\
    \phi^{(4)}(v_{n,+}) & = 2 (\beta - v_{n,+}^2) (\beta - 3 v_{n,+}^2)\\
    & = 2 \beta_n^2 \sech^4(\sqrt{\beta_n}v_{n,+}) - 4 \beta_n^2 \sech^2(\sqrt{\beta_n}v_{n,+})\tanh^2(\sqrt{\beta_n}v_{n,+}) \\
    & = 2 (1 + O(c n^{-1/2})).
\end{align*}
Take $\mathsf{W}_n = n^{1/4}\mathsf{V}_n = n^{-1/4}\mathsf{U}_n$, $\mca w_+ = n^{1/4}v_{n,+} = \sqrt{3 c} + O(n^{-1/4})$, and $\mca w_- = n^{1/4} v_{n,-}$. Define 
\begin{align*}
    & h_{c,n}(w) \\
    = & -\frac{\sqrt{n}\phi^{(2)}(v_{n,+})}{2} (w - \mca w_{\sign(w)})^2 - \frac{n^{1/4}\phi^{(3)}(v_{n,+})}{6}  (w - \mca w_{\sign(w)})^3 - \frac{\phi^{(4)}(v_{n,+})}{24}(w - \mca w_{\sign(w)})^4.
\end{align*}
By a change of variable and Taylor expansion, the density for $\mathsf{W}_n$ satisfies
\begin{align}\label{eq: Wn approax}
    & f_{\mathsf{W}_n}(w) 
    \propto g_{c,\gamma}(w) = \exp\Big( h_{c,n}(w) + O(\infnorm{\phi^{(6)}}/6!)\frac{(w-\mca w_{\sign(w)})^6}{\sqrt{n}} \Big).
\end{align}

By Lemma~\ref{sa-lem:sub-gaussian}, for $\ell \in \{-,+\}$, condition on $\mathsf{W}_n \in \ca I_{c,n,\ell}$, $\mathsf{W}_n - \mca w_\ell$ is sub-Gaussian with $\psi_2$-norm bounded by $\mathtt{C}$. Let $\mathsf{W}_{c,n}$ be a random variable with density at $w$ proportional to $\exp(h_{c,n}(w))$. By similar argument as Equations~\ref{eq: drift high dem 1} and \ref{eq: drift high dem 2}, $$\dKS(\mathsf{W}_n| \mathsf{W}_n \in \ca I_{c,n,\ell}, \mathsf{W}_{c,n}| \mathsf{W}_{c,n} \in \ca I_{c,n,\ell}) \leq \mathtt{C} (\log n)^3 n^{-1/2}).$$

The other steps, \textit{conditional Berry-Esseen}, \textit{reduction through TV-distance inequality}, and \textit{non-Gaussian approximation for $n^{\frac{1}{4}}e(n^{\frac{1}{4}}\mathsf{W}_{c,n})$} can be proceeded in the same way as in the proof for Lemma~\ref{sa-lem:fixed-temp-be}, with $\mathsf{W}_n - \mca w_\ell$ sub-Gaussian condition on $\mathsf{W}_n \in \ca I_{c,n,\ell}$ with $\psi_2$-norm bounded by $\mathtt{C}$, and respectively for $\mathsf{W}_{c,n}$.

\subsection{Proof for Lemma~\ref{sa-lem: knife-edge} Knife-Edge Representation}

Again we take $\mathsf{U}_n$ to be the latent variable from Lemma~\ref{sa-lem: definetti}, and $\mathsf{W}_n = n^{-1/4}\mathsf{U}_n$. From Step 2 in the proof of Lemma~\ref{sa-lem: localization to singularity}, $f_{\mathsf{W}_n}(w) = I_n(c)^{-1} h_n(w)$, with $I_n(c) = \int_{-\infty}^{\infty}h_n(w)dw$, and
\begin{align*}
    h_n(w) = \exp ( - \frac{\sqrt{n}}{2} w^2 + n \log \cosh (n^{-\frac{1}{4}} \sqrt{\beta_n} w ) ) = \exp (-\frac{c_n}{2} w^2 - \frac{\beta_n^2}{12} w^4 + g(w)\beta_n^3 n^{-\frac{1}{2}}w^6),
\end{align*}
where by smoothness of $\log(\cosh(\cdot))$, $\infnorm{\theta} \leq \mathtt{K}$. 

\smallskip

\noindent \textbf{Case 1: When $\sqrt{n}(\beta_n - 1) = o(1)$.} We can apply Berry-Esseen conditional on $\mathsf{U}_n$ the same way as in the proof of Lemma~\ref{sa-lem: localization to singularity}, and its Step 2 can also be applied here to show that if we take $\widetilde{\mathsf{W}}_c$ to be a random variable with density proportional to $\exp(-c_n^2/2 w^2 - \beta_n^2/12 w^4)$, then $\dKS(\mathsf{W}_n,\widetilde{\mathsf{W}}_c) = O((\log n)^3 n^{-1/2})$. Moreover, $c_n = o(1)$ and $\beta_n = 1 - o(1)$. Hence $\dKS(\mathsf{W}_n,\mathsf{W}_0) = o(1)$. The rest of the proof then follows from Step 3 to Step 5 in the proof for the critical regime case in Lemma~\ref{sa-lem:fixed-temp-be}.

\smallskip

\noindent \textbf{Case 2: When $\sqrt{n}(1 - \beta_n) \gg 1$.} Again we still have $\norm{\mathsf{U}_n}_{\psi_2} = O(n^{1/4})$. And we take $v_+ > 0$, $v_- < 0$ to be the two solutions of $v - \sqrt{\beta_n}\tanh(\sqrt{\beta_n}v) = 0$. Similarly as in the previous case, the first two steps in the proof of Lemma~\ref{sa-lem: localization to singularity} implies $\dKS(\mathsf{W}_n,\widetilde{\mathsf{W}}_c) = o(1)$, where the density of $\mathsf{W}_c$ is proportional to $\exp(-c_n^2/2w^2 - \beta_n^2/12 w^4)$. Since $c_n \gg 1$, the first term in the exponent dominates, and we can show $\dKS(\mathsf{W}_n, \mathsf{W}_c^{\dag}) = o(1)$, where $\mathsf{W}_c^\dag$ has density proportional to $\exp(-c_n^2/2 w^2)$. Again, we can Taylor expand to get $n^{1/4}e(n^{1/4}\mathsf{W}))
=\E[X_i] n^{\frac{1}{4}}\tanh\left(n^{-\frac{1}{4}}\mathsf{W}\right) = \E[X_i] [\mathsf{W} -  O (\frac{\mathsf{W}^2}{3\sqrt{n}})]$, and show $\dKS(n^{1/4}e(n^{1/4}\mathsf{W}_c^\dag), \E[X_i]\mathsf{W}_c^\dag) = o(1)$. Combining with stablization of variance as in the proof of Lemma~\ref{sa-lem: fixed temp MPLE} (high temperature case), we can show $$\dKS(\mca g_n, n^{-1/4}\E[X_i^2]^{1/2}\mathsf{Z} + \E[X_i]\mathsf{W}_c^\dag) = o(1).$$ Since $\mathsf{Z}$ and $\mathsf{W}_c^\dag$ are independent Gaussian random variables, we also have $\dKS (\mca g_n/\sqrt{\V[\mca g_n]}, \mathsf{Z}) = o(1)$.

\smallskip

\noindent \textbf{Case 3: When $\sqrt{n}(\beta_n - 1) \gg 1$.} By Lemma~\ref{sa-lem: localization to singularity} (2), 
\begin{align}\label{eq:low-knife-edge}
    \sup_{t \in \reals}\bigg|\P_{\beta_n, h} (n^{\frac{1}{4}} \mca g_n \leq t| \mca m \in \ca I_{c,\ell}) - \P ( n^{-\frac{1}{4}}\E[X_i^2]^{\frac{1}{2}}\mathsf{Z} +\beta_n^{\frac{1}{2}}\E[X_i]\mathsf{W}_{c_n,n} \leq t & |\mathsf{W}_{c_n,n} \in \ca I_{c,\ell})\bigg| = o(1),
\end{align}
where $\mathsf{W}_{c,n}$ has density proportional to $\exp(h_{c,n}(w))$, with
\begin{align*}
    & h_{c,n}(w) \\
    = & -\frac{\sqrt{n}\phi^{(2)}(v_+)}{2} (w - \mca w_{\sign(w)})^2 - \frac{n^{1/4}\phi^{(3)}(v_+)}{6}  (w - \mca w_{\sign(w)})^3 - \frac{\phi^{(4)}(v_+)}{24}(w - \mca w_{\sign(w)})^4,
\end{align*}
and $\ca I_{c,n,-} = (-\infty,K_{c,n,-})$ and $\ca I_{c,n,+} = (K_{c,n,+},\infty)$ such that $\E[\mathsf{W}_{c,n}|\mathsf{W}_{c,n} \in \ca I_{c,n,\ell}] = w_{c,n,\ell}$ for $\ell \in \{-,+\}$. Now we calculate the order of the coefficients under $\sqrt{n}(\beta_n - 1) \gg 1$. First, suppose $\beta_n = 1 + c n^{\gamma}$ for some $\gamma \in (0,\infty)$ and $c$ not depending on $n$. Then $v_+ = \sqrt{\frac{3(\beta_n - 1)}{\beta_n^2}} + O(\beta_n - 1) = \sqrt{3 c}n^{-\gamma/2} +  O(n^{-\gamma})$. Taylor expand $\tanh$ and $\sech$ at $0$, 
\begin{align*}
    \phi^{(2)}(v_+) & = 1 - \beta_n + v_+^2 = - c n^{-\gamma} + c n^{-\gamma}3(1 + c n^{-\gamma})^{-2} + O((c n^{-\gamma})^{5/2}) \\
    & = 2 c n^{-\gamma}(1 + O(cn^{-\gamma})), \\
    \phi^{(3)}(v_+) & = 2 \beta_n^{3/2} \sech^2(\sqrt{\beta_n} v_+) \tanh(\sqrt{\beta_n} v_+) \\
    & = 2 (1 + O(c n^{-\gamma}))(1 + O(v_+^2))(\sqrt{\beta_n}v_+ + O(v_+^3)) \\
    & = 2 \sqrt{3 c} n^{-\gamma/2} (1 + O(c n^{-\gamma})), \\
    \phi^{(4)}(v_+) & = - 2 \beta_n^4 \sech^4(\sqrt{\beta_n}v) + 4 \sech^2(\sqrt{\beta_n}v)\tanh^2(\sqrt{\beta_n}v) \\
    & = -2 (1 + O(c n^{-\gamma})).
\end{align*}
We see when $\gamma = 1/2$, all of $\sqrt{n}\phi^{(2)}(v_+)$, $n^{1/4}\phi^{(3)}(v_+)$ and $\phi^{(4)}(v_+)$ are of order 1. And when $c_n = \sqrt{n}(\beta_n - 1) \gg 1$, we have $\sqrt{n}\phi^{(2)}(v_+) \gg n^{1/4}\phi^{(3)}(v_+) \gg \phi^{(4)}(v_+)$. Since $w_+ = n^{1/4} v_+ = \sqrt{3 c_n} \gg 1$, and similarly, $|w_-| \gg 1$, condition on $\mathsf{W}_{c,n} \in [n]$, $\mathsf{W}_{c,n} - \E[\mathsf{W}_{c,n}|\mathsf{W}_{c,n} \in [n]]$ is $\mathtt{C}$-sub-Gaussian, $\ell \in \{-,+\}$. By similar concentration arguments as in the proof for Step 2 in Lemma~\ref{sa-lem: localization to singularity} (1), we can show the second order term in $h_{c,n}$ dominates, and for $\ell \in \{-,+\}$,
\begin{align*}
    \sup_{t \in \reals}|\P_{\beta_n,h}(\mathsf{W}_{c,n} - \E[\mathsf{W}_{c,n}|\mathsf{W}_{c,n} \in \ca I_\ell] \leq t| \mathsf{W}_{c,n} \in \ca I_\ell) - \Phi(\sqrt{n (1 - \beta_n + v_\ell^2)} t)| = o(1).
\end{align*}
The conclusion then follows from pluggin the (conditional) Gaussian approximation for $\mathsf{W}_{c_n,n}$ back into Equation~\eqref{eq:low-knife-edge}, and the fact that $\mathsf{Z}$ is independent to $\mathsf{W}_{c,n}$ and also Gaussian.

\subsection{Proof of Lemma~\ref{lem: mult-be}}
Throughout the proof, the Ising spins $\bW=(W_i)_{i=1}^n$ are distributed according to Assumption~\ref{assump-one-block} with parameters $(\beta,h)$. For brevity, we write $\P$ in place of $\P_{\beta,h}$.

\begin{center}
\textbf{I. High Temperature or Nonzero External Field}
\end{center}
Let $\mathsf{U}_n$ be the latent random variable from Lemma~\ref{sa-lem:sub-gaussian}. Condition on $\mathsf{U}_n$, $\bX_i W_i$'s are i.i.d random vectors. For $u \in \reals$, define
\begin{align*}
    \Sigma(u) & = \operatorname{Cov}[\bX_i (W_i - \pi)|\mathsf{U}_n = u] \\
    & = \E[\bX_i \bX_i^{\top}]\E[(W_i - \pi)^2|\mathsf{U}_n = u] - \E[\bX_i] \E[\bX_i]^{\top}(\E[W_i|\mathsf{U}_n] - \pi)^2 \\
    & \gtrsim \Cov[\bX_i] \E[(W_i - \pi)^2|\mathsf{U}_n = u] \\
    & \gtrsim \Cov[\bX_i] \min\{(1 - \pi)^2, (1 + \pi)^2\}, \\
    e(u) & = \E[\bX_i (W_i - \pi)|\mathsf{U}_n = u] = \E[\bX_i] (\tanh(\sqrt{\beta/n}u + h) - \pi), 
\end{align*}
and to save notations, we denote
\begin{align*}
    t(u) = \sqrt{n} (\tanh(\sqrt{\beta/n}u + h) - \pi).
\end{align*}
Suppose $\mathsf{Z}_d \sim \mathsf{N}(\mathbf{0}, \mathbf{I}_{d \times d})$ independent to $\mathsf{U}_n$. By \cite[Theorem 2.1]{chernozhukov2017central}
\begin{equation}\label{eq: cond multi}
\begin{aligned}
    & \sup_{u \in \reals}\sup_{A \in \mathcal{R}} \bigg|\P \bigg(\frac{1}{\sqrt{n}} \sum_{i = 1}^n X_i (W_i - \pi) \in A \bigg| \mathsf{U}_n = u \bigg) - \P \bigg(\Sigma(u)^{1/2} \mathsf{Z}_d +  \E[\bX_i] t(\mathsf{U}_n) \in A \bigg) \bigg| \\
    \leq & \Big(\frac{B_n^2 \log (n)^7}{n}\Big)^{1/6}.
\end{aligned}
\end{equation}
From the proofs of Lemma~\ref{sa-lem:fixed-temp-be}, we know the term $t(\mathsf{U}_n)$ stabilizes, $$\dKS \Big(t(\mathsf{U}_n), \sigma \mathsf{Z}\Big) = O(n^{-1/2}), \qquad \sigma = \Big(\frac{\beta(1 - \pi^2)^2}{1 - \beta(1 - \pi^2)}\Big)^{1/2}.$$ 
By Lemma~\ref{sa-lem:sub-gaussian}, 
\begin{align*}
    \norm{\Sigma(\mathsf{U}_n) - \Sigma} = O_{\psi,2}(n^{-1/2}), \qquad \Sigma = (1 - \pi^2) \E[\bX_i \bX_i^{\top}].
\end{align*}
For each $\varepsilon$, define $A_{\varepsilon}$ to be the event $\{\norm{\Sigma(\mathsf{U}_n)^{1/2} - \Sigma^{1/2})\mathsf{Z}_d} \leq \varepsilon\}$. Since $d$ is fixed, we can work with each dimension to get
\begin{equation}\label{eq: sigma-conc}
\begin{aligned}
    & \sup_{\mathbf{t} \in \reals^{2d}} |\P (\Sigma(\mathsf{U}_n)^{1/2}\mathsf{Z}_d + \E[\bX_i] t(\mathsf{U}_n) \leq \mathbf{t}) - \P(\Sigma^{1/2}\mathsf{Z}_d +\E[\bX_i] t(\mathsf{U}_n) \leq \mathbf{t})| \\
    \leq & \sup_{\varepsilon > 0} \sup_{\mathbf{t} \in \reals^{2d}} |\P (\Sigma(\mathsf{U}_n)^{1/2}\mathsf{Z}_d + \E[\bX_i] t(\mathsf{U}_n) \leq \mathbf{t}, A_{\varepsilon}) \\
    & \qquad \qquad \qquad - \P(\Sigma^{1/2}\mathsf{Z}_d+\E[\bX_i]t(\mathsf{U}_n) \leq \mathbf{t}, A_{\varepsilon})| + \P(A_{\varepsilon}^c)\\
    \leq & \sup_{\varepsilon> 0} 2\P (A_{\varepsilon}^c) + \sup_{\mathbf{t} \in \reals^{2d}}\sup_{\boldsymbol{\varepsilon}\in \reals^{2d},\norm{\boldsymbol{\varepsilon}}\leq \varepsilon}\P(\Sigma^{1/2} \mathsf{Z}_d\in (\mathbf{t} - \boldsymbol{\varepsilon}, \mathbf{t} + \boldsymbol{\varepsilon})) \\
    \lesssim & \sup_{\varepsilon > 0}\exp(-n\varepsilon^2) +  \sup_{\mathbf{t} \in \reals^{2d},}\sup_{\boldsymbol{\varepsilon}\in \reals^{2d},\norm{\boldsymbol{\varepsilon}}\leq \varepsilon}\P(\Sigma^{1/2} \mathsf{Z}_d \in (\mathbf{t} - \boldsymbol{\varepsilon}, \mathbf{t} + \boldsymbol{\varepsilon})) \\
    = & O(n^{-1/2} \sqrt{\log n}),
\end{aligned}
\end{equation}
where in the last line, we have chosen $\varepsilon = n^{-1/2}\sqrt{\log n}$ and used Nazarov's inequality (Lemma A.1 in \cite{chernozhukov2017central}). Since $\mathsf{Z}_d$ and $\mathsf{U}_n$ are independent, we can show via data processing inequality that
\begin{align*}
    & \sup_{\mathbf{t} \in \reals^{2d}}|\P(\Sigma^{1/2}\mathsf{Z}_d +  \E[\bX_i] t(\mathsf{U}_n)) \leq \mathbf{t}) - \P((\Sigma^{1/2}\mathsf{Z}_d) + \E[\bX_i] \sigma \mathsf{Z}) \leq \mathbf{t})| \\
    \leq & \sup_{\mathbf{t} \in \reals^{2d}} |\P( t(\mathsf{U}_n) \leq \mathbf{t}) - \P(\E[\bX_i] \sigma \mathsf{Z} \leq \mathbf{t})| 
    = O(n^{-1/2}).
\end{align*}
Combining the previous results, 
\begin{align*}
    \sup_{A \in \mathcal{R}} \bigg|\P\bigg(\frac{1}{\sqrt{n}}\sum_{i = 1}^n X_i W_i \in A\bigg) - \P(\Sigma^{1/2}\mathsf{Z}_d + \E[\bX_i] \sigma \mathsf{Z}_1 \in A) \bigg| = O(n^{-1/2}\sqrt{\log n}).
\end{align*}

\begin{center}
\textbf{II. Critical Temperature}
\end{center}

We still have conditional Berry-Esseen as in Equation~\eqref{eq: cond multi}. The proof of Lemma~\ref{sa-lem:fixed-temp-be} implies
$$\dKS(n^{-1/4}t(\mathsf{U}_n),\mathsf{R}) = O(n^{ -1/2}).$$ 
Hence $\norm{\Sigma(\mathsf{U}_n) - \E[\Sigma(\mathsf{U}_n)]}_{\operatorname{max}} = O_{\psi,2}(n^{-1/4})$. By concentration of $\mathsf{U}_n$, approximation of $n^{-1/4}t(\mathsf{U}_n)$ by $\mathsf{R}$, and anti-concentration of $\mathsf{R}$, we can use similar arguments as Equation~\eqref{eq: sigma-conc} to get
\begin{align*}
    \sup_{\mathbf{t} \in \reals^{2d}} |\P (\Sigma(\mathsf{U}_n)^{1/2}\mathsf{Z}_d + \E[\bX_i] t(\mathsf{U}_n) \leq \mathbf{t}) - \P(\Sigma^{1/2}\mathsf{Z}_d +\E[\bX_i]
    t(\mathsf{U}_n)  \leq \mathbf{t})| = O(n^{-1/2}(\log n)^{1/4}).
\end{align*}
By independence between $\mathsf{Z}_d$ and $\mathsf{U}_n$, and approximation of $n^{-1/4}t(\mathsf{U}_n)$ by $\mathsf{R}$, we can use data processing inequality to get
\begin{align*}
        \sup_{\mathbf{t} \in \reals^{2d}} | \P(\Sigma^{1/2}\mathsf{Z}_d +\E[\bX_i] t(\mathsf{U}_n)  \leq \mathbf{t}) - \P(n^{-\frac{1}{4}} \Sigma^{1/2}\mathsf{Z}_d + \E[\bX_i] \mathsf{R})| = O(n^{-1/2}).
\end{align*}

\begin{figure}
    \centering
    \includegraphics[width=8cm]{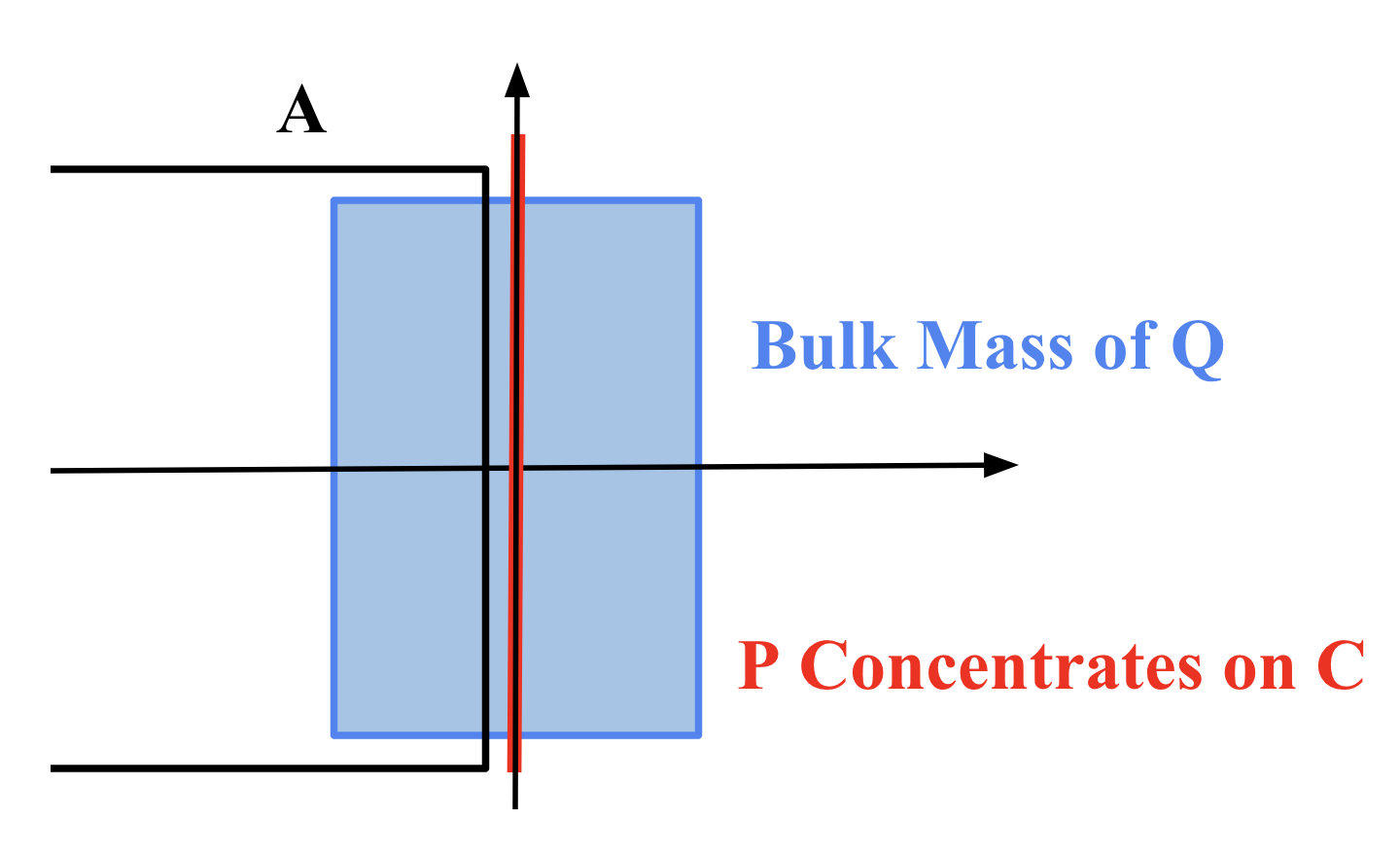}
    \caption{The law $\P$ of $\E[\bX_i]W$, concentrates on $\mathcal{C} = \{s\E[\bX_i]: s \in \reals\}$, while the law $\Q$ of $\E[\bX_i]\mathsf{W} + n^{-\frac{1}{4}}\Sigma^{-\frac{1}{2}}\mathsf{Z}_d$ is degenerate. The bulk mass of $\Q$ lies in a cylinder with axis $\mathcal{C}$ and width of order $\sqrt{\log n}n^{-\frac{1}{4}}$. Consider $\Sigma = I_2$, $\E[\bX_i] = \mathbf{e}_2$, $\mathcal{C} = \{s \be_2: s \in \reals\}$, and $A = \{(x,y) \in \reals^2: -M \leq x \leq -\varepsilon, -M \leq y \leq M\}$ for some small $\varepsilon > 0$ and large $M > 0$. Then $\P(A) = 0$ while $\Q(A)$ is close to $\frac{1}{2}$.}
    \label{fig: multi-be}
\end{figure}
It follows that
\begin{align*}
    \sup_{A \in \mathcal{R}} \bigg|\P\bigg(n^{-1/4}\sum_{i = 1}^n X_i W_i \in A\bigg) - \P(n^{-\frac{1}{4}}\Sigma^{1/2}\mathsf{Z}_d + \E[\bX_i] \mathsf{R} \in A) \bigg| = O(n^{-1/2}\sqrt{\log n}).
\end{align*}

\begin{center}
\textbf{III. Low Temperature}
\end{center}
We still have conditional Berry-Esseen as in Equation~\eqref{eq: cond multi}. From the proof of Lemma~\ref{sa-lem:fixed-temp-be} (3) and Remark~\ref{sa-remarK: conditional concentration under low temp}, for $\ell = -,+$,
$$\dKS(t(\mathsf{U}_n) - \sqrt{n} \pi_{\ell}|\sign(\mca m) = \ell, \sigma \mathsf{Z}) = O(n^{-1/2}),$$
where $\sigma^2 = \frac{\beta(1 - \pi_+^2)^2}{1 - \beta(1 - \pi_+^2)}$. The rest of the proof follows from the arguments for \textit{I. High Temperature or Nonzero External Field}, using conditional concentration of $\mathsf{U}_n$ given $\sign(\mca m)$.

\section{Proofs: Section~\ref{sa-sec:mple}}

\subsection{Proof of Lemma~\ref{sa-lem:var-inconsist}}

Our proof is constructive. We show that consistent estimate of $n\V[\widehat{\tau}_n]$ would imply that one can distinguish between two constructed hypotheses easily. Let $\mathcal{P}_n$ be the class of distributions of random vectors $(\bW = (W_1, \cdots, W_n), \bY = (Y_1, \cdots, Y_n))$ taking values in $\reals^{2n}$ that satisfies Assumptions 1,2,3. Consider the following two data generating processes:
\begin{align*}
    \text{DGP}_0: & \quad \beta = 0, \quad G(\cdot,\cdot) \equiv 1, \quad \rho_n = 1, \quad Y_i(\cdot, \cdot) = f_i(\cdot,\cdot) +  \varepsilon_i, \quad f_i(\cdot,\cdot) \equiv 1, \\
    \text{DGP}_1: & \quad \beta = u, \quad G(\cdot,\cdot) \equiv 1, \quad \rho_n = 1, \quad Y_i(\cdot, \cdot) = f_i(\cdot,\cdot) + \varepsilon_i, \quad f_i(\cdot,\cdot) \equiv 1,
\end{align*}
where $0 < u < 1$, and in both cases $(\varepsilon_i: 1 \leq i \leq n)$ are i.i.d $\mathsf{N}(0,1)$ random variables, independent to $\bW$. Denote by $\P_{0,n}$ and $\P_{1,n}$ the laws of $(\bW,\bY)$ under $\text{DGP}_0$ and $\text{DGP}_1$. Then
\sm{\begin{align*}
    d_{\text{KL}}(\P_{0,n}(\bW,\bY), \P_{1,n}(\bW, \bY)) & = d_{\text{KL}}(\P_{0,n}(\bW),\P_{1,n}(\bW)) + d_{\text{KL}}(\P_{0,n}(\bY|\bW),\P_{1,n}(\bY|\bW))\\
    & = d_{\text{KL}}(\P_{0,n}(\bW),\P_{1,n}(\bW)),
\end{align*}}
the first line uses chain rule of $d_{\text{KL}}$, the second line uses $$d_{\text{KL}}(\P_{0,n}(\bY|\bW),\P_{1,n}(\bY|\bW)) = d_{\text{KL}}(\P_{0,n}(\bY),\P_{1,n}(\bY)) = 0.$$ From Theorem 2.3 (and its proof) in \cite{bhattacharya2018inference}, $$M := \lim_{n \rightarrow \infty} d_{\text{KL}}(\P_{0,n}(\bW),\P_{1,n}(\bW)) < \infty.$$ Hence for large enough $n$, 
\begin{align*}
    d_{\text{TV}}(\P_{0,n}(\bW,\bY), \P_{1,n}(\bW, \bY)) & \leq 1 - \frac{1}{2}\exp(-d_{\text{KL}}(\P_{0,n}(\bW,\bY), \P_{1,n}(\bW, \bY))) \\
    &\leq 1 - \frac{1}{2}\exp(-M).
\end{align*}
Le Cam's method (Section 15.2.1 in \cite{wainwright2019high}) gives for large enough $n$, 
\begin{align*}
     \inf_{\widehat{\V}} \sup_{\P_n \in \mathcal{P}_n}&\E_{\P_n}[n(\widehat{\V}[\widehat{\tau} - \tau] - \V[\widehat{\tau} - \tau])] \\
    \geq & n|\V_{\P_{n,0}}[\widehat{\tau} - \tau]-\V_{\P_{n,1}}[\widehat{\tau} - \tau]| (1 -  d_{\text{TV}}(\P_{0,n}(\bW,\bY), \P_{1,n}(\bW, \bY))) \\
    \geq & \varepsilon \exp(-M)/2,
\end{align*}
in the last line we used Theorem 2 (1) to get $n \V_{\P_{n,0}}[\widehat \tau - \tau] - n \V_{\P_{n,1}}[\widehat \tau - \tau] = \varepsilon (1 + o(1))$.

\subsection{Proof of Lemma~\ref{sa-lem: fixed temp MPLE}}

The following discussions will be organized according to the three different cases: (1) When $\beta<1$. (2) When $\beta\geq 1$, $\mca m$ concentrates around $0$. (3) When $\beta\geq 1$ and $\mca m$ concentrates around two symmetric locations $w_{+}>0$ and $w_{-}<0$ with $|w_{+}|=|w_{-}|$.

We have required $\wh \beta \in [0,1]$. For analysis, consider an unrestricted pseud-likelihood estimator,
\begin{align*}
   \nonumber \betaur = \argmax_{\beta \in \reals} l(\beta;\bW),
\end{align*}
where $l(\beta;\bW)$ is the pseudo log-likelihood given by
\begin{align*}
    l(\beta;\bW) = \sum_{i \in [n]} \log \mathbb{P}_{\beta} \left( W_i \mid \mathbf{W}_{-i} \right) = \sum_{i \in [n]} -\log \bigg( \frac{1}{2}W_i \tanh(\beta \mca m_i) + \frac{1}{2}\bigg).
\end{align*}
We show that $l(\beta;\bW)$ is concave. 
\begin{align*}
    \frac{\pta}{\pta \beta} l(\beta;\bW)&=-\frac{1}{n}
        \sum_{i=1}^n\frac{(n^{-1}\sum_{j\neq i}W_j)W_i\sech^2(\beta n^{-1}\sum_{j\neq i}W_j)}{W_i\tanh(\beta n^{-1}\sum_{j\neq i }W_j)+1}\\
    &=-\frac{1}{n}\sum_{i=1}^n\bl \frac{1}{n}\sum_{j\neq i}W_j\br(W_i-\tanh(\beta n^{-1}\sum_{j\neq i }W_j)),
\end{align*}
and
\begin{align*}
    l^{(2)}(\beta;\bW)= \frac{1}{n}\sum_{i=1}^n\bl \frac{1}{n}\sum_{j\neq i}W_j\br^2\sech^2\bl\frac{\beta}{n}\sum_{j\neq i}W_j\br > 0.
\end{align*}
Hence $l(\cdot;\bW)$ is concave everywhere in $\reals$. This shows $\wh \beta = \min\{\max\{\betaur,0\},1\}$. Now we study limiting distribution of $\betaur$
\begin{center}
    \textbf{1. High and critical temperature regime.}
\end{center}
To obtain a more precise distribution for $\betaur$, we use Fermat's condition to obtain that 
\begin{align*}
    0 &=\frac{1}{n}\sum_{i=1}^n\bl\frac{1}{n}\sum_{j\neq i}W_j\br\bl W_i-\tanh\bl\betaur n^{-1}\sum_{j\neq i}W_j\br\br\\
    &=\frac{1}{n}\sum_{i=1}^n\bl\mca m-\frac{W_i}{n}\br\bl W_i-\tanh(\betaur\mca m)+\sech^2(\betaur\mca m)\frac{\betaur W_i}{n}+O(n^{-2})\br\\
    &=\frac{1}{n}\sum_{i=1}^n\bl\mca m-\frac{W_i}{n}\br\bl\bl 1+\sech^2(\betaur\mca m)\frac{\betaur}{n}\br W_i-\tanh(\betaur\mca m)+O(n^{-2})\br\\
    &=\bl 1+\frac{\betaur}{n}\sech^2(\betaur\mca m)\br\bl\mca m^2-\frac{1}{n}\br-\frac{n-1}{n}\mca m\tanh(\betaur\mca m)+O(n^{-2})\mca m,
\end{align*}
here $O(\cdot)$'s are all up to an absolute constant. By Lemma~\ref{sa-lem: localization to singularity} with $X_i = 1$, we can show $\E[|(n \mca m)^{-1}|] \leq \mathtt{C} n^{-1/2}$. By Markov inequality, $(n \mca m)^{-1} = O_\P(n^{-1/2})$. Taylor expanding $\tanh$, we have
\begin{align}\label{sa-eq:beta-hat-expansion}
    \nonumber \betaur&=\frac{n}{(n-1)\mca m}\tanh^{-1}\lef(\mca m-\frac{1}{n\mca m}\rig) \\
    \nonumber &=\frac{n}{(n-1)\mca m}\bl \mca m-\frac{1}{n\mca m}+\frac{1}{3}\bl\mca m-\frac{1}{n\mca m}\br^3+O\bl\bl\mca m-\frac{1}{n\mca m}\br^5\br\br\\
    &= 1-\frac{1}{n\mca m^2}+\frac{\mca m^2}{3} + O_\P(n^{-1}),
\end{align}
where in the above equation, both $O(\cdot)$ and $O_\P(\cdot)$ are up to absolute constants.
The rest of the results are given according to the different temperature regimes.

\textbf{(1) The High Temperature Regime.} Using Lemma~\ref{sa-lem: fixed temp MPLE} with $X_i = 1$, our result for the high temperature regime with $\beta<1$ implies that
$n^{\frac{1}{2}}\mca m\overset{d}{\to}\mathsf{N}( 0,\frac{1}{1-\beta})  \Rightarrow (1-\beta)n\mca m^2\overset{d}{\to} \chi^2(1)$. Therefore we conclude that $\frac{1-\beta}{1-\betaur}\overset{d}{\to}\chi^2(1)$. The conclusion then follows from $\wh \beta = \min\{\max\{\betaur,0\},1\}$.

\textbf{(2) The Critical Temperature Regime.}
Using Lemma~\ref{sa-lem: fixed temp MPLE} with $X_i = 1$, we have $\dKS(n^{\frac{1}{4}}\mca m, \mathsf{W}_0) = o(1)$. This implies $n^{\frac{1}{2}}(\betaur-1) \overset{d}{\to} Law(\frac{\mathsf{W}_0^2}{3}-\frac{1}{\mathsf{W}_0^2}).$ Since $\mathsf{W}_0 = O_\P(1)$, $\P(\betaur < 0) = o(1)$. The conclusion then follows from $\wh \beta = \min\{\max\{\betaur,0\},1\}$.

\begin{center}
    \textbf{2. The low temperature regime.} 
\end{center}
When $\mca m$ concentrates around $\pi_+$ and $\pi_-$ we have when $\mca m>0$, use the fact  that $\pi_{\ell}=\tanh(\beta\pi_{\ell})$ for $\ell\in\{+,-\}$,
\begin{align*}
    \betaur&-\beta=\frac{(1-O(n^{-1}))(\mca m-\tanh(\beta\mca m))}{\mca m\sech^2(\beta\mca m)}+\mca mO(\delta^2)+O(n^{-1})\\
    &=\frac{(1-O(n^{-1}))\lef((\mca m-\pi_{\ell})-(\tanh(\beta\mca m)-\tanh(\beta\pi_{\ell}))\rig)}{\pi_{\ell}\lef(\sech^2(\beta\pi_{\ell})-2(\mca m-\pi_{\ell})\tanh(\beta\pi_{\ell})\sech^2(\beta\pi_{\ell})+O(\mca m-\pi_{\ell})^2\rig)\lef(1+\frac{\mca m-\pi_{\ell}}{\pi_{\ell}}\rig)} \\
    & \qquad +\mca mO(\delta^2)+O(n^{-1})\\
    &=(1-O(n^{-1}))\frac{(1-\beta\sech^2(\beta\pi_{\ell}))(\mca m-\pi_{\ell})}{\pi_{\ell}\sech^2(\beta\pi_{\ell})}(1+O(\mca m-\pi_{\ell}))+\mca mO(\delta^2)+O(n^{-1}).
\end{align*}
and the similar argument gives
\begin{align*}
    \mca m(\betaur - \beta^{\ast}) = \frac{1 - \beta^{\ast} \sech^2(\beta^{\ast} \pi_{\ell})}{\sech^2(\beta^{\ast} \pi_{\ell})}(\mca m - \pi_{\ell}) + O_{\psi_1}(n^{-1}).
\end{align*}
The conclusion then Lemma~\ref{sa-lem:fixed-temp-be} (3) and the convergence of $\mca m$ to $\pi_+$ or $\pi_-$.

\subsection{Proof of Lemma~\ref{sa-lem: drifting MPLE}}
Again we consider the unrestricted PMLE given by
\begin{align*}
   \nonumber \betaur = \argmax_{\beta \in \reals} l(\beta;\bW),
\end{align*}
where $l(\beta;\bW)$ is the pseudo log-likelihood given by
\begin{align*}
    l(\beta;\bW) = \sum_{i \in [n]} \log \mathbb{P}_{\beta} \left( W_i \mid \mathbf{W}_{-i} \right) = \sum_{i \in [n]} -\log \bigg( \frac{1}{2}W_i \tanh(\beta \mca m_i) + \frac{1}{2}\bigg).
\end{align*}

For $\beta \in [0,1]$, that is $c_\beta = \sqrt{n}(\beta - 1) \leq 0$, Equation~\eqref{sa-eq:beta-hat-expansion} and the approximation of $\mca m$ by $n^{-1/2}\mathsf{Z} + n^{-1/4}\mathsf{W}_c$ from Lemma~\ref{sa-lem: localization to singularity} gives
\begin{align*}
    \sup_{\beta \in [0,1]}\sup_{t \in \reals}|\P (1 - \wh \beta \leq t) - \P (z_{\beta,n}^{-2} - \frac{3}{n} z_{\beta,n}^2  \leq t)| = o(1).
\end{align*}
The conclusion follows from the fact that $x \mapsto \max\{\min\{x,0\},1\}$ is $1$-Lipschitz.

\section{Proofs: Section~\ref{sa-sec:stochlin}}

\subsection{Preliminary Lemmas}

\begin{lemma}\label{pi fixed point}
Recall $\bW = (W_i)_{1 \leq i \leq n}$ takes value in $\{-1,1\}^n$ with
\begin{align*}
    \P \left( \bW = \bw \right) = \frac{1}{Z} \exp \bigg( \frac{\beta}{n} \sum_{i < j} W_i W_j + h \sum_{i = 1}^n W_i \bigg), \quad h \neq 0 \text{ or } h = 0, 0 \leq \beta \leq 1.
\end{align*}
Recall $\pi$ is the unique solution to $x = \tanh(\beta x + h)$. Then $\E[W_i] = \pi + O(n^{-1})$.
\end{lemma}

\begin{proof}
If $h = 0$, then $\pi = \E[W_i] = 0$. If $h \neq 0$, then the concentration of $\mca m = n^{-1} \sum_{i = 1}^n W_i$ towards $\pi$ in Lemma~\ref{sa-lem:fixed-temp-be} implies,
\begin{align*}
    \E[W_i] = & \E[\E[W_i|W_{-i}]]
    = \E[\tanh(\beta \mca m_i + h)] \nnb\\
    = &  \E[\tanh(\beta \pi + h) + \sech^2(\beta \pi + h)(\mca m_i - \pi) - \sech^2(\beta m^{\ast} + h)\tanh(\beta m^{\ast} + h)(\mca m_i - \pi)^2] \nnb\\
    = & \tanh(\beta \pi + h) + O(n^{-1}) \\
    = & \pi + O(n^{-1}),
\end{align*}
where $\mca m^{\ast}$ is a number between $\mca m$ and $\pi$, and we have used boundedness of $\sech$.
\end{proof}

\begin{lemma}\label{lem: concentration of M_i/N_i}
Suppose Assumption~\ref{assump-one-block}, 2, and 3 hold with $h = 0, 0 \leq \beta \leq 1$ or $h \neq 0$: (1)
\begin{align*} 
    & \max_{i \in [n]} \left|\frac{M_i}{N_i} - \pi \right| = O_{\psi_{\beta, \gamma}}(n^{-\bgamma}) + O_{\psi_2}(N_i^{-1/2}).
\end{align*}

(2) Define $A(\mathbf{U}) = (G(U_i,U_j))_{1 \leq i,j \leq n}$. Condition on $\mathbf{U}$ such that $A(\mathbf{U}) \in \mathcal{A} = \{A \in \bb R^{n \times n}: \min_{i \in [n]} \sum_{j \neq i}A_{ij} \geq 32 \log n\}$, for large enough $n$, for each $i \in [n]$ and $t > 0$,
\begin{align*}
    \P \left( \left|\frac{M_i}{N_i} - \pi\right| \geq 4 \E[N_i | \mathbf{U}]^{-1/2}t^{1/2} + C_{\beta,h}n^{-\bgamma} t^{\pgamma}\middle| \mathbf{U} \right) \leq 2 \exp(-t) + n^{-98},
\end{align*}
where $C_{\beta,h}$ is some constant that only depends on $\beta,h$.

(3) When $h = 0$, and $\beta \in [0,1]$, then there exists a constant $\mathtt{K}$ that does not depend on $\beta$, such that for large enough $n$, for each $i \in [n]$ and $t > 0$,
\begin{align*}
    \P \left( \left|\frac{M_i}{N_i} - \pi\right| \geq 4 \E[N_i | \mathbf{U}]^{-1/2}t^{1/2} + \mathtt{K} n^{-\bgamma} t \middle| \mathbf{U} \right) \leq 2 \exp(-t) + n^{-98}.
\end{align*}
\end{lemma}

\begin{proof}
Take $\mathsf{U}_n$ to be a random variable with density
\begin{align*}
    f_{\mathsf{U}_n}(u) = \frac{\exp \left(- \frac{1}{2} u^2 + n \log \cosh \left( \sqrt{\frac{\beta}{n}} u + h\right) \right)}{\int_{-\infty}^{\infty} \exp \left(- \frac{1}{2} v^2 + n \log \cosh \left( \sqrt{\frac{\beta}{n}} v + h\right) \right) d v}.
\end{align*} 
Condition on $\mathsf{U}_n$, $W_i$'s are i.i.d. Decompose by
\begin{align*}
    \frac{M_i}{N_i} - \pi = \sum_{j \neq i} \frac{E_{ij}}{N_i} \left(W_j - \E[W_j|\mathsf{U}_n] \right) + \E[W_j|\mathsf{U}_n] - \pi.
\end{align*}
Condition on $\mathsf{U}_n$, $W_i$'s are i.i.d. Berry-Esseen theorem condition on $\mathsf{U}_n$ and $\mathbf{E}$ gives, 
\begin{align}\label{sa-eq: nbh avg decompose}
    \sup_{t \in \reals} \bigg|\P\Big(\frac{M_i}{N_i} - \pi  \leq t\Big|\mathbf{E} \Big) - \P \Big(\sqrt{\frac{v(\mathsf{U}_n)}{N_i}} Z + e(\mathsf{U}_n) \leq t \Big|\mathbf{E}\Big) \bigg| = O(n^{-\frac{1}{2}}),
\end{align}
where $e(\mathsf{U}_n) := \E[W_i|\mathsf{U}_n] - \pi = \tanh(\sqrt{\beta/n}\mathsf{U}_n + h) - \pi$, and $v(\mathsf{U}_n) := \V[W_i - \pi|\mathsf{U}_n]$. By McDiarmid's inequality, 
\begin{align*}
    \P \bigg( |\sum_{j \neq i} \frac{E_{ij}}{N_i} \left(W_j - \E[W_j|\mathsf{U}_n] \right)| \geq 2 N_i^{-1/2} t\bigg | \mathbf{E} \bigg) \leq 2 \exp(-t^2).
\end{align*}
Plugging into Equation~\eqref{sa-eq: nbh avg decompose}, we can show (1) holds.

Next, we want to show condition on $\mathbf{U}$ such that $A(\mathbf{U}) \in \mathcal{A}$, $\P(N_i \leq \E[N_i|\mathbf{U}]/3|\mathbf{U}) \leq n^{-100}$:

Notice that for any $\mathbf{U}$ such that $\rho_n \min_{i \in [n]} \sum_{j \neq i}A_{ij}(\mathbf{U}) \rightarrow \infty$, 
Condition on $A$ such that $A \in \mathcal{A}$, $E_{ij} = \rho A_{ij}\iota_{ij}$, $1 \leq i \leq j \leq n$ are i.i.d Bernouli random variables, and for each $i,j$, $\sum_{k \neq i,j}A_{ki} \geq 32 \log n - 1 \geq 31 \log n$ for $n \geq 3$. By bounded difference inequality, for all $t > 0$,
\begin{align*}
    \P \bigg( \bigg|\sum_{k \neq i,j}E_{ki} - \sum_{k \neq i,j}\rho_n A_{ki}\bigg| \geq \rho_n \sqrt{\sum_{k \neq i,j}A_{i,j}^2}t \bigg) \leq 2 \exp(-2t^2).
\end{align*}
Hence condition on $A$, with probability at least $1 - n^{-100}$, 
\begin{align}\label{N_i lower bound}
    \nonumber \sum_{k \neq i,j}E_{ki} & \geq \sum_{k \neq i,j}\rho_n A_{ki} - 8 \sqrt{\log n} \rho_n \sqrt{\sum_{k \neq i,j}A_{ij}^2}
    \geq \rho_n \sum_{k\neq i,j} A_{ki} - 8 \sqrt{\log n} \rho_n \sqrt{\sum_{k \neq i,j}A_{ki}} \\
    \nonumber & \geq \rho_n \sqrt{\sum_{k \neq i,j}A_{ki}}\left(\sqrt{\sum_{k \neq i,j}A_{ki}} - 8 \sqrt{\log n}\right) \\
    \nonumber & \geq \rho_n \sqrt{\sum_{k \neq i,j}A_{ki}} \left(\sqrt{\sum_{k \neq i,j}A_{ki}} - 8 \sqrt{31^{-1} \sum_{k \neq i,j}A_{ij}}\right)\\
    & \geq \rho_n \sum_{k \neq i,j}A_{ij}/3 \geq \frac{31}{3}\log n,
\end{align}
and since $\rho_n A_{i,j} = \E[E_{ij}|\mathbf{U}] \in [0,1]$, $\sum_{k \neq i,j}E_{ki} + 1 \geq \E[N_j|\mathbf{A}]/3$. By Equation~\ref{N_i lower bound}, condition on $\mathbf{U}$ such that $A(\mathbf{U}) \in \mathcal{A}$, $\P(N_i \leq \E[N_i|\mathbf{U}]/3|\mathbf{U}) \leq n^{-100}$. 

Hence we can disintegrate over the distribution of $\mathbf{E}$ to get
\begin{align*}
    \P \left( |\sum_{j \neq i} \frac{E_{ij}}{N_i} \left(W_j - \E[W_j|\mathsf{U}_n] \right)| \geq 4 \E[N_i|\mathbf{U}]^{-1/2} t \middle | \mathbf{U} \right) \leq 2 \exp(-t^2) + n^{-100}.
\end{align*}
By Equation~\ref{eq: conditional mean and var on Un} and Lemma~\ref{sa-lem:sub-gaussian}, and the Lipschitzness of $\tanh$ that 
\begin{align*}
    \E[W_i|\mathsf{U}_n] - \pi = O_{\psi_{\beta,h}}\left(n^{-\bgamma}\right).
\end{align*}
Plugging into Equation~\eqref{sa-eq: nbh avg decompose}, we can show (2) holds.

Under the setting of (3), the only part that depends on $\beta$ in our proof is $\mathsf{U}_n$. Since we show in Lemma~\ref{sa-lem:sub-gaussian} $\norm{\mathsf{U}_n}_{\psi_1} \leq \mathtt{K} n^{1/4}$ for some absolute constant $\mathtt{K}$, which is essentially the $\beta = 1$ rate, the conclusion of (3) then follows.
\end{proof}

\subsection{Proof of Lemma~\ref{sa-lem:unbiased}}
Since we use the conditional probability $p_i$ in the inverse probability weight, we have
\begin{align*}
    \E[\wh\tau_{n,\text{UB}}|(f_i)_{i \in [n]}, \mathbf{E
    }] & = \frac{1}{n} \sum_{i = 1}^n \E \bigg[ \frac{T_i Y_i}{p_i} - \frac{(1 - T_i) Y_i}{1 - p_i} \bigg|(f_i)_{i \in [n]}, \mathbf{E
    } \bigg] \\
    & = \frac{1}{n} \sum_{i = 1}^n \E \bigg[\E \bigg[ \frac{T_i Y_i}{p_i} - \frac{(1 - T_i) Y_i}{1 - p_i} \bigg|\bT_{-i},(f_i)_{i \in [n]}, \mathbf{E
    } \bigg] \bigg| (f_i)_{i \in [n]}, \mathbf{E
    } \bigg],
\end{align*}
and the conclusion follows from $\E[T_i|\bT_{-i},(f_i)_{i \in [n]}, \mathbf{E}] = p_i$.

\subsection{Proof of Lemma~\ref{sa-lem: approx delta_1}}

First consider the treatment part.
\begin{align*}
    n^{-\balpha} \sum_{i = 1}^n \frac{T_i}{p_i} g_i \left(1, \pi \right) 
    = n^{-\balpha} \sum_{i = 1}^n  g_i \left(1, \pi \right) + n^{-\balpha} \sum_{i = 1}^n \frac{T_i - p_i}{p_i} g_i \left(1, \pi \right).
\end{align*}
For the second term, taylor expand $p_i^{-1}, p_i$ as follows:
\begin{equation}\label{taylor}
\begin{aligned}
    p_i^{-1} = & 1 + \exp \left(-2 \beta \mca m_i -  2 h\right) = 1 + \exp \left(-2\beta \frac{n-1}{n}\pi - 2 h\right) \\
    & -  \exp \left(-2\beta \frac{n-1}{n} \pi - 2 h\right) 2 \beta \left(\mca m_i - \frac{n-1}{n}\pi\right) + \frac{1}{2}\exp(-\xi_i^{\ast}) 4 \beta^2 \left(\mca m_i - \frac{n-1}{n}\pi \right)^2,
\end{aligned}
\end{equation}
where $\xi_i^{\ast}$ is some random quantity that lies between $4 \frac{\beta}{n} \sum_{j \neq i} W_j$ and $4 \frac{\beta}{n} \sum_{j \neq i} \pi$. Taking the parameters $c_i^+ = g_i \left(1, \pi \right) \left(1 + \exp(-2 \beta \pi - 2 h) \right)$, $d^+ = \beta(1 - \tanh(\beta \pi + h))\E[g_i(1,\pi)].$ Then
\begin{align*}
     & n^{-\balpha} \sum_{i = 1}^n \frac{T_i - p_i}{p_i} g_i \left(1, \pi \right) \\
    \stackrel{(1)}{=} & n^{-\balpha} \sum_{i = 1}^n (T_i - p_i)g_i \left(1,\pi \right) (1 + \exp \left(-2 \beta \pi - 2 h\right)  -  \exp \left(-2\beta \pi - 2 h\right) 2 \beta (\mca m_i - \pi)) \\ 
    & \qquad + O_{\psi_{\beta,h},tc}(n^{-\bgamma}) \\ 
    \stackrel{(2)}{=} & n^{-\balpha} \sum_{i = 1}^n c_i(T_i - p_i) + O_{\psi_{\beta,h},tc}((\log n)^{1/2}n^{-\bgamma}) \\
    \stackrel{(3)}{=} & n^{-\balpha} \sum_{i = 1}^n c_i^+ \left[T_i - \frac{1}{1 + \exp(-2 \beta \pi - 2 h)} - \frac{2 \beta \exp(2\beta \pi + 2 h)}{(1 + \exp(2 \beta \pi + 2 h))^2}(\mca m_i - \pi)\right] \\
    & \qquad + O_{\psi_{\beta,h},tc}((\log n)^{1/2}n^{-\bgamma}) \\
    = & n^{-\balpha} \sum_{i = 1}^n \frac{c_i^+}{2} \left(W_i - \tanh( \beta \pi + h)\right)  \\
    & \quad - n^{-\balpha} \sum_{i = 1}^n \frac{2 \beta \exp(2\beta \pi + 2 h)}{(1 + \exp(2 \beta \pi + 2 h))^2}(\frac{1}{n} \sum_{j \neq i} c_j^+)\left(W_i - \pi \right) + O_{\psi_{\beta,h},tc}((\log n)^{1/2}n^{-\bgamma}) \\
    \stackrel{(4)}{=} & n^{-\balpha}\sum_{i = 1}^n \left[g_i \left(1,\pi \right) +  (c_i^+/2 - d^+) \left(W_i - \pi\right)\right] + O_{\psi_{\beta,h},tc}((\log n)^{1/2}n^{-\bgamma}).
\end{align*}
\paragraph*{Proof of (1):} By Lemma~\ref{sa-lem:fixed-temp-be}, $\mca m - \pi = O_{\psi_{\beta, h}}(n^{-\bgamma})$. The claim follows from Equation~\ref{taylor} and a union bound argument. 
\paragraph*{Proof of (2):} 
\begin{align*}
    n^{-\balpha} \sum_{i = 1}^n (T_i - p_i) g_i(1,\pi)(\mca m_i - \pi)
    = & \frac{1}{2} (\mca m - \pi) n^{-\balpha} \sum_{i = 1}^n (W_i - \tanh(\beta \mca m + h)) g_i(1,\pi) \\
    & \qquad + O(n^{-\balpha}).
\end{align*}
By Lemma~\ref{sa-lem:fixed-temp-be},
\begin{align*}
    \mca m - \pi = O_{\psi_{\beta,h}, tc}(n^{-\bgamma}).
\end{align*}
Taylor expand $\tanh(x)$ at $x = \beta \pi + h$, we have
\begin{align*}
    & n^{-\balpha} \sum_{i = 1}^n g_i(1,\pi) (W_i - \tanh(\beta \mca m + h)) \\
    = & n^{-\balpha} \sum_{i = 1}^n g_i(1,\pi) (W_i - \tanh(\beta \pi + h) - \beta \sech^2(\beta \pi + h)(\mca m - \pi) + \tanh(\beta \pi + h)\sech^2(\beta \pi + h)(\mca m - \pi)^2 \\
    & \qquad + O((\mca m - \pi)^3)) \\
    = & O_{\psi_{\beta,h,tc}}(1).
\end{align*}
hence
\begin{align*}
    n^{-\balpha} \sum_{i = 1}^n (T_i - p_i)g_i(1,\pi)(\mca m_i - \pi) = O_{\psi_{\beta,h}, tc}((\log n)^{1/2}n^{-\bgamma}).
\end{align*}

\paragraph*{Proof of (3):} The first line follows from a Taylor expansion of $p_i = (1 + \exp(2 \beta \mca m_i + 2 h))^{-1}$ at $\pi$, and $\mca m_i - \pi = O_{\psi_{\beta,h}}(n^{-\bgamma})$, noticing that $c_i$, $\infnorm{\psi^{\prime \prime}}$ are bounded. The second line follows by reordering the terms.

\paragraph*{Proof of (4):} By Lemma~\ref{pi fixed point}, $\tanh(\beta \pi + h) = \pi + O(n^{-1})$. By boundedness and i.i.d of $g_i(1,\pi)$, $\frac{1}{n} \sum_{j \neq i} c_j = \overline{c} + O(n^{-1}) = \E[c_i] + O_{\P}(n^{-1/2}) + O(n^{-1})$. Similarly, for the control part, taking the parameters $c_i^- = g_i \left(-1, \pi \right) \left(1 + \exp(2 \beta \pi + 2 h) \right)$, $d^- = \beta(1 - \tanh(-\beta \pi - h))\E[g_i(-1,\pi)].$
\begin{align*}
    & - n^{-\balpha} \sum_{i = 1}^n \frac{1 - T_i}{1 - p_i} g_i(-1,\pi) \\
    = & - n^{-\balpha} \sum_{i = 1}^n g_i(-1,\pi) + n^{-\balpha} \sum_{i = 1}^n (c_i^{-}/2 - d^-)(W_i - \pi) + O_{\psi_{\beta,h},tc}((\log n)^{1/2}n^{-\bgamma}).
\end{align*}
Using Lemma~\ref{pi fixed point} again, we can show $(1 +\exp(- 2 \beta \pi - 2 h))/2 = 1/\pi + O(n^{-1})$ and $(1 + \exp(2 \beta \pi + 2 h))/2 = 1/(1 - \pi) + O(n^{-1})$, $\tanh(- \beta \pi - h) = - \pi + O(n^{-1})$. The result then follows from replacing these quantities in $c_i^{+}, c_i^{-}, d^+, d^-$ by corresponding ones using $\pi$.

% Under the setting of $h = 0$, $\beta \in [0,1]$, (which implies $\pi = 0$), we replace $n^{-\bgamma}$ in the rates by $n^{-1/4}$. The only part where the rate of convergence depends $\beta$ in our proof is the rate of convergence of $\mca m - \pi$ and $\mca m - \tanh(\beta \mca m  + h )$. Taking $X_i \equiv 1$ in Lemma~\ref{sa-lem: localization to singularity}, the sum $\frac{1}{n} \sum_{i = 1}^n X_i W_i$ equals $\mca m$, we get from the uniform approximation $\norm{\mca m - \pi}_{\psi_1} \leq \mathtt{C} n^{-1/4}$ and $\norm{\mca m - \tanh(\beta \mca m  + h )}_{\psi_1} \leq \mathtt{C} n^{-1/4}$, where $\mathtt{C}$ is an absolute constant. The second claim in the statement then follows.

\subsection{Proof of Lemma~\ref{lem: delta_2,2}}

We decompose by $\Delta_{2,2} = \Delta_{2,2,1} + \Delta_{2,2,2}$, where
\begin{align*}
    \Delta_{2,2,1} = n^{-\balpha} \sum_{i = 1}^n \frac{T_i - \E[p_i]}{\E[p_i]} g_i^{\prime}(1,\pi) \left(\frac{M_i}{N_i} - \pi \right), \\
    \Delta_{2,2,2} = n^{-\balpha} \sum_{i = 1}^n T_i \left(p_i^{-1} - \E[p_i]^{-1} \right) g_i^{\prime}(1,\pi) \left(\frac{M_i}{N_i} - \pi \right).
\end{align*}
Notice that the first term is a quadractic form. Define $\mathbf{H}$ such that $H_{ij} = \frac{g_i^{\prime}(1,\pi) E_{ij}}{2 \E[p_i] N_i}$. Then $\Delta_{2,2,1} = n^{-\balpha} (\bW - \pi)^{\trans} \mathbf{H} (\bW - \pi)$. Take $\mathsf{U}_n$ to be the latent variable from Lemma~\ref{sa-lem: definetti}. Then we decompose
\begin{align*}
    \Delta_{2,2,1} = \Delta_{2,2,1,a} + \Delta_{2,2,1,b} + \Delta_{2,2,1,c} + \Delta_{2,2,1,d},
\end{align*}
where
\begin{align*}
    \Delta_{2,2,1,a} & = n^{-\balpha}(\bW - \E[\bW|\mathsf{U}_n])^{\trans} \bH  (\bW - \E[\bW|\mathsf{U}_n]), \\
    \Delta_{2,2,1,b} & = n^{-\balpha} (\E[\bW|\mathsf{U}_n] - \pi)^{\trans} \bH (\bW - \E[\bW|\mathsf{U}_n]), \\
    \Delta_{2,2,1,c} & = n^{-\balpha} (\bW - \E[\bW|\mathsf{U}_n])^{\trans} \bH (\E[\bW|\mathsf{U}_n] - \pi), \\
    \Delta_{2,2,1,d} & = n^{-\balpha} (\E[\bW|\mathsf{U}_n] - \pi)^{\trans} \bH (\E[\bW|\mathsf{U}_n] - \pi).
\end{align*}
Since $\norm{\mathbf{H}}_2 \leq \norm{\bH}_F \leq \frac{B}{2 \pi} \sqrt{n}(\min_i N_i)^{-1/2}$, we can apply Hanson-Wright inequality conditional on $\mathsf{U}_n, \mathbf{E}$,
\begin{align*}
    \Delta_{2,2,1,a} = O_{\psi_1}(n^{\frac{1}{2} - \balpha}(\min_i N_i)^{-1/2}).
\end{align*}
Since $g_i^{\prime}(1,\pi)$'s are independent to $W_i$, by Lemma~\ref{sa-lem:fixed-temp-be}, $$n^{-\balpha}\sum_{i = 1}^n (W_i - \pi) g_i^{\prime}(1,\pi) = O_{\psi_{\beta,h},tc}(1).$$ By Equation~\ref{eq: conditional mean and var on Un}, Lipschitzness of $\tanh$ and Lemma~\ref{sa-lem:sub-gaussian}, $\E[W_i|\mathsf{U}_n] - \pi = O_{\psi_{\beta,h}}(n^{-\bgamma})$, hence  
\begin{align*}
    \Delta_{2,2,1,b} & = \left(\E[W_i|\mathsf{U}_n] - \pi \right) n^{-\balpha} \sum_{i = 1}^n  \frac{T_i - \E[p_i]}{\E[p_i]} g_i^{\prime}(1,\pi)
    = O_{\psi_{\beta,h}, tc} \left((\log n)^{-1/2} n^{-\bgamma}\right).
\end{align*}
Then by concentration of $\frac{M_i}{N_i}$ from Lemma~\ref{lem: concentration of M_i/N_i}, we have
\begin{align*}
    |\Delta_{2,2,1,c}| & = \left|\frac{\E[W_i|\mathsf{U}_n] - \pi}{2 \E[p_i]} n^{-\balpha} \sum_{i = 1}^n g_i^{\prime}(1,\pi) \left(\frac{M_i}{N_i} - \pi\right)\right| \\
    & \leq n^{\bgamma} \left|\frac{\E[W_i|\mathsf{U}_n] - \pi}{2 \E[p_i]}\right| \cdot \max_{i \in [n]} \left|\frac{M_i}{N_i} - \pi \right| \\
    & = O_{\psi_{2},tc}\left(\log n \max_{i \in [n]}\E[N_i|\mathbf{U}]^{-1/2}\right) + O_{\psi_{\beta,\gamma},tc}(n^{-\bgamma}).
\end{align*}
The bound for $\Delta_{2,2,1,d}$ follows from the definition of $\bH$ and $\mathsf{U}_n$,
\begin{align*}
    \Delta_{2,2,1,d} = n^{\bgamma} \left(\tanh\left( \sqrt{\frac{\beta}{n}} \mathsf{U}_n + h\right) - \E \left[\tanh\left( \sqrt{\frac{\beta}{n}} \mathsf{U}_n + h\right)\right] \right)^2 = O_{\psi_{\beta, \gamma}}(n^{-\bgamma}).
\end{align*}
For $\Delta_{2,2,2}$, a Taylor expansion of $p_i$ in terms of $\mca m_i$, and the concentration of $\frac{M_i}{N_i}$ in Lemma~\ref{lem: concentration of M_i/N_i} implies that
\begin{align*}
    |\Delta_{2,2,2}| & \lesssim n^{-\balpha} \sum_{i = 1}^n |p_i^{-1} - \E[p_i]^{-1}| \cdot \Big|\frac{M_i}{N_i} - \pi \Big| \\
    & \lesssim n^{\bgamma} \max_{1 \leq i \leq n} |\exp(- 2 \beta \mca m_i - 2h) - \E[\exp( - 2 \beta \mca m_i - 2 h)]| \cdot \Big|\frac{M_i}{N_i} - \pi \Big| \\
    & \lesssim n^{\bgamma} |\mca m - \pi| \cdot \max_{1 \leq i \leq n} \Big|\frac{M_i}{N_i} - \pi \Big| + O(n^{-1}) \\
    & = O_{\psi_{2},tc}\left(\log n \max_{i \in [n]}\E[N_i|\mathbf{U}]^{-1/2}\right) + O_{\psi_{2},tc}(n^{-1/2}).
\end{align*}

% In the case $h = 0$ and $\beta \in [0,1]$, we replace $n^{-\bgamma}$ in the rates by $n^{-1/4}$. We will show that we can make the constants in error terms not depending on $\beta$ and $n$: For $\Delta_{2,2,1,a}$, it is because $W_i \in \{-1,1\}$, hence the constants in Hanson-Wright does not depend on $n$. For $\Delta_{2,2,1,b}$, $\Delta_{2,2,1,c}$ and $\Delta_{2,2,1,d}$, it is because Lemma~\ref{sa-lem: localization to singularity} and Lemma~\ref{sa-lem:sub-gaussian} imply $\norm{\mca m - \pi}_{\psi_1} \leq \mathtt{C} n^{-1/4}$ and $\norm{\mathsf{U}_n}_{\psi_1} \leq \mathtt{C} n^{1/4}$, where $\mathtt{C}$ is an absolute constant.

\subsection{Proof of Lemma~\ref{lem:delta_231}}

Throughout the proof, the Ising spins $\bW=(W_i)_{i=1}^n$ are distributed according to Assumption~\ref{assump-one-block} with parameters $(\beta,h)$. For brevity, we write $\P$ in place of $\P_{\beta,h}$.

Take $\mathsf{U}_n$ to be the latent variable given in Lemma~\ref{sa-lem: definetti}. We further decompose by
\begin{align*}
    & \Delta_{2,3,1} = n^{-\balpha} \sum_{i = 1}^n \frac{1}{2} g_i^{(2)}(1, \eta_i^{\ast}) \left( \sum_{j \neq i} \frac{E_{ij}}{N_i}(W_i - \pi)\right)^2 
    = \Delta_{2,3,1,a} + \Delta_{2,3,1,b} + \Delta_{2,3,1,c},
\end{align*}
where $\eta_i^{\ast}$ is some value between $\pi$ and $M_i/N_i$, and 
\begin{align}\label{sa-eq:delta-231-decomp}
    \nonumber \Delta_{2,3,1,a} & = n^{-\balpha} \sum_{i = 1}^n \frac{1}{2} g_i^{(2)}(1, \eta_i^{\ast}) \left( \sum_{j \neq i} \frac{E_{ij}}{N_i}(W_j - \E[W_j|\mathsf{U}_n])\right)^2, \\
    \nonumber \Delta_{2,3,1,b} & = n^{-\balpha} \sum_{i = 1}^n \frac{1}{2} g_i^{(2)}(1, \eta_i^{\ast}) \left( \sum_{j \neq i} \frac{E_{ij}}{N_i}(W_j - \E[W_j|\mathsf{U}_n])\right)\left(\E[W_j|\mathsf{U}_n] - \pi\right), \\
    \Delta_{2,3,1,c} & =n^{-\balpha} \sum_{i = 1}^n \frac{1}{2} g_i^{(2)}(1, \eta_i^{\ast}) \left(\E[W_j|\mathsf{U}_n] - \pi\right)^2.
\end{align}
\begin{center}
    \textbf{Part I: $\Delta_{2,3,1,c}$.}
\end{center}
Since $\E[W_i|\mathsf{U}_n,\mathbf{U}] = \tanh\left(\sqrt{\frac{\beta}{n}}\mathsf{U}_n + h \right)$, we have $\E[W_i|\mathsf{U}_n] - \pi = O_{\psi_{\beta,h}}(n^{-\bgamma})$ and $(\E[W_i|\mathsf{U}_n] - \pi)^2 = O_{\psi_{\pgamma/2}}(n^{-2\bgamma})$. It then follows from boundness of $g_i^{(2)}(1, \eta_i^{\ast})$ that
\begin{align*}
    \Delta_{2,3,1,c} = O_{\psi_{\pgamma/2}}(n^{-\bgamma}).
\end{align*}

\begin{center}
    \textbf{Part II: $\Delta_{2,3,1,b}$.}
\end{center}
Condition on $\mathsf{U}_n$, $W_i$'s are i.i.d. By Mc-Diarmid inequality conditional on $\mathsf{U}_n$ for each $\sum_{j \neq i}\frac{E_{ij}}{N_i}(W_j - \E[W_j|\mathsf{U}_n])$ and using a union bound over $i \in [n]$, for all $i \in [n]$, for all $t > 0$,
\begin{align*}
    \P \left(\left|\Delta_{2,3,1,b}\right| \geq 2 \max_i N_i^{-1/2} n^{\bgamma} |\E[W_j|\mathsf{U}_n] - \pi| \sqrt{t} \middle| \mathsf{U}_n, \mathbf{E} \right) \leq 2 n \exp(-t).
\end{align*}
The tails for $n^{\bgamma}(\E[W_j|\mathsf{U}_n] - \pi)$ are also controlled,
\begin{align*}
    \P \left( n^{\bgamma}\left|\E[W_j|\mathsf{U}_n] - \pi \right| \geq C_{\beta,h} (\log n)^{1/\pgamma}\right) \leq n^{-1/2}.
\end{align*}
Integrate over the distribution of $\mathsf{U}_n$ and using a union bound, for large $n$, for all $t > 0$,
\begin{align*}
    \P \left( |\Delta_{2,3,1,b}| \geq 2 C_{\beta,h} \max_i N_i^{-1/2} t^{ 1/\pgamma}\middle| \mathbf{E} \right) \leq 2 n \exp(-t) + C_{\beta,h} n^{-1/2}.
\end{align*}
By Equation~\ref{N_i lower bound}, condition on $\mathbf{U}$ such that $A(\mathbf{U}) \in \mathcal{A}$, $\P(N_i \leq \E[N_i|\mathbf{U}]/3|\mathbf{U}) \leq n^{-100}$. Hence for such $\mathbf{U}$,
\begin{align*}
    \P \left(|\Delta_{2,3,1,b}| \geq 4 C_{\beta,h} \max_i \E[N_i|\mathbf{U}]^{-1/2}t^{ 1/\pgamma}\middle| \mathbf{U} \right) \leq 2 n \exp(-t) + C_{\beta,h}n^{-1/2}.
\end{align*}
In other words, conditional on $\mathbf{U}$ s.t. $A(\mathbf{U}) \in \mathcal{A}$,
\begin{align*}
    \Delta_{2,3,1,b} = O_{\psi_{\beta,h},tc}(\max_i \E[N_i|\mathbf{U}]^{-1/2}).
\end{align*}

% There exists an absolute constant $C$ such that  $\left|\E[W_i|U = u,\mathbf{U}] - \E[W_i|U = u^{\prime},\mathbf{U}] \right| = \left|\tanh\left(\sqrt{\frac{2 \beta}{n}}u + h \right) - \tanh\left(\sqrt{\frac{2 \beta}{n}}u^{\prime} + h \right) \right| \leq C \sqrt{\frac{2 \beta}{n}}|u - u^{\prime}|$. Hence $\left| \sum_{j \neq i} \frac{E_{ij}}{N_i}(\E[W_i|U,\mathbf{U}] - \pi)\right| \leq C \sqrt{\frac{2 \beta}{n}} |U - \E[U]|$. Contidion on $\bU$, $\bW_i$'s are i.i.d, hence by the same argument as in the proof of Lemma~\ref{lem: delta_2,2} with $\beta  = 0$, 
% \begin{align}\label{moment i.i.d}
%     \E \left[\left(\sum_{j \neq i} \frac{E_{ij}}{N_i}(W_i - \E[W_i|U, \mathbf{U}]) \right)^2 \middle | U, \mathbf{E}\right] = O(N_i^{-1}).
% \end{align} 
% Hence by Cauchy-Schwarz inequality,
% \begin{align*}
%     & \E \left[|\Delta_{2,3,1,b}| \middle| \mathbf{E} \right] \\
%     \lesssim & n^{-\balpha} \sum_{i = 1}^n  B \E \left[\left(\sum_{j \neq i} \frac{E_{ij}}{N_i}(W_i - \E[W_i|U, \mathbf{U}]) \right)^2 \middle | U, \mathbf{E}\right]^{\frac{1}{2}} \E \left[\left( \sum_{j \neq i} \frac{E_{ij}}{N_i}(\E[W_i|U,\mathbf{U}] - \pi)\right)^2\right]^{\frac{1}{2}} \\
%     \lesssim & B n^{\frac{1}{2} - \balpha} (\min_i N_i)^{-\frac{1}{2}},
% \end{align*}
% and hence $\Delta_{2,3,1,b} - \E[\Delta_{2,3,1,b}|\mathbf{E}] = O_{\P} \left(B n^{\frac{1}{2} - \balpha} (\min_i N_i)^{-\frac{1}{2}}\right)$. 
% \comment{Conditional i.i.d and sub-gaussianity of latent $\mathsf{U}_n$}

\begin{center}
    \textbf{Part III: $\Delta_{2,3,1,a}$.}
\end{center}
For notational simplicity, we will denote
\begin{align*}
    B_i & = \frac{1}{2} g_i^{(2)}(1, \eta_i^{\ast}) \left( \sum_{j \neq i} \frac{E_{ij}}{N_i}(W_j - \E[W_j|\mathsf{U}_n])\right)^2 \\
    & = \frac{1}{2} \theta \left(\frac{M_i}{N_i}\right) \left( \sum_{j \neq i} \frac{E_{ij}}{N_i}(W_j - \E[W_j|\mathsf{U}_n])\right)^2=: F(\bW,\mathsf{U}_n),
\end{align*}
and since we assume $g_i(\ell,\cdot)$ is $C^4$ for $\ell \in \{-1,1\}$, we know $\theta(\ell,\cdot)$ is $C^2$ for $\ell \in \{-1,1\}$. Then we can decompose $\Delta_{2,3,1,a} - \E[\Delta_{2,3,1,a}|\mathbf{E}]$ as
\begin{align*}
    \Delta_{2,3,1,a} - \E[\Delta_{2,3,1,a}|\mathbf{E}] & = n^{-\balpha} \sum_{i = 1}^n \left( B_i - \E[B_i|\mathsf{U}_n,\mathbf{E}]\right) + n^{-\balpha} \sum_{i = 1}^n \left(\E[B_i|\mathsf{U}_n,\mathbf{E}] - \E[B_i|\mathbf{E}]\right).
\end{align*}
where $F$ is a function that possibly depends on $\beta(\mathbf{U})$ and $\mathbf{E}$. 

\smallskip

\noindent \textbf{First part of $\Delta_{2,3,1,a}$: }The first two terms have a quadratic form in $W_j - \E[W_j|\mathsf{U}_n]$, except for the term $\theta(M_i/N_i)$.  We will handle it via a generalized version of Hanson-Wright inequality. Fix $\mathsf{U}_n$ and $\mathbf{E}$, consider
\begin{align*}
    H(\bW) = n^{-1/2} \sum_{i = 1}^n \frac{1}{2} \theta \left(\frac{M_i}{N_i}\right) \left( \sum_{j \neq i} \frac{E_{ij}}{N_i}(W_j - \E[W_j|\mathsf{U}_n])\right)^2.
\end{align*}
Denoting by $D_k H$ the partial derivative of $H$ w.r.p to $W_k$ and $D_{k,l}$ the mixed partials, then
\begin{align*}
    D_k H(\bW) = & n^{-1/2} \sum_{i \neq k}^n \frac{1}{2} \theta^{\prime} \left(\frac{M_i}{N_i}\right) \frac{E_{ik}}{N_i} \left(\sum_{j \neq i} \frac{E_{ij}}{N_i}(W_j - \E[W_j|\mathsf{U}_n])\right)^2 \\
    & + n^{-1/2} \sum_{i \neq k}^n \theta \left(\frac{M_i}{N_i}\right) 2 \left(\sum_{j \neq i} \frac{E_{ij}}{N_i} (W_j - \E[W_j|\mathsf{U}_n]) \right) \frac{E_{ik}}{N_i}.
\end{align*}
Since we have assumed $f$ is at least $4$-times continuously differentiable, we can apply standard concentration inequalities for $\sum_{j \neq i}\frac{E_{ij}}{N_i}(W_j - \E[W_j|\mathsf{U}_n])$ to get
\begin{align*}
    |\E[D_k H(\bW)|\mathsf{U}_n, \mathbf{E}]| \lesssim n^{-1/2} \sum_{i =1}^n E_{ik} N_i^{-3/2}.
\end{align*}
Hence the gradient of $H$ is bounded by
\begin{align*}
    \norm{\E[D H(\bW)|\mathsf{U}_n, \mathbf{E}]}_2^2 \lesssim & \sum_{k = 1}^n n^{-1} \left( \sum_{i = 1}^n E_{ik} N_i^{-3/2} \right)^2 \\
    \lesssim & \sum_{k = 1}^n n^{-1} \left(\sum_{i = 1}^n E_{ik}N_i^{-3} + \sum_{j_1 = 1}\sum_{j_2 \neq j_1} \frac{E_{j_1 k} E_{j_2 k}}{N_{j_1}^{3/2} N_{j_2}^{3/2}} \right) \\
    \lesssim & \frac{\max_i N_i^2}{\min_i N_i^3}.
\end{align*}
Moreover, the mix partials are
\begin{align*}
    D_{k,l}H(\bW) = & n^{-1/2} \sum_{i \neq k,l}^n \theta^{\prime \prime} \left(\frac{M_i}{N_i}\right) \left(\sum_{j\neq i} \frac{E_{ij}}{N_i}(W_j - \E[W_j|\mathsf{U}_n]) \right)^2 \frac{E_{ik} E_{il}}{N_i^2} \\
    & + 2 n^{-1/2} \sum_{i =1}^n \theta^{\prime}\left(\frac{M_i}{N_i}\right) 2 \left(\sum_{j\neq i} \frac{E_{ij}}{N_i}(W_j - \E[W_j|\mathsf{U}_n]) \right) \frac{E_{ik} E_{il}}{N_i^2}  \\
    & \qquad + n^{-1/2} \sum_{i =1}^n \theta \left(\frac{M_i}{N_i}\right)\frac{E_{ik} E_{il}}{N_i^2}.
\end{align*}
Hence $\infnorm{D_{k,l}H(\bW)} \lesssim n^{-1/2}\sum_{i =1}^n \frac{E_{ik} E_{il}}{N_i^2}$. Hence
\begin{align*}
    \infnorm{\norm{H F}_{F}^2} \lesssim & \sum_{k = 1}^n \sum_{l = 1}^n \left(n^{-1/2} \sum_{i = 1}^n \frac{E_{ik} E_{il}}{N_i^2} \right)^2 
    \lesssim  n^{-1} \sum_{i_1 = 1}^n \sum_{l =1}^n \frac{E_{i_1 l}}{N_{i_1}} \sum_{k = 1}^n \frac{E_{i_1 k}}{N_{i_1}} \sum_{i_2 = 1}^n \frac{E_{i_2 k}}{N_{i_2}} \frac{1}{N_{i_2}}
    \lesssim \frac{\max_i N_i}{\min_i N_i^2}.
\end{align*}
Moreover, since $H F$ is symmetric,
\begin{align*}
    \infnorm{\norm{H F}_2} \leq \infnorm{\norm{H F}_1} \lesssim \max_{k} \sum_{l =1}^n n^{-1/2} \sum_{i = 1}^n \frac{E_{ik}E_{il}}{N_i^2} \lesssim n^{-1/2}\frac{\max_i N_i}{\min_i N_i}.
\end{align*}
Hence by Theorem 3 from \cite{dagan2021learning}, for all $t > 0$,
\begin{align*}  
    \P \bigg( \bigg|n^{-1/2} \sum_{i = 1}^n (B_i & - \E[B_i|\mathsf{U}_n, \mathbf{E}]) \bigg| \geq t \bigg | \mathsf{U}_n, \mathbf{E} \bigg) \\
    & \leq \exp \left(-c \min \left(\frac{t^2}{\frac{\max_i N_i^2}{\min_i N_i^3} + \frac{\max_i N_i}{\min_i N_i^2}}, \frac{t}{n^{-1/2} \frac{\max_i N_i}{\min_i N_i}}\right) \right).
\end{align*}
By Equation~\ref{N_i lower bound} and a similar argument for upper bound, for each $i \in [n]$, conditional on $\mathbf{U}$ such that $A(\mathbf{U}) \in \mathcal{A}$, with probability at least $1 - n^{-100}$, $\E[N_i|\mathbf{U}]/2 \leq N_i \leq 2 \E[N_i|\mathbf{U}]$. Hence for each $t > 0$,
\begin{align*}
    \P \left(\left|n^{-1/2} \sum_{i = 1}^n (B_i - \E[B_i|\mathsf{U}_n, \mathbf{E}]) \right| \geq 8 \max_i \E[N_i|\mathbf{U}]^{-1/2}\sqrt{t} + 8 C_{\beta,h} n^{-1/2}t \middle | \mathbf{U}\right) \leq & \exp(-t) \\
    & \qquad + n^{-99},
\end{align*}
that is
\begin{align}\label{231a I}
    n^{-\balpha} \sum_{i = 1}^n \left(B_i - \E[B_i|\mathsf{U}_n, \mathbf{E}] \right) = O_{\psi_2, tc}\left(n^{\frac{1}{2} - \balpha} \max \E[N_i|\mathbf{U}]^{-1/2} \right) + O_{\psi_1, tc} \left(n^{-1/2}\right).
\end{align}

\smallskip

\noindent \textbf{Second part of $\Delta_{2,3,1,a}$: } Next, we will show $n^{1 - \balpha} \left(\E \left[B_i \middle| \mathsf{U}_n, \mathbf{U}, \mathbf{E} \right] - \E \left[B_i | \mathbf{E} \right]\right)$, is small. There exists a function $F$ that possibly depends on $\beta$ and $\mathbf{E}$ such that
\begin{align*}
    F(\mathbf{W},\mathsf{U}_n) = \frac{1}{2} \theta \left(\frac{M_i}{N_i}\right) \left( \sum_{j \neq i} \frac{E_{ij}}{N_i}(W_j - \E[W_j|\mathsf{U}_n])\right)^2.
\end{align*}
Define $p(u) = \P(W_j = 1|\mathsf{U}_n,\mathbf{U})$. Then
\begin{align*}
\E[B_i|\mathsf{U}_n = u, \mathbf{U}, \mathbf{E}] = \E [F(\bW,\mathsf{U}_n)|\mathsf{U}_n = u] = \sum_{\bw \in \{-1,1\}^n} \prod_{l = 1}^n p(u)^{w_l}(1 - p(u))^{1 - w_l} F(\bw,u).
\end{align*}
Using chain rule and product rule for derivatives,
\begin{align*}
    & \partial_u \E \left[B_i | \mathsf{U}_n = u, \mathbf{U} \right] \\
    = & \sum_{\bw \in \{-1,1\}^n} \Bigg[ \sum_{l = 1}^n \prod_{s \neq l} p(u)^{w_s}(1 - p(u))^{1 - w_s} \left(F((\bw_{-l}, w_l = 1), u) - F((\bw_{-l}, w_l = -1), u) \right) \\
    & \qquad + \prod_{i = 1}^n p(u)^{w_i}(1 - p(u))^{1- w_i} \partial_u F(\bw, u)\Bigg] p^{\prime}(u)\\
    = & \sum_{l = 1}^n \E_{\bW_{-l}} \left[ F((\bW_{-l}, W_l = 1), u) - F((\bW_{-l}, W_l = -1), u) \right] p^{\prime}(u) + \E_{\bW} \left[ \partial_u F(\bW, u) \right] p^{\prime}(u) \\
    = & \sum_{l = 1}^n O_{\P} \left( \frac{1}{\sqrt{N_i}} \frac{E_{il}}{N_i}\right) \infnorm{p^{\prime}} + O_{\P} \left( \frac{1}{\sqrt{N_i}} \infnorm{p^{\prime}} \right) \infnorm{p^{\prime}}
    = O_{\P}((n N_i)^{-0.5}),
\end{align*}
where in the last line, we have used
\begin{align*}
    & |D_{W_l} F(\bw, u)| \lesssim \infnorm{\theta^{\prime}} \frac{E_{il}}{N_i} \left( \sum_{j \neq i} \frac{E_{ij}}{N_i}(W_j - \E[W_j|U,\mathbf{U}])\right)^2 + \infnorm{\theta} \left|\sum_{j \neq i} \frac{E_{ij}}{N_i}(W_j - \E[W_j|U,\mathbf{U}])\right| \frac{E_{il}}{N_i}, \\
    & |\partial_u F(\bw, u)| \lesssim  \infnorm{\theta} \infnorm{p^{\prime}} \left|\sum_{j \neq i} \frac{E_{ij}}{N_i}(W_j - \E[W_j|U,\mathbf{U}])\right|,
\end{align*}
and that fact that $\infnorm{p^{\prime}} = O((2 \beta/n)^{0.5})$ and Hoeffiding's inequality for $\sum_{j \neq i}\frac{E_{ij}}{N_i}(W_j - \E[W_j|\mathsf{U}_n])$, 
\begin{align}\label{sa-eq:F lipschitz}
    \left|\partial_{u}\E\left[F(\bw,\mathsf{U}_n)\middle|\mathsf{U}_n = u, \mathbf{E} \right] \right| & \leq \E \left[\left|\partial_u F(\bw,\mathsf{U}_n) \right| \middle| \mathsf{U}_n = u\right]
    = O\left(n^{-1/2} \min_i N_i^{-1/2} \right).
\end{align}
Since $\mathsf{U}_n = O_{\psi_{\beta, h}}(n^{\balpha - 1/2})$, we have
\begin{align}\label{231a II}
    \nonumber n^{-\balpha} \sum_{i = 1}^n \left(\E[B_i| \mathsf{U}_n,\mathbf{U}] - \E[B_i|\mathbf{U}] \right) & = O_{\psi_{\beta,h}} \left(n^{1-\balpha} n^{-1/2} \min_i N_i^{-1/2} n^{\balpha - 1/2} \right) \\
    & = O_{\psi_{\beta,h}} \left(\min_i N_i^{-1/2}\right).
\end{align}
Combining Equations~\ref{231a I} and \ref{231a II}, conditional on $\mathbf{U}$ such that $A(\mathbf{U}) \in \mathcal{A}$, 
\begin{align*}
    n^{-\balpha} \sum_{i = 1}^n \left(B_i - \E[B_i|\mathbf{E}]\right) & = O_{\psi_2, tc}\left(n^{\frac{1}{2} - \balpha} \max \E[N_i|\mathbf{U}]^{-1/2} \right) + O_{\psi_1,tc}\left(n^{-1/2}\right) \\
    & \qquad + O_{\psi_{\beta,h},tc}\left(\max_i \E[N_i]^{-1/2} \right).
\end{align*}
Combining the bounds for $\Delta_{2,3,1,a}, \Delta_{2,3,1,b}, \Delta_{2,3,1,c}$, we get the desired result.

\subsection{Proof of Lemma~\ref{lem: delta_2,3,2}}

Throughout the proof, the Ising spins $\bW=(W_i)_{i=1}^n$ are distributed according to Assumption~\ref{assump-one-block} with parameters $(\beta,h)$. For brevity, we write $\P$ in place of $\P_{\beta,h}$.

Recall
\begin{align*}
    \Delta_{2,3,2} & = n^{-\balpha} \sum_{i = 1}^n \frac{1}{2} \frac{W_i - \E[W_i | \bW_{-i}]}{p_i}\left[g_i \left(1, \frac{M_i}{N_i} \right) - g_i \left(1, \pi \right) - g_i^{\prime} \left(1, \pi \right) \left(\frac{M_i}{N_i} - \pi \right)\right].    
\end{align*}
First, we will consider the effect of fluctuation of $p_i$ and $\E[W_i|\bW_{-i}]$. Recall
\begin{align*}
    \E[W_i|\bW_{-i}]  = \tanh \left(\beta \mca m_i + h \right),  \quad
    p_i  = \left(1 + \exp\left( - 2 \beta \mca m_i - 2h \right)\right)^{-1}.
\end{align*}
It follows from the boundeness of $\beta \mca m_i + h$, $\mca m_i - \pi = O_{\psi_{\beta,h}}(n^{-\bgamma})$ that for each $i \in [n]$, 
\begin{align*}
    \frac{W_i - \E[W_i|\bW_{-i}]}{p_i} = 2 \frac{W_i - \pi}{\pi + 1} + O_{\psi_{\beta,h}}(n^{-\bgamma}).
\end{align*}
Moreover for some $\eta_i^{\ast}$ between $M_i/N_i$ and $\pi$, using Lemma~\ref{lem: concentration of M_i/N_i} we have
\begin{align*}
    & g_i \left(1, \frac{M_i}{N_i} \right) - g_i \left(1, \pi \right) - g_i^{\prime} \left(1, \pi \right) \left(\frac{M_i}{N_i} - \pi \right) \\
    = & \frac{1}{2} g_i^{\prime \prime}(1, \eta_i^{\ast})\left(\frac{M_i}{N_i} - \pi\right)^2 = O_{\psi_{\pgamma/2},tc}(n^{-2\bgamma}) + O_{\psi_1,tc}(N_i^{-1}).
\end{align*}
Using a union bound over $i$ and an argument for the product of two terms with bounded Orlicz norm with tail control, we have
\begin{align*}
    \Delta_{2,3,2} = &  n^{-\balpha} \sum_{i = 1}^n \frac{W_i - \pi}{\pi + 1} \left[ g_i \left(1, \frac{M_i}{N_i} \right)  - g_i(1, \pi) - g_i^{\prime}(1,\pi) \left(\frac{M_i}{N_i} - \pi \right)\right] \\
    & + O_{\psi_{\pgamma/2},tc}((\log n)^{-1/\pgamma}n^{-2\bgamma}) + O_{\psi_{1},tc}((\log n)^{-1/\pgamma} N_i^{-1}).
\end{align*}
Next, we will show $n^{-\balpha} \sum_{i = 1}^n \frac{W_i - \pi}{\pi + 1} \left[ g_i \left(1, \frac{M_i}{N_i} \right)  - g_i(1, \pi) - g_i^{\prime}(1,\pi) \left(\frac{M_i}{N_i} - \pi \right)\right]$ is small. Suppose $g_i(1,\cdot)$ is $p$-times continuously differentiable. Define 
\begin{align*}
    \delta_p = n^{-\balpha} \sum_{i = 1}^n \frac{W_i - \pi}{\pi + 1} g_i^{(p)}\left( 1,\pi\right) \left( \frac{M_i}{N_i} - \pi\right)^p.
\end{align*}
We will use the conditioning strategy to analyse $\delta_p$: Decompse by
\begin{align*}
    \delta_p = \delta_{p,1} + \delta_{p,2} + \delta_{p,3},
\end{align*}
with
\begin{align*}
    \delta_{p,1} & = n^{-\balpha} \sum_{i = 1}^n \frac{W_i - \E[W_i|\mathsf{U}_n]}{\pi + 1} g_i^{(p)}\left( 1,\pi\right) \left( \frac{M_i}{N_i} - \E[W_i|\mathsf{U}_n]\right)^p, \\
    \delta_{p,2} & = n^{-\balpha} \sum_{i = 1}^n \frac{\E[W_i|\mathsf{U}_n] - \pi}{\pi + 1} g_i^{(p)}\left( 1,\pi\right) \left( \frac{M_i}{N_i} - \E[W_i|\mathsf{U}_n]\right)^p, \\
    \delta_{p,3} & = n^{-\balpha} \sum_{i = 1}^n \frac{W_i - \pi}{\pi + 1} g_i^{(p)}(1,\pi)\left[\left(\frac{M_i}{N_i} - \E[W_i|\mathsf{U}_n]\right)^p - \left(\frac{M_i}{N_i} - \pi\right)^p \right].
\end{align*}
First, we will show $\delta_{p,2}$ and $\delta_{p,3}$ are small. By Hoeffding inequality, $M_i/N_i - \E[W_i|\mathsf{U}_n] = O_{\psi_2}(N_i^{-1/2})$. Moreover, $\E[W_i|\mathsf{U}_n] - \pi = O_{\psi_{\beta,h}}(n^{-\bgamma})$. Hence 
\begin{align*}
    \delta_{p,2} = O_{\psi_{\beta,h},tc}(\max_i N_i^{-1/2}).
\end{align*}
For $\delta_{p,3}$, we have 
\begin{align*}
    \left(\frac{M_i}{N_i} - \E[W_i|\mathsf{U}_n] \right)^p - \left(\frac{M_i}{N_i} - \pi \right)^p 
    & = p \left(\frac{M_i}{N_i} - \xi^{\ast} \right)^{p-1}\left(\E[W_i|\mathsf{U}_n] - \pi\right),
\end{align*}
where $\xi^{\ast}$ is some quantity between $\E[W_i|\mathsf{U}_n]$ and $\pi$. Since $x \mapsto x^{p-1}$ is either monotone or convex and none-negative, condition on $\mathbf{E}$,
\begin{align*}
    \left|\frac{M_i}{N_i} - \xi^{\ast}\right|^{p-1} & \leq \max \left\{\left|\frac{M_i}{N_i} - \E[W_i|\mathsf{U}_n]\right|^{p-1}, \left|\frac{M_i}{N_i} - \pi \right|^{p-1}\right\} \\
    & = O_{\psi_{\frac{\pgamma}{p-1}}}(n^{-(p-1)\bgamma}) + O_{\psi_{\frac{2}{p-1}}}(N_i^{-\frac{p-1}{2}}).
\end{align*}
Combining with boundedness of $g_i^{(p)}(1,\pi)$ and tail control of $\E[W_i|\mathsf{U}_n]$, we have
\begin{align*}
    \delta_{p,3} & = O_{\psi_{\frac{\pgamma}{p-1}}}\left((\log n)^{\frac{1}{\pgamma}}n^{-(p-1)\bgamma}\right) + O_{\psi_{\frac{2}{p-1}}}\left((\log n)^{\frac{1}{\pgamma}} N_i^{-\frac{p-1}{2}}\right).
\end{align*}
For $\delta_{p,1}$, we will again use the generalized version of Hanson-Wright inequality. For each $k \in [n]$,
\begin{align*}
    \partial_k \delta_{p,1} = &  n^{-\balpha} \sum_{i \neq k} \frac{W_i - \E[W_i|\mathsf{U}_n]}{\pi + 1} g_i^{(p)}(1,\pi) p \left(\frac{M_i}{N_i} - \E[W_i|\mathsf{U}_n] \right)^{p-1}\frac{E_{ik}}{N_i} \\
    & \qquad + n^{-\balpha} g_k^{(p)}(1,\pi) \left(\frac{M_k}{N_k} - \E[W_i|\mathsf{U}_n]\right)^p.
\end{align*}
Hence condition on $\mathbf{E}$,
\begin{align*}
    \norm{\E \left[\nabla \delta_{p,1} \right]} = O\left(n^{1/2 - \balpha} N_i^{-(p-1)/2} \right).
\end{align*}
Taking mixed partials w.r.p $\delta_{p,1}$ and using boundedness of $g_i^{(p)}$, we have
\begin{align*}
    \infnorm{\partial_k \partial_l \delta_{p,1}} \lesssim n^{-\balpha} \sum_{i \neq k,l} \frac{E_{ik}E_{il}}{N_i^2} + n^{-\balpha}\frac{E_{lk}}{N_l} + n^{-\balpha}\frac{E_{kl}}{N_k}.
\end{align*}
It follows that
\begin{align*}
    \infnorm{\norm{\operatorname{Hess}(\delta_{p,1})}_2}\lesssim \infnorm{\norm{\operatorname{Hess}(\delta_{p,1})}_{F}} \lesssim n^{1/2 - \balpha} \left(\frac{\max_i N_i^3}{\min_i N_i^4}\right)^{1/2}.
\end{align*}
It then follows from Equation~\ref{N_i lower bound} and Theorem 3 in \cite{dagan2021learning} that conditional on $\mathbf{U}$ such that $A(\mathbf{U}) \in \mathcal{A}$,
\begin{align*}
    \delta_{p,1} - \E[\delta_{p,1}|\mathbf{E}] = O_{\psi_1, tc} \left(n^{1/2 - \balpha} \left( \frac{\max_i \E[N_i|\mathbf{U}]^3}{\min_i \E[N_i|\mathbf{U}]^4}\right)^{1/2} \right).
\end{align*}

\paragraph*{Trade-off Between Smoothness of $g_i(1, \cdot)$ and Sparsity of Graph}
Assume $g_i(1,\cdot)$ is $p+1$-times continuously differentiable. Then by the decomposition of $\Delta_{2,3,2}$, condition on $\mathbf{U}$ such that $A(\mathbf{U}) \in \mathcal{A}$,
\begin{align*}
    & \Delta_{2,3,2} - \E[\Delta_{2,3,2}|\mathbf{E}] \\
    = & \sum_{l =2}^p \delta_l - \E[\delta_l|\mathbf{E}] + n^{-\balpha}\sum_{i =1}^n\left[\frac{Y_i^{(p+1)}(1,\xi_i^{\ast})}{(p+1)!} \left(\frac{M_i}{N_i} - \pi\right)^{p+1}- \E \left[\frac{Y_i^{(p+1)}(1,\xi_i^{\ast})}{(p+1)!} \left(\frac{M_i}{N_i} - \pi\right)^{p+1}\middle| \mathbf{E} \right]\right] \\
    & +  O_{\psi_{\pgamma/2},tc}((\log n)^{-1/\pgamma}n^{-2\bgamma}) + O_{\psi_{1},tc}((\log n)^{-1/\pgamma} (\min_i\E[N_i|\mathbf{U}])^{-1}).
\end{align*}
Then by the concentration of $M_i/N_i - \pi$ given in Lemma~\ref{lem: concentration of M_i/N_i}, we have
\begin{align*}
    & \Delta_{2,3,2} - \E[\Delta_{2,3,2}|\mathbf{E}] \\
    = &  O_{\psi_{\pgamma/2},tc}((\log n)^{-1/\pgamma}n^{-2\bgamma}) + O_{\psi_{1},tc}((\log n)^{-1/\pgamma} (\min_i\E[N_i|\mathbf{U}])^{-1}) \\
    & + O_{\psi_1, tc} \left(n^{1/2 - \balpha} \left( \frac{\max_i \E[N_i|\mathbf{U}]^3}{\min_i \E[N_i|\mathbf{U}]^4}\right)^{1/2} \right) + O_{\psi_{2/(p+1)},tc}\left(n^{\bgamma} (\min_i \E[N_i|\mathbf{U}]^{-(p+1)/2}) \right).
\end{align*}

\subsection{Proof of Lemma~\ref{sa-lem: hajek}}
For notational simplicity, denote $\wh{\mca p} = \frac{1}{n}\sum_{i = 1}^n T_i$ and $\mca p = \frac{1}{2}\tanh(\beta \pi + h) + \frac{1}{2} = \frac{1}{2} \pi + \frac{1}{2}$. Then
\begin{align*}
    & \frac{1}{n} \sum_{i = 1}^n \frac{T_i Y_i}{\wh{\mca p}} - \frac{1}{n} \sum_{i =1}^n \frac{T_i Y_i}{\mca p} 
    = \frac{1}{n}\sum_{i=1}^n \frac{T_i Y_i}{\wh{\mca p}} \frac{\mca p - \wh{\mca p}}{\mca p}.
\end{align*}
Taylor expand $x \mapsto \tanh(\beta x + h)$ at $x = \pi$, we have
\begin{align*}
    2(\wh{\mca p} - \mca p) = & \mca m - \tanh(\beta \mca m + h)\\
    = & \pi + \mca m  - \pi - \tanh(\beta \pi + h) - \beta \sech^2(\beta \pi + h) (\mca m - \pi) + O((\mca m - \pi)^2)\\
    = & (1 - \beta \sech^2(\beta \pi + h))(\mca m - \pi) + O((\mca m - \pi)^2),
\end{align*}
where $O(\cdot)$ is up to a universal constant. Together with concentration of $\frac{1}{n}\sum_{i = 1}^n T_i Y_i$ towards $p \E[Y_i]$, we have
\begin{align*}
    \frac{1}{n} \sum_{i = 1}^n \frac{T_i Y_i}{\wh{\mca p}} - \frac{1}{n} \sum_{i =1}^n \frac{T_i Y_i}{\mca p} 
    = - \frac{1 -\beta(1 - \pi^2)}{1 + \pi} \E[g_i(1,\frac{M_i}{N_i})] + O_{\psi_1}(n^{-2\bgamma}).
\end{align*}
A Taylor expansion of $g_i$ and concentration of $M_i / N_i$ then implies
\begin{align*}
    \E[g_i(1,\frac{M_i}{N_i})] 
    = \E[g_i(1,\pi)] + \E[g_i^{(1)}(1,\pi)(\frac{M_i}{N_i} - \pi)] + \frac{1}{2}\E[g_i^{(2)}(1,\pi^{\ast})(\frac{M_i}{N_i} - \pi)^2] = O(n^{-2\bgamma}).
\end{align*}
The conclusion then follows.

\subsection{Proof of Lemma~\ref{sa-lem:lin-fixed-temp}}

Throughout the proof, the Ising spins $\bW=(W_i)_{i=1}^n$ are distributed according to Assumption~\ref{assump-one-block} with parameters $(\beta,h)$. For brevity, we write $\P$ in place of $\P_{\beta,h}$.

By Lemma~\ref{sa-lem: approx delta_1} to Lemma~\ref{sa-lem: hajek}, we show
\begin{align}\label{linearization I}          
&n^{\bgamma}(\wh\tau_n - \tau_n) \\
=& n^{-\balpha}\sum_{i = 1}^n (R_i - \E[R_i] + b_i)(W_i - \pi)  + \varepsilon,
\end{align}
where $R_i = \frac{g_i(1,\frac{M_i}{N_i})}{1 + \pi} + \frac{g_i(-1,\frac{M_i}{N_i})}{1 - \pi}$, and $b_i  = \sum_{j \neq i} \frac{E_{ij}}{N_j} g_j^{\prime}\left(1, \pi \right)$, and $\varepsilon$ is such that condition on $\mathbf{U}$ such that $A(\mathbf{U}) \in \mathcal{A} = \{A \in \bb R^{n \times n}: \min_{i \in [n]} \sum_{j \neq i}A_{ij} \geq 32 \log n\}$, 
\begin{align}\label{linearization II}
    \nonumber & \varepsilon =  O_{\psi_{1},tc}\left(\log n \max_{i \in [n]}\E[N_i|\mathbf{U}]^{-1/2}\right) + O_{\psi_1,tc}(\sqrt{\log n} n^{-\bgamma})  \\
    & + O_{\psi_1, tc} \bigg(n^{1/2 - \balpha} \bigg( \frac{\max_i \E[N_i|\mathbf{U}]^3}{\min_i \E[N_i|\mathbf{U}]^4}\bigg)^{1/2} \bigg) + O_{\psi_{2/(p+1)},tc}\left(n^{\bgamma} (\min_i \E[N_i|\mathbf{U}]^{-(p+1)/2}) \right).
\end{align}
Following the strategy as in the proof of Theorem 4 in \cite{li2022random}, we will show $b_i$ is close to $R_i$: First, decompose by
\begin{align*}
    & \sum_{j\neq i}\frac{E_{ij}}{N_j}g_j^{\prime}(1,\pi) - R_i \\
    = & \sum_{j \neq i} \frac{E_{ij}}{N_j}g_j^{\prime}(1,\pi) - \sum_{j\neq i}\frac{E_{ij}}{n \E[G(U_i,U_j)|U_j]}g_j^{\prime}(1,\pi) + \sum_{j\neq i}\frac{E_{ij}}{n \E[G(U_i,U_j)|U_j]}g_j^{\prime}(1,\pi) - R_i.
\end{align*}
By Equation~\ref{N_i lower bound}, condition on $\mathbf{U}$ such that $A(\mathbf{U}) \in \mathcal{A}$, $$|\sum_{j \neq i} \frac{E_{ij}}{N_j}g_j^{\prime}(1,\pi) - \sum_{j\neq i}\frac{E_{ij}}{n \E[G(U_i,U_j)|U_j]}g_j^{\prime}(1,\pi)| \leq C n^{-1/2}$$ with probability at least $1 - n^{-99}$. Moreover, $\frac{E_{ij}}{\E[G(U_i,U_j)|U_j]}g_j^{\prime}(1,\pi), j \neq i$ are i.i.d condition on $U_i$, hence $\sum_{j \neq i}\frac{E_{ij}}{n \E[G(U_i,U_j)|U_j]}g_j^{\prime}(1,\pi) - R_i = O_{\psi_2}((n \E[G(U_i,U_j)|U_j]^{-1/2}) = O_{\psi_2}(\E[N_j|X]^{-1/2})$. It follows that conditional on $\mathbf{U}$ such that $A(\mathbf{U}) \in \mathcal{A}$,
\begin{align}\label{Qni}
    \max_i |\sum_{j \neq i}\frac{E_{ij}}{N_j}g_j^{\prime}(1,\pi) - R_i | = O_{\psi_2,tc}(\max_i \E[N_i|\mathbf{U}]^{-1/2}).
\end{align}
Again using the conditional i.i.d decomposition, Hoeffiding inequality and $\mathsf{U}_n$'s concentration for the two terms respectively,
\begin{align}\label{sa-eq: lin error network}
    \nonumber & |n^{-\balpha}\sum_{i =1}^n [\sum_{j \neq i}\frac{E_{ij}}{N_j}g_j^{\prime}(1,\pi) - R_i](W_i - \pi)|\\
    \nonumber \leq &  |n^{-\balpha}\sum_{i =1}^n [\sum_{j \neq i}\frac{E_{ij}}{N_j}g_j^{\prime}(1,\pi) - R_i](W_i - \E[W_i|\mathsf{U}_n])|  \\
    \nonumber & + n^{\bgamma} |\E[W_i|\mathsf{U}_n] - \pi|\max_i |\sum_{j \neq i}\frac{E_{ij}}{N_j}g_j^{\prime}(1,\pi) - R_i | \\
    = & O_{\psi_{2}} (n^{\frac{1}{2}- \balpha}\max_i \E[N_i|\mathbf{U}]^{-1/2}) + O_{\psi_{\beta,h},tc} ((\log n)^{1/\pgamma}\max_i \E[N_i|\mathbf{U}]^{-1/2}) = \varepsilon^{\prime}.
\end{align}
Hence denote the term of stochastic linearization by $G_n$, i.e.
\begin{align*}
    G_n = n^{-\balpha}\sum_{i =1}^n (R_i - \E[R_i] + Q_i)(W_i - \pi).
\end{align*}
Since $R_i - \E[R_i] + Q_i$'s are i.i.d independent to $W_i$'s with bounded third moment, we know from Lemma~\ref{sa-lem:fixed-temp-be} that $G_n$ can be approximated by either a Gaussian or non-Gaussian law, that is order $1$, this gives
\begin{align*}
    & \sup_{t \in \reals}\P \left(\wh \tau_n - \tau_n |\mathbf{U}) \leq t \right) - \P\left(G_n \leq t|\mathbf{U}\right) \\
    \leq & \sup_{t \in \reals} \min_{u > 0} \P \left(G_n \leq t + u \right) - \P\left(G_n \leq t\right) + \P(\varepsilon + \varepsilon^{\prime} \geq u) \\
    \leq & \sup_{t \in \reals}\min_{u > 0} \P \left(G_n \leq t + u \right) - \P\left(G_n \leq t + u\right) + \P(\varepsilon + \varepsilon^{\prime} \geq u) + \P \left(t \leq G_n \leq t + u\right) \\
    \leq & O(n^{-1/2}) + \min_{u > 0}\exp(-(u/\mathtt{r})^{\mathtt{a}}) + \mathtt{c} u \\
    = & O((\log n)^{\mathtt{a}}\mathtt{r}(\mathbf{U})),
\end{align*}
where $O(\cdot)$ does not depend on the value of $\mathbf{U}$ and
\begin{align*}
    \mathtt{r}(\mathbf{U}) = &  n^{-\bgamma} + \max_i \E[N_i|\mathbf{U}]^{-1/2} + n^{1/2- \balpha}\left(\frac{\max_i\E[N_i|\mathbf{U}]^3}{\min \E[N_i|\mathbf{U}]^4} \right)^{1/2} \\
    & + n^{\bgamma}\max_i \E[N_i|\mathbf{U}]^{-(p+1)/2}.
\end{align*}
To analyse the second term, recall $\E[N_i|\mathbf{U}]  = \rho_n \sum_{j \neq i}G(U_i,U_j)$. Hence
\begin{align*}
    & \E \left[\max_i \left(\E[N_i|\mathbf{U}]\right)^{-1/2}\Indicator(A(\mathbf{U}) \in \mathcal{A})\right] \\
    = & (n \rho_n)^{-1/2}\E \left[\max_i\left(\frac{1}{n}\sum_{j \neq i} G(U_i,U_j) \right)^{-1/2} \Indicator(A(\mathbf{U}) \in \mathcal{A})\right] \\
    = & O(\sqrt{\log n}(n \rho_n)^{-1/2}),
\end{align*}
the last line is because with probability at least $1 - n^{-98}$, $E = \{\frac{1}{2}g(U_i) \leq \frac{1}{n}\sum_{j \neq i}G(U_i, U_j) \leq 2 g(U_i), \forall 1 \leq i \leq n\}$ happens, and by maximal inequality, $\max_i |g(U_i)|^{-1/2} = O_{\psi_2}(\sqrt{\log n})$. And on $\{A(\mathbf{U}) \in \mathcal{A}\} \cap E$, $\max_i (\frac{1}{n}\sum_{j \neq i}G(U_i,U_j))^{-1/2} \leq (32 \log n/n)^{-1/2}$, since we assume $G$ is positive. By similar argument for the last two terms in $\mathtt{r}(\mathbf{U})$, we have
\begin{align*}
    \E \left[r(\mathbf{U}) \Indicator(A(\mathbf{U}) \in \mathcal{A})\right] & \leq n^{-\bgamma} + \sqrt{\log n} (n \rho_n)^{-1/2} + \sqrt{\log n} n^{\bgamma}(n \rho_n)^{-(p+1)/2}.
\end{align*}
Recall that $\mathcal{A} = \{A(\mathbf{U}): \min_i \sum_{j \neq i} A_{ij}(\mathbf{U})\geq 32 \log n \}$. Since $\sum_{j \neq i} A_{ij}(\mathbf{U}) \sim \operatorname{Bin}(n-1,\E[G(X_1,X_2)])$, we know from Chernoff bound for Binomials and union bound over $i$ that $\P(A(\mathbf{U}) \notin \mathcal{A}) \leq n^{-99}$. The conclusion then follows.

\subsection{Proof of Lemma~\ref{sa-lem:lin-stoch-unif}}
Our proof for Lemma~\ref{sa-lem: approx delta_1} to Lemma~\ref{sa-lem: hajek} relies on the following devices:

(1) Taylor expansion of $\tanh(\cdot)$ in the inverse probability weighting for unbiased estimator, and taylor expansion of $Y_i(\ell,\cdot)$ at $\E[T_i]$ for $\ell \in \{0,1\}$. 
Then the higher order terms are in terms of $\mca m - \pi$ and $\frac{M_i}{N_i} - \pi$. In Lemma~\ref{sa-lem: localization to singularity} (taking $X_i \equiv 1$), we show
\begin{align*}
    \norm{\mca m}_{\psi_1} \leq \mathtt{K} n^{-1/4},
\end{align*}
and in Lemma~\ref{lem: concentration of M_i/N_i}, we show
\begin{align*}
    \norm{\frac{M_i}{N_i}}_{\psi_1} \leq \mathtt{K} n^{-1/4} + \mathtt{K} (n \rho_n)^{-1/2},
\end{align*}
where $\mathtt{K}$ is some constant that does not depend on $\beta$. This shows for the higher order terms, we always have
\begin{align*}
    \mca m^2 = \mca m (1 + o_\P(1)), \qquad (M_i/N_i)^2 = (M_i/N_i)(1 + o_\P(1)),
\end{align*}
where the $o_\P(\cdot)$ terms does not depend on $\beta$.

(2) Condition i.i.d decomposition based on the de-Finetti's lemma (Lemma \ref{sa-lem: definetti}). Suppose $\mathsf{U}_n$ is the latent variable from Lemma~\ref{sa-lem: definetti}, we use decompositions based on $\mathsf{U}_n$: For Lemma~\ref{lem: delta_2,2} to Lemma~\ref{lem: delta_2,3,2}, we break down higher order terms in the form
\begin{align*}
    & F(\mathbf{W},\mathbf{E}) - \E[F(\mathbf{W},\mathbf{E})|\mathbf{E}] \\
    = & F(\mathbf{W},\mathbf{E}) - \E[F(\mathbf{W},\mathbf{E})|\mathbf{E},\mathsf{U}_n] + \E[F(\mathbf{W},\mathbf{E})|\mathbf{E},\mathsf{U}_n] - \E[F(\mathbf{W},\mathbf{E})|\mathbf{E}].
\end{align*}
For the first part $F(\mathbf{W},\mathbf{E}) - \E[F(\mathbf{W},\mathbf{E})|\mathbf{E},\mathsf{U}_n]$, we use the conditional i.i.d of $W_i$'s given $\mathsf{U}_n$. For the second part, we use concentration from Lemma~\ref{sa-lem:sub-gaussian} that there exists a constant $\mathtt{K}$ not depending on $\beta$ or $n$, such that $\norm{\mathsf{U}_n}_{\psi_1} \leq \mathtt{K} n^{1/4}$ and the effective term $\norm{\tanh(\sqrt{\frac{\beta}{n}} \mathsf{U}_n)}_{\psi_1} \leq \mathtt{K} n^{-1/4}$. In particular, the rate of concentration for conditional i.i.d Berry-Esseen and concentration of $\tanh(\sqrt{\frac{\beta}{n}}\mathsf{U}_n)$ does not depend on $\beta$.

By the same proof from Lemma~\ref{sa-lem: approx delta_1} to Lemma~\ref{sa-lem: hajek}, we can show in $\wh \tau_n - \tau_n$, the second and higher order terms in terms of $W_i - \pi$ can always be dominated by the first order terms, with a rate that does not depend on $\beta$.

The conclusion then follows from the two devices and the same proof logic of Lemma~\ref{sa-lem: approx delta_1} to Lemma~\ref{sa-lem: hajek}.

\section{Proofs: Section~\ref{sec: jacknife}}

\subsection{Proof of Lemma~\ref{sa-lem:jacknife}}
Define $g(U_j) =\E[G(U_i,U_j)|U_j]$, for $i \neq j$. Reordering the terms,
\begin{align*}
    \overline{\tau}^a = \frac{n - 1}{n^2} \sum_{j \in [n]} \frac{T_j}{1/2}h_j(1,M_j/N_j) - \frac{1 - T_j}{1 - 1/2} h_j(-1,M_j/N_j).
\end{align*}
Hence $\tau^a_{(i)} - \overline{\tau}^a$ has the representation given by
\begin{align}\label{JACK-0}
    \nonumber & \tau^a_{(i)} - \overline{\tau}^a \\
     \nnb = & - \frac{1}{n} \frac{T_i}{1/2} h_i \Big(1,\frac{M_i}{N_i}\Big) + \frac{1}{n^2} \sum_{j \in [n]} \frac{T_j}{1/2} h_j\Big(1, \frac{M_j}{N_j}\Big)
    + \frac{1}{n} \frac{1 - T_i}{1 - 1/2} h_i \Big(1,\frac{M_i}{N_i}\Big) \\
    & \qquad - \frac{1}{n^2} \sum_{j \in [n]} \frac{1 - T_j}{1 - 1/2} h_j\Big(1, \frac{M_j}{N_j}\Big)\\
    \nonumber = & - \frac{1}{n} \Big(\frac{T_i}{1/2} h_i(1,0) - 1/2 \E[h_i(1,0)] \Big) 
    + \frac{1}{n} \Big(\frac{1 - T_i}{1 - 1/2} h_i(-1,0) - (1 - 1/2) \E[h_i(-1,0)] \Big) \\
    & \qquad + O_{\psi_{2,tc}}(n^{-1}(n \rho_n)^{-\frac{1}{2}}) \\
    = & -\frac{1}{n} \Big(\frac{h_i(1,0)}{1/2} + \frac{h_i(-1,0)}{1 - 1/2}\Big)(T_i - 1/2) + O_{\psi_{2,tc}}(n^{-1}(n \rho_n)^{-\frac{1}{2}}) \\
    = & -\frac{1}{n} \Big(\frac{f_i(1,0)}{1/2} + \frac{f_i(-1,0)}{1 - 1/2}\Big)(T_i - 1/2) + O_{\psi_{2,tc}}(n^{-1}(n \rho_n)^{-\frac{1}{2}}) + o_{\P}(n^{-1}),
\end{align}
where the second to last line is due to $-\frac{1}{n} \frac{1}{1/2} 1/2(h_i(1,0) - \E[h_i(1,0)]) + \frac{1}{n} \frac{1}{1 - 1/2} (1 - 1/2) (h_i(-1,0) - \E[h_i(-1,0)]) = - \frac{2}{n} \varepsilon_i + \frac{2}{n} \varepsilon_i = 0$.

Now we look at $b$-part. For representation purpose, we look at only the treatment part. The control part can be analysized by in the same way. Reordering the terms, 
\begin{align*}
    \overline{\tau}^b & = \frac{1}{n} \sum_{i \in [n]} \tau_{(i)}^b = \frac{1}{n} \sum_{i \in [n]} \frac{1}{n} \sum_{j \in [n]} \frac{T_j}{1/2} \bigg[h_j \Big(1, \frac{M_j}{N_j}_{(i)}\Big) - h_j \Big(1, \frac{M_j}{N_j} \Big)\bigg] \\
    & = \frac{1}{n} \sum_{j \in [n]} \frac{T_j}{1/2} \frac{1}{n}\sum_{i \in [n]} \bigg[h_j \Big(1, \frac{M_j}{N_j}_{(i)}\Big) - h_j \Big(1, \frac{M_j}{N_j} \Big)\bigg].
\end{align*}
Hence $\tau_{(i)}^b - \overline{\tau}^b$ has the representation given by
\begin{align}\label{JACK-1}
    \tau_{(i)}^b - \overline{\tau}^b = \frac{1}{n} \sum_{j \in [n]} \frac{T_j}{1/2} \bigg[h_j \Big(1,\frac{M_j}{N_j}_{(i)} \Big) - \frac{1}{n}\sum_{\iota \in [n]} h_j \Big(1,\frac{M_j}{N_j}_{(\iota)} \Big)\bigg].
\end{align}
The analysis follows from a Taylor expansion of $h_j(1,\cdot)$. For some $\xi_{j,i}^{\ast}$ between $\frac{M_j}{N_j}_{(i)}$ and $0$ for each $j,i$,
\begin{align}\label{JACK-Taylor}
    h_j \Big(1, \frac{M_j}{N_j}_{(i)} \Big) = &  h_j (1,0) + \partial_2 h(1,0) \Big(\frac{M_j}{N_j}_{(i)} - 0\Big) + \frac{1}{2} \partial_{2,2} h(1,0) \Big(\frac{M_j}{N_j}_{(i)} - 0\Big)^2 \\
    & + \frac{1}{6} \partial_{2,2,2} h(1,\xi_{j,i}^{\ast}) \Big(\frac{M_j}{N_j}_{(i)} - 0\Big)^3,
\end{align}
where we have used $\partial_2 h_j(1,\cdot) = \partial_2 [h(1,\cdot) + \varepsilon_j] = \partial_2 h(1,\cdot)$.

\paragraph*{Part 1: Linear Terms}
\begin{equation}\label{JACK-2}
\begin{aligned}
    \frac{M_j}{N_j}_{(i)} - \frac{1}{n}\sum_{\iota \in [n]} \frac{M_j}{N_j}_{(\iota)} 
    & = \sum_{l \neq i} \frac{E_{lj}}{N_j^{(i)}}W_l - \frac{1}{n}\sum_{\iota \in [n]}  \sum_{l \neq \iota} \frac{E_{lj}}{N_j^{(\iota)}}W_l\\
    & = \sum_{l = 1}^n E_{lj} W_l \bigg(\frac{1}{N_j^{(i)}} - \frac{1}{n} \sum_{\iota \in [n],\iota \neq l} \frac{1}{N_j^{(\iota)}} \bigg) - \frac{E_{ij}}{N_j^{(i)}}W_i.
\end{aligned}
\end{equation}
By a decomposition argument,
\begin{align*}
    \frac{1}{N_j^{(i)}} - \frac{1}{n}\sum_{\iota \in [n],\iota \neq l} \frac{1}{N_j^{(\iota)}} 
    & = \frac{1}{N_j^{(i)}} - \frac{1}{n-1}\sum_{\iota \in [n],\iota \neq l} \frac{1}{N_j^{(\iota)}} + \frac{1}{(n - 1)n} \sum_{\iota \in [n],\iota \neq l} \frac{1}{N_j^{(\iota)}} \\
    & = \frac{1}{n -  1} \sum_{\iota \in [n], \iota \neq l} \frac{E_{ji} - E_{j \iota}}{N_j^{(i)}N_j^{(\iota)}} + \frac{1}{(n - 1)n} \sum_{\iota \in [n],\iota \neq l} \frac{1}{N_j^{(\iota)}} \\
    & = n^{-1} (n \rho_n)^{-1} \frac{E_{ij} - \rho_n g(U_j)}{\rho_n g(U_j)^2} +  \frac{1}{(n - 1)n} \sum_{\iota \in [n],\iota \neq l} \frac{1}{N_j^{(\iota)}}.
\end{align*}
Hence
\begin{align*}
    & \sum_{l = 1}^n E_{lj} W_l \bigg(\frac{1}{N_j^{(i)}} - \frac{1}{n} \sum_{\iota \in [n],\iota \neq l} \frac{1}{N_j^{(\iota)}} \bigg) \\
    = & (n \rho_n)^{-1} \frac{E_{ij} - \rho_n g(U_j)}{\rho_n g(U_j)^2} \frac{1}{n}\sum_{l = 1}^n E_{lj} W_l + \frac{\sum_{l = 1}^n E_{lj}W_l}{N_j^{(i)}} O_{\psi_{2,tc}}((n \rho_n)^{-\frac{3}{2}}) \\
    & \quad + \frac{1}{n - 1} \sum_{\iota \in [n], \iota \neq l} \frac{\sum_{l = 1}^n E_{lj} W_l}{n N_j^{(\iota)}}.
\end{align*}
Condition on $U_j$, $(E_{lj} W_l: l \neq j)$ are i.i.d mean-zero, hence Bernstein inequality gives $\frac{1}{n} \sum_{l = 1}^n E_{lj} W_l = O_{\psi_2}(\sqrt{n^{-1}\rho_n}) + O_{\psi_1}(n^{-1})$, which implies
\begin{align*}
    (n \rho_n)^{-1} \frac{E_{ij} - \rho_n g(U_j)}{\rho_n g(U_j)^2} \frac{1}{n}\sum_{l = 1}^n E_{lj} W_l & = O_{\psi_2}((n \rho_n)^{-\frac{3}{2}}) + O_{\psi_1}((n \rho_n)^{-2}), \\
    \frac{1}{n - 1} \sum_{\iota \in [n], \iota \neq l} \frac{\sum_{l = 1}^n E_{lj} W_l}{n N_j^{(\iota)}} & = O_{\psi_2}(n^{-\frac{3}{2}} \rho_n^{-\frac{1}{2}}) + O_{\psi_1}(n^{-2}).
\end{align*}
Putting back into Equation~\eqref{JACK-2},
\begin{align*}
     \frac{M_j}{N_j}_{(i)} - \frac{1}{n}\sum_{\iota \in [n]} \frac{M_j}{N_j}_{(\iota)} &  = - \frac{E_{ij}}{N_j^{(i)}} W_i + O_{\psi_1}((n \rho_n)^{-\frac{3}{2}}).
\end{align*}
Looking at contribution from the first order term in Taylor expanding $h_j(1,\cdot)$ to $\tau_{(i)}^b - \overline{\tau}^b$ in Equation~\eqref{JACK-1},
\begin{align*}
    & \frac{1}{n} \sum_{j \in [n]} \partial_2 h(1,0) \frac{T_j}{1/2} \bigg[\frac{M_j}{N_j}_{(i)} - \frac{1}{n}\sum_{\iota \in [n]} \frac{M_j}{N_j}_{(\iota)}\bigg] \\
    = & - \sum_{j \in [n]} \partial_2 h(1,0) W_i \frac{1}{n} \frac{E_{ij}}{N_j^{(i)}} \frac{T_j}{1/2} + O_{\psi_{1,tc}}((n \rho_n)^{-\frac{3}{2}}) \\
    = & -  W_i \frac{1}{n} \sum_{j \in [n]} \partial_2 h(1,0) \frac{E_{ij}}{n \rho_n g(U_j)} \frac{T_j}{1/2} - W_i \frac{1}{n} \sum_{j \in [n]} \partial_2 h(1,0) \frac{E_{ij}}{N_j^{(i)}} \frac{n \rho_n g(U_j) - N_j}{n \rho_n g(U_j)} \frac{T_j}{1/2} \\
    & \qquad + O_{\psi_{1,tc}}((n \rho_n)^{-\frac{3}{2}}) \\
    = & - W_i \frac{1}{n} \sum_{j \in [n]} \partial_2 h(1,0) \frac{E_{ij}}{n \rho_n g(U_j)} \frac{T_j}{1/2} + O_{\psi_{1,tc}}((n \rho_n)^{-\frac{3}{2}}).
\end{align*}
Since $(E_{ij} T_j/g(U_j): j \in [n])$ are independent condition on $U_i$, standard concentration inequality gives
\begin{align*}
    & \frac{1}{n} \sum_{j \in [n]} \partial_2 h(1,0) \frac{T_j}{1/2} \bigg[\frac{M_j}{N_j}_{(i)} - \frac{1}{n}\sum_{\iota \in [n]} \frac{M_j}{N_j}_{(\iota)}\bigg] \\
    = & - W_i  \frac{1}{n} \sum_{j \in [n]} \partial_2 h(1,0) \frac{E_{ij}}{n \rho_n g(U_j)}\frac{T_j}{1/2} + O_{\psi_{1,tc}}((n \rho_n)^{-\frac{3}{2}}) \\
    = & - W_i \partial_2 h(1,0) \frac{1}{n} \sum_{j \in [n]}  \frac{E_{ij}}{n \rho_n g(U_j)}\frac{T_j}{1/2} + O_{\psi_{1,tc}}((n \rho_n)^{-\frac{3}{2}}) \\
    = & - \partial_2 h(1,0) \frac{W_i}{n} \E \bigg[\frac{E_{ij}}{\rho_n g(U_j)}\bigg|U_i\bigg] + O_{\psi_{1,tc}}((n \rho_n)^{-\frac{3}{2}}).
\end{align*}
Since we assumed $\partial_2 h(1,0) = \partial_2 f(1,0) + o_{\P}(1) = \partial_2 f_j(1,0) + o_{\P}(1)$ where
\begin{align*}
    & \frac{1}{n} \sum_{j \in [n]} \partial_2 h(1,0) \frac{T_j}{1/2} \bigg[\frac{M_j}{N_j}_{(i)} - \frac{1}{n}\sum_{\iota \in [n]} \frac{M_j}{N_j}_{(\iota)}\bigg] \\
    = & - \frac{W_i}{n} \E \bigg[\frac{E_{ij} \partial_2 f_j(1,0)}{\rho_n g(U_j)}\bigg|U_i\bigg] + O_{\psi_{1,tc}}((n \rho_n)^{-\frac{3}{2}}) + o_{\P}(n^{-1}).
\end{align*}
Together with the leading term in  Equation~\eqref{JACK-1}, we have
\begin{align*}
    & n \sum_{i \in [n]} \bigg( \frac{1}{n} \sum_{j \in [n]} \partial_2 h_j(1,0) \frac{T_j}{1/2}  \bigg[\frac{M_j}{N_j}_{(i)} - \frac{1}{n}\sum_{\iota \in [n]} \frac{M_j}{N_j}_{(\iota)}\bigg] + \tau_{(i)}^a - \overline{\tau}^a \bigg) \cdot \\
    & \qquad  \bigg(\frac{2}{n_q} \sum_{j \in \ca I_q} \partial_2 h_j(1,0) \frac{T_j}{\theta_q} \bigg[\frac{M_j}{N_j}_{(i)} - \frac{1}{n}\sum_{\iota \in [n]} \frac{M_j}{N_j}_{(\iota)}\bigg] + \tau_{(i)}^a - \overline{\tau}^a \bigg) \\
    = & \frac{n}{n^2} \sum_{i \in [n]} \bigg( \E \bigg[\frac{E_{ij} \partial_2 f_j(1,0)}{\rho_n g(U_j)}\bigg|U_i\bigg] + \frac{f_i(1,0)}{1/2} (T_i - 1/2)\bigg) \cdot \\
    & \qquad  \bigg( \E \bigg[\frac{E_{ij} \partial_2 f_j(1,0)}{\rho_n g(U_j)}\bigg|U_i\bigg] + \frac{f_i(1,0)}{1/2} (T_i - 1/2)\bigg) + O_{\psi_{1,tc}}((n \rho_n^3)^{-1}) + o_{\P}(1)\\
    = & \frac{n_l^2}{n^2}  \E\bigg[\Big(\E \bigg[\frac{E_{ij} \partial_2 f_j(1,0)}{\rho_n g(U_j)}\bigg|U_i\bigg] + \frac{f_i(1,0)}{1/2} (T_i - 1/2)\Big) \cdot \\
    & \phantom{\E \bigg[\bigg|U_i\bigg]} \Big(\E \bigg[\frac{E_{ij} \partial_2 f_j(1,0)}{\rho_n g(U_j)}\bigg|U_i\bigg] +  \frac{f_i(1,0)}{1/2} (T_i - 1/2)\Big)\bigg] + O_{\psi_{1,tc}}((n \rho_n^3)^{-1}) + o_{\P}(1) \\
    = & \be_s^{\top} \E[\mathbf{S}_{\ell} \mathbf{S}_{\ell}^{\top}] \be_q + O_{\psi_{1,tc}}((n \rho_n^3)^{-1}) + o_{\P}(1).
\end{align*}
\paragraph*{Part 2: Higher Order Terms}
For the second order terms, first notice that if $l \notin [n]$, then
\begin{align*}  
    & \bigg(\frac{M_j}{N_j}_{(i)}\bigg)^2 - \frac{1}{n}\sum_{\iota \in [n], \iota \neq l} \bigg(\frac{M_j}{N_j}_{(\iota)}\bigg)^2 \\
    = & \frac{1}{n} \sum_{\iota \in [n], \iota \neq l} \bigg(\frac{M_j}{N_j}_{(i)} + \frac{M_j}{N_j}_{(\iota)}\bigg)\frac{M_j(E_{ij} - E_{\iota j}) - (E_{ij}W_i - E_{\iota j}W_{\iota})N_j + E_{ij}E_{\iota j}(W_i - W_{\iota})}{N_j^{(i)} N_j^{(\iota)}} \\
    = & O_{\psi_{2,tc}}((n \rho_n)^{-\frac{3}{2}}),
\end{align*}
where we have used $(M_j/N_j)_{\iota} = O_{\psi_2}((n \rho_n)^{-\frac{1}{2}})$ and $N_j^{-1} = O_{\psi_2}((n \rho_n)^{-1})$. If $l \in [n]$, then again
\begin{align*}
    & \bigg(\frac{M_j}{N_j}_{(i)}\bigg)^2 - \frac{1}{n}\sum_{\iota \in [n], \iota \neq l} \bigg(\frac{M_j}{N_j}_{(\iota)}\bigg)^2 \\
     = & \bigg(\frac{M_j}{N_j}_{(i)}\bigg)^2 - \frac{1}{n - 1}\sum_{\iota \in [n], \iota \neq l} \bigg(\frac{M_j}{N_j}_{(\iota)}\bigg)^2 + \frac{1}{(n - 1)n}\sum_{\iota \in [n], \iota \neq l} \bigg(\frac{M_j}{N_j}_{(\iota)}\bigg)^2 \\
    = & O_{\psi_{2,tc}}((n \rho_n)^{-\frac{3}{2}}).
\end{align*}
Hence
\begin{align*}
    n \sum_{i \in [n]} \bigg(\partial_{2,2} h(1,0) & \frac{2}{n} \sum_{j \in [n]} T_j \bigg[\Big(\frac{M_j}{N_j}_{(i)}\Big)^2 - \frac{1}{n}\sum_{\iota \in [n]} \Big(\frac{M_j}{N_j}_{(\iota)}\Big)^2\bigg]\bigg) \cdot \\
     & \bigg(\partial_{2,2} h(1,0) \frac{2}{n_q} \sum_{j \in \ca I_q} T_j \bigg[\Big(\frac{M_j}{N_j}_{(i)}\Big)^2 - \frac{1}{n}\sum_{\iota \in [n]} \Big(\frac{M_j}{N_j}_{(\iota)}\Big)^2\bigg]\bigg) 
     = O_{\psi_{2,tc}}((n \rho_n^3)^{-1}).
\end{align*}
For the third order residual, observe that $(\frac{M_j}{N_j}_{(\iota)})^3 = O_{\psi_2}((n \rho_n)^{-3/2})$. Then
\begin{align*}
    n \sum_{i \in [n]} \bigg(\frac{2}{n} & \sum_{j \in [n]} T_j \bigg[\partial_{2,2,2} h\Big(1,\xi_{j,i}^{\ast}\Big)\Big(\frac{M_j}{N_j}_{(i)}\Big)^3- \frac{1}{n}\sum_{\iota \in [n]} \partial_{2,2,2} h\Big(1,\xi_{j,\iota}^{\ast}\Big) \Big(\frac{M_j}{N_j}_{(\iota)}\Big)^3\bigg]\bigg) \cdot \\
    & \bigg(\frac{2}{n_q} \sum_{j \in \ca I_q} T_j \bigg[\partial_{2,2,2} h\Big(1,\xi_{j,i}^{\ast}\Big)\Big(\frac{M_j}{N_j}_{(i)}\Big)^3- \frac{1}{n}\sum_{\iota \in [n]} \partial_{2,2,2} h\Big(1,\xi_{j,\iota}^{\ast}\Big) \Big(\frac{M_j}{N_j}_{(\iota)}\Big)^3\bigg]\bigg) \\\
    & = O_{\psi_{2,tc}}((n \rho_n^3)^{-1}).
\end{align*}
The conclusion then follows from Equations \eqref{JACK-0}, \eqref{JACK-1} and \eqref{JACK-Taylor}.

\subsection{Proof of Lemma~\ref{sa-lem:lp}}

Define $\mathtt{r}(x) = (1,x)^\top$. Denote $\pi = \E[W_i] = 2 \E[T_i]  - 1$. Then

\begin{center}
    \textbf{Case 1: $\beta < 1$}
\end{center}
First, consider the gram-matrix.
Take $\zeta_i := \sqrt{n \rho_n} (\frac{M_i}{N_i} - \pi)$. Then $1 \lesssim \V[\zeta_i] \lesssim 1$. Take $b_n = \sqrt{n \rho_n} h_n$. Take
\begin{align*}
    \mathbf{B}_n := \frac{1}{n b_n} \sum_{i = 1}^n \mathtt{r} \Big(\frac{\zeta_i}{b_n}\Big) \mathtt{r} \Big(\frac{\zeta_i}{b_n}\Big)^{\top} K \Big(\frac{\zeta_i}{b_n}\Big),
\end{align*}
where $\mathtt{r}: \reals \rightarrow \reals^2$ is given by $\mathtt{r}(u) = (1,u)^{\top}$. Take $Q$ to be the probability measure of $\zeta_i$ given $\mathbf{E}$. Then
\begin{align*}
    \mathbf{B} := \E[\mathbf{B}_n|\mathbf{E}] 
    = \begin{bmatrix}
        \int_{-\infty}^{\infty} \frac{1}{b_n} K(\frac{x}{b_n}) d Q(x) & \int_{-\infty}^{\infty} \frac{x}{b_n}\frac{1}{b_n} K(\frac{x}{b_n}) d Q(x) \\
        \int_{-\infty}^{\infty} \frac{x}{b_n}\frac{1}{b_n} K(\frac{x}{b_n}) d Q(x) & \int_{-\infty}^{\infty} (\frac{x}{b_n})^2 \frac{1}{b_n} K(\frac{x}{b_n}) d Q(x)
    \end{bmatrix}.
\end{align*}
In particular, $\lambda_{\min}(\mathbf{B}) \gtrsim 1$. Now we want to show each entry of $\mathbf{B}_n$ converge to those of $\mathbf{B}$. Take 
\begin{align*}
    F_{p,q}(\bW) := \be_p^{\top} \mathbf{B}_n \be_q
    = \frac{1}{n b_n} \sum_{i = 1}^n \Big(\frac{\zeta_i}{b_n}\Big)^{p + q} K \Big(\frac{\zeta_i}{b_n}\Big), \qquad p,q \in \{0,1\}.
\end{align*}
Denote $\partial_j$ to be the partial derivative w.r.p to $W_j$. Since $K$ is Lipschitz with bounded support,
\begin{align}\label{lp deriv}
    |\partial_j F_{p,q}(\bW)| 
    \lesssim \frac{1}{b_n^2} \frac{1}{n}\sum_{i = 1}^n \Big|\partial_j \Big(\frac{M_i}{N_i} - \pi\Big)\Big|
    \lesssim \frac{1}{b_n^2} \frac{1}{n}\sum_{i = 1}^n \frac{E_{ij}}{N_i}.
\end{align}
Condition on $\mathbf{E}$,
\begin{align*}      
    F_{p,q}(\bW) & = \E[F_{p,q}(\bW)|\mathbf{E}] + O_{\psi_2}\Big(\sum_{j = 1}^n |\partial_j F_{p,q}(\bW)|^2\Big)
    = \be_p^{\top} \mathbf{B} \be_q + O_{\psi_2}\Big(\frac{1}{n b_n^4} \frac{1}{n}\sum_{j = 1}^n \Big( \sum_{i = 1}^n \frac{E_{ij}}{N_i}\Big)^2\Big).
\end{align*}
Hence for all $p,q \in \{0,1\}$, 
\begin{align*}
    \be_p^{\top} \mathbf{B}_n \be_q = \be_p^{\top} \mathbf{B} \be_q + O_{\psi_2}((n b_n^4)^{-1}).
\end{align*}
Since both $\mathbf{B}_n$ and $\mathbf{B}$ are two by two matrices, $\norm{\mathbf{B}_n - \mathbf{B}}_{\operatorname{op}} \lesssim O_{\psi_2}((n b_n^4)^{-1})$. By Weyl's Theorem,
\begin{align}\label{lp weyl}
    |\lambda_{\min}(\mathbf{B}_n) - \lambda_{\min}(\mathbf{B})| 
    \leq \norm{\mathbf{B}_n - \mathbf{B}}_{\operatorname{op}}
    \lesssim (n b_n^4)^{-1},
\end{align}
and together with $\lambda_{\min}(\mathbf{B}) \gtrsim 1$, implies $\lambda_{\min}(\mathbf{B}_n) \gtrsim 1$. Take
\begin{align*}
    \bSigma_n := \frac{1}{n b_n^2} \sum_{i = 1}^n \mathtt{r} \Big(\frac{\zeta_i}{b_n}\Big) \mathtt{r} \Big(\frac{\zeta_i}{b_n}\Big)^{\top} K^2 \Big(\frac{\zeta_i}{b_n}\Big)\V[Y_i|\zeta_i].
\end{align*}
Hence variance can be bounded by
\begin{align}\label{variance}
    \V[\widehat{\gamma}_0|\mathbf{E},\mathbf{W}] & = \be_0^{\Trans} \mathbf{B}_n^{-1} \mathbf{\Sigma}_n \mathbf{B}_n^{-1} \be_0 \lesssim (n b_n)^{-1}, \\
    \V[\widehat{\gamma}_1|\mathbf{E},\mathbf{W}] & = n \rho_n \be_1^{\Trans} \mathbf{B}_n^{-1} \mathbf{\Sigma}_n \mathbf{B}_n^{-1} \be_1 \lesssim (n \rho_n)(n b_n^3)^{-1} = \rho_n b_n^{-3}.
\end{align}
Next, consider the bias term. Since $f(1,\cdot) \in C^2$, whenever $|\frac{M_i}{N_i} - \pi| \leq h_n = (n \rho_n)^{-1/2} b_n$, 
\begin{align*}
    f(1,M_i/N_i) & = f(1,\pi) + \partial_2 f(1,\pi) \Big(\frac{M_i}{N_i} - \pi \Big) + O \Big(\Big(\frac{M_i}{N_i} - \pi \Big)^2\Big) \\
    & = f(1,\pi) +  \partial_2 f(1,\pi) \Big(\frac{M_i}{N_i} - \pi \Big) + O ((n \rho_n)^{-1}b_n^2).
\end{align*}
Hence using the fourth and third lines above respectively,
\begin{equation}\label{bias}
\begin{aligned}
\E[\widehat{\gamma}_0|\mathbf{E}, \bW] 
& = \be_0^{\Trans} \mathbf{B}_n^{-1} \bigg[\frac{1}{n b_n} \sum_{i = 1}^n \mathtt{r}\Big(\frac{\zeta_i}{b_n}\Big)K \Big( \frac{\zeta_i}{b_n}\Big)f \Big(1,\frac{M_i}{N_i} \Big) \bigg] \\
& = \be_0^{\Trans} \mathbf{B}_n^{-1} \bigg[\frac{1}{n b_n} \sum_{i = 1}^n \mathtt{r}\Big(\frac{\zeta_i}{b_n}\Big)K \Big( \frac{\zeta_i}{b_n}\Big) \bigg(\mathtt{r} \Big(\frac{\zeta_i}{b_n}\Big)^{\top}(f(1,\pi),\frac{1}{\sqrt{n\rho_n}}\partial_2 f(1,\pi))^{\top} + O_{\psi_2}((n \rho_n)^{-\frac{1}{2}})\bigg) \bigg]\\
& = f(1,\pi) + O_{\psi_2}((n \rho_n)^{-\frac{1}{2}}),\\
\E[\widehat{\gamma}_1|\mathbf{E}, \bW] 
& = \sqrt{n \rho_n} \be_1^{\Trans} \mathbf{B}_n^{-1} \bigg[\frac{1}{n b_n} \sum_{i = 1}^n \mathtt{r}\Big(\frac{\zeta_i}{b_n}\Big)K \Big( \frac{\zeta_i}{b_n}\Big)f \Big(1,\frac{M_i}{N_i} \Big) \bigg] \\
& = \sqrt{n \rho_n} \be_1^{\Trans} \mathbf{B}_n^{-1} \bigg[\frac{1}{n b_n} \sum_{i = 1}^n \mathtt{r}\Big(\frac{\zeta_i}{b_n}\Big)K \Big( \frac{\zeta_i}{b_n}\Big) \bigg(\mathtt{r} \Big(\frac{\zeta_i}{b_n}\Big)^{\top}(f(1,\pi),\frac{1}{\sqrt{n\rho_n}}\partial_2 f(1,\pi))^{\top} + O_{\psi_2}((n \rho_n)^{-1})\bigg) \bigg]\\
& = \partial_2 f(1,\pi) + O_{\psi_2}((n \rho_n)^{-\frac{1}{2}}),
\end{aligned}
\end{equation}
Putting together Equations~\eqref{variance} and \eqref{bias},
\begin{align*}
    \widehat{\gamma}_0 - \gamma_0 = O_{\P}((n \rho_n)^{-\frac{1}{2}} + (n b_n)^{-\frac{1}{2}}), \quad 
    \widehat{\gamma}_1 - \gamma_1 = O_{\P}((n \rho_n)^{-\frac{1}{2}} + \rho_n b_n^{-3}).
\end{align*}
Hence any $b_n$ such that $b_n = \Omega(n^{-1/4} + \rho_n^{1/3})$ will make $(\widehat{\gamma}_0, \widehat{\gamma}_1)$ a consistent estimator for $(\gamma_0,\gamma_1)$. For any $0 \leq \rho_n \leq 1$ such that $n \rho_n \rightarrow \infty$, such a sequence $b_n$ exists.

\begin{center}
    \textbf{Case 2: $\beta = 1$}
\end{center}
The order $\frac{M_i}{N_i}$ is $n^{-1/4}$ if $\liminf_{n \rightarrow \infty} n \rho_n^2 > c$ for some $c > 0$; and is $(n \rho_n)^{-1/2}$ if $n \rho_n^2 = o(1)$. We consider these two cases separately.

\paragraph*{Case 2.1: $\liminf_{n\rightarrow \infty}n \rho_n^2 > c$ for some $c > 0$} Take $\eta_i = n^{\frac{1}{4}}(\frac{M_i}{N_i} - \pi)$. Take $d_n = n^{1/4}h_n$. And with the same $\mathtt{r}$ defined in Case 1,
\begin{align*}
    \mathbf{D}_n := \frac{1}{n d_n} \sum_{i = 1}^n \mathtt{r} \Big(\frac{\eta_i}{d_n}\Big) \mathtt{r} \Big(\frac{\eta_i}{d_n}\Big)^{\top} K \Big(\frac{\eta_i}{d_n}\Big), \qquad \mathbf{D} = \E[\mathbf{D}_n].
\end{align*}
Under the assumption $\liminf_{n \rightarrow \infty }n \rho_n^2 \leq c$ for some $c > 0$, we have $1 \lesssim \V[\eta_i] \lesssim 1$. Hence $\lambda_{\min}(\mathbf{D}) \gtrsim 1$. To study the convergence between $\mathbf{D}_n$ and $\mathbf{D}$, again consider for $p, q \in \{0,1\}$,
\begin{align*}
    G_{p,q}(\mathbf{W}) := \be_p^{\top} \mathbf{D}_n \be_q 
    = \frac{1}{n d_n} \sum_{i=1}^n \Big(\frac{\eta_i}{d_n}\Big)^{p+q} K\Big(\frac{\eta_i}{d_n}\Big)
    = \frac{1}{n^{5/4} h_n} \sum_{i = 1}^n \Big( h_n^{-1}(\frac{M_i}{N_i} - \pi)\Big)^{p+q} K \Big(h_n^{-1}(\frac{M_i}{N_i} - \pi)\Big). 
\end{align*}
Still let $\mathsf{U}_n$ be the latent variable from Lemma~\ref{sa-lem: definetti}, $W_i$'s are independent conditional on $\mathsf{U}_n$. Hence by similar argument as Equation~\eqref{lp deriv}, we can show
\begin{align*}
    G_{p,q}(\bW) = \E[G_{p,q}(\bW)|\mathsf{U}_n,\mathbf{E}] + O_{\psi_2}((n d_n^4)^{-1}).
\end{align*}
Moreover, recall we denote by $\omega_i \in [k]$ the block unit $i$ belongs to, then
\begin{align*}
    \E[G_{p,q}(\bW)|\mathsf{U}_n,\mathbf{E}]
    = & \sum_{\bW \in \{-1,1\}^n} \prod_{i = 1}^n p(\mathsf{U}_{n, \omega_i})^{W_s}(1 - p(\mathsf{U}_{n, \omega_i}))^{1 - W_s} G_{p,q}(\bW),
\end{align*}
$p(U_l)= \P(W_i = 1|U_{\ell}) = \frac{1}{2}(\tanh(\sqrt{\beta_{\ell}/n}\mathsf{U}_n + h_{\ell}) + 1)$, $i \in \ca I_{\ell}$. Take the derivative term by term,
\begin{align*}
    \partial_{U_{\ell}}\E[G_{p,q}(\bW)|\mathsf{U}_n,\mathbf{E}] 
    = \sum_{j \in \ca I_{\ell}} \E_{\bW_{-j}}[G_{p,q}(W_j = 1, W_{-j}) - G_{p,q}(W_j = -1, W_{-j})] p^{\prime}(U_{\ell}).
\end{align*}
Using Lipschitz property of $x \mapsto (x/h_n)^{p+q} K(x/h_n)$, 
\begin{align*}
     |G_{p,q}(W_j = 1, W_{-j}) - G_{p,q}(W_j = -1, W_{-j})|
     \lesssim \frac{1}{n^{5/4}h_n} \sum_{i =1 }^n \frac{1}{h_n} \frac{E_{ij}}{N_i}.
\end{align*}
Hence for all $\ell \in \mathscr{C}$,
\begin{align*}
    |\partial_{U_{\ell}}\E[G_{p,q}(\bW)|\mathsf{U}_n,\mathbf{E}] |
    \lesssim \sum_{j \in \ca I_{\ell}} \frac{1}{n^{5/4}h_n} \sum_{i =1 }^n \frac{1}{h_n} \frac{E_{ij}}{N_i} \infnorm{p^{\prime}} 
    \lesssim \frac{1}{n^{3/4}h_n^2}.
\end{align*}
Moreover, for all $\ell \in \mathscr{C}$, $\norm{U_{\ell}}_{\varphi_2} \lesssim n^{1/4}$. Together, this gives
\begin{align*}
    \E[G_{p,q}(\bW)|\mathsf{U}_n, \mathbf{E}] - \E[G_{p,q}(\bW)|\mathbf{E}] = O_{\P}((n^{1/2}h_n^2)^{-1}) = O_{\P}(d_n^{-2}).
\end{align*}
Hence if we take $d_n \gg 1$ (which implies $n d_n^4 \gg 1$), then $G_{p,q}(\bW) = \E[G_{p,q}(\bW)|\mathbf{E}] + o_{\P}(1)$, implying $\norm{\mathbf{D}_n - \mathbf{D}}_2 = o_{\P}(1)$ and $\lambda_{\min}(\mathbf{D}_n) - \lambda_{\min}(\mathbf{D}) = o_{\P}(1)$, making $\lambda_{\min}(\mathbf{D}_n) \gtrsim_{\P} 1$. Take
\begin{align*}
    \boldsymbol{\Upsilon}_n := \frac{1}{n d_n^2} \sum_{i = 1}^n \mathtt{r} \Big(\frac{\eta_i}{d_n}\Big) \mathtt{r} \Big(\frac{\eta_i}{d_n}\Big)^{\top} K^2 \Big(\frac{\eta_i}{d_n}\Big)\V[Y_i|\eta_i].
\end{align*} 
Hence variance can be bounded by
\begin{align}\label{variance 2}
    \V[\widehat{\gamma}_0|\mathbf{E},\mathbf{W}] & = \be_0^{\Trans} \mathbf{D}_n^{-1} \mathbf{\Upsilon}_n \mathbf{D}_n^{-1} \be_0 \lesssim (n d_n)^{-1}, \\
    \V[\widehat{\gamma}_1|\mathbf{E},\mathbf{W}] & = n^{1/2}\be_1^{\Trans} \mathbf{D}_n^{-1} \mathbf{\Upsilon}_n \mathbf{D}_n^{-1} \be_1 \lesssim n^{1/2} (n d_n^3)^{-1} = n^{-1/2} d_n^{-3}.
\end{align}
By similar argument as in Case 1, assume $d_n \gg 1$, we can show
\begin{align*}
    \E[\widehat{\gamma}_0|\mathbf{E}] - \gamma_0 = O(n^{-1/4} + n^{-1/2}d_n^2), \qquad \E[\widehat{\gamma_1}|\mathbf{E}] - \gamma_1 = O(n^{-1/4}d_n^2).
\end{align*}
Hence if we choose $d_n$ such that $1 \ll d_n \ll n^{1/8}$, then
$(\widehat{\gamma}_0, \widehat{\gamma}_1)$ is a consistent estimator for $(\gamma_0, \gamma_1)$. The only assumption we made for the existence of such a $d_n$ is $\liminf_{n \rightarrow \infty} n \rho_n^2 \geq c$ for some $c > 0$.

\paragraph*{Case 2.2: $n \rho_n^2 = o(1)$} Take $\eta_i := \sqrt{n \rho_n}(\frac{M_i}{N_i} - \pi)$, $d_n = \sqrt{n \rho_n} h_n$. By similar decomposition based on latent variables, we can show if $n \rho_n \rightarrow \infty$ as $n \rightarrow \infty$, then there exists $h_n$ such that $(\widehat{\gamma}_0, \widehat{\gamma}_1)$ is a consistent estimator for $(\gamma_0, \gamma_1)$.

\section{Proofs: Section~\ref{sa-sec: add}}

\subsection{Preliminary Lemmas}

\begin{lemma}\label{pi fixed point low temp}
Recall $\bW = (W_i)_{1 \leq i \leq n}$ takes value in $\{-1,1\}^n$ with
\begin{align*}
    \P \left( \bW = \bw \right) = \frac{1}{Z} \exp \bigg( \frac{\beta}{n} \sum_{i < j} W_i W_j\bigg), \quad \beta > 1.
\end{align*}
Recall $\pi_+$ and $\pi_-$ are the positive and negative solutions to $x = \tanh(\beta x + h)$, respectively, and $\mca m = n^{-1} \sum_{i = 1}^n W_i$. Then $\E[W_i|\sign(\mca m) = \ell] = \pi_{\ell} + O(n^{-1})$ for $\ell = -, +$.
\end{lemma}

\begin{proof}
The conditional concentration of $\mca m = n^{-1} \sum_{i = 1}^n W_i$ towards $\pi_{\ell}$ in Lemma~\ref{sa-lem:fixed-temp-be} implies,
\begin{align*}
    & \E[W_i|\sign(\mca m) = \ell] \\
    =  & \E[\E[W_i|W_{-i}, \sign(\mca m) = \ell]| \sign(\mca m) = \ell] \\
    = & \E[\tanh(\beta \mca m_i + h) \Indicator(\sign(\mca m_i) = \ell)|\sign(\mca m) = \ell] + O(n^{-1})\\
    = &  \E[\tanh(\beta \pi + h) + \sech^2(\beta \pi + h)(\mca m_i - \pi) - \sech^2(\beta m^{\ast} + h)\tanh(\beta m^{\ast} + h)(\mca m_i - \pi)^2] + O(n^{-1})\\
    = & \tanh(\beta \pi + h) + O(n^{-1}) \\
    = & \pi + O(n^{-1}),
\end{align*}
where $\mca m^{\ast}$ is a number between $\mca m$ and $\pi$, and we have used boundedness of $\sech$.
\end{proof}

\begin{lemma}\label{lem: concentration of M_i/N_i low temp}
Suppose Assumption~\ref{assump-one-block}, and Assumption 2, 3 hold with $h = 0$ and $\beta > 1$. Then for $\ell = _, +$: (1) Condition on $\sign(\mca m) = \ell$,
\begin{align*} 
    & \max_{i \in [n]} \left|\frac{M_i}{N_i} - \pi_{\ell} \right| = O_{\psi_1}(n^{-1/2}) + O_{\psi_2}(N_i^{-1/2}).
\end{align*}
(2) Define $A(\mathbf{U}) = (G(U_i,U_j))_{1 \leq i,j \leq n}$. Condition on $\mathbf{U}$ such that $$A(\mathbf{U}) \in \mathcal{A} = \{A \in \bb R^{n \times n}: \min_{i \in [n]} \sum_{j \neq i}A_{ij} \geq 32 \log n\}$$ for large enough $n$, for each $i \in [n]$ and $t > 0$,
\begin{align*}
    \P_{\beta,h}\left( \left|\frac{M_i}{N_i} - \pi_{\ell}\right| \geq 4 \E[N_i | \mathbf{U}]^{-1/2}t^{1/2} + C n^{-1/2} t^{1/2}\middle| \mathbf{U}, \sign(\mca m) = \ell \right) \leq 2 \exp(-t) + n^{-98},
\end{align*}
where $C$ is some absolute constant.
\end{lemma}

\begin{proof}
Throughout the proof, the Ising spins $\bW=(W_i)_{i=1}^n$ are distributed according to Assumption~\ref{assump-one-block} with parameters $(\beta,h)$. For brevity, we write $\P$ in place of $\P_{\beta,h}$.

Let $\mathsf{U}_n$ to be the latent variable defined in Lemma~\ref{sa-lem:sub-gaussian}. Decompose by
\begin{align*}
    \frac{M_i}{N_i} - \pi_{\ell} = \sum_{j \neq i} \frac{E_{ij}}{N_i} \left(W_j - \E[W_j|\mathsf{U}_n] \right) + \E[W_j|\mathsf{U}_n] - \pi_{\ell}.
\end{align*}
Condition on $\mathsf{U}_n$, $W_i$'s are i.i.d. Berry-Esseen theorem gives that with $\mathsf{Z} \sim \mathsf{N}(0,1)$ independent to $\mathsf{U}_n$, we have
\begin{align}\label{sa-eq: nbh avg decompose low temp}
    \sup_{t \in \reals} \bigg|\P\Big(\frac{M_i}{N_i} \leq t\Big|\mathbf{E}, \mathsf{U}_n \Big) - \P \Big(\sqrt{\frac{v(\mathsf{U}_n)}{N_i}} \mathsf{Z} + e(\mathsf{U}_n) \leq t \Big|\mathbf{E}, \mathsf{U}_n\Big) \bigg| = O(n^{-\frac{1}{2}}),
\end{align}
where $e(\mathsf{U}_n) = \E[W_i|\mathsf{U}_n] - \pi = \tanh(\sqrt{\beta/n}\mathsf{U}_n + h) - \pi$, and $v(\mathsf{U}_n) = \V[W_i - \pi|\mathsf{U}_n]$. By McDiarmid's inequality, 
\begin{align*}
    \P \bigg( |\sum_{j \neq i} \frac{E_{ij}}{N_i} \left(W_j - \E[W_j|\mathsf{U}_n] \right)| \geq 2 N_i^{-1/2} t\bigg | \mathbf{E} \bigg) \leq 2 \exp(-t^2).
\end{align*}
Conclusion (1) then follows from the conditional concentration of $\mathsf{U}_n$ in Remark~\ref{sa-remarK: conditional concentration under low temp}.

Notice that $\bW$ and $\mathsf{U}_n$ are independent to the random graph. Conclusion (1) and the same analysis as in Lemma~\ref{lem: concentration of M_i/N_i} give conclusion (2). 
\end{proof}

\subsection{Proof of Lemma~\ref{sa-lem:lin-h-nonzero}}

The result is a special case of Lemma~\ref{sa-lem:lin-fixed-temp} in Section~\ref{sa-sec:stochlin} when $h \neq 0$. 

\subsection{Proof of Theorem~\ref{sa-thm:dist-nonzero}}

The result follows from Lemma~\ref{sa-lem:fixed-temp-be}, Lemma~\ref{sa-lem:lin-h-nonzero}, and the same anti-concentration argument as in the proof of Lemma~\ref{sa-lem:fixed-temp-be}.

\subsection{Proof of Lemma~\ref{sa-lem:lin-low}}

\begin{center}
\textbf{I. The Unbiased Estimator}
\end{center}
First, we consider the unbiased estimator 
\begin{align*}
    \wh\tau_{n,\text{UB}} = \frac{1}{n} \sum_{i = 1}^n \bigg[\frac{T_i Y_i}{p_i} - \frac{(1 - T_i) Y_i}{1 - p_i}\bigg],
\end{align*}
with $p_i = \bb P(W_i = 1| \bW_{-i}) = \left(\exp \left(-2\beta \mca m_i\right) + 1\right)^{-1}$. Our analysis will be similar to the proofs in Section~\ref{sa-sec:unbiased}, but using the concentration of $n^{-1} \sum_{i = 1}^n W_i$ conditional on $\sign(\mca m)$ shown in Lemma~\ref{sa-lem:fixed-temp-be} instead of the unconditional concentration of $n^{-1} \sum_{i = 1}^n W_i$. We decompose by
\begin{align*}
    n^{-1}\sum_{i = 1}^n \frac{T_i}{p_i} g_i\Big(1,\frac{M_i}{N_i}\Big)  = n^{-1}\sum_{i = 1}^n \frac{T_i}{p_i} g_i(1,\pi_{\ell}) + n^{-1}\sum_{i = 1}^n \frac{T_i}{p_i} \bigg[g_i\Big(1, \frac{M_i}{N_i}\Big) - g_i(1,\pi_{\ell}) \bigg].
\end{align*}
For the first term, we Taylor expand the expression for $p_i^{-1}$ in terms of $\mca m_i$, and get
\begin{align*}
    n^{-1} \sum_{i = 1}^n \frac{T_i}{p_i} g_i(1,\pi_{\ell}) & = n^{-1} \sum_{i = 1}^n g_i(1, \pi_{\ell}) + n^{-1} \sum_{i = 1}^n \frac{T_i - p_i}{p_i} g_i(1,\pi_{\ell}) \\
    & = n^{-1}\sum_{i = 1}^n (Y_i(1, \pi_{\ell}) + (c_{i,l}/2 + d_l)(W_i - \pi_{\ell})) + O_{\psi_2,tc}(\sqrt{\log n} n^{-1/2}),
\end{align*}
condition on $\sign(\mca m) = \ell$, where
\begin{align*}
    c_{i,l} = g_i(1,\pi_{\ell})(1 + \exp(2 \beta \pi_{\ell})), \qquad d_l = \frac{\beta(1 + \exp(2 \beta \pi_{\ell}))}{1 + \cosh(2 \beta \pi_{\ell})}\E[g_i(1, \pi_{\ell})].
\end{align*} 
For the second term, we Taylor expand $g_i(1, \cdot)$ at $\pi_{\ell}$: For some $\eta_i^{\ast}$ between $\pi_{\ell}$ and $\frac{M_i}{N_i}$, 
\begin{align*}
    & \frac{1}{n}\sum_{i = 1}^n \frac{T_i}{p_i} \bigg[g_i\bigg(1, \frac{M_i}{N_i}\bigg) - g_i(1, \pi_{\ell}) \bigg] = \Delta_{2,1}^{\prime} + \Delta_{2,2}^{\prime} + \Delta_{2,3}^{\prime},
\end{align*}
 where
 \begin{align*}
    & \Delta_{2,1}^{\prime} =  \frac{1}{n} \sum_{i =1}^n g_i^{\prime} \left(1, \pi_{\ell}\right) \left(\frac{M_i}{N_i} - \pi_{\ell}\right), \\
    & \Delta_{2,2}^{\prime} = \frac{1}{n} \sum_{i = 1}^n \frac{T_i - p_i}{p_i} g_i^{\prime}\left( 1, \pi_{\ell}\right) \left(\frac{M_i}{N_i} - \pi_{\ell} \right), \\
    & \Delta_{2,3}^{\prime} = \frac{1}{n}\sum_{i = 1}^n  \frac{T_i g_i^{\prime \prime} \left(1, \eta_i^{\ast}\right)}{2 p_i} \left(\frac{M_i}{N_i} - \pi_{\ell}\right)^2.
\end{align*}

\smallskip

\noindent \textbf{Term $\Delta_{2,1}^{\prime}$:} Denote $\bg = (g_i)_{1 \leq i \leq n}$. Rearranging the terms,
\begin{align*}
    \Delta_{2,1}^{\prime} - \E[\Delta_{2,1}^{\prime}|\mathbf{E}, \bg, \sign(\mca m) = \ell] & = \frac{1}{n}\sum_{i = 1}^n \bigg[ \sum_{j \neq i}\frac{E_{ij}}{N_j} g_j^{\prime}(1,\pi_{\ell})\bigg](W_i - \pi_{\ell}).
\end{align*}

\smallskip

\noindent \textbf{Term $\Delta_{2,2}^{\prime}$:} Take $u_{\ell} = (\pi_{\ell} + 1)/2$ for $\ell \in \{-,+\}$. Decompose by $$\Delta_{2,2}^{\prime} = \Delta_{2,2,1}^{\prime} + \Delta_{2,2,2}^{\prime},$$
where
\begin{align*}
    \Delta_{2,2,1}^{\prime} & = \frac{1}{n} \sum_{i = 1}^n \frac{T_i - u_{\ell}}{u_{\ell}} g_i^{\prime}(1, \pi_{\ell}) \Big(\frac{M_i}{N_i} - \pi_{\ell} \Big), \\
    \Delta_{2,2,2}^{\prime} & = \frac{1}{n} \sum_{i = 1}^n T_i (p_i^{-1} - u_{\ell}^{-1}) g_i^{\prime}(1, \pi_{\ell}) \Big(\frac{M_i}{N_i} - \pi_{\ell} \Big).
\end{align*}
Since $\beta < \infty$ and $v_{-}$, $u_-$ and $u_+$ are bounded away from $0$ and $1$. Rearranging the terms, we get
\begin{align*}
    \Delta_{2,2,1}^{\prime} = n^{-1} (\mathbf{W} - \pi_{\ell} \mathbf{1})^{\Trans} \mathbf{H}^{\ell}(\mathbf{W} - \pi_{\ell} \mathbf{1})
\end{align*}
where $\mathbf{H}^{\ell}$ is the $n \times n$ matrix with $H^{\ell}_{ij} = g_i^{\prime}(1, \pi_{\ell}) E_{ij} (2 u_{\ell} N_i)^{-1}$ and $\mathbf{1}$ is the $n$-dimensional vector with all entries $1$. To analyze the quadratic form, we use the same strategy as in the proof of Lemma~\ref{lem: delta_2,2}: Let $\mathsf{U}_n$ be the one defined in Lemma~\ref{sa-lem:sub-gaussian}, and we know $W_1, \cdots, W_n$ are conditional i.i.d given $\mathsf{U}_n$. Then we can decompose $\Delta_{2,2,1}^{\prime}$ into four terms based on 
\begin{align*}
    \bW - \pi_{\ell}\mathbf{1} = (\bW - \E[\bW|\mathsf{U}_n]) + (\E[\bW|\mathsf{U}_n] - \pi_{\ell} \mathbf{1}).
\end{align*}
Conditional Berry-Esseen given $\mathsf{U}_n$, conditional concentration of $\mathsf{U}_n$, $\mca m$ and $\frac{M_i}{N_i}$ given $\sign(\mca m)$ in Remark~\ref{sa-remarK: conditional concentration under low temp}, Lemma~\ref{sa-lem:fixed-temp-be} and Lemma~\ref{lem: concentration of M_i/N_i low temp}, and the same argument as in the proof for Lemma~\ref{lem: delta_2,2} implies that condition on $\bg, \bE$ and $\sign(\mca m)$, 
\begin{align*}
    \norm{\Delta_{2,2,j}^{\prime} - \E[\Delta_{2,2,j}^{\prime}|\bg, \mathbf{E}, \sign(\mca m)]}_{\psi_2} &  = \log (n) n^{-1/4} (\min_i N_i)^{-1/2} +  n^{-1/2}, \qquad j = 1,2.
\end{align*}

\smallskip

\noindent \textbf{Term $\Delta_{2,3}^{\prime}$:} Now we proceed to $\Delta_{2,3}^{\prime}$. Decompose by $\Delta_{2,3}^{\prime} = \Delta_{2,3,1}^{\prime} + \Delta_{2,3,2}^{\prime}$, where
\begin{align*}
    \Delta_{2,3,1}^{\prime} & = \frac{1}{n}\sum_{i = 1}^n \Big[g_i \Big( 1, \frac{M_i}{N_i}\Big) - g_i(1, \pi_{\ell}) - g_i^{\prime}(1, \pi_{\ell})\Big(\frac{M_i}{N_i} - \pi_{\ell}\Big) \Big], \\
    \Delta_{2,3,2}^{\prime} & = \frac{1}{n}\sum_{i = 1}^n \frac{T_i - p_i}{p_i}\Big[g_i \Big( 1, \frac{M_i}{N_i}\Big) - g_i(1, \pi_{\ell}) - g_i^{\prime}(1, \pi_{\ell})\Big(\frac{M_i}{N_i} - \pi_{\ell}\Big) \Big].
\end{align*}
Define $\Delta_{2,3,1,l}^{\prime}$ to be the counterparts of $\Delta_{2,3,1,l}$ in Equation~\ref{sa-eq:delta-231-decomp} with $\pi$ by replaced by $\pi_{\ell}$ for $l \in \{a,b,c\}$, the same argument in the proof of Lemma~\ref{lem:delta_231} shows 
\begin{align*}
    &\norm{\Delta_{2,3,1,a}^{\prime} - \E[\Delta_{2,3,1,a}^{\prime}|\bg, \bE, \sign(\mca m) = \ell]}_{\psi_2,tc} = O((\min_i N_i)^{-1/2} n^{-1/2}), \\ & \norm{\Delta_{2,3,1,b}^{\prime}}_{\psi_2,tc} = O((\min_i N_i)^{-1/2} n^{-1/2}), \\ & \norm{\Delta_{2,3,1,c}^{\prime}}_{\psi_2,tc} = O(n^{-1}).
\end{align*}
condition on $\sign(\mca m) = \ell$ for $\ell = -, +$. Combining the three parts,
\begin{align*}
    \norm{\Delta_{2,3,1}^{\prime}}_{\psi_2, tc} = O((\min_i N_i)^{-1/2} n^{-1/2}),
\end{align*}
condition on $\sign(\mca m) = \ell$ for $\ell = -, +$. Taylor expanding $p_i = (1 + \exp(- 2 \beta \mca m_i))^{-1}$ as a function of $\mca m_i$ at $\pi_{\ell}$, the same argument as in Lemma~\ref{lem: delta_2,3,2} shows
\begin{align*}
    \Delta_{2,3,2}^{\prime} = & \frac{1}{n}\sum_{i =1}^n \frac{W_i - \pi_{\ell}}{\pi_{\ell} + 1} \Big[g_i \Big( 1, \frac{M_i}{N_i}\Big) - g_i(1, \pi_{\ell}) - g_i^{\prime}(1, \pi_{\ell})\Big(\frac{M_i}{N_i} - \pi_{\ell}\Big) \Big] + O_{\psi_2,tc}((\min_i N_i)^{-1/2} n^{-1/2}).
\end{align*}
Conditional concentration of $\mathsf{U}_n$, $\mca m$ and $\frac{M_i}{N_i}$ given $\sign(\mca m)$ in Remark~\ref{sa-remarK: conditional concentration under low temp}, Lemma~\ref{sa-lem:fixed-temp-be} and Lemma~\ref{lem: concentration of M_i/N_i low temp}, and the same argument as in the proof for Lemma~\ref{lem: delta_2,3,2} implies that condition on $\bg, \bE$ and $\sign(\mca m)$,
\begin{align*}
    \norm{\Delta_{2,3,2}^{\prime}}_{\psi_2,tc} = O((\min_i N_i)^{-1/2} n^{-1/2} + (\min_i N_i)^{-(p+1)/2})
\end{align*}

\smallskip 

\noindent \textbf{Putting together.} Putting together the decompositions, condition on $\bE$ and $\sign(\mca m) = \ell$, 
\begin{align*}
    \wh\tau_{n,\text{UB}} - \E[\wh\tau_{n,\text{UB}}|\mathbf{E}, \bg, \sign(\mca m)] - \frac{1}{n}\sum_{i = 1}^n L_{n,i,\ell}(W_i - \pi_{\ell}) = O_{\psi_2,tc}((\min_i N_i)^{-\frac{1}{2}} n^{-\frac{1}{2}} + (\min_i N_i)^{-\frac{p + 1}{2}}),
\end{align*}
where with $c_{i,l} = g_i(1,\pi_{\ell})(1 + \exp(2 \beta \pi_{\ell}))$, and $d_l = \frac{\beta(1 + \exp(2 \beta \pi_{\ell}))}{1 + \cosh(2 \beta \pi_{\ell})}\E[g_i(1, \pi_{\ell})]$, 
\begin{align*}
    L_{n,i,\ell} & = Y_i(1, \pi_{\ell}) + \Big(c_{i,l}/2 + d_l + \sum_{j \neq i}\frac{E_{ij}}{N_j} g_j^{\prime}(1,\pi_{\ell})\Big)(W_i - \pi_{\ell}).
\end{align*}
Consider the event $\Omega_i = \{\sign(\mca m) = \ell, |\sum_{j \neq i}W_j| \leq 1\}$ and $\Omega = \cup_{1 \leq i \leq n} \Omega_i$. We then have
\begin{align*}
    \bb P\bl\sum_{j\neq i}W_j=1\br+\bb P\bl\sum_{j\neq i}W_j=-1\br\leq C\exp(-nC).
\end{align*}
implying $\P(\Omega_i) \leq C \exp(-nC)$, $1 \leq i \leq n$. Hence
\begin{align}\label{sa-eq: unbiased estimator low temp}
    \nonumber & \E[\wh\tau_{n,\text{UB}}|\mathbf{E}, \bg, \sign(\mca m)] \\
    \nonumber = & \frac{1}{n} \sum_{i = 1}^n \E\bigg[\bigg(\frac{T_i Y_i(1,M_i/N_i)}{\P(W_i = 1|W_{-i})} - \frac{(1 - T_i) Y_i(-1,M_i/N_i)}{\P(W_i = -1|W_{-i})}\bigg) \Indicator(\Omega_i^c)\bigg|\mathbf{E}, \bg, \sign(\mca m) \bigg] + O(\P(\Omega))\\
    \nonumber = & \frac{1}{n} \sum_{i = 1}^n \E\bigg[\bigg(\frac{T_i Y_i(1,M_i/N_i)}{\P(W_i = 1|W_{-i}, \sign(\mca m))} - \frac{(1 - T_i) Y_i(-1,M_i/N_i)}{\P(W_i = -1|W_{-i}, \sign(\mca m))}\bigg) \Indicator(\Omega_i^c)\bigg|\mathbf{E}, \bg, \sign(\mca m) \bigg] + O(\P(\Omega))\\
    \nonumber = & \frac{1}{n} \sum_{i = 1}^n \E\bigg[\frac{T_i Y_i(1,M_i/N_i)}{\P(W_i = 1|W_{-i}, \sign(\mca m))} - \frac{(1 - T_i) Y_i(-1,M_i/N_i)}{\P(W_i = -1|W_{-i}, \sign(\mca m))}\bigg|\mathbf{E}, \bg, \sign(\mca m) \bigg] + O(\P(\Omega)) \\
    = & \tau_l + O(C \exp(-Cn)).
\end{align}
Hence condition on $\bE$ and $\sign(\mca m) = \ell$,
\begin{align*}
    \wh\tau_{n,\text{UB}} - \tau_{n,\ell}= \frac{1}{n}\sum_{i = 1}^n L_{n,i,\ell}(W_i - \pi_{\ell}) + O_{\psi_2,tc}((\min_i N_i)^{-\frac{1}{2}} n^{-\frac{1}{2}} + (\min_i N_i)^{-(p+1)/2}),
\end{align*}
\begin{center}
\textbf{II. The Hajek Estimator}
\end{center}
Now, we consider the difference between the unbiased estimator and the Hajek estimator. For notational simplicity, denote $\wh{\mca p} = \frac{1}{n}\sum_{i = 1}^n T_i$ and $\mca p_{\ell} = \frac{1}{2}\tanh(\beta \pi_{\ell} + h) + \frac{1}{2} = \frac{1}{2} \pi_{\ell} + \frac{1}{2}$. Then
\begin{align*}
    & \frac{1}{n} \sum_{i = 1}^n \frac{T_i Y_i}{\wh{\mca p}} - \frac{1}{n} \sum_{i =1}^n \frac{T_i Y_i}{\mca p_{\ell}} 
    = \frac{1}{n}\sum_{i=1}^n \frac{T_i Y_i}{\wh{\mca p}} \frac{\mca p_{\ell} - \wh{\mca p}}{\mca p_{\ell}}.
\end{align*}
Taylor expand $x \mapsto \tanh(\beta x + h)$ at $x = \pi_{\ell}$, we have
\begin{align*}
    2(\wh{\mca p} - \mca p_{\ell}) = & \mca m - \tanh(\beta \mca m + h)\\
    = & \pi_{\ell} + \mca m  - \pi_{\ell} - \tanh(\beta \pi_{\ell} + h) - \beta \sech^2(\beta \pi_{\ell} + h) (\mca m - \pi_{\ell}) + O((\mca m - \pi_{\ell})^2)\\
    = & (1 - \beta \sech^2(\beta \pi_{\ell} + h))(\mca m - \pi_{\ell}) + O((\mca m - \pi_{\ell})^2),
\end{align*}
where $O(\cdot)$ is up to a universal constant. Together with the fact that condition on $\sign(\mca m) = \ell$, $\frac{1}{n}\sum_{i = 1}^n T_i Y_i$ concentrates towards $\mca m \E[Y_i|\sign(\mca m) = \ell]$, we have
\begin{align*}
    \frac{1}{n} \sum_{i = 1}^n \frac{T_i Y_i}{\wh{\mca p}} - \frac{1}{n} \sum_{i =1}^n \frac{T_i Y_i}{\mca p_{\ell}} 
    = - \frac{1 -\beta(1 - \pi_{\ell}^2)}{1 + \pi_{\ell}} \E\Big[g_i\Big(1,\frac{M_i}{N_i}\Big)\Big|\sign(\mca m) = \ell\Big] + O_{\psi_1}(n^{-1}),
\end{align*}
 condition on $\sign(\mca m) = \ell$. A Taylor expansion of $g_i$ and concentration of $M_i / N_i$ then implies
\begin{align*}
    & \E \Big[g_i\Big(1,\frac{M_i}{N_i}\Big) \Big| \sign(\mca m) = \ell \Big] \\
    = & \E[g_i(1,\pi_{\ell})] + \E\Big[g_i^{(1)}(1,\pi_{\ell})\Big(\frac{M_i}{N_i} - \pi_{\ell}\Big)\Big|\sign(\mca m) = \ell\Big] + \frac{1}{2}\E\Big[g_i^{(2)}(1,\pi^{\ast})\Big(\frac{M_i}{N_i} - \pi_{\ell} \Big)^2 \Big|\sign(\mca m) = \ell\Big] \\
    = & O(n^{-1}),
\end{align*}
where $\pi^{\ast}$ is some number between $\pi_{\ell}$ and $M_i/N_i$. The conclusion then follows.

\subsection{Proof of Lemma~\ref{sa-lem:clt-low}}

The result follows from Lemma~\ref{sa-lem:fixed-temp-be} (3), Lemma~\ref{sa-lem:lin-low}, and the same anti-concentration argument as in the proof of Lemma~\ref{sa-lem:fixed-temp-be}.

\subsection{Proof of Lemma~\ref{sa-lem: mult-stoch-lin}}

As in the case of one block analyzed in Section~\ref{sa-sec:stochlin}, $\wh \btau_n$ is not an unbiased estimate of $\btau_n$. We first consider an unbiased estimator to $\btau_n$ and then consider the difference.

\begin{center}
\textbf{I. The Unbiased Estimator}
\end{center}
Consider $\wh{\btau}_{n,UB} = (\wh{\tau}_{n,UB,1}, \cdots, \wh{\tau}_{n,UB,K})$, where
\begin{align*}
    \wh{\tau}_{n,UB,k} & = \frac{1}{n_k} \sum_{i \in \scrC_k} \frac{D_i Y_i}{p_i} - \frac{(1 - D_i) Y_i}{1 - p_i}, \quad k \in [K].
\end{align*}
Here $p_i = \sum_{k = 1}^K \Indicator(i \in \scrC_k) (1 + \exp(2 \beta_k \mca m_{i,k} + 2 h_k))^{-1}$ and $\mca m_{i,k} = n_k^{-1} \sum_{j \in \scrC_k, j \neq i}W_j$. 

Denote $\mca m = n^{-1}\sum_{i =1}^n W_i$, $\mca m_k = n_k^{-1} \sum_{i \in \scrC_k} W_i$. For notational simplicity, we denote $\pi_{l,\sign(\mca m_l)}$ by $\pi_l$ for low temperature blocks $l \in \mathscr{L}$, and omit the index by $(\bs)$ with $\bs = \bsign$. As in the one-block case, we decompose by 
\begin{align*}
    \wh{\tau}_{n,UB,k} & = \Delta_{1,k} + \Delta_{2,k}, 
\end{align*}
where
\begin{align*}
    \Delta_{1,k} & = \frac{1}{n_k} \sum_{i \in \scrC_k} \frac{D_i}{p_i} f_i\left(1,\zeta_i\right), \quad \Delta_{2,k} = \frac{1}{n_k} \sum_{i \in \scrC_k} \frac{D_i}{p_i} \bigg(f_i\bigg(1, \frac{M_i}{N_i}\bigg) - f_i(1,\zeta_i)\bigg),
\end{align*}
and $\zeta_i = \frac{\sum_{k =1}^K N_{i,k} \pi_k}{\sum_{k = 1}^K N_{i,k}}$. 

\begin{center}
    \textbf{I.1: Term $\Delta_{1,k}$}
\end{center}

Condition on $\bE, \bg = \{g_i: i \in [n]\}$ and $\bsign$, the randomness of $\Delta_{1,k}$ only comes from $(W_i)_{i \in \scrC_k}$, that is, the Ising bits from the same block. Hence 
\begin{align*}
    \Delta_{1,k} - \E[\Delta_{1,k}|\bE,\bg,\bsign] = \frac{1}{n_k}\sum_{i \in \scrC_k} \frac{D_i - p_i}{p_i} Y_i(1,\zeta_i) - \E \bigg[ \frac{1}{n_k}\sum_{i \in \scrC_k} \frac{D_i - p_i}{p_i} Y_i(1,\zeta_i)\bigg| \bE, \bg, \bsign \bigg].
\end{align*}
The analysis in Lemma~\ref{sa-lem: approx delta_1} and Lemma~\ref{sa-lem:lin-low} with $g_i(1,\zeta_k)\Indicator(i \in \scrC_k)$ replacing $g_i(1,\pi)$ implies
\begin{align*}
    & \frac{1}{n_k}\sum_{i \in \scrC_k} \frac{D_i - p_i}{p_i} Y_i(1,\zeta_i) \\
    & = \frac{n}{n_k} \cdot \frac{1}{n} \sum_{i = 1}^n \frac{D_i - p_i}{p_i} Y_i(1,\zeta_i) \Indicator(i \in \scrC_k) \\
    & = \frac{1}{n_k} \sum_{i \in \scrC_k}\left(\mathtt{c}_k Y_i(1,\zeta_i) + \mathtt{d}_k \E[Y_i(1,\zeta_i)] \right)(W_i - \pi_k) + O_{\psi_{\beta_k,h_k},tc}(n^{-2\bgammak}).
\end{align*}
condition on $\bE, \bg, \bsign$, where $\mathtt{c}_k  
= (1 + \exp(2\beta_k \pi_k + 2 h_k))/2$ and $\mathtt{d}_k = \beta_k(1 + \exp(2 \beta_k \pi_k + 2 h_k))/(1 + \cosh(2 \beta_k \pi_k + 2 h_k))$. 

\begin{center}
    \textbf{I.2: Term $\Delta_{2,k}$}
\end{center}

The linearization of $\Delta_{2,k}$ involves $M_i/N_i$, which depends all blocks even if the estimator is for block $k$. We will find its stochastic linearization in terms of units in all blocks.

% \begin{lemma}\label{lem: multiblock -- delta_2}
% Suppose Assumption~\ref{Graph-1}, \ref{Pout-1}, \ref{assump: ising block} and \ref{Ising-2} hold. Assume $Y_i(1,\cdot)$ and $Y_i(-1,\cdot)$ are $C^4$. Then
% \begin{equation*}
% \begin{split}
%     \Delta_{2,k} - \E[\Delta_{2,k}|\bE, \bg,\bsign] = \sum_{l = 1}^K \frac{1}{n_k} \sum_{i \in \scrC_l}  \left[\sum_{j \in \scrC_k, j \neq i}\frac{E_{ij}}{N_j}\left(Y_j^{\prime}(1,\zeta_j) - Y_j^{\prime}(-1,\zeta_j)\right)\right](W_i - v_l) \\
%     + O_{\psi_1}\bigg(\sqrt{\log n} \bigg(n^{-\bgammak} \max_i N_i^{-\frac{1}{2}} + n^{-\bgammak}\max_{1 \leq l \leq K}n^{-\bgammal} + n^{-\frac{1}{2}}\frac{\max_i N_i^3}{\min_i N_i^4}\bigg)\bigg).
% \end{split}
% \end{equation*}
% \end{lemma}

By a Taylor expansion of $g_i(1,\cdot)$ at $\zeta_i$
\begin{align*}
    \Delta_2 & = \frac{1}{n_k} \sum_{i \in \scrC_k} \frac{D_i}{p_i}\bigg[g_i^{\prime}(1,\zeta_i)\bigg(\frac{M_i}{N_i} - \zeta_i\bigg) + g_i\left( \frac{M_i}{N_i}\right)\bigg(\frac{M_i}{N_i} - \zeta_i \bigg)^2\bigg] \\
    & = \Delta_{2,1} + \Delta_{2,2} + \Delta_{2,3},
\end{align*}
where
\begin{align*}
    \Delta_{2,1} & = \frac{1}{n_k} \sum_{i \in \scrC_k} g_i^{\prime}(1,\zeta_i)\left(\frac{M_i}{N_i} - \zeta_i\right), \\
    \Delta_{2,2} & = \frac{1}{n_k} \sum_{i \in \scrC_k} \frac{D_i - p_i}{p_i} g_i^{\prime}(1,\zeta_i)\left(\frac{M_i}{N_i} - \zeta_i\right), \\
    \Delta_{2,3} & = \frac{1}{n_k} \sum_{i \in \scrC_k} \frac{D_i}{p_i} g_i \left(\frac{M_i}{N_i}\right)\left(\frac{M_i}{N_i} - \zeta_i\right)^2, 
\end{align*}
with $g_i(x) = \int_0^1 (1 - t) Y_i^{(2)}(1,\zeta_i + t (x - \zeta_i))dt$. In particular, $g_i$ is $C^2$. 
\paragraph*{Term $\Delta_{2,1}$:} Rearranging $\Delta_{2,1}$, we get the effective term in the stochastic linearization. 
\begin{align*}
    \Delta_{2,1} & = \frac{1}{n_k}\sum_{i \in \scrC_k} g_i^{\prime}(1,\zeta_i)\bigg[\sum_{l = 1}^K \sum_{j \in \scrC_l, j \neq i}\frac{E_{ij}}{N_i}(W_j - \pi_l)\bigg] \\
    & = \sum_{l = 1}^K \frac{1}{n_k} \sum_{i \in \scrC_l}  \bigg[\sum_{j \in \scrC_k, j \neq i}\frac{E_{ij}}{N_j}Y_j^{\prime}(1,\zeta_j)\bigg](W_i - \pi_l).
\end{align*}
\paragraph*{Term $\Delta_{2,2}$:} We want to show $\Delta_{2,2}$ is negligible. Consider the effect from each block separately. We claim that condition on $\bg$, $\bE$, $\bsign$,
\begin{align*}
    \Delta_{2,2} & = \frac{1}{n_k} \sum_{i \in \scrC_k} \frac{D_i - p_i}{p_i} g_i^{\prime}(1,\zeta_i)\bigg(\sum_{l = 1}^K \sum_{j \in \scrC_l, j \neq i} \frac{E_{ij}}{N_i}(W_j - \pi_l)\bigg) \\
    & = \sum_{l = 1}^K \frac{1}{2 n_k} \sum_{i \in \scrC_k} \frac{W_i - \pi_k}{(\pi_k + 1)/2} g_i^{\prime}(1,\zeta_i) \bigg(\sum_{j \in \scrC_l, j \neq i} \frac{E_{ij}}{N_i}(W_j - \pi_l)\bigg) \\
    & \qquad \qquad  + O_{\psi_2, tc}\bl\sqrt{\log n} n^{-\bgammak} \bl\max_{1 \leq i \leq n} N_i^{-1/2} + \max_{1 \leq l \leq K} n^{-\bgammal}\br\br.
\end{align*}
To get the second line, notice that $p_i = (1 + \exp(2 \beta_k \mca m_{i,k} + 2 h_k))^{-1}$ is Lipschitz in $\mca m_{i,k}$, and since  $(W_i: i \in \scrC_k), 1 \leq k \leq K$ form independent Ising models, we can use Lemma~\ref{sa-lem:fixed-temp-be} to get $\mca m_{i,k} - \pi_k = O_{\psi_{\beta_k,h_k}}(n_k^{-\bgammak})$ for $k \in \mathscr{H} \cup \mathscr{C}$, and condition on $\sign(\mca m_{k})$, $\mca m_{i,k} - \pi_k = O_{\psi_{\beta_k,h_k}}(n_k^{-\bgammak})$ for $k \in \mathscr{L}$. Hence for each $k \in [K]$,
\begin{align*}
    \left|\frac{D_i - p_i}{p_i} - \frac{W_i - \pi_k}{(\pi_k + 1)/2}\right| & = O_{\psi_{\beta_k,h_k}}(n^{-\bgammak}), \quad \text{condition on } \bsign.
\end{align*}
Suppose $\mathsf{U}_{n,l}$ is the latent variable underlining the distribution of $(W_i: i \in \scrC_l), l \in [K]$ as in Lemma~\ref{sa-lem:sub-gaussian}. Conditional on $\bE$ and $\bsign$, using Hoeffiding's inequality and the concentration of $\mathsf{U}_{n,l}$, we have
\begin{align*}
    \sum_{j \in \scrC_l, j \neq i}\frac{E_{ij}}{N_i} (W_j - \pi_l)
    & = \sum_{j \in \scrC_l, j \neq i}\frac{E_{ij}}{N_i} (W_j - \E[W_j|\mathsf{U}_{n,l}]) + \sum_{j \in \scrC_l, j \neq i}\frac{E_{ij}}{N_i} (\E[W_j|\mathsf{U}_{n,l}] - \pi_l) \\
    & = O_{\psi_2}(N_i^{-1/2}) + O_{\psi_{\beta_l,h_l}}(n^{-\bgammal}).
\end{align*}
From the fact that $\P(|Z_1 Z_2| \geq t) \leq \P(\sqrt{\log n} |Z_2| \geq t) + \P(|Z_1| \geq \sqrt{\log n})$ for any two random variables $Z_1$ and $Z_2$, and using a union bound over the summation over $i \in \scrC_k$, we get the second line for $\Delta_{2,2}$. 

Now consider the first term of $\Delta_{2,2}$. With the help of the latent variables $\mathsf{U}_{n,k}, 1 \leq k \leq K$, decompose by
\begin{align*}
    \Gamma_{k,l} = \Gamma_{k,l,a} + \Gamma_{k,l,b} + \Gamma_{k,l,c} + \Gamma_{k,l,d},
\end{align*}
where
\begin{align*}
    \Gamma_{k,l,a} = & \frac{1}{n_k} \sum_{i \in \scrC_k} \sum_{j \in \scrC_l}(W_i - \E[W_i|\mathsf{U}_{n,k}])(W_j - \E[W_i|\mathsf{U}_{n,l}]) \frac{g_i^{\prime}(1,\zeta_i)}{\pi_k + 1} \frac{E_{ij}}{N_i} \Indicator(i \neq j), \\
    \Gamma_{k,l,b} = & \frac{1}{n_k}\sum_{i \in \scrC_k}(W_i - \E[W_i|\mathsf{U}_{n,k}])g_i^{\prime}(1,\zeta_i) \frac{N_{i,l}}{N_i}(\E[W_i|\mathsf{U}_{n,l}] - \pi_l), \\
    \Gamma_{k,l,c} = & \frac{1}{n_k} \sum_{i \in \scrC_k} (\E[W_i|\mathsf{U}_{n,k}] - \pi_k) g_i^{\prime}(1,\zeta_i) \bigg(\sum_{j \in \scrC_l, j \neq i} \frac{E_{ij}}{N_i}(W_j - \pi_l)\bigg) \\
    \Gamma_{k,l,d} = & \frac{1}{n_k}\sum_{i \in \scrC_k}(\E[W_i|\mathsf{U}_{n,k}] - \pi_k)g_i^{\prime}(1,\zeta_i) \frac{N_{i,l}}{N_i}(\E[W_i|\mathsf{U}_{n,l}] - \pi_l).
\end{align*}
Since conditional on $\mathsf{U}_{n,k}$ and $\mathsf{U}_{n,l}$, $(W_i: i \in \scrC_k \cup \scrC_l)$ are i.i.d., we can use Hoeffding's inequality and boundedness of $g_i^{\prime}(1,\zeta_i)$ to get conditional on $\bE$ and $\bsign$,
\begin{align*}
    & \frac{1}{n_k}\sum_{i \in \scrC_k} (W_i - \E[W_i|\mathsf{U}_{n,k}]) g_i^{\prime}(1,\zeta_i)\frac{N_{i,l}}{N_i} = O_{\psi_2}(n_k^{-1/2}), 
\end{align*}
and
\begin{align*}
    \sum_{j \in \scrC_l, j \neq i} \frac{E_{ij}}{N_i}(W_j - \pi_l) & = \sum_{j \in \scrC_l, j \neq i} \frac{E_{ij}}{N_i}(W_j - \E[W_j|\mathsf{U}_{n,l}]) + \frac{N_{i,l}}{N_i}(\E[W_i|\mathsf{U}_{n,l}] - \pi_l) \\
    & = O_{\psi_2}(N_i^{-1/2}) + O_{\psi_{\beta_l, h_l}}(n_l^{-\bgammal}).
\end{align*}
It follows that condition on $\bE$ and $\bsign$, 
\begin{align*}
    \Gamma_{k,l,b} - \E[\Gamma_{k,l,b}|\bE, \bsign] & = O_{\psi_1,tc}(\sqrt{\log n} n_k^{-\frac{1}{2}} n_l^{-\bgammal}), \\
    \Gamma_{k,l,c} - \E[\Gamma_{k,l,c}|\bE, \bsign] & = O_{\psi_{2},tc}(\sqrt{\log n} n_k^{-\bgammak} N_i^{-\frac{1}{2}}) + O_{\psi_1,tc}(\sqrt{\log n} n_k^{-\bgammak} n_l^{-\bgammal}), \\
    \Gamma_{k,l,d} - \E[\Gamma_{k,l,d}|\bE, \bsign] & = O_{\psi_1,tc}(\sqrt{\log n}n_k^{-\bgammak} n_l^{-\bgammal}).
\end{align*}
For $\Gamma_{k,l,a}$, observe that with $\omega_i = \sum_{k = 1}^K k \Indicator(i \in \scrC_k)$, 
\begin{align*}
    \Gamma_{k,l,a} & = \frac{1}{n_k}\sum_{i \in \scrC_k \sqcup \scrC_l} \sum_{j \in \scrC_k \sqcup \scrC_l}(W_i - \E[W_i|\mathsf{U}_{n, \omega_i}])(W_j - \E[W_j|\mathsf{U}_{n, \omega_j}])H_{ij}, \\
    H_{ij} & = \frac{g_i^{\prime}(1,\zeta_i)}{v_k + 1}\frac{E_{ij}}{N_i} \Indicator(i \in \scrC_k, j \in \scrC_l, i \neq j).
\end{align*}
Apply Hanson-Wright inequality conditional on $\bE$, $\mathsf{U}_{n,l}$ and $\mathsf{U}_{n,k}$, we get $$\Gamma_{k,l,a} - \E[\Gamma_{k,l,a}|\bE, \bsign] = O_{\psi_1}((n_k \min_i N_i)^{-\frac{1}{2}}).$$ Put together, conditional on $\bE$ and $\bsign$,
\begin{align*}
    \Delta_{2,2} - \E[\Delta_{2,2}|\bE, \bsign] = O_{\psi_1, tc}\bigg(\sqrt{\log n} n^{-\bgammak} \bigg(\max_{1 \leq i \leq n} N_i^{-1/2} + \max_{1 \leq l \leq K} n^{-\bgammal}\bigg)\bigg).
\end{align*}
\paragraph*{Term $\Delta_{2,3}$:} Similar to the analysis in Section~\ref{sa-sec:stochlin}, we decompose $\Delta_{2,3} = \Delta_{2,3,1} + \Delta_{2,3,2}$ where
\begin{align*}
    \Delta_{2,3,1} = \frac{1}{n_k} \sum_{i \in \scrC_k} g_i \bigg(\frac{M_i}{N_i}\bigg) \left(\frac{M_i}{N_i} - \zeta_i \right)^2, \quad \Delta_{2,3,2} = \frac{1}{n_k} \sum_{i \in \scrC_k} \frac{D_i - p_i}{p_i} g_i \left(\frac{M_i}{N_i}\right) \left(\frac{M_i}{N_i} - \zeta_i \right)^2,
\end{align*}
where $\Delta_{2,3,1}$ is further decomposed based on latent variables $\mathsf{U}_{n,l}, 1 \leq l \leq K$, that is, $$\Delta_{2,3,1} = \Delta_{2,3,1,a} + \Delta_{2,3,1,b} + \Delta_{2,3,1,c},$$ where
\begin{align*}
    \Delta_{2,3,1,a} & = \frac{1}{n_k} \sum_{i \in \scrC_k} g_i \bigg(\frac{M_i}{N_i}\bigg)\bigg(\sum_{l = 1}^K \sum_{j \in \scrC_l, j\neq i} \frac{E_{ij}}{N_i}(W_j - \E[W_j|\mathsf{U}_{n,l}]) \bigg)^2, \\
    \Delta_{2,3,1,b} & = \frac{1}{n_k} \sum_{i \in \scrC_k} g_i \bigg(\frac{M_i}{N_i}\bigg)\bigg(\sum_{l = 1}^K \sum_{j \in \scrC_l, j\neq i} \frac{E_{ij}}{N_i}(W_j - \E[W_j|\mathsf{U}_{n,l}]) \bigg) \bigg(\sum_{l = 1}^K \frac{N_{i,l}}{N_i}(\E[W_j|\mathsf{U}_{n,l}] - \pi_l) \bigg), \\
    \Delta_{2,3,1,c} & = \frac{1}{n_k} \sum_{i \in \scrC_k} g_i \left(\frac{M_i}{N_i}\right)\bigg(\sum_{l = 1}^K \frac{N_{i,l}}{N_i}(\E[W_j|\mathsf{U}_{n,l}] - \pi_l) \bigg)^2.
\end{align*}
\textit{Term $\Delta_{2,3,1,a}$:}  Consider the $\Delta_{2,3,1,a}$ as a (random) function on $\bW$ and $\mathsf{U}_{n,1}, \cdots, \mathsf{U}_{n,K}$. Let
\begin{align*}
     F_i(\bW, \mathsf{U}_{n,1}, \cdots, \mathsf{U}_{n,k}) = g_i \bigg(\frac{M_i}{N_i}\bigg)\bigg(\sum_{l = 1}^K \sum_{j \in \scrC_l, j\neq i} \frac{E_{ij}}{N_i}(W_j - \E[W_j|\mathsf{U}_{n,l}]) \bigg)^2
\end{align*}
Notice that conditional on $\mathsf{U}_{n,l}, 1 \leq l \leq K$, $W_j$'s are independent random variables, and we can rewrite $$\Delta_{2,3,1,a} = \frac{n}{n_k} \frac{1}{n} \sum_{i = 1}^n  g_i \left(\frac{M_i}{N_i}\right) \Indicator(i \in \scrC_k)(\sum_{l =1}^K \sum_{j \in \scrC_l, j\neq i} \frac{E_{ij}}{N_i}(W_j - \E[W_j|\mathsf{U}_{n,l}]))^2.$$ It follows from the same concentration argument for $\Delta_{2,3,1,a}$ in the proof for Lemma~\ref{lem:delta_231} that conditional on $\bE$ and $\bsign$,
\begin{align*}
    \Delta_{2,3,1,a} - \E[\Delta_{2,3,1,a}|\bE, \mathsf{U}_{n,1}, \cdots, \mathsf{U}_{n,K}]& = O_{\psi_2, tc}(n^{-\frac{1}{2}}\max_i N_i^{-1/2}) + O_{\psi_1, tc}(n^{-1}).
\end{align*}
Define $p_l(u) = \E[W_i = 1| \mathsf{U}_{n,l} = u, i \in \scrC_l]$. Then we can write
\begin{align*}
    & \E[F_i(\bW, \mathsf{U}_{n,1}, \cdots, \mathsf{U}_{n,k})|\bE, \mathsf{U}_{n,1} = u_1,\cdots, \mathsf{U}_{n,k} = u_k] \\
    = & \sum_{\bw \in \{-1,1\}^n} \prod_{l= 1}^K \prod_{i \in \scrC_l}p_l(u_l)^{w_i}(1 - p_l(u_l))^{1-w_i}F_i(\bW,u_1, \cdots, u_k).
\end{align*}
By the same argument as in the proof for Lemma~\ref{lem: delta_2,3,2}, 
\begin{align*}
    \partial_{u_l} \E[F_i(\bW, \mathsf{U}_{n,1}, \cdots, \mathsf{U}_{n,k})| \bE, \mathsf{U}_{n,1} = u_1, \cdots, \mathsf{U}_{n,k} = u_k] = O((n N_i)^{-\frac{1}{2}}).
\end{align*}
It then follows from the concentration of $\mathsf{U}_{n,1}$ to $\mathsf{U}_{n,K}$ that condition on $\bE$ and $\bsign$,
\begin{align*}
    \E[\Delta_{2,3,1,a}|\bE, \mathsf{U}_{n,1}, \cdots, \mathsf{U}_{n,K}] - \E[\Delta_{2,3,1,a}|\bE] & = O_{\psi_2}((n N_i)^{-\frac{1}{2}}\max_{1 \leq l \leq K} n^{-\bgammal}).
\end{align*}
Moreover, since $\sum_{j \in \scrC_l, j \neq i} \frac{E_{ij}}{N_i}(W_j - \E[W_j|\mathsf{U}_{n,l}]) = O_{\psi_2}(N_i^{-\frac{1}{2}})$ and $\E[W_j|\mathsf{U}_{n,l}] - \pi_l = O_{\psi_{\beta_l,h_l}}(n^{-\bgammal})$, we have conditional on $\bE$ and $\bsign$,
\begin{align*}
    \Delta_{2,3,1,b} = O_{\psi_2, tc}(\sqrt{\log n} \max_i N_i^{-\frac{1}{2}} \max_{1 \leq l \leq K}n^{-\bgammal}), \quad \Delta_{2,3,1,c} = O_{\psi_2,tc}(\max_{1 \leq l \leq K} n^{-2 \bgammal}).
\end{align*}
Putting together, $$\Delta_{2,3,1} - \E[\Delta_{2,3,1}|\bE] = O_{\psi_2, tc}\left(\sqrt{\log n} \max_i N_i^{-\frac{1}{2}} \max_{1 \leq l \leq K}n^{-\bgammal} + \max_{1 \leq l \leq K} n^{-2 \bgammal}\right).$$
Consider the $p$-th order term in the expansion of  $\Delta_{2,3,2}$, $$\delta_p = \frac{1}{n_k} \sum_{i \in \scrC_k} \frac{T_i - \pi_k}{\pi_k} g_i^{(p)}(1, \zeta_i)\bigg(\frac{M_i}{N_i} - \zeta_i\bigg).$$ Following conditional i.i.d argument as in Lemma~\ref{lem: delta_2,3,2}, we can show condition on $\bE$ and $\bsign$,
\begin{align*}
    \delta_p - \E[\delta_p|\bE, \bsign]& = O_{\psi_2, tc} \left(\sqrt{\log n} \bigg(n^{-\bgammak} \max_i N_i^{-\frac{1}{2}} + \max_{1 \leq l \leq K}n^{-p \bgammal} + n^{-\frac{1}{2}}\frac{\max_i N_i^3}{\min_i N_i^4}\bigg)\right).
\end{align*}
Hence assuming $g_i(1, \cdot)$ is $C^{p+1}$. Taylor expand $g_i(1,\cdot)$ up to the $p$-th order, we get
\begin{align*}
    & \Delta_{2,3,2} - \E[\Delta_{2,3,2}|\bE, \bsign] \\
    & = \sum_{j = 1}^p \delta_l - \E[\delta_l|\bE, \bsign] + \frac{1}{n_k} \sum_{i \in \scrC_k} \frac{D_i - v_k}{v_k} Y_i^{(p+1)}(1, \eta_i^{\ast})\left(\frac{M_i}{N_i} - \zeta_i \right)^{p+1} + \text{error} \\
    & = O_{\psi_2, tc} \bigg(\sqrt{\log n} \bigg(n^{-\bgammak} \max_i N_i^{-\frac{1}{2}} + \max_{1 \leq l \leq K}n^{-2 \bgammal} + n^{-\frac{1}{2}}\frac{\max_i N_i^3}{\min_i N_i^4}\bigg)\bigg).
\end{align*}

\begin{center}
\textbf{I.3 Averaging Over $\bE$}
\end{center}
From the previous steps, condition on $\bE$ and $\bsign$,
\begin{align*}
    \bigg\lVert \wh{\btau}_{n,UB} - \E[\wh{\btau}_{n,UB}|\bE,\bg,\bsign] - n^{-1}\sum_{l = 1}^K \sum_{i \in \scrC_l} \bar{\bS}_{l,i}(W_i - \pi_l)\bigg\rVert_2 = O_{\psi_2}(\bar{\mathtt{r}}_n),
\end{align*}
where $\bar{\mathtt{r}}_n = \max_{k \in [K]}n^{-2 \bgammak} + \sqrt{\log n} \max_{k \in [K]} n^{-\bgammak}(n \rho_n)^{-1} + \sqrt{\log n} n^{-1/2} + (n \rho_n)^{-(p+1)/2}$, and $\bar{\bS}_{l,i}$ is the vector $(\bar{S}_{1,l,i}, \cdots, \bar{S}_{K,l,i})^{\Trans}$, where
\begin{align*}  
        \bar{S}_{k,l,i} = \frac{n}{n_k} \bigg[\sum_{j \in \scrC_k, j \neq i}\frac{E_{ij}}{N_{i,k}} (Y_j^{\prime}\left(1,\zeta_j\right) - Y_j^{\prime}\left(-1,\zeta_j\right)) + \Indicator(k = l)(\mathtt{c}_k g_i(1,\zeta_i) + \mathtt{d}_k \E[g_i(1,\zeta_i)])\bigg],
\end{align*}
with $\zeta_i = \frac{\sum_{\ell = 1}^k N_{i,\ell} \pi_{\ell}}{\sum_{\ell = 1}^k N_{i,\ell}}$. Condition on $U_i$, $E_{i,j}$ for all $1 \leq j \leq n$ are independent with $|E_{ij}| \leq 1$ and $\V[E_{ij}|U_i] \lesssim \rho_n$. Hence using Bernstein's inequality,
\begin{align*}
    \frac{\sum_{\ell = 1}^k N_{i,\ell}\pi_{\ell}}{\sum_{\ell = 1}^k N_{i,\ell}}
     = \frac{\frac{1}{n \rho_n}\sum_{\ell = 1}^k N_{i,\ell}\pi_{\ell}}{\frac{1}{n \rho_n}\sum_{\ell = 1}^k N_{i,\ell}}
     = \frac{\frac{1}{n}\sum_{\ell = 1}^k n_{\ell}\pi_{\ell}g(U_i) + O_{\psi_2}((n \rho_n)^{-\frac{1}{2}})}{g(U_i) + O_{\psi_2}((n \rho_n)^{-\frac{1}{2}})}
     = \overline{\pi} + O_{\psi_2}((n \rho_n)^{-\frac{1}{2}}),
\end{align*}
with $\overline{\pi} = \sum_{k = 1}^K p_k \pi_k$. The same argument as the proof for Lemma~\ref{sa-lem:lin-fixed-temp} implies
\begin{align*}
    \max_{1 \leq i \leq n} \bigg|\sum_{j \in \scrC_k, j \neq i}\frac{E_{ij}}{N_{i}} (Y_j^{\prime}\left(1,\zeta_j\right) - Y_j^{\prime}\left(-1,\zeta_j\right)) - p_k Q_i \bigg| = O_{\psi_2, tc}((n \rho_n)^{-1/2}),
\end{align*}
with
\begin{align*}
    Q_i = \E \Big[\frac{G(U_i, U_j)}{\E[G(U_i, U_j)|U_j]} (g_j^{\prime}(1, \overline{\pi}) - g_j^{\prime}(-1, \overline{\pi}))\Big| U_i, \bsign \Big].
\end{align*}
Hence with $\bS_{l,i} = (S_{1,l,i}, \cdots, S_{K,l,i})^{\Trans}$, where
\begin{align*}
    S_{k,l,i} = Q_i + \Indicator(k = l) p_k^{-1}(\mathtt{c}_k g_i(1,\zeta_i) + \mathtt{d}_k \E[g_i(1,\zeta_i)]).
\end{align*}
Hence by the same analysis as Equation~\eqref{sa-eq: lin error network} in the proof of Lemma~\ref{sa-lem:lin-fixed-temp},
\begin{align*}
    & \bigg\lVert \wh{\btau}_{n,UB} - \E[\wh{\btau}_{n,UB}|\bE,\bg,\bsign] - \frac{1}{n} \sum_{l = 1}^K \sum_{i \in \scrC_l} \bS_{l,i}(W_i - \pi_l) \bigg\rVert_2 \\
    & = O_{\psi_1,tc}((n \rho_n)^{-1/2} \max_{k \in [K]} n^{-\bgammak} + (n \rho_n)^{-(p+1)/2}).
\end{align*}
We already know $\wh{\btau}_{n,UB}$ is the unbiased estimator. The same argument as Equation~\eqref{sa-eq: unbiased estimator low temp} shows that condition on $\sign$, 
\begin{align*}
    \norm{\E[\wh{\btau}_{n,UB}|\bE,\bg,\bsign] - \btau_n}_2 = O(\exp(-n)).
\end{align*}
This finishes the proof for the unbiased estimator.

\begin{center}
\textbf{II. The Hajek Estimator}
\end{center}
The analysis will be the same as those for Lemma~\ref{sa-lem: hajek}. For simplicity, denote $\wh{\mca p}_k = n_k^{-1} \sum_{i \in \scrC_k} W_i$ and $\mca p_k = \frac{1}{2}\tanh(\beta_k \mca m_k + h_k) + \frac{1}{2} = \frac{1}{2} \mca m_k + \frac{1}{2}$. Then
\begin{align*}
    & \frac{1}{n_k} \sum_{i \in \scrC_k} \frac{W_i Y_i}{\wh{\mca p}_k} - \frac{1}{n_k} \sum_{i \in \scrC_k} \frac{W_i Y_i}{\mca p_k} 
    = \frac{1}{n_k}\sum_{i \in \scrC_k} \frac{T_i Y_i}{\wh{\mca p}_k} \frac{\mca p_k - \wh{\mca p}_k}{\mca p_k}.
\end{align*}
The analysis in Lemma~\ref{sa-lem: hajek} implies
\begin{align*}
    2(\wh{\mca p}_k - \mca p_k) = (1 - \beta \sech^2(\beta \pi + h))(\mca m_k - \pi_k) + O((\mca m_k - \pi_k)^2),
\end{align*}
and 
\begin{align*}
    \E[g_i(1,\frac{M_i}{N_i})] 
    = \E[g_i(1,\pi)] + \E[g_i^{(1)}(1,\overline{\pi})(\frac{M_i}{N_i} - \overline{\pi})] + \frac{1}{2}\E[g_i^{(2)}(1,\pi^{\ast})(\frac{M_i}{N_i} - \overline{\pi})^2] = O( \max_{1 \leq l \leq K}n^{-2\bgammal}).
\end{align*}
Hence
\begin{align*}
    \frac{1}{n_k} \sum_{i \in \scrC_k} \frac{W_i Y_i}{\wh{\mca p}_k} - \frac{1}{n_k} \sum_{i \in \scrC_k} \frac{W_i Y_i}{\mca p_k} 
    = - \frac{1 -\beta_k(1 - \pi_k^2)}{1 + \pi_k} \E[g_i(1,\overline{\pi})|\bsign] + O_{\psi_1}( \max_{1 \leq l \leq K}n^{-\bgammal}).
\end{align*}
The conclusion then follows from step \textit{I. The Unbiased Estimator.}

\subsection{Proof of Lemma~\ref{sa-lem:clt-block}}

We want to apply Lemma~\ref{lem: mult-be} to the stochastic linearizations obtained from Lemma~\ref{sa-lem: mult-stoch-lin}, $$n_l^{-1} \sum_{i \in \scrC_l} \bS_{l,i,(\bs)}(W_i - \pi_{l,(\mathbf{s})}),$$ for different blocks separately. First, we need to check if $\bS_{l,i,(\bs)}$ satisfies the covariate constraints in Lemma~\ref{lem: mult-be}. Recall $\bS_{l,i,(\bs)} = (S_{1,l,i,(\bs)}, \cdots, S_{K,l,i,(\bs)})^{\Trans}$, where
\begin{align*}
    S_{k,l,i,(\bs)} = Q_{i,(\bs)} + \Indicator(k = l) p_k^{-1}(R_{i,l,(\bs)} - \E[R_{i,l,(\bs)}]), \qquad 1 \leq k, l \leq K, 1 \leq i \leq n,
\end{align*}
Definitions of $Q_{i,(\bs)}$ and $R_{i,l,(\bs)}$ imply that $\min_{k,l \in [K]}\E[S_{k,l,i,(\bs)}^2] > 0$ and $\max_{k,l}|S_{k,l,i}| < \infty$ almost surely, satisfying the conditions in Lemma~\ref{lem: mult-be}. Hence
\begin{align*}
    \sup_{A \in \mathscr{R}} | \P_{\boldsymbol{\beta}, \boldsymbol{h}}( n_l^{-1} \sum_{i \in \scrC_l} \bS_{l,i,(\bs)} & (W_i - \pi_{l,(\mathbf{s})}) \in A | sgn(\mca m_l)  \Indicator(l \in \mathcal{L}) = s_l) - \\
    & \P(n^{-1/2} \bSigma_{l,(\bs)}^{1/2} \mathsf{Z}_K + \Indicator(l \in \mathscr{H} \cup \mathscr{L}) n^{-1/2} \sigma_{l,(\bs)} \E[\bS_{l,i,(\bs)}] \mathsf{Z}_{(l)} \\
    & \qquad \qquad + n^{-1/4} \Indicator(l \in \mathscr{C}) \E[\bS_{l,i,(\bs)}] \mathsf{R}_{(l)} \in A)| = O\Big(\Big( \frac{\log(n)^7}{n}\Big)^{1/6} \Big),
\end{align*}
where 
\begin{align*}
    \bSigma_{l,(\bs)} = \E[\bS_{l,i,(\bs)} \bS_{l,i,(\bs)}^{\top}](1 - \pi_{l,(\bs)}^2).
\end{align*}
Now replacing $n_l$ by $n p_l$. The assumption that $n_l/n = p_l + O(n^{-1/2})$ and the Nazarov inequality implies
\begin{align}\label{sa-eq:approax of stoch lin}
    \nonumber \sup_{A \in \mathscr{R}} | \P_{\boldsymbol{\beta}, \boldsymbol{h}}( n^{-1} \sum_{i \in \scrC_l} \bS_{l,i,(\bs)} & (W_i - \pi_{l,(\mathbf{s})}) \in A | sgn(\mca m_l)  \Indicator(l \in \mathcal{L}) = s_l) - \\
    \nonumber & \P(n^{-1/2} p_l \bSigma_{l,(\bs)}^{1/2} \mathsf{Z}_K + \Indicator(l \in \mathscr{H} \cup \mathscr{L}) n^{-1/2} p_l \sigma_{l,(\bs)} \E[\bS_{l,i,(\bs)}] \mathsf{Z}_{(l)} \\
    & \qquad \qquad + n^{-1/4} \Indicator(l \in \mathscr{C}) p_l \E[\bS_{l,i,(\bs)}] \mathsf{R}_{(l)} \in A)| = O\Big(\Big( \frac{\log(n)^7}{n}\Big)^{1/6} \Big),
\end{align}
The independence between $\bS_{l,i,(\bs)}$ for different $i$'s and the independence between Ising-spins across blocks then imply the stochastic linearization from Lemma~\ref{sa-lem: mult-stoch-lin} can be approximated by summation of right hand sides of Equation~\eqref{sa-eq:approax of stoch lin}. Lemma~\ref{sa-lem: mult-stoch-lin} and Nazarov inequality applied on the $n^{-1/2} \bSigma^{1/2} \mathsf{Z}_K$ part then imply the conclusion.

\bibliographystyle{plain}
\bibliography{bib.bib}